\documentclass[11pt]{report}
\usepackage{etex}
\reserveinserts{28} % might require another inclusion before use!

\usepackage{silence} % Allow filtering console warnings
\usepackage{graphicx} % support inclusion of graphics files in various formats
\usepackage{tcolorbox} % for graphics placeholders
\usepackage{latexsym}
\usepackage{amsfonts}
\usepackage{amssymb} % for nmid
\usepackage{epsfig}
\usepackage{xcolor}
\usepackage{mathtools} % for DeclarePairedDelimiter
\usepackage{listings} % for listing program code
\usepackage{enumitem} % control spacing and labels of enumerate and itemize environments
\usepackage{framed} % for putting frames around figures and text, or shading them
\definecolor{shadecolor}{rgb}{0.9,0.9,0.9}
\usepackage{comment}
\usepackage{suffix} % For starred commands
\usepackage{longtable} % not sure we actually use this one anywhere
\usepackage{booktabs} % for nicer rules in tables
\usepackage{tabu} % More table stuff... maybe we don't even need longtable with it
\usepackage{siunitx}
\usepackage{array}
\usepackage{soul} % for strikeout and underline
\usepackage{include/bibalias} % alias one bibliography key by another
\usepackage[export]{adjustbox} % for framed \includegraphics
\usepackage{lipsum}
\usepackage{multicol} % for multi-columned lists (within a single document column)
\usepackage{multirow} % create table cells spanning several rows. Not related to the multicol package!
\usepackage{abbrevs}
\usepackage{tablefootnote}
\usepackage{xfrac} % For split-level fractions
\usepackage{xspace}
\usepackage{relsize} % Change font size relative to current size
\usepackage[shortcuts]{extdash} % Non-breakable hyphens
\usepackage{etoolbox}

\newrobustcmd{\LocalAtBeginEnvironment}[1]{\csappto{@begin@#1@hook}}
\newrobustcmd{\LocalAtEndEnvironment}[1]{\csappto{@end@#1@hook}}

\usepackage[autosize]{dot2texi}
\usepackage{tikz}
\usetikzlibrary{shapes,arrows}
\usepackage{xparse} % for expl3 syntax

\usepackage{datetime} % for currenttime

\makeatletter
\newcommand{\fornonempty}[2]{
  \def\@tmp@a{#1}%
  \ifx\@tmp@a\@empty
  \else
      #2
  \fi
}
\newcommand{\emptyornot}[3]{
  \def\@tmp@a{#1}%
  \ifx\@tmp@a\@empty
      #2
  \else
      #3
  \fi
}
\makeatother

\newcommand{\inlineordisplay}[2]{\ifinner {#1} \else {#2} \fi}
\newcommand{\displayorinline}[2]{\ifinner {#2} \else {#1} \fi}

\definecolor{DarkRedERC}{rgb}{0.7,0.1,0.1}
\definecolor{DarkGreenDMC}{rgb}{0.1,0.5,0.1}

\providecommand\erc[2][self]{\noindent\textsf{\color{DarkRedERC} [Eyal to {#1}: {#2}]}}
\WithSuffix\newcommand\erc*[1]{\textsf{\color{DarkRedERC} [{#1}]}}
\providecommand\strong{\textbf}

\DeclarePairedDelimiter\paren{(}{)}      % (parentheses)
\DeclarePairedDelimiter\sbrace{[}{]}     % [square brackets]
\DeclarePairedDelimiter\ceil{\lceil}{\rceil}       
\DeclarePairedDelimiter\floor{\lfloor}{\rfloor}       % |_ floor _|
\let\ceiling\ceil

\newcommand{\breakingparen}[1]{
 \mathord{\left(  {\protect\vphantom {#1}} \right. \kern-\nulldelimiterspace}
 #1
 \mathord{\kern-\nulldelimiterspace \left.  {\protect\vphantom {#1}} \right) }
}
\newcommand{\breakingcparen}[1]{
 \mathord{\left\{  {\protect\vphantom {#1}} \right. \kern-\nulldelimiterspace}
 #1
 \mathord{\kern-\nulldelimiterspace \left.  {\protect\vphantom {#1}} \right\} }
}
\newcommand{\breakingcondset}[2]{
 \mathord{\left\{  {\protect\vphantom{{#1}{#2}}} \right. \kern-\nulldelimiterspace}
 #1
 \mathord{\left. \middle|  {\protect\vphantom{{#1}{#2}}} \right. \kern-\nulldelimiterspace}
 #2
 \mathord{\kern-\nulldelimiterspace \left.  {\protect\vphantom{{#1}{#2}}} \right\} }
}

\newcommand{\splitatcommas}[1]{%
  \begingroup
  \begingroup\lccode`~=`, \lowercase{\endgroup
    \edef~{\mathchar\the\mathcode`, \penalty0 \noexpand\hspace{0pt plus 1em}}%
  }\mathcode`,="8000 #1%
  \endgroup
}

\let\sparen\sbrace

\newcommand{\eps}{\varepsilon}
\newcommand{\half}{\displayorinline{\frac{1}{2}}{\sfrac{1}{2}}}

\newcommand{\vphi}{\varphi}

\renewcommand{\*}{\cdot}

\newcommand{\wloss}{without loss of generality\xspace}

\newabbrev\xth{\textsuperscript{th}}
\newabbrev\xrd{\textsuperscript{rd}}
\newabbrev\xst{\textsuperscript{st}}
\newabbrev\xnd{\textsuperscript{nd}}

\newabbrev\etc{{etc.}}
\newabbrev\cf{{cf.}}

\newcommand{\graphicsplaceholder}[3][]{%
  \ifx\\#1\\%
  \begin{tcolorbox}[valign=center,width=#2,height=#3,arc=0.5mm,auto outer arc]\centering \sf missing graphic\end{tcolorbox}%
  \else%
  \begin{tcolorbox}[valign=center,width=#2,height=#3,arc=0.5mm,auto outer arc]\centering \sf missing: {#1}\end{tcolorbox}%
  \fi%
}

\newcommand{\includegraphicsorplaceholder}[4][]{%
\IfFileExists{#2}{
  \includegraphics[width=#3]{#2}
}{
  \graphicsplaceholder[#1]{#3}{#4}
}
}

\let\hfrac=\horizontalfraction

\ExplSyntaxOn
\NewDocumentCommand{\replacein}{mmm}
 {
  \tl_set:Nn \l_replacein_tl { #3 }
  \tl_replace_all:Nnn \l_replacein_tl { #1 } { #2 }
  \tl_use:N \l_replacein_tl
 }
\tl_new:N \l_replacein_tl
\ExplSyntaxOff

\newcommand{\textbsf}[1]{\textbf{\textsf{#1}}}

\usepackage{abbrevs}

\newcommand{\kavmafrid}{\lower0.17em\hbox{-\kern-0.13em-}}

\let\abouteq\approx

\newcommand{\complexityclass}[1]{{\bf #1}\xspace}

\let\prob\probability

\let\expect\expectation

\newabbrev{\Whp}{With high probability}

\makeatletter
\def\overbracket#1{\mathop{\vbox{\ialign{##\crcr\noalign{\kern3\p@}
\downbracketfill\crcr\noalign{\kern3\p@\nointerlineskip}
$\hfil\displaystyle{#1}\hfil$\crcr}}}\limits}
\def\underbracket#1{\mathop{\vtop{\ialign{##\crcr
$\hfil\displaystyle{#1}\hfil$\crcr\noalign{\kern3\p@\nointerlineskip}
\upbracketfill\crcr\noalign{\kern3\p@}}}}\limits}
\def\overparenthesis#1{\mathop{\vbox{\ialign{##\crcr\noalign{\kern3\p@}
\downparenthfill\crcr\noalign{\kern3\p@\nointerlineskip}
$\hfil\displaystyle{#1}\hfil$\crcr}}}\limits}
\def\underparenthesis#1{\mathop{\vtop{\ialign{##\crcr
$\hfil\displaystyle{#1}\hfil$\crcr\noalign{\kern3\p@\nointerlineskip}
\upparenthfill\crcr\noalign{\kern3\p@}}}}\limits}
\def\downparenthfill{$\m@th\braceld\leaders\vrule\hfill\bracerd$}
\def\upparenthfill{$\m@th\bracelu\leaders\vrule\hfill\braceru$}
\def\upbracketfill{$\m@th\makesm@sh{\llap{\vrule\@height3\p@\@width.7\p@}}%
\leaders\vrule\@height.7\p@\hfill
\makesm@sh{\rlap{\vrule\@height3\p@\@width.7\p@}}$}
\def\downbracketfill{$\m@th
\makesm@sh{\llap{\vrule\@height.7\p@\@depth2.3\p@\@width.7\p@}}%
\leaders\vrule\@height.7\p@\hfill
\makesm@sh{\rlap{\vrule\@height.7\p@\@depth2.3\p@\@width.7\p@}}$}
\makeatother

\newcommand{\bigOmega}[1]{{\mathrm{\Omega}\mathopen{}\left( #1 \right)}}

\newcommand{\bigO}[1]{{{O}\mathopen{}\left( #1 \right)}}

\newcommand{\littleo}[1]{{o\mathopen{}\left( #1 \right)}}

\let\Otilde=\aboutO

\newcommand{\setsize}[1]{{\left| #1 \right|}}

\newcommand{\conditionalset}[2]{{\left\{ {#1} \: \middle| \: {#2} \right\}}}
\let\condset\conditionalset

\newcommand{\tuple}[1]{{\left( {#1} \right)}}

\newcommand{\breakingtuple}[1]{\breakingparen{\splitatcommas{#1}}}
\newcommand{\itemrange}[2]{#1 , \, \ldots , \, #2}
\let\itemrng\itemrange
\newcommand{\idxrange}[3]{{\itemrange{{#1}_{#2}}{{#1}_{#3}}}}
\let\idxrng\idxrange

\let\rngset\rangeset

\let\rngseq\rangesequence
\let\rngtuple\rangesequence

\newcommand{\idxrngtuple}[3]{{\left( \idxrange{#1}{#2}{#3} \right)}}

\newcommand{\setof}[1]{{\left\{ {#1} \right\}}}
\newcommand{\inlinesetof}[1]{\{ #1 \}}
\newcommand{\singletonset}[1]{{\left\{ {#1} \right\}}}

\newcommand{\familyof}[2]{\setof{#1}_{#2}}

\let\emptyset\varnothing

\let\included\subseteq
\let\inc\subseteq

\let\pincluded\subsetneq

\newcommand{\union}{\cup}
\newcommand{\Union}{\bigcup}
\newcommand{\intersect}{\cap}

\newcommand{\dunion}{\mathrel{\mathaccent\cdot{\cup}}}
\newcommand{\Dunion}{{\mathaccent\cdot{\bigcup}}}

\newcommand{\setpoweringoperator}{\wp}
\newcommand{\powerset}[1]{{ \setpoweringoperator \paren*{#1}}}

\newcommand{\naturals}{\mathbb{N}}

\newcommand{\integers}{{\mathbb{Z}}}

\newcommand{\positivereals}{{\mathbb{R}^{+}}}

\newcommand{\true}{\textsc{true}}
\newcommand{\false}{\textsc{false}}

\newcommand{\genericoneargmop}[2]{{{\rm {#1}}\paren*{#2}}}

\newcommand{\genoneargsubscriptmop}[3]{{\mathop{{\rm {#1}}}_{#2}\paren*{#3}}}

\newcommand{\genoneargsubscriptsetmop}[4]{{\mathop{{\rm {#1}}}_{#2}\setof{#3}}}

\newcommand{\setmax}[2][]{\genoneargsubscriptsetmop{max}{#1}{#2}{}}

\newcommand{\opsup}[1]{\genericoneargmop{sup}{#1}}

\newcommand{\opmax}[1]{\genericoneargmop{max}{#1}}

\renewcommand{\lim}[2][]{\genoneargsubscriptmop{lim}{#1}{#2}{}}
\let\abs\absolutevalue

\renewcommand\log[2][]{{{\rm log}_{#1}}\mathopen{}\left(#2\right)}
\newcommand{\fnrngdom}[2]{#1\!\rightarrow #2}
\let\fndef\functiondefinition

\newcommand{\domrestrict}[2]{{{#1} {\big |}_{#2}}}

\let\fcompose\functioncompose

\let\fdomain\functiondomain
\let\fdom\functiondomain
\let\domainof\functiondomain
\let\frange\functionrange

\newcommand{\functionsupport}[1]{\genericoneargmop{supp}{#1}}
\let\fsupport\functionsupport

\let\fimage\functionimage

\newcommand{\inversefunction}[1]{{#1}^{-1}}
\let\finverse\inversefunction

\let\fapply\applyfunction

\let\fapplyinv\applyinversefunction

\let\finverseapply\applyinversefunction

\let\fpreimageof\applyinversefunction

\let\gedges\graphedges

\let\gverts\graphvertices

\let\gdegree\graphvertexdegree

\let\gvindeg\graphvertexindegree

\let\gvoutdeg\graphvertexoutdegree

\let\gmaxindeg\graphmaxindegree

\newcommand{\ndivides}[2]{\left. {#1} \hspace{1pt} \middle| \hspace{-4.3pt} \slash \hspace{0.5pt} {#2} \right.}
\newcommand{\norm}[2][]{%
\ifx\hfuzz#1\hfuzz
  {\lVert #2 \rVert}%
\else
  {\lVert #2 \rVert}_{#1}%
\fi
}

\makeatletter
\def\imod#1{\allowbreak\mkern10mu({\operator@font mod}\,\,#1)}
\makeatother

\newcommand{\factorial}[1]{\ensuremath{{#1}!}}

\newcommand{\UnindexedPlanOperator}[1]{\textsf{#1}}
\newcommand{\indexPlanOperator}[1]{{\ifdefined\index\index[operators]{\uplanop{#1}}\fi}}

\let\planop\PlanOperator

\let\planOp\PlanOperator

\let\uplanop\UnindexedPlanOperator

\newcommand{\indexPlanOperatorAlias}[2]{\ifdefined\index\index[operators]{\textsf{#1}|see {\textsf{#2}}}\fi}
\newcommand{\IntroducePlanOperatorAlias}[2]{\uplanop{#1}\indexPlanOperatorAlias{#1}{#2}}

\let\planopalias\PlanOperatorAlias

\newcommand{\UnindexedCompressionScheme}[1]{\textrm{\textsc{#1}}}
\newcommand{\indexCompressionScheme}[1]{\ifdefined\index\index[schemes]{\textsc{#1}}\fi}

\let\cscheme\CompressionScheme

\let\ucscheme\UnindexedCompressionScheme

\newcommand{\indexCompressionSchemeShorthand}[2]{\ifdefined\index\index[schemes]{\textsc{#1}|see {\textsc{#2}}}\fi}
\newcommand{\IntroduceCompressionSchemeShorthand}[2]{\ucscheme{#1}\indexCompressionSchemeShorthand{#1}{#2}}
\let\cschemesh\CompressionSchemeShorthand
\let\cssh\CompressionSchemeShorthand

\newcommand{\sqlcode}[1]{\ifmmode\text{\ttfamily \detokenize{#1}}\else\texttt{\detokenize{#1}}\fi}

\newcommand{\identifier}[1]{\ifmmode\text{\ttfamily \detokenize{#1}}\else\texttt{\detokenize{#1}}\fi}

\let\colname\identifier

\let\colspace\columnspace

\let\idxcolname\indexedcolumnname

\let\colsize\columnlength
\let\collength\columnlength
\let\collen\columnlength

\newcommand{\columnelement}[2]{\colname{#1}\left\{#2\right\}}

\let\colval\columnelement

\WithSuffix\newcommand\columnelement*[2]{#1\left\{#2\right\}}
\newcommand{\colvalstar}[2]{#1\sbrace{#2}}

\let\colval\columnvalue

\let\idxcolval\indexedcolumnvalue

\let\coldom\columndomain

\let\colsupp\columnsupport

\let\colimage\columnimage

\let\colfreq\columnfrequencydistribution

\let\abouteq\cong
\let\eps\varepsilon

\newcommand{\sqlnull}{\texttt{NULL}\xspace}

\usepackage{enumitem}
\usepackage{listings}
\usepackage{tabu}

\usepackage{include/tabto}
\usepackage{listings}
\usepackage{tcolorbox}

\usepackage{makecell}
\usepackage[symbol]{footmisc}

\definecolor{codegreen}{rgb}{0,0.6,0}
\definecolor{codegray}{rgb}{0.5,0.5,0.5}
\definecolor{codepurple}{rgb}{0.58,0,0.82}
\definecolor{backcolour}{rgb}{0.95,0.95,0.92}
 
\lstdefinestyle{mystyle}{
    language=bash,
    backgroundcolor=\color{backcolour},   
    commentstyle=\color{codegreen},
    keywordstyle=\color{black},
    numberstyle=\tiny\color{codegray},
    stringstyle=\color{codepurple},
    basicstyle=\footnotesize\ttfamily,
    breakatwhitespace=false,         
    breaklines=true,                 
    captionpos=b,                    
    keepspaces=true,                 
    numbers=none,
    numbersep=5pt,                  
    showspaces=false,                
    showstringspaces=false,
    showtabs=false,                  
}
 
\newcommand{\portdigraphedge}[4]{\ensuremath{\tuple{\tuple{{#1},{#2}} , \tuple{{#3}, {#4}}}}}

\let\cconnected\columnarcircuitconnected

\let\graphportlabelssymbol\Lambda

\let\gportlssym\graphportlabelssymbol

\let\gportls\graphportlabels

\let\gportlsin\graphinportlabels

\let\gportlsout\graphoutportlabels

\let\gports\graphports

\newcommand{\circuitports}[1][]{P\fornonempty{#1}{\paren*{#1}}}

\newcommand{\circuitinlabels}[1]{\mathop{\mathsf{labels}_{\fornonempty{#1}{#1,}\text{in}}}}
\newcommand{\circuitoutlabels}[1]{\mathop{\mathsf{labels}_{\fornonempty{#1}{#1,}\text{out}}}}

\let\pginports\directedportgraphinports

\let\gportsof\graphportsof

\let\csignature\circuitsignature

\let\csig\circuitsignature

\let\csigin\circuitsignaturein

\let\csigout\circuitsignatureout

\let\opsig\operatorsignature

\let\opsigin\operatorinsignature

\newcommand{\types}{\mathop{\mathsf{types}}}

\let\typesofsig\typesofsignature
\newcommand{\intypesofsignature}[1]{\mathop{\mathsf{types}_{C,\text{in}}}}

\newcommand{\outtypesofsignature}[1]{\mathop{\mathsf{types}_{C,\text{out}}}}

\newcommand{\labels}{\mathop{\mathsf{labels}}}
\newcommand{\labelsofsignature}[1]{\mathop{\mathsf{labels}}\paren*{#1}}
\let\labelsofsig\labelsofsignature

\newcommand{\inlabelsofsignature}[1]{\mathop{\mathsf{labels}_\text{in}}\paren*{#1}}
\let\inlabelsofsig\inlabelsofsignature
\newcommand{\outlabelsofsignature}[1]{\mathop{\mathsf{labels}_\text{out}}\paren*{#1}}
\let\outlabelsofsig\outlabelsofsignature

\let\opsigout\operatoroutsignature

\let\circins\columnarcircuitinputs

\let\circouts\columnarcircuitoutputs

\let\subcolof\subseteq

\let\subcolumnof\subseteq

\newcommand{\zeroupto}[1]{\rngset{0}{{#1 - 1}}}

\newlist{operators}{description}{1}
\setlist[operators]{style=nextline, leftmargin=1cm, font=\normalfont}
\newcommand{\opitem}[1][]{\item[\planop{#1} \hspace{\leftmargin} \nopagebreak\vspace{\smallskipamount}]}
\WithSuffix\newcommand\opitem*[1][]{\item[$#1$ \hspace{\leftmargin} \nopagebreak\vspace{\smallskipamount}]}

\newlist{cschemes}{description}{1}
\setlist[cschemes]{style=nextline, leftmargin=1cm, font=\normalfont}
\newcommand{\csitem}[1][]{\item[\cscheme{#1} \hspace{\leftmargin} \nopagebreak\vspace{\smallskipamount}]}
\WithSuffix\newcommand\csitem*[1][]{\item[$#1$ \hspace{\leftmargin} \nopagebreak\vspace{\smallskipamount}]}

\newcommand{\graphedgebetween}[2]{\tuple{#1,#2}}

\newcommand{\undecodedset}{\mathcal{I}}

\newcommand{\infinitynorm}{L^\infty}

\newenvironment{oplisting}{

\begingroup
\small
\begin{tabular}{lllll}
  \smaller{\textbsf{In/out?}}
& \smaller{\textbsf{Label}}
& \smaller{\textbsf{Type}}
& \smaller{\textbsf{Length}}
& \smaller{\textbsf{Description}} \\
}  {
\end{tabular}
\endgroup
\\[8pt]
}

\newenvironment{cslisting}{%
\nopagebreak
\begingroup
\small
\nopagebreak
\begin{tabular}{llll}
  \smaller{\textbsf{Label}}
& \smaller{\textbsf{Type}}
& \smaller{\textbsf{Length}}
& \smaller{\textbsf{Description}} \\
}  {
\end{tabular}
\endgroup
\\[6pt]
}

\newcommand{\unittype}{\ensuremath{\tau_\text{unit}}}
\newcommand{\bittype}{\ensuremath{\tau_\text{bit}}}
\newcommand{\bottomtype}{\ensuremath{\tau_\bot}}

\let\booltype\booleantype
\let\bittype\booleantype
\newcommand{\inttype}{\ensuremath{\tau_\text{int}}}

\let\opof\operatorof

\let\copof\circuitoperatorof

\let\gedges\graphedges

\let\gverts\graphvertices

\newcommand{\cppcode}[1]{\ifmmode\text{\ttfamily \detokenize{#1}}\else\texttt{\detokenize{#1}}\fi}

\let\opfunc\operatorfunction
\newcommand{\operatorfunctionproject}[2]{\vphi_{\uplanop{#1}, {#2}}}

\let\ofp\opfuncproject

\let\ofpof\operatorfunctionatportof

\newcommand{\circuitfunction}[1]{\vphi_{#1}}
\let\circfunc\circuitfunction

\newcommand{\circuitfunctionatport}[2]{\vphi_{\emptyornot{#1}{}{#1, }{#2}}}
\let\cfp\circuitfunctionatport

\let\circfuncof\circuitfunctionof

\newcommand{\circuitfunctionatportof}[3]{\circuitfunctionatport{#1}{#2}\paren*{#3}}
\let\cfpof\circuitfunctionatportof

\newcommand{\circuitsignaturemapping}[1][]{\emptyornot{#1}{\pi}{\pi_{#1}}}
\let\csigmap\circuitsignaturemapping

\let\csigmapof\circuitsignaturemappingof

\newcommand{\circuittuple}[1][]{\tuple{\csignature[#1], \gverts{#1}, \copof[#1], \gedges{#1}, \csigmap[#1]}}

\let\opsiginlabels\inlabelsofoperator

\let\opsiginlabels\inlabelsofoperator

\let\opsigoutlabels\outlabelsofoperator

\let\opsigoutlabels\outlabelsofoperator

\let\opsigintypes\intypesofoperator

\let\opsigintypes\intypesofoperator

\let\opsigouttypes\outtypesofoperator

\let\opsigouttypes\outtypesofoperator

\let\circuitinlabels\inlabelsofcircuit
\let\circuitinlabels\inlabelsofcircuit

\let\circuitsiglabels\labelsofcircuit

\let\circuitsiglabels\labelsofcircuit

\let\typedom\typedomain

\newcommand{\smashtilde}[1]{\smash{\tilde{#1}}}

\newcommand{\circuitliftingofoperator}[2]{C_\uplanop{Op}}
\let\circuitofop\circuitliftingofoperator

\newcommand{\capucscheme}[1]{{\normalfont \ucscheme{#1}}}

\usepackage{amsmath}
\usepackage{algorithmicx}
\usepackage{algorithm}
\usepackage{algpseudocode}
\algnewcommand\algorithmicinput{\textbf{Input:}}
\algnewcommand\algorithmicoutput{\textbf{Output:}}
\algnewcommand\Input{\item[\algorithmicinput]}
\algnewcommand\Output{\item[\algorithmicoutput]}

\usepackage[T1]{fontenc}
\usepackage[utf8]{inputenc}

\usepackage[nottoc]{tocbibind} % Adds entries for bibliography, indices, etc to the table of contents

\usepackage{imakeidx} % for end-of-document indices; 
\usepackage{titlesec} % Allows changing title formats using the \titleformat commands
\titleformat*{\section}{\LARGE\bfseries}
\titleformat*{\subsection}{\Large\bfseries}
\titleformat*{\subsubsection}{\large\bfseries}

\usepackage{silence}
\usepackage{tabularx}
\usepackage{etoolbox}
\usepackage{amsmath}

\usepackage{nth} % for saying things like "\nth{3} edition" in the bibliography
\usepackage[hyphens]{url} % Bibliographies usually have  URLs...

\usepackage{subcaption}
\usepackage{float}

\usepackage{varioref} % For above/below references

\usepackage{xcolor}

\usepackage[hyperref,thmmarks]{ntheorem}
\usepackage{aliascnt}

\usepackage{changebar}

\usepackage{titling}

\newsavebox{\abstractbox}

\renewenvironment{abstract}
 {%
  \global\setbox\abstractbox=\vtop\bgroup
  \begin{center}\bfseries\abstractname\end{center}%
 }
 {\par\egroup}

\usepackage{hyperref}
\usepackage{cleveref}

\makeatletter
  \hypersetup{% try hidelinks=true ...
              hypertexnames=false, % unless this is false, links get fscked up
              colorlinks=true,
              pdfstartview=FitH,
              linkcolor=.,
              anchorcolor=.,
              citecolor=.,
              filecolor=.,
              urlcolor=.,
              menucolor=.}

\crefname{subsubsection}{subsubsection}{subsubsections}
\crefname{subsection}{subsection}{subsections}

\makeatother
\makeatletter
\def\newaliasedtheorem#1[#2]#3{%
  \newaliascnt{#1@alt}{#2}
  \newtheorem{#1}[#1@alt]{#3}
  \@namedef{#1@altname}{#3}
}

 \newtheoremstyle{ams-theorem}%
   {\item[\hskip\labelsep \theorem@headerfont ##1\ ##2\theorem@separator]}%
   {\item[\hskip\labelsep {\theorem@headerfont ##1\ ##2}{\theorem@headerfont\ (##3)}{\theorem@headerfont
   \theorem@separator}]}
 \newtheoremstyle{nonumberams-theorem}%
   {\item[\theorem@headerfont \hskip\labelsep ##1\theorem@separator]}%
   {\item[{\theorem@headerfont\hskip \labelsep ##1}{\theorem@headerfont\ (##3)}{\theorem@headerfont
   \theorem@separator}]}
 \newtheoremstyle{ams-remark}%
   {\item[\hskip\labelsep \theorem@headerfont ##1\ ##2\theorem@separator]}%
   {\item[\hskip\labelsep {\theorem@headerfont ##3\ ##2}\theorem@separator]}
 \newtheoremstyle{nonumberams-remark}%
   {\item[\theorem@headerfont \hskip\labelsep ##1\theorem@separator]}%
   {\item[{\theorem@headerfont\hskip \labelsep ##3}\theorem@separator]}

 \theoremstyle{ams-theorem}
 \theoremheaderfont{\normalfont\bfseries}
 \theorembodyfont{\itshape}
 \theoremseparator{.}
 \theoremnumbering{arabic}
 \theoremsymbol{}

 \renewtheorem*{theorem*}{Theorem}
 
 \newaliasedtheorem{corollary}[theorem]{Corollary}
 \renewtheorem*{corollary*}{Corollary}
 \newaliasedtheorem{proposition}[lemma]{Proposition}
 \renewtheorem*{proposition*}{Proposition}
 \newaliasedtheorem{claim}[lemma]{Claim}

 \theorembodyfont{\normalfont}

 \newaliasedtheorem{definition}[lemma]{Definition}
 \newaliasedtheorem{observation}[lemma]{Observation}

 \theoremstyle{ams-remark}
 \theoremheaderfont{\normalfont\itshape}
 \theorembodyfont{\normalfont}

\newaliasedtheorem{numbered-example}[lemma]{Example}

 \newtheorem*{proofsketch}{Proof Sketch}
 \newtheorem*{proof-attempt}{Proof Attempt}
 \newaliasedtheorem{fact}[lemma]{Fact}
 \newaliasedtheorem{assumption}[lemma]{Assumption}
 
 \newtheorem*{note}{Note}
 
 \newaliasedtheorem{question}[lemma]{Question}
 \newaliasedtheorem{conjecture}[lemma]{Conjecture}
 \newaliasedtheorem{hypothesis}[lemma]{Hypothesis}

\makeatother
\newcommand{\appendixref}[1]{\hyperref[#1]{Appendix~\ref*{#1}}}
\newcommand{\wrt}{w.r.t.\xspace}
\makeatother

\makeatletter
{\end{proof} }

\newenvironment{proofsketchof}[1][]
{%
 \def\firstargument{#1}
 \def\noargumentspecified{}
 \ifx\firstargument\noargumentspecified
  \begin{proofsketch}
 \else
  \begin{proofsketch}[Proof Sketch for \autoref{#1}]
  \label{proofsketch:#1}
 \fi
 {}
} %
{\end{proofsketch} }

\makeatother
\usepackage{algorithmicx}
\usepackage{algcompatible}

\usepackage{algpseudocode}

\usepackage{array}
\usepackage{eqparbox}
\makeatletter
\let\@tmp\@xfloat
\begingroup\expandafter\expandafter\expandafter\endgroup
\expandafter\ifx\csname IncludeInRelease\endcsname\relax
  \usepackage{fixltx2e}
\fi
\let\@xfloat\@tmp
\makeatother
\usepackage{stfloats}
\usepackage{url}
\usepackage{microtype} % Micro-typography; may help in squeezing
\usepackage{enumitem}
\usepackage{mdframed} % for placeholder frames

\usepackage[T1]{fontenc}
\usepackage{lmodern}

\usepackage{fullpage}

\makeatletter

\def\newaliasedtheorem#1[#2]#3{%
  \newaliascnt{#1@alt}{#2}
  \newtheorem{#1}[#1@alt]{#3}
  \expandafter\newcommand\csname #1@altname\endcsname{#3}
}

\newcommand\Autoref[1]{\@first@ref#1,@}
\def\@throw@dot#1.#2@{#1}% discard everything after the dot
\def\@set@refname#1{%    % set \@refname to autoefname+s using \getrefbykeydefault
    \edef\@tmp{\getrefbykeydefault{#1}{anchor}{}}%
    \xdef\@tmp{\expandafter\@throw@dot\@tmp.@}%
    \def\@refname{\@nameuse{\@tmp autorefnameplural}}%
}
\def\@first@ref#1,#2{%
  \ifx#2@\autoref{#1}\let\@nextref\@gobble% only one ref, revert to normal \autoref
  \else%
    \@set@refname{#1}%  set \@refname to autoref name
    \@refname~\ref{#1}% add autoefname and first reference
    \let\@nextref\@next@ref% push processing to \@next@ref
  \fi%
  \@nextref#2%
}
\def\@next@ref#1,#2{%
   \ifx#2@ and~\ref{#1}\let\@nextref\@gobble% at end: print and+\ref and stop
   \else, \ref{#1}% print  ,+\ref and continue
   \fi%
   \@nextref#2%
}

\makeatletter
\renewcommand{\@chapapp}{}% Not necessary...

\makeatother

\newcommand{\emailaddress}[1]{\href{#1}{\nolinkurl{#1}}}

\let\email\emailaddress

\makeindex[name=operators,title={Index of computational operators\label{idx:operators}},intoc=true]
\makeindex[name=schemes,title={Index of representation and compression schemes\label{idx:schemes}},intoc=true]
\indexsetup{noclearpage}

\title{A model of computation for analytic column stores
\\
\large with focus on columnar representations of data and composite compression schemes
}
\author{Eyal Rozenberg\thanks{Bikurim Street 11/2, Haifa 3457609, Israel/Palestine. E-mail: \email{eyalroz@technion.ac.il}. Work done in part while at the DB architectures group at CWI, Science Park 123, Amsterdam 1098 XG, The Netherlands.}}
\date{}

\begin{document}
% \begin{minipage}{\textwidth}
%\layout
\begin{titlingpage}
\begin{abstract}
This work presents an abstract model for the computations performed by analytic column stores or columnar query processors. The model is based on circuits whose wires carry columns rather than scalar values, and whose nodes apply operators with column inputs and outputs. This model allows expression of most of the architectural features of existing column-store DBMSes through columnar execution plans, rather than such features being implemented sui-generis, and without the column store maintaining significant out-of-plan data. A strict adherence to columnarity allows for a relatively simple and robust model; enabling extensive and intensive optimization of almost all aspects of query processing; and also enabling massive uniform parallelization of query process on modern hardware. Moreover, the model's expressivity makes it useful also as an \emph{analytical} tool for considering design aspects and features of existing column stores, individually and comparatively.

To achieve the model's wide expressiveness, much of this work develops representation schemes of relevant data structures as combinations of plain columns, with columnar circuits used as scheme encoders and decoders. A particular focus is given to schemes which also compress the data, and their use in query execution --- as an integral part of the computation: Subcircuits of larger columnar circuits, not black boxes. Decoder and encoder circuits are thus also composed to form more elaborate schemes. Such formulation allows both for an alternative view of well-known compression schemes, and for the development of new columnar compression schemes with useful features; these should be of independent interest irrespective of column store systems.
\end{abstract}
%\begin{acknowledgements}
%Thanks mum.
%\end{acknowledgements}
\maketitle
% \end{minipage}

\end{titlingpage}

% \begin{titlingpage}
%     \maketitle
%     \begin{abstract}
%         For over a century, fingerprints have been an undisputed
%         personal identifier.  Recent court rulings have sparked
%         interest in verifying unique techniques to make the current
%         methods even more reliable. Ducks, as they do not have
%         fingers, play a key role in the development of new methods to
%         protect the innocent of our society.
%     \end{abstract}
% \end{titlingpage}

\tableofcontents
\clearpage

% ---------------------------------------
% ---------------------------------------
% ---------------------------------------
% ---------------------------------------
% ---------------------------------------

\chapter*{Introduction}
\label{chap:introduction}
\addcontentsline{toc}{chapter}{\nameref{chap:introduction}}

%\begin{chapquote}{Alan Perlis}
%It is better to have 100 functions operate on one data structure than 10 functions on 10 data structures.
%\end{chapquote}

Why bother spending time on formalizing a model of computation for columnar data processing systems? Column stores have, after all, already become well-established and widely-used. Isn't this pursuit several decades too late, and thus mostly irrelevant?

It is the author's belief that nothing could be further from the truth, and that the dearth of \emph{theoretical} underpinnings is an obstacle to the progress of more \emph{practical} research into column stores or columnar DBMSes and their further acceleration. The reason for this, in a nutshell, is that column stores designers have failed to realize and exploit the full potential of \emph{the column as a fundamental concept and focus of computation}. A formal model of computation would:

\begin{itemize}
 \item make the plethora of architectural features column stores typically support \cite[\S 1]{ABHIM2013} be taken care of as integral parts of the execution of a plan, without special treatment by feature-specific code. Consequently,

 \item bring as much of a column store's computational work when processing a query into the sphere of consideration and influence by execution plan optimization phases.

 \item facilitate the conception and specification of a rich space of composite compression schemes, fully and seamlessly integrated into execution plans.

 \item form a basis for a further theoretical endeavor: A grammatic formalism for execution plan transformation/optimization rules --- rather than such rules constituting mostly ad-hoc unconstrained code.

% \item Provide a framework in which to consider vectorization-vs-compilation \cite{SZB2011} dilemmata, at plan generation time or at run-time.

% \item Provide a conceptual framework in which to consider the vectorization-vs-compilation dilemma \cite{SZB2011}, conducive to its potential resolution.

 \item inspire a more streamlined, flexible architecture for future column store software systems, especially those focused on emerging parallel-execution hardware, lending itself to better utilization of their capabilities as well as easier extensibility.

% \item The particular model presented in this work provides inherent motivation for execution plan fragment compilation --- more so than, say, the consideration of the relational algebra tree for an SQL query.
\end{itemize}

%Let us flesh out these statements somewhat.

\subsection*{The non-history of column store models of computation}

It is somewhat peculiar that research into column stores has not produced any formal models of computation long ago already. An (essentially) columnar storage model for DBMSes was proposed as such already in 1985, in the form of the Decomposition Storage Model (DSM) \cite{CK1985}; columnar file storage format (``transposed files'') was suggested even earlier, in \cite{Batory1979}; and work in which a columnar model is implicit goes as far back as half a century ago, to the late 1960s \cite{EB1969}.
Columnar DBMSes as full-fledged products have also long been available: KDB and Sybase IQ \cite{French1997} were released in the 1990s; MonetDB \cite{Boncz2002} and C\=/Store \cite{SABCCFLLMOo2005} were made availble in the 2000s as free software --- the former is in active development under a free license even today. In fact, by the time of writing, all major commercial DBMSes which consider analytics to be a use-case sport columnarity in some form or another --- even if only as an auxiliary index for a row store: See \cite[\S 2.2, \S 3.3]{ABHIM2013} for the state of affairs as of 2013, which has since progressed further with systems by Amazon \cite{GATKPSS2015}, Microsoft \cite{LBHHNP2015}, Oracle \cite{Oracle2015} and so on. But these advances in practice have not manifested themselves in theoretical research; that is still wholly dominated by efforts inspired by, and geared towards, row-oriented DBMSes, unordered relations, and/or transactional work. The author was unable to find a single paper evaluating the models of computation --- formal or informal --- used by the prominent column stores of recent years.

%Still, while theoretical discussions inspired by and related to row-oriented (and transactional) DBMSes abound, the author could not find a single published paper putting forward a model of computation inspired by (or implemented in) any of the above-mentioned columnar systems.

The more abstract fields of the theory of computation generally, and parallel computation theory particularly, also offer no panacea in terms of a relevant model. The closest models of computation formalized and studied in the literature (see, e.g., \cite[Chapter II]{S1998} for a textbook overview and \cite{KSS2018} for a recent and more domain-specific publication) seem to fall into one of the following categories:
\begin{enumerate}
 \item A network of many computing nodes, acting independently and communicating with each other, where the parallelism is expressed by multiple nodes working on potentially different elements of a column/array).
 \item A data-parallel computer, supporting complex control flows, exhibiting either SIMD (Single-instruction multiple-data) or SPMD (Single-program, multiple-data) behavior.
 \item Circuits / circuit families, carrying scalar values on the wires, using a small fixed set of very-simple operator nodes (boolean algebra, or arithmetic operations on rational numbers). Parallelism may expressed by multiple nodes processing multiple elements of the input (in our case, the column).
\end{enumerate}

The network abstraction of the first category is not useful for us to adopt: Networking may be a significant part of implementing a system as it scales, but the parallelism does not inherently depend on this network structure, nor is the parallel nature based on it. The second model is too powerful for our use: Allowing loops and such control structures would make it rather non-meaningful to distinguish between individual operators and execution plans more generally. Finally, the circuit model, in itself, does not express the concept of the column, or of the processing of an entire column at once. What we would be after is some kind of middle-ground between the two last options, as we shall elaborate in this monograph.

\smallskip

Yet another subfield of computer science in which we find some context for our desired model of computation is that of programming languages. Several languages have been devised over the years intended to manipulate arrays as fundamental types; and some of these involve ``dataflow graphs'' in one form or another (e.g. \cite{FCO1990}). Unfortunately, the author is not well-versed in the scholarly work in this subfield enough to authoritatively draw on such work; but it would seem previous research efforts have not fleshed out or formalized a proper model of computation which one may adapt for use with column stores. Some inspiration can be drawn from the the monadic and diadic operators of the APL language \cite{Iverson1962}, formalized with vector and matrix arguments in mind (as per the relatively late but more comprehensive set of definitions in \cite{APLX5Manual}).

\medskip

Finally, the reader should bear in mind that existing column stores do not actually share a single identical model of computation or common set of structural features it relates to. Differences can be quite significant: Proper-DSM-based column stores couple each column of data with a keying column relating it to the rest of its table; while others operates on tuples in multi-column projections. Or, taking another aspect of the model, some column stores support execution plans with strictly acyclic dependencies between operators --- circuit or DAG-like plans; others allow full-fledged imperative programming, with loops and all (even if these are rarely used in practice).

\subsubsection*{What existing column stores have left out}

Typically (if not universally),the columns appearing a column store's execution plan are those found in a database schema --- columns within tables to which users can make queries --- and intermediate results of operators within the plan. A wealth of auxiliary data is kept separate from the execution plans: \sqlnull indication; column statistics; index structures; partitioning, sharding or cracking information, and more yet. The column store code involving this auxiliary data is not reflected in the execution plans, but rather hard-coded to apply in certain cases and in certain ways. Also, such data is held in structures of idiosyncratic design; and while they might be well-designed for their specific function, they are not themselves columnar --- another reason for their necessitating out-of-plan code.

Whatever is done outside of an execution plan cannot benefit from repatedly-applied and contextually-applied plan optimizations and transformations; it can also hardly benefit, if at all, from the use of just-in-time compilation. Also, such idiosyncratic use of auxiliary data may impose constraints on the actual plan, such as additional materialization points of intermediate results, or less flexibility in scheduling.

\subsubsection*{Compression, data layout, and `pushed-down' execution}

Recently, the author was working on implementing lightweight decompression on a GPU. True to the principle mentioned above, several well-known schemes --- \ucscheme{Dictionary}, \ucscheme{FrameOfReference} and so on --- were concretized using distinct, fixed-width-type columns in memory. As implemention progressed, it turned out that large parts of the code were used by more than one scheme; and that operators originally written for executing queries on uncompressed data were being used as-is in decompression.

These occurrences in practice naturally induce two theoretical questions:

\begin{itemize}
 \item Perhaps it would be useful to decompose compression schemes into simpler transformations (which do not, individually, decompress anything), and consider them separately as well as in combination?
 \item Is there really a distinction between the computational operators we employ during query execution ``proper'', and the computational work performed for decompression?
\end{itemize}

The potential for \emph{composition} of compression schemes, as opposed to their \emph{decomposition} is rather obvious, both to a query processing engine designer as we as to the end user. For example, if we have a time series with points at intervals whose lengths only have few possible values, then an effective compression scheme could be: Storing the differences between samples rather than the values (\ucscheme{Delta} compression), and using less bits for the differences (\ucscheme{NullSuppression}), or perhaps storing indices into a dictionary of possible interval lengths (\ucscheme{Dict}).

The \emph{decomposition} of compression schemes becomes useful mostly in the larger context of executing an actual query, when we can sometimes avoid some of the decompression work, and act on a partially-decompressed form, or on partial data. Let us take the previous example of time-series data: Suppose we wish to perform some aggregation on a feature of events in the time-series which occur after long intervals. In this case, we would like to be able to skip the reconstitution of timestamps from differences, followed by taking differences again to filter out the irrelevant data. But to do so, we must be able to do away with the \ucscheme{Delta} decompression of the composite scheme.

Existing column stores supporting compressed data do not support composition or decomposition of compression schemes (e.g. Vectorwise \cite{ZHNB2006} and C\=/Store/Vertica \cite{LFVTVDB2012}); and the same seems to be true for high-performance row-stores (e.g. HyPerDB \cite{Neumann2011, LMFBNK2016}). Some of them do have a few useful composite schemes baked-in; and some have some scan functionality `pushed-down' to operate on the compressed data as it is being decompressed. But the use of each of these is idiosyncratic; and query-specific optimizations and work-avoidance be applied. Specifically, the two examples given above cannot be realized these systems.

\paragraph{Structure of this monograph} This work progressively presents concepts, constructs, definitions, based on previously-exposited ones; but as far as motivation is concerned, the document's direction is mostly in reverse. We thus begins with the formulation of a model of computation (\autoref{chap:columnar-computational-model} with rather limited justification for the choices of its specifics; we continue with an exploration of the expressivity of combinatorial structures within the model of computation, as columns (\autoref{chap:columnar-representation}; it is only then that we ``apply'' these representation schemes, in formulating common and less-common compression schemes in columnar terms (\autoref{chap:columnar-compression}). A final chapter \autoref{chap:applying-the-model-notes}, prospectively sketches out aspects of the possible design of an analytic column store implementing the model of computation fully and directly.

\section*{Acknowledgements} Part of the work on this monograph was conducted while the author was a post-doctoral researcher at CWI Amsterdam (\url{https://www.cwi.nl}), in the Database Architectures group; the author wishes to thank all group members, and the group's founder, Prof. Emeritus Dr. Martin Kersten, for an enlightening stay. The author particularly wishes to thank Prof. Dr. Peter Boncz, for his guidance and insights during that time. Finally, the author wishes to thank Profs. Drs. Boncz and Urbani , as well as Aris Koning of MonetDB Solutions B.V., for useful discussion and comments on earlier versions of this work.

% ---------------------------------------
% ---------------------------------------
% ---------------------------------------
% ---------------------------------------
% ---------------------------------------

\chapter{A model for columnar computation}
\label{chap:columnar-computational-model}

In this chapter we formalize a model of the computations performed by column stores. In more concrete terms, we extend the folklore model of computation of \emph{combinatorial boolean logic circuits} to columns.

A boolean logic circuit consists of nodes, or vertices, which are logic gates, such as \uplanop{And} or \uplanop{Not}; and of directed edges, corresponding to physical wires between logic gate outputs and inputs, each ``carrying'' a one-bit (scalar) value. The vertices and edges form a DAG (directed acyclic graph). Circuit inputs are conceputalized as wires with only their destination attached to gate input, while outputs are wires with only their source attached to a gate output. The function computed by a circuit is computed inductively, starting from an assignment of a sequence of bits to the inputs, and propagated through the vertices to deeper and deeper internal edges --- until the final gate or gates which compute the circuit's outputs --- in correspondence to how setting 0/1 values on the input wire propagates through the physical nodes over the wire, eventually setting the output wire charges. We will not restate any of this formally; the reader may consult a detailed textbook treatment in \cite{KJ2010}.

In the context of a column store, with most activity being (ideally) uniform over large columns, instead of forming gigantic circuits with single-bit or single-scalar-value edges, we instead have each input, and each edge, carry a \emph{column} with elements of some data type. This is where the similarity between the model we formalize and circuits reaches its end, both in the technical details of the definition and with respect to the richness and expressivity of the model.
%Some of these will, of necessity, be presented with the definitions aspects of the model itself; but most of them will await \Autoref{chap:columnar-representation,chap:columnar-compression}.

\paragraph{Notational conventions} In the following, $\fimage{f},\fdomain{f}, \frange{f}$ denote the image, domain and range, respectively, of a function $f$. Composition of functions is denoted $\fcompose{f}{g}$ where $\fapply{\paren{\fcompose{f}{g}}}{x} = \fapply{f}{\fapply{g}{x}}$. The natural numbers, $\naturals$ are considered to include 0. %$\divides{a}{b}$ indicates $a$ divides $b$ ($b$ is an integer multiple of $a$).
All combinatorial structures we define are finite unless otherwise stated. Diagrams presenting two-dimensional arrays are column-major unless otherwise stated or indicated by element indices. We shall refer to ``types'' in the sense of programming language type systems: Types have domains --- sets of possible values; they have a representation in physical computer memory; and they have associated operators, which may be applied to values in their domains. We'll assume a fixed set of allowed types, which includes various integers and floating-point types with representations of fixed sizes in bits. This set of types will specifically include $\inttype$, an integer type large enough to represent the indices of the column being considered; as well as $\unittype$, the unit type (with a single possible value); and $\bottomtype$, is a type with an empty domain (having no values). When we refer to ``any type'' or to an ``arbitrary type'' --- this is with respect to the global set of allowed types (which will will otherwise not be concerned with).

\section{Model building blocks}
\label{sec:computationa-model-building-blocks}

\subsection{Columns}
\label{subsec:columns}

We begin with a choice of a column store's fundamental, most basic, data structure:

\begin{definition}[Column]
\label{def:column}
A \emph{(plain) column} is a function $\fndef{\colname{c}}{\rngset{0}{n-1}}{\fdomain{\tau}}$ for some fixed-representation-size type $\tau$ and $n \in \naturals$. $n$ is the column's \emph{length} (denoted $\setsize{\colname{c}}$), $\tau$ is its \emph{element type} and the size of $\tau$'s physical representation is the column's \emph{width}.
\end{definition}
A column may be \emph{empty} in case $n$ is 0, or it may interpreted as a \emph{scalar}, a single value of type $\tau$, for the case of $n = 1$. Both cases are valid, though the latter will be explicitly useful in this work, while the former will be mostly ignored. The set of all columns of element type $\tau$ is denoted $\colspace{\tau}$.

\medskip

Our definition of a column is not entirely concrete: It does not specify a particular physical layout for column data. In our model of computation, columns will in fact have multiple possible physical layouts, developed at length in subsequent chapters. Still, as a concrete starting point, we choose a default physical layout in which columns manifest: The \emph{standard representation} of a column is the sequence of its evaluations over its domain, by the natural order of integers, i.e. $\rngseq{\colname{c}(0)}{\colname{c}(n-1)}$. While also defined abstractly, this representation translates immediately into the physical world --- a sequence of values in a computer's memory.

In this work we often conflate columns in the abstract with their standard representation. Also, in light of this physical expression, this work uses the array subscripting syntax of C-like programming languages to denote column elements, e.g. for some $i \in \naturals$, $\colval{c}{i}$ denotes $\colname{c}(i)$, the function \colname{c} applied to $i$.

\begin{definition}[Element frequency distribution / probability mass function]
\label{def:column-frequency-distribution}
Let $\colname{c}$ be a column of element type $\tau$. The \emph{(element) frequency distribution} or \emph{frequency function} of $\colname{c}$ is the function $\fndef{\colfreq{c}}{\fdomain{\tau}}{\naturals}$ counting $\tau$-value appearances in \colname{col}, i.e. $\fapply{\colfreq{c}}{x} = \frac{1}{\colsize{c}} \setsize{\fapplyinv{\colname{c}}{x}}$. Normalizing this function by $\colsize{c}$, we obtain the \emph{probability mass function} for the uniform distribution over the elements of $\colname{c}$.
\end{definition}

Now, the support of  $\colfreq{c}$ (i.e. $\fpreimageof{\colfreq{c}}{\positivereals}$) is of significance --- the set of all $\tau$ values appearing in $\colname{c}$ --- while the support of $\colname{c}$ as a function is simply $\zeroupto{\colsize{c}}$ regardless of the column's contents. Thus, abusing notation, we refer to the support of  $\colfreq{c}$ as ``the support of column $\colname{c}$'', denoting it $\colsupp{c}$.

% \begin{definition}[Column support set]
% Consider the support of $\colfreq{c}$: The set of $\fdomain{\tau}$ values on which $\colfreq{c}$ takes a non-zero value. This is really just $\fimage{\colname{c}}$, when considering $\colname{c}$ as a function; abusing notation, however, we will also refer to it as the \emph{support} or \emph{support set} of column $\colname{c}$.
% \end{definition}

% \begin{definition}[Column concatenation]
% \label{def:column-concatenation}
% Let $\colname{c}_1$, $\colname{c}_2$ be columns of lengths $n_1, n_2$ respectively. Their \emph{concatenation}, denoted $\colname{c}_1 . \colname{c}_2$, is the function:
% \[
% \paren{\colname{c}_1 . \colname{c}_2}(x) =
% \begin{cases}
%   \colname{c}_1(x)           & \phantom{n_1 \leq \;} x < n_1 \\
%   \colname{c}_2(x - n_1)     & n_1 \leq x < n_1 + n_2
% \end{cases}
% \]
% \end{definition}

\subsection{Circuit nodes: Computational operators}
\label{subsec:operators}

\begin{figure}[ht]
  \centering
  \includegraphics[scale=1]{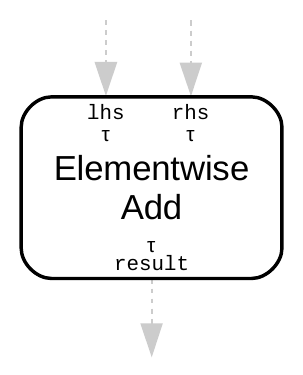}
  \caption[Illustration of an operator as a single vertex within a columnar circuit] {Illustration of a computational operator as a circuit node: Elementwise addition of two (equal-length) columns of type $\tau$. The (grayed-out) arrows indicate how potential graph edges may connect to its ports.}
  \label{fig:operator:example}
\end{figure}

In a boolean logic circuit, and with nodes limited to zero, one or two inputs and one output, there can be no more than $2^{2^0} + 2^{2^1} + 2^{2^2}$ possible kinds of nodes. In fact, only a small subset of them is necessary to define the model of computation, since each of the functions can be computed using a small gadget-circuit using only the small subset: Just \uplanop{And}, \uplanop{Or}, and \uplanop{Not} suffice, or even just \uplanop{Nand} or \uplanop{Nor}, alone \cite[\S 3.2]{KJ2010}. The situation is entirely different when circuit node inputs are \emph{columns} --- the variety of computational operators becomes infinite. We do not, therefore , limit our discussion to some fixed set of operators, and instead consider arbitrary operators inspecifically. A column store system may limit its operators to a finite set, or support defining them dynamically (as many DBMSes, including column stores, do).

\begin{definition}[Signature]
\label{def:signature}
A \emph{(single-direction) computational signature} is a tuple $\tuple{\labels, \types}$: A set of labels, and a family of types
$\typesofsig{\opsig[]} = \familyof{ \tau_\ell }{\ell \in \labelsofsig{\opsig[]}}$ corresponding to each of them. A \emph{bidirectional/in-out computational signature} is a pair $\tuple{\opsigin[],\opsigout[]}$ of an input and an output single-direction signatures, with disjoint sets of labels.
\end{definition}
For an in-out signature $\sigma$, we denote $\inlabelsofsignature{\sigma}$ for $\labelsofsignature{\opsigin[]}$ and $\outlabelsofsignature{\sigma}$ for $\labelsofsignature{\opsigout[]}$ and similarly $\intypesofsignature{\sigma}$ and $\outtypesofsignature{\sigma}$.
When some entity $X$ has an associated (in-out) signature $\sigma_X$, we denote ${\labels}_{X,\text{in}} = \inlabelsofsignature{\sigma_X}$ and similarly for output labels and for types. Finally, we occasionally assume an implicit order on $\labels$ sets.

\begin{definition}[Operator]
\label{def:operator}
A \emph{computational operator} is a tuple $\tuple{M, \sigma}$ comprising some computational automaton $M$ with finite description and a corresponding signature $\sigma$. Omitting details, the automaton --- a deterministic Turing machine \cite[\S 3.1]{Sipser2012} unless otherwise specified --- takes as input the standard representations of columns of types $\typesofsig{\opsigin[]}$, and if it halts, produces the standard representations of columns of types $\typesofsig{\opsigout[]}$.
\end{definition}
We will abuse this definition by referring to operators performing the ``same'' computation for different input types, as though they were the exact same operator (e.g. elementwise addition and multiplication, defined for multiple possible column element types).% (e.g. elementwise addition of signed integers vs. 32-bit IEEE 754 floating-point values).

\smallskip

For an operator \uplanop{Op}, we denote by $\opfunc{Op}$ the partial function computed by \uplanop{Op}: A partial function from (a subset of) $\prod_{\ell \in \inlabelsofsig{\opsig[{\uplanop{Op}}]}}\colspace{\tau_\ell}$ to $\prod_{\ell \in \outlabelsofsig{\opsig[{\uplanop{Op}}]}}\colspace{\tau_\ell}$. We denote by $\ofp{\uplanop{Op}}{\ell}$ the output column of $\opfunc{Op}$ with label $\ell$ (the \emph{projection} of the output on $\ell$, if you will). If \uplanop{Op} has a single output labeled $\ell$, we may conflate $\opfunc{Op}$ with $\ofp{Op}{\ell}$. The arity of $\opfunc{Op}$ is also referred to as the arity of \uplanop{Op} itself.

As the first few examples of actual computational operators, suppose we wish to have the ``same'' logic operators of traditional circuits in columnar circuits as well --- by having them apply to multiple tuple of inputs, elementwise. Choosing \uplanop{And} as an example, we define the following example operator:

\vbox{
\begin{operators}
\opitem*[\uplanop{ElementwiseAnd}\indexPlanOperator{Elementwise}]
  \begin{oplisting}
     Input  & $\colname{lhs}$    & $\bittype$ & $n$ & \\
     Input  & $\colname{rhs}$    & $\bittype$ & $n$ & \\
     Output & $\colname{result}$ & $\bittype$ & $n$ & \\
  \end{oplisting}
  The output satisfies $\colval{result}{i} = \colval{lhs}{i} \land \colval{rhs}{i}$.
\end{operators}
}
The $\sigma_\text{in}$ and $\sigma_\text{out}$ signatures of this operator are the labels and types on the lines marked ``Input'' and ``Output'' above, respectively. As for the length designations, these are actually a constraint on the lengths for which the operator produces an output --- as, unlike in boolean circuits, operators are partial rather than complete functions. In this case, \uplanop{ElementwiseAnd} will only produce output for pairs of input columns having the same length.

\begin{note} In practice, storing bits often involves issue such as machine byte-alignment. We do not account for alignment issues in this text, except to note that they can be overcome using mechanisms developed in \autoref{chap:columnar-representation} below.
\end{note}

Our example elementwise operator immediately generalizes to any function $\fndef{f}{\vec{\tau}_\text{in}}{\vec{\tau}_\text{out}}$ with $\vec{\tau}_\text{in} = \idxrngtuple{\tau}{1}{k_\text{in}}$ and $\vec{\tau}_\text{out} = \rngtuple{\tau'_1}{\tau'_{k_\text{out}}}$:
\begin{operators}
\opitem*[\planop{Elementwise}_f]
  \begin{oplisting}
     Input   & $\colname{arguments}_1$              & $\tau_1$                 & $n$ & \\
     \vdots  & $\vdots$                             & \vdots                   & \vdots \\
     Input   & $\colname{arguments}_{k_\text{in}}$  & $\tau_{k_\text{in}}$     & $n$ &  \\
     Output  & $\colname{results}_1$                & $\tau'_1$                & $n$ &  \\
     \vdots  & $\vdots$                             & \vdots                   & \vdots \\
     Output  & $\colname{results}_{k_\text{out}}$   & $\tau'_{k_\text{out}}$   & $n$ &  \\
  \end{oplisting}
  Letting $x_j = \colvalstar{\colname{arguments}_j}{i}$ and $\vec{y} = \fapply{f}{\vec{x}}$, the output satisfies  $\colvalstar{\colname{results}_j}{i} = y_j$ for all $j \in \zeroupto{k_\text{out}}$. When $k_\text{in} = 1$ (or, respectively, $k_\text{out} = 1$, the single argument (respectively, result) column is referred to as \colname{arguments} (respectively, \colname{results}), dropping the index. (When $f$ is a unary function, one can think of $\uplanop{Elementwise}_f$ this as the $[\IntroducePlanOperatorAlias{Map}{Elementwise}$ higher-order function applied to $f$.)
\end{operators}
this kind of elementwise operators will be put to considerable use throughout this monograph.

\medskip

We present two additional examples of simple operator on columns, with forethought of a columnar computation example later on, in \Autoref{sec:columnar-circuits,sec:circuit-composition-and-transformation}.

\begin{operators}
\opitem[Replicate]
  \begin{oplisting}
     Input  & $\colname{value}$        & $\inttype$  & $1$                & A scalar value to replicate\\
     Input  & $\colname{factor}$       & $\inttype$  & $1$                & The number of times to replicate\\
     Output & $\colname{replicated}$   & $\tau$      & $\colname{factor}$ & Copies of $\colname{value}$ \\
  \end{oplisting}
  Produces a column with the uniform value $\colname{value}$ at each index $\zeroupto{\colname{factor}}$.

\opitem[Select]
  \begin{oplisting}
     Input  & $\colname{data}$       & $\tau$       & $n$                & A column from which to select \\
     Input  & $\colname{selection}$  & $\booltype$  & $n$                & Indication of which elements are selected \\
     Output & $\colname{selected}$   & $\tau$       & $\setsize{\finverseapply{\colname{data}}{\true}}$ & The selected elements of \colname{data} \\
  \end{oplisting}
  Keeps only those elements of a column which are ``selected'' by a $\true$ value in a selection column. Note that this operator has two variants: One which maintains the relative order of elements in \colname{selected} as in \colname{data}, and a more relaxed variant in which any permutation of the selected elements is a valid output; the difference is quite significant for parallel implementations. %(Also known as \IntroducePlanOperatorAlias{Filter}{Select})
\end{operators}
In the description of this operator we've used $\colname{value}$ and $\colname{factor}$ as scalars rather than columns; this should be interpreted as a use of $\colval{value}{0}$ and $\colval{factor}{0}$ --- the single values in these length-1 column.

%See also \cite[\S 4.1]{GBDPS2018} --- but there the \colname{selection} is implicit, produced by an additioanl operator). 

\subsubsection*{Side note: On the strength of circuit operators}
\label{subsubsec:strength-of-operators}

The choice of Turing machines for \autoref{def:operator} above is motivated by actual column stores our model of computation abstracts from --- principally MonetDB \cite{IGNMMK12}, C\=/Store \cite{SABCCFLLMOo2005} and Vectorwise \cite{ZB2012}: Most column stores can and do run user-defined functions (UDFs) in one or more high-level language; and these functions may well be Turing-complete. Even if one only considers the operators built-in to the column store --- they may involve arbitrary code, and are not restricted apriori in their complexity.

On the other hand, in a typical column store (including the abovementioned ones), UDFs are used sparingly if at all (depending on the application domain); and most of the built-in operators, require no more $\Otilde{n}$ time for length-$n$ inputs-plus-outputs, and $\Otilde{1}$ space. Even the more involved built-in operators are not terribly complex (e.g. $\bigO{n^2}$ time for a \uplanop{Join} in the most unforgiving figuring). Furthermore --- and recalling our model of computation being formalized with a mind to enable parallelization and distribution of computation --- we expect most (not all) operators to fall into an even weaker complexity class: Computations which may be extremely-effectively paralellized to $n$ machines, or cores, or threads. Even $\complexityclass{NC}$ \cite[\S 6.7.1]{AB2009} would be far too much. Without delving into details, these should have constant parallel running time and be decomposable to very small fixed-size traditional (non-columnar) circuits. This requirement is similar to the ``pipelinability'' property of operators defined by HyperDB \cite{Neumann2011}, but not quite identical, as may become apparent to a careful of subsequent sections and chapters. Further discussion of this point is beyond the scope of this monograph.
%Regardless of whether in actual execution on a computer such compositions are used as-is, their existence should prove useful for reasoning about execution and effecting execution plan optimization. These aspirations, however, are beyond the scope of this monograph.

%Still, one can note that most operators presented in this monograph do meet this parallelizability requirement, and, in fact, involve either at most a single element in each of the input columns or at most a single element in each of the output columns.

\subsection{Circuit layout: Port graphs}
\label{subsec:port-graphs}

Next, a combinatorial structure for our circuits. Boolean logic circuits could make do with a directed (acyclic) graph, as the \uplanop{Or} and \uplanop{And} binary operators are symmetric, so there is no need to distinguish between their two inputs; and the edge directionality distinguishes inputs from outputs. For operators on columns (and even for asymmetric operators on scalars), inputs are not interchangeable (and neither are outputs). Thus, if we think of them as vertices in a DAG, each such vertex must have different `ports' which edges start from or arrive at. This inspires the following definition.

\begin{figure}[ht]
  \centering
%   \begin{subfigure}[b]{0.45\linewidth}
    \centering
    \includegraphics[scale=0.4]{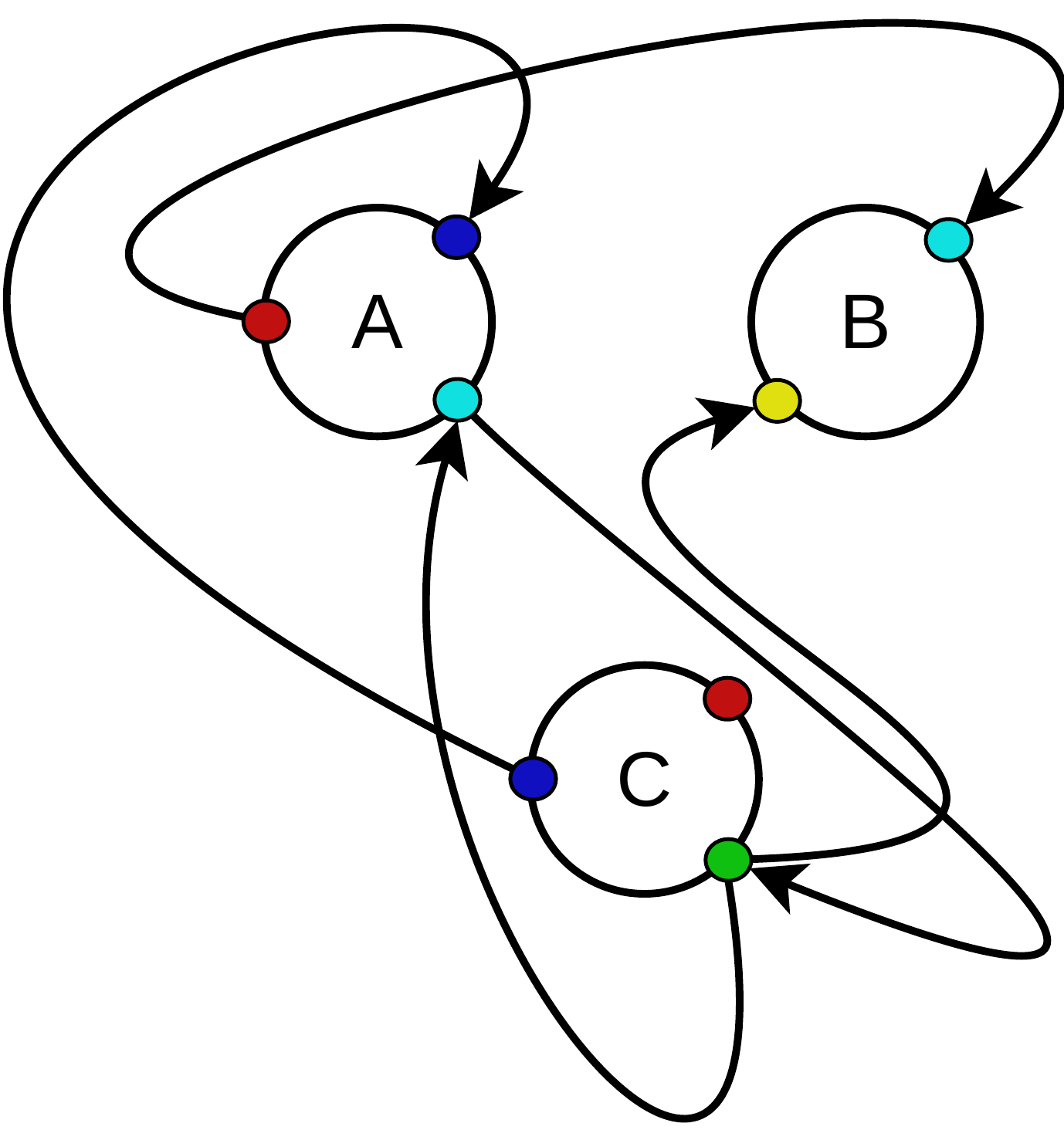}
%     \label{subfig:port-graph:example}
%     \subcaption{The port digraph itself}
%   \end{subfigure}
%   \begin{subfigure}[b]{0.30\linewidth}
%     \centering
%     \includegraphics[scale=0.25]{graphics/columnar_circuits/vertex_connectivity_digraph_example}
%     \label{subfig:vertex-connectivity-digraph:example}
%     \subcaption{The (induced) vertex connectivity digraph}
%   \end{subfigure}
  \begin{minipage}[b]{0.8\linewidth}\footnotesize
  \vspace{10pt}
  \begin{description}
    \item[Vertex set] $V = \setof{\mathsf{A}, \mathsf{B}, \mathsf{C}}$
    \item[Port label sets] $\gportls{A} = \setof{\text{blue},\allowbreak\text{cyan},\allowbreak\text{red}}$, $\gportls{B} = \setof{\text{yellow}, \allowbreak\text{cyan}}$, $\gportls{C} = \setof{\text{red},\allowbreak\text{green},\allowbreak\text{blue}}$
    \item[Edge set] $\inlinesetof{
      \cconnected{\mathsf{A}}{\text{red}}{\mathsf{B}}{\text{cyan}},\allowbreak
      \cconnected{\mathsf{A}}{\text{cyan}}{\mathsf{C}}{\text{green}},\allowbreak
      \cconnected{\mathsf{C}}{\text{blue}}{\mathsf{A}}{\text{blue}},\allowbreak
      \cconnected{\mathsf{C}}{\text{green}}{\mathsf{A}}{\text{cyan}},\allowbreak
      \cconnected{\mathsf{C}}{\text{green}}{\mathsf{B}}{\text{yellow}} }$.
  \end{description}
  \end{minipage}
  \caption[A port digraph]{A 3-node port digraph}
  \label{fig:port-digraph:example}
\end{figure}

\begin{definition}[Port graph]
A \emph{(simple) directed port graph} (or \emph{(simple) port digraph} for short) is a tuple $G = \tuple{V, \gportls{}, E}$ of vertices $V$, per-vertex port label sets $\gportls{}$ and edges $E$, such that:
\begin{itemize}
 \item  $\gportlssym$ is a family of per-vertex port label sets: $\gportlssym = \familyof{\gportls{v}}{v \in V}$.
 \item  The edges in $E$ connect ports (i.e vertex-label tuples) rather than vertices. In other words: Denoting the ports of each vertex $v$ by $\gportsof{G}{v} = \setof{v} \times \gportls{v}$, and the overall \emph{port set} of $G$ by $\gports[G] = \gportsof{G}{V} = \Union_{v \in V} \gportsof{G}{v}$ --- we have $E \inc \gports[G] \times \gports[G]$.
\end{itemize}
Non-simple port digraphs and undirected port graphs are defined similarly, mutatis mutandis.
\end{definition}

A port $p$ in a port digraph is said to be \emph{engaged} if it is an endpoint of an edge in $G$ (and disengaged otherwise); if the origin port of its engaging edge is $p'$, $p$ is said to be \emph{engaged by $p'$}. Two vertices $u,v$ in a port digraph are said to be \emph{(directly) connected} if the graph has an edge between a port of $u$ and a port of $v$. A port $p'$ is said to be \emph{reachable} from port $p$ if there is a path of edges beginning at $p$ and ending at $p'$; reachability from a vertex means port reachability from any of its ports; the reachability of a vertex means port reachability of any of its ports.

\begin{definition}[Induced sub-port-digraph]
Let $G = \tuple{\gverts{}, \gportls{}, \gedges{}}$ be port digraph with $\gverts{}' \inc \gverts{}$. The \emph{subgraph of $G$ induced by $\gverts{}'$} is the port graph $\domrestrict{G}{\gverts{}'} = \tuple{\gverts{}', \gportls{}', \gedges{}' }$ where $\gportls{}' = \domrestrict{\gportls{}}{V'}$ and $\gedges{}' = \gedges{} \intersect \paren{\gportsof{G}{V'} \times \gportsof{G}{V'}}$.
\end{definition}

We denote by $\gvindeg{p}$ (respectively, $\gvoutdeg{p}$) the input (respectively output) degree of port $v$ in a port graph $G$. The input (respectively output) degree $\gvindeg{v}$ (respectively $\gvoutdeg{v}$) of a vertex $v$ in a port graph is the sum of all of its ports input (respectively output) degrees. For a pair of port sets $P_1, P_2 \inc \gports[G]$, $\gdegree{P_1, P_2}$ denotes the number of edges from ports in the first set to ports in the second; and for a pair of vertex sets $V_1, V_2 \inc \gverts{G}$, $\gdegree{V_1, V_2}$ denotes $\gdegree{\gportsof{G}{V_1}, \gportsof{G}{V_2}}$.

\medskip

Generally, a port graph vertex can have edges coming in and going out via the same port. As the ports are a method of qualifying incoming and outgoing edges, it makes sense to consider those cases in which a given port, a given label, may only be associated with one direction --- either incoming or outgoing. More formally,
same port in a port graph can have edges coming in and going out

\begin{definition}[Port orientation]
\label{def:port-orientation}
Let $G$ be a port digraph. A partition $\Pi = \tuple{P_s, P_t}$ of $G$'s ports is an \emph{orientation} (\wrt $G$) if $G$'s edges only connect ports in $P_s$ to ports in $P_t$. If $G$ is port graph for which such a partition exists, it is said to have \emph{orientable ports}.
\end{definition}

\medskip

In conclusion of this section, bear in mind that the port (di)graph is an abstraction of mere convenience. One could very well `encode' ports as gadgets within a single binary relation, either more trivially with an auxiliary unary relation, or with some gadgetry and higher overhead in graph size, but no auxiliary relations --- all using trivial model-theoretical techniques. It would also be possible to have labels as special vertices, and use oriented 4-uniform hyperedges with two regular vertices and two port labels. Our choice is geared towards the definition of circuits, below, since we'll want to maintain the mental picture of a physical circuit layout, where a graph edge is a physical wire, a graph vertex is a component placed on a PCB (printed circuit board), and a port is one of the holes into which a components' pins fit.

\section{Columnar circuits}
\label{sec:columnar-circuits}

\begin{figure}[ht]
\begin{subfigure}[b]{0.475\linewidth}
    \centering
    \includegraphics[scale=0.75]{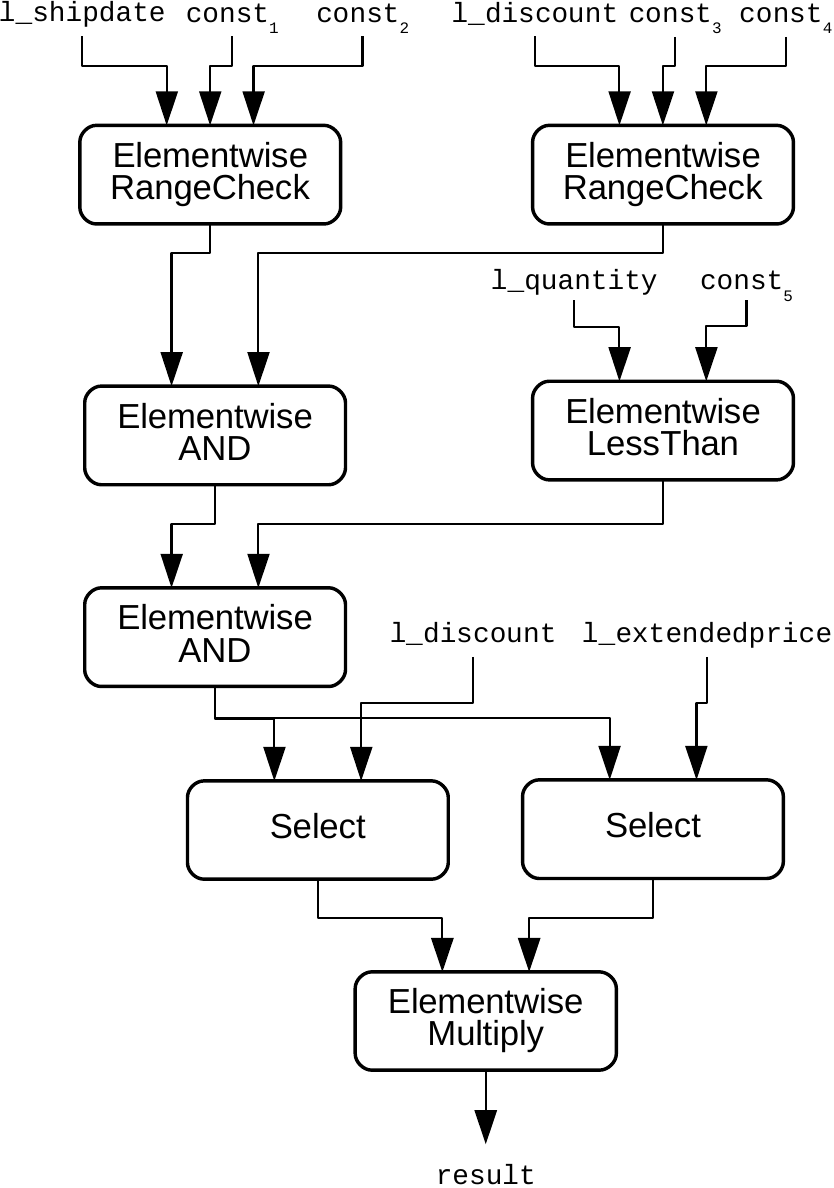}
    \indexPlanOperator{Elementwise}
    \indexPlanOperator{Select}
    \subcaption{A columnar circuit for the query plan, drawn with height determined by dependency depth. The constants are conidered to be inputs although they aren't quite that.}
    \label{subfig:columnar-circuit:example}
  \end{subfigure}
  \hspace{0.03\linewidth}
  \begin{subfigure}[b]{0.50\linewidth}
    \centering
    \begin{lstlisting}
X_1:date := load_column("l_shipdate")
X_2:date := c1 # min shipdate
X_3:date := c2 # max shipdate
X_4:decimal := load_column("l_discount")
X_5:decimal := c3 # min discount
X_6:decimal := c4 # max discount
X_7:int := load_column("l_quantity")
X_8:int := c5 # quantity threshold
X_9 := elementwise.in_range(X_1,X_2,X_3)
X_10 := elementwise.in_range(X_4,X_5,X_6)
X_11 := elementwise.less_than(X_7,X_8)
X_12 := elementwise.and(X_9,X_10)
X_13 := elementwise.and(X_12,X_11)
X_14 := load_column("l_extended_price")
X_15 := load_column("l_discount")
X_16 := select(X_13, X_14)
X_17 := select(X_13, X_15)
X_18 := elementwise.multiply(X_16,X_17)
X_19 := reduction.sum(X_18)
# use X_19 (the "result" output)

\end{lstlisting}
    \subcaption{A program in a MonetDB-MAL-like static-single-assignment language, corresponding to the columnar circuit (with element type information).}
    \label{subfig:columnar-circuit:monetdb-equivalent}
  \end{subfigure}

  \caption[A possible query execution plan for query Q6 of the TPC-H benchmark]{A possible query execution plan for query Q6 of the TPC-H benchmark \cite{TPCH} }
  \label{fig:columnar-circuit:example}

\end{figure}

We have already indicated columnar circuits are to be similar to boolean logic circuits, except for having columns instead of bits on the wires. Still, the orientation of the edges will necessitate a proper definition, somewhat more rigorous and verbose. Before presenting it, however, let us consider example circuit, \autoref{fig:columnar-circuit:example}: It counterposes a program-listing-style execution plan (with syntax and structure similar to those used in MonetDB) to a columnar circuit for processing the same query. The program-listing-style plan cannot itself be the equivalent of a circuit, as its instructions/operators are in a total order, while in the circuit there is merely a partial order imposed by the data dependencies. One may thus think of the circuit as being counterposed to the equivalence class of all reorderings of the program-listing-style plan, in which instructions only use variables computed in previous instructions.

\medskip

Not every port digraph can serve as the underlying structure of a columnar circuit. A port digraph $G$ is said to \emph{have circuit layout} \wrt a partition $\Pi$ of its ports if $G$ is acyclic, $\Pi$ is a port orientation for $G$ (as per \autoref{def:port-orientation}), and no port of $G$ is the target of more than one edge (i.e. $\gmaxindeg{G} \leq 1$). If, given a port digraph, such a partition $\Pi$ exists, $G$ is said to \emph{admit circuit layout} if there exists a partition $\Pi$ with respect to which it has circuit layout. (If it seems odd that a graph in ``circuit layout'' cannot actually have circuits, i.e. cycles --- recall that boolean logic circuits are also acyclic and not literally circuits; it's the electricity driving their operator physically that runs in circuits).

\medskip

A last stepping-stone before finally defining a columnar circuit is the induction of the ports. Indeed, given a finite set $V$, a function $\copof[]$ from $V$ into some set of available operators would induce a family of input and of output ports for each element of $V$: \[\gportlsin{V, \copof[]} = \familyof{ \setof{v} \times \opsiginlabels[{\opof[v]}]}{v \in V} \qquad
\gportlsout{} = \familyof{ \setof{v} \times \opsigoutlabels[{\opof[v]}]}{v \in V}\]and with these two defined, we denote by $\gportls{V, \copof[]}$ their elementwise union, i.e. $\gportls{V, \copof[]}$ are all the induced ports for each vertex. This completes the machinery necessary for:

\begin{definition}[Columnar circuit]
A tuple $C = \circuittuple[]$ --- a computational signature $\sigma$ (as per \autoref{def:signature}), a set of vertices $V$, a mapping of vertices to operators $\copof[]$ (which induces the ports of each vertex), a set of edges $E$, and an ``interface mapping'' $\pi$ of circuit signature labels to ports --- constitutes a \emph{columnar circuit} if the following hold:

\begin{enumerate}
 \item The tuple $G_C = \tuple{V, \gportls{V,\copof[]}, E}$ constitutes a port digraph --- the \emph{layout graph} or \emph{structure graph} of $C$.
 \item The structure graph $G_C$ has circuit layout with port orientation $\Pi = \breakingtuple{\Union \gportlsin{V,\copof[]},\Union \gportlsout{V,\copof[]}}$.
 \item Edges only connect ports with the same associated element type, i.e. if $\portdigraphedge{u}{a}{v}{b} \in \gedges{}$ then $\paren{\opsigouttypes[{\fapply{\copof[]}{u}}]}_a = \paren{\opsigintypes[{\fapply{\copof[]}{v}}]}_b$.
 \item The interface $\csigmap[]$ maps of the circuit input ports to disengaged vertex input ports, and every disengaged input port has exactly one circuit input port mapped to it; in other words, $\domrestrict{\csigmap[]}{\csigin[]}$ is a bijection between $\circuitinlabels[C]$ and the disengaged input ports of $\gportlsin{G_C}$. %The rationale is that the input is not assigned to internally, it must receive some external assignment for the operator to execute.
 \item The interface $\csigmap[]$  maps circuit output labels to output ports in $\Union \gportlsout{G_C}$.

\end{enumerate}

%\item The port graph induced by the circuit (with each vertex having its corresponding operator's ports) --- has circuit layout, where
%circuit inputs (resp. outputs) are ports with operator-input (resp. operator-output) labels, and circuit outputs are ports with operator-output labels. Ths if we denote $\gportls{C} = \familyof{\inlabelsofsig{\opsig[{\copof[C](v)}]} \union \outlabelsofsig{\opsig[{\copof[C](v)}]} }{v \in \gverts{}}$, the tuple $G(C) = \breakingtuple{\gverts{}, \gportls{C}, \gedges{}}$ constitutes a port graph in circuit layout \wrt in.

%\item Edges only connect operator output ports to other operators' input ports (and thus $C$ has circuit-layout with respect to this particular partition);%, i.e. $\graphedgebetween{\tuple{u,a}}{\tuple{v,b}} \in \gedges{}$ satisfy $a \in \outlabelsofsig{\opsig[{\copof[C](u)}]}$ and  $b \in \inlabelsofsig{\opsig[{\copof[C](v)}]}$.
%\item The in-degree of all input ports in $G(C)$ is either $0$ or $1$, i.e. an operator's input port can either be disengaged, or engaged by a single output port of another operator (= be using a single intermediate column as input).
%\end{itemize}
\end{definition}

\begin{note}
This definition does not support ``pass-through'' wires in a circuit, connecting an input to an output directly with no operator used; without a pass-through capability, circuits may require the use of an idempotent (Elementwise-identity-applying) operator to achieve the same functionality --- relaying its input to its output. Similarly, unused input ports are not supported, but could be simulated  using \uplanop{NoOp} nodes (essentially, \uplanop{Elementwise} with the unary identity function): If a circuit input is mapped by $\pi$ to the input of one of these, it can be considered unused.
\end{note}

\begin{definition}[Circuit input and output ports]
The \emph{input ports} of a computational circuit $C$ is the set $\circins[C] = \csigmapof{C}{\circuitinlabels[C]}$; as per the constraint above, these are exactly all disengaged ports in $C$. The \emph{output ports} of $C$ are $\circouts[C] = \csigmapof{C}{\circuitoutlabels{C}}$.
\end{definition}

We come now to the central definition of our model of computation, relating the combinatorial objects to computed functions:

\begin{definition}[Circuit-computed function]
Let $C = \circuittuple[]$ be a columnar circuit let $v \in \gverts{}$ with corresponding operator $\uplanop{Op} = \copof[](v)$ and input arity $r = \setsize{\opsigin[\uplanop{Op}]}$ (which may be 0) and let $p = \tuple{v,a} \in \gportsof{C}{v}$. We now inductively define a family of functions, $\familyof{\circuitfunctionatport{}{p}}{p \in \circuitports[C]}$, each taking input variables $X = \familyof{x_{\ell}}{\ell \in \circuitinlabels[C]}$ (the circuit $C$'s overall inputs); our definition is inductive:
\[
\circuitfunctionatportof{}{p}{X} =
\begin{cases}
 x_\finverseapply{\circuitsignaturemapping}{p}
   &\text{$p$ is a disengaged input port} \\
 \circuitfunctionatportof{}{p'}{X}
   & \text{$p$ is an engaged input port with $\tuple{p',p} \in \gedges{}$} \\
 \ofp{Op}{a} \paren*{ Y }
   & \text{$p = \tuple{v,a}$ is an output port and $Y = \familyof{ \circuitfunctionatportof{}{\tuple{v,\ell}}{X} }{\ell \in \inlabelsofsig{\opsig[{\uplanop{Op}}]}}$}
\end{cases}
\]
In other words: The inputs are cascaded through the columnar circuit, just like in a boolean logic circuit; nodes (operators) apply their associated function to their inputs, and the resulting outputs are ``carried'' by wires on to input ports of subsequent nodes.

Finally, \emph{The function computed by $C$}, denoted $\circfunc{C}$, is $X \mapsto \familyof{\circuitfunctionatportof{}{ \csigmapof{}{\ell} }{X}}{\ell \in \circuitoutlabels{C}}{}$.
\end{definition}

\begin{figure}[ht]
%\begin{subfigure}[b]{0.475\linewidth}
    \centering
    \includegraphics[scale=0.75]{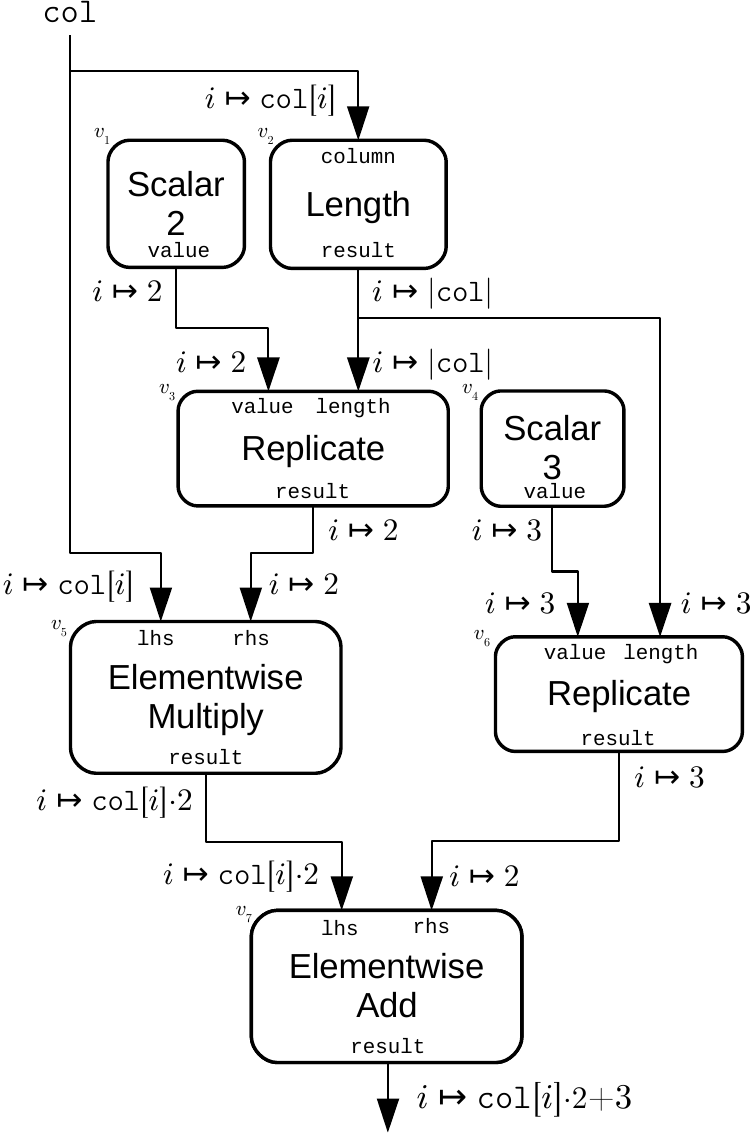}
    \indexPlanOperator{Scalar}
    \indexPlanOperator{Length}
    \indexPlanOperator{Replicate}
    \indexPlanOperator{Elementwise}
%    \subcaption{A columnar circuit, with the port-computed functions indicated in $x \mapsto f(x)$ form}
%    \label{subfig:columnar-circuit:function-resolution:example:diagram}
%   \end{subfigure}
%   \hspace{0.03\linewidth}
%   \begin{subfigure}[b]{0.50\linewidth}
%   \begin{minipage}[b]{0.50\linewidth}\footnotesize
%     \end{minipage}
%
%     \subcaption{Recursive application of the inductive resolution --- equation form}
%     \label{subfig:columnar-circuit:function-resolution:example:equation}
%   \end{subfigure}
%
  \caption{Resolution of the function computed by a columnar circuit}
  \label{fig:columnar-circuit:function-resolution:example}
\end{figure}

To make the above definition more concrete, let us work out the function computed by a specific circuit: The one appearing in \autoref{fig:columnar-circuit:function-resolution:example}. For brevity, we use $\uplanop{EA}$, $\uplanop{EM}$, $\uplanop{R}$, $\uplanop{L}$, $\uplanop{S2}$ and $\uplanop{S3}$ in the following instead of $\uplanop{Elementwise}_{+}$ (Elementwise add), $\uplanop{Elementwise}_{\*}$ (Elementwise multiply), $\uplanop{Replicate}$, $\uplanop{Length}$, $\uplanop{Scalar 2}$ and $\uplanop{Scalar 3}$ respectively, and the column names $\colname{res}$ and $\colname{val}$ instead of $\colname{result}$ and $\colname{value}$.

\begin{align*}
    \circfuncof{C}{\colname{col}}
    &= \cfpof{C}{\tuple{v_7,\colname{res}}}{\colname{col}} \\
    &= \ofpof{EA}{\colname{res}}{\cfp{C}{\tuple{v_7, \colname{lhs}}},\cfp{C}{\tuple{v_7, \colname{rhs}}}} \\
    &= \ofpof{EA}{\colname{res}}{
         \cfp{C}{\tuple{v_5, \colname{res}}},
         \cfp{C}{\tuple{v_6, \colname{res}}}
       } \\
    &= \ofpof{EA}{\colname{res}}{
         \ofpof{EM}{\colname{res}}{ \cfp{C}{\tuple{v_5, \colname{lhs}}} , \cfp{C}{\tuple{v_5,\colname{rhs}}} },
         \ofpof{R}{\colname{res}}{ \cfp{C}{\tuple{v_6, \colname{val}}} , \cfp{C}{\tuple{v_6, \colname{length}}} }
       }  \\
    &= \ofpof{EA}{\colname{res}}{
         \ofpof{EM}{\colname{res}}{
           \colname{col},
           \cfp{C}{\tuple{v_3, \colname{res}}}
         },
         \ofpof{R}{\colname{res}}{
           \cfp{C}{\tuple{v_4, \colname{val}}} ,
           \cfp{C}{\tuple{v_2, \colname{length}}}
         }
       }  \\
    &= \ofpof{EA}{\colname{res}}{
         \begin{aligned}
         &\ofpof{EM}{\colname{res}}{
           \colname{col},
           \ofpof{R}{\colname{res}}{
             \cfp{C}{\tuple{v_3, \colname{val}}},
             \cfp{C}{\tuple{v_3, \colname{length}}}
           }
         },\\
         &\ofpof{R}{\colname{res}}{
           \ofpof{S3}{\colname{val}}{\*},
           \ofpof{L}{\colname{res}}{\cfp{C}{\tuple{v_2, \colname{column}}}}
         }
         \end{aligned}
       }  \\
    &= \ofpof{EA}{\colname{res}}{
         \begin{aligned}
         &\ofpof{EM}{\colname{res}}{
	   \colname{col},
	   \ofpof{R}{\colname{res}}{
	     \cfp{C}{\tuple{v_1, \colname{val}}},
	     \cfp{C}{\tuple{v_2, \colname{length}}}
	   }
         },\\
         &\ofpof{R}{\colname{res}}{
           \ofpof{S3}{\colname{val}}{\*},
           \ofpof{L}{\colname{res}}{\colname{col}}
         }
         \end{aligned}
       }  \\
    &= \ofpof{EA}{\colname{res}}{
         \ofpof{EM}{\colname{res}}{
           \colname{col},
           \ofpof{R}{\colname{res}}{
             \ofpof{S2}{\colname{val}}{\*},
             \ofpof{L}{\colname{res}}{\cfp{C}{\tuple{v_2, \colname{column}}}}
           }
         },
         \ofpof{R}{\colname{res}}{
           3,
           \collength{col}
         }
       }  \\
    &= \ofpof{EA}{\colname{res}}{
         \ofpof{EM}{\colname{res}}{
           \colname{col},
           \ofpof{R}{\colname{res}}{
           2,
           \ofpof{L}{\colname{res}}{\colname{col}}
           }
         },
         \ofpof{R}{\colname{res}}{
           3,
           \collength{col}
         }
       }  \\
    &= \ofpof{EA}{\colname{res}}{
         \ofpof{EM}{\colname{res}}{
           \colname{col},
           \ofpof{R}{\colname{res}}{
           2,
           \collength{col}
           }
         },
         \ofpof{R}{\colname{res}}{
           3,
           \collength{col}
         }
       }  \\
\intertext{and applying this to a single arbitrary column element, we get}
    \circfuncof{C}{\colname{col}} \sbrace{i}
    &=
       \ofpof{EA}{\colname{res}}{
         \ofpof{EM}{\colname{res}}{
           \colname{col},
           \ofpof{R}{\colname{res}}{
           2,
           \collength{col}
           }
         },
         \ofpof{R}{\colname{res}}{
           3,
           \collength{col}
         }
       } \sbrace{i} \\
    &= \ofpof{EM}{\colname{res}}{
         \colname{col},
         \ofpof{R}{\colname{res}}{
         2,
         \collength{col}
         }
       }\sbrace{i}  +
       \ofpof{R}{\colname{res}}{
         3,
         \collength{col}
       }\sbrace{i} \\
    &= \colval{col}{i} \*
       \ofpof{R}{\colname{res}}{
         2,
         \collength{col}
       }\sbrace{i}  +
       3 \\
    &= \colval{col}{i} \* 2 + 3
\end{align*}
this functions computed on each port are also noted in \autoref{fig:columnar-circuit:function-resolution:example} itself.

\medskip

An implementation of the model of computation we've defined is a system (a program, a machine or a multi-node cluster of machines) which, given a columnar circuit and an assignment to its variables, evaluates $\circfunc{G}$ as in the above example, on its entire domain, i.e. materializes the columns produced by circuit $C$ when provided with columns as assignments to its inputs.

\begin{definition}[Decision circuit]
\label{def:decision-circuit}
A (columnar) decision circuit is a columnar circuit $C$  with a single output type $\booltype$, whose output column never has length other than 1 (i.e. $\frange{\circfunc{C}} \inc \setof{\true, \false}$). A decision circuit is said to \emph{accept} input columns $X = \familyof{\colname{c}_\tau}{\tau \in \intypesofsignature{C}}$ if $\circfuncof{C}{X} = \true$, and to \emph{reject} the input if $\circfuncof{C}{X} = \false$. If $\circfunc{C}$ is a complete function (accepting or rejecting every input), $C$ is said to be a \emph{complete} decision circuit.
\end{definition}

\subsubsection*{Side note: Circuits vs instruction sequences}
\label{subsubsec:circuits-vs-instruction-sequences}

The circuit model is the more fitting characterization of the kind of computation column stores execute, as there is no ordering among the vertices except through their data dependencies. Instruction sequences, the much more commonly used formalism for specifying computations, are ordered; and are also amenable to control flow denotations such as loops, conditionals and jumps --- none of which are supported by the model of computation we have presented. This being said, in sequential, non-vectorized/columnar execution of programs, out-of-order execution which does not interfere with data dependencies is common practice. And we could just constrain instructions to merely Single-Static-Assignment applications of columnar operators, defining a program to be an equivalence class of instruction sequences under permutations ensuring that all instructions producing the inputs of an operator application appear before this application.

\section{Circuit composition and transformation}
\label{sec:circuit-composition-and-transformation}

A traditional boolean circuits is not typically generated from scratch, but rather composed of subcircuits, computing simpler functions. The same should be the case with columnar circuits as well, for column store constructing execution plans. Such plans undergo repeated transformations after their initial generation --- involving removal, addition and replacement of individual operators and entire subplans. This section presents some formal machinery for applying such transformations to columnar circuits. We will also be making extensive use of them for further conceptualization work in later chapters.

\begin{definition}[circuit union]
Let $C_1 = \circuittuple[1]$, $C_2 = \circuittuple[2]$ be two circuit, and assume $\gverts{}_1$, $\gverts{}_2$ are disjoint, as are $\circuitsiglabels[1]$ and $\circuitsiglabels[2]$). The (disjoint) \emph{union}, or \emph{union circuit}, \emph{of $C_1$ and $C_2$} is the circuit $C_\cup =
\breakingtuple{\csignature[]_\union, \gverts{}_\union, \gedges{}_\union, \copof[]_\union, \csigmap[]_\union}$ where
\[
\begin{aligned}
\csignature[]_\union &= \tuple{\csigin[1] \union \csigin[2]\, , \, \csigout[1] \union \csigout[2]} \\
\gverts{}_\union     &= \gverts{1} \union \gverts{2} \\
\gedges{}_\union     &= \gedges{1} \union \gedges{2} \\
\copof[]_\union      &= \copof[1] \union \copof[2] \\
\csigmap[]_\union    &= \csigmap[1] \union \csigmap[2]
\end{aligned}
\]
For the case of non-disjoint vertex or signature label sets, one may differentiate them by applying $x \mapsto \tuple{1, x}$ for one circuit and $x \mapsto \tuple{2, x}$ for the other.
\end{definition}
If we order the input and output labels of circuits $C_1$ and $C_2$, the latter after the former, their union can also be thought of as a concatenation: It takes a concatenation of inputs to each of them and produces a concatenation of outputs from each of them (when the outputs are defined, of course).

\medskip

%This next transformation involves connecting ports within a circuit. We first note, though, that not any pair of ports in a circuit's layout can be connected while maintain the circuit constraints: A port $p_1$ is \emph{connectible} to port $p_2$ (and $p_2$ connectible from $p_1$) if $p_1$ is an output port; $p_2$ is a disengaged input port; the two ports share the same element type (via $\copof$); and $p_1$ cannot be reachable from $p_2$ in the circuit structure $G(C)$ (in which case the connection would form a cycle). We can now make

\begin{definition}[Circuit input assignment]
\label{def:circuit-input-assignment}
Let $C = \circuittuple[]$ be a columnar circuit whose signature contains the label $\ell$, let $p_1 = \tuple{v, b} = \fapply{\csigmap[C]}{\ell}$ be the vertex input port to which $\ell$ is mapped in $C$, and let $p_2 = \tuple{u,a}$ be a vertex output port within $C$, with the same associated element type as $p_1$. The circuit resulting from \emph{the assignment of vertex output $p_1$ to input $\ell$ in $C$}, or \emph{the assignment of the $a$ output of $u$ to the $\ell$ input in $C$} is the circuit
$C' = \breakingtuple{\csignature[]', \gverts{}', \gedges{}', \copof[]', \csigmap[]'}$
where
\[
\begin{aligned}
\csignature[]' &= \tuple{\csigin[]', \csigout[]} \\
\csigin[]'     &= \tuple{\inlabelsofsignature{\csig[]'} \, , \, \intypesofsignature{\csig[]'}} \\
\inlabelsofsignature{\csig[]'}
               &= \inlabelsofsignature{\csig[]} \setminus \setof{a} \\
\intypesofsignature{\csig[]'}
               &= \domrestrict{\intypesofsignature{\csig[]} }{\inlabelsofsignature{\csig[]'} } \\
\gverts{}'     &= \gverts{} \\
\gedges{}'     &= \gedges{} \union \singletonset{ \tuple{p_2,p_1} } \\
\copof[]'      &= \copof[] \\
\csigmap[]'    &= \csigmap[] \setminus \tuple{a,p_1}
\end{aligned}
\]
\end{definition}

The above definition is ``intra-circuit''; but one may wish to engage a circuit's input with the output of another circuit, rather than another vertex in the same circuit. This can be done by reducing the inter-circuit case to the intra-circuit case by taking the union circuit of the two circuits to be connected.

\begin{definition}[Induced subcircuit]
Let $C = \circuittuple[]$ be columnar circuit and let $V' \inc \gverts{}$ (with $\gverts{}'' = \gverts{} \setminus \gverts{}'$). The \emph{subcircuit of $C$ induced by $\gverts{}'$} is a circuit with the layout of the sub-port-graph of $G(C)$ induced by $\gverts{}'$, the ports of $C$ mapped to $\gverts{}'$ by $\csigmap[]$), plus an artificial circuit port for every port $p$ of a $\gverts{}'$ vertex whose connection from/to $\gverts{}''$ is broken by taking the induced subgraph. We label these new artifical circuit ports using the $\gports{C}$ ports to which they had been connected, assuming without loss of generality that these labels will be disjoint from $\labelsofsignature{\csignature[]}$.
More formally, the induced subcircuit, denoted $\domrestrict{C}{\gverts{}'}$ is the tuple $\breakingtuple{\csignature[]', \gverts{}', E', \copof[]', \csigmap[]' }$ with
\begin{align*}
\gedges{}'      &= \domrestrict{\gedges{}}{\gverts{}'} \\
\copof[]'       &= \domrestrict{\copof[]}{\gverts{}'} \\
\csignature[]'  &= \tuple{\csigin[]',\csigout[]'} \\
\csigin[]'      &= \tuple{\inlabelsofsignature{\csig[]'}, \intypesofsignature{\csig[]'}} \\
\inlabelsofsignature{\csig[]'}
                &= L_\text{retained} \dunion L_\text{exposed} \\
L_\text{retained}  &= \condset{ \ell \in \inlabelsofsignature{\csig[]} }{ \csigmapof{}{\ell} \in \pginports[C]{\gverts{}'}}  \\
L_\text{exposed}  &= \condset{ \ell_\tuple{v,a} }{\in \pginports[C]{\gverts{}'} \text{ and } \gdegree{\gverts{}'', \tuple{v,a}} > 0} \\
 \intypesofsignature{\csig[]'}_{\ell}
                 &= \begin{cases}
                      \intypesofsignature{\csig[]'}_{\ell} & \ell \in L_\text{retained} \\
                      \intypesofsignature{\opsig[\uplanop{Op}]}_a & %\ell \in L_\text{exposed} \text{ i.e. }
                                                                     \ell = \ell_\tuple{v,a}\text{ for some }\tuple{v,a} \in \pginports[C]{\gverts{}'}\\
                    \end{cases} \\
 \csigout[]'  &\phantom{= \vphantom{1}}\text{defined similarly to $\csigin[]'$} \\
 \csigmap[]'\paren{\ell}
                 &= \begin{cases}
                      \csigmap[]\paren{\ell} & \ell \in L_\text{retained} \\
                      p & %\ell \in L_\text{exposed} \text{ i.e. }
                          \ell = \ell_p \text{ for some } p \in \pginports[C]{\gverts{}'} \\
                    \end{cases}
\end{align*}
\end{definition}

\begin{figure}[H]
  % ----------------------------------------------------
  % Examples of an induced subcircuit
  % ----------------------------------------------------
  \centering
  \begin{subfigure}[b]{0.25\linewidth}
    \centering
    \includegraphics[scale=0.65]{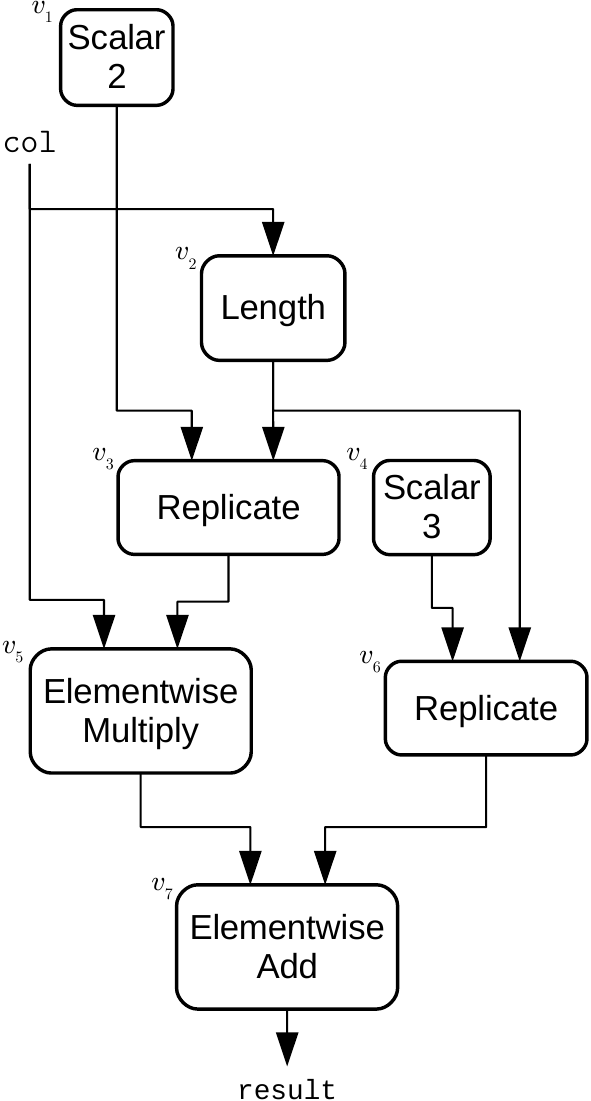}
    \indexPlanOperator{Scalar}
    \indexPlanOperator{Length}
    \indexPlanOperator{Replicate}
    \indexPlanOperator{Elementwise}
    \subcaption{The full original circuit (port names not shown)\\ \mbox{}\\ \mbox{}}
    \label{subfig:columnar-circuit:induced-subcircuit:example:simplified-complete-circuit}
  \end{subfigure}
  \hspace{0.25cm}
  \begin{subfigure}[b]{0.25\linewidth}
    \centering
    \includegraphics[scale=0.65]{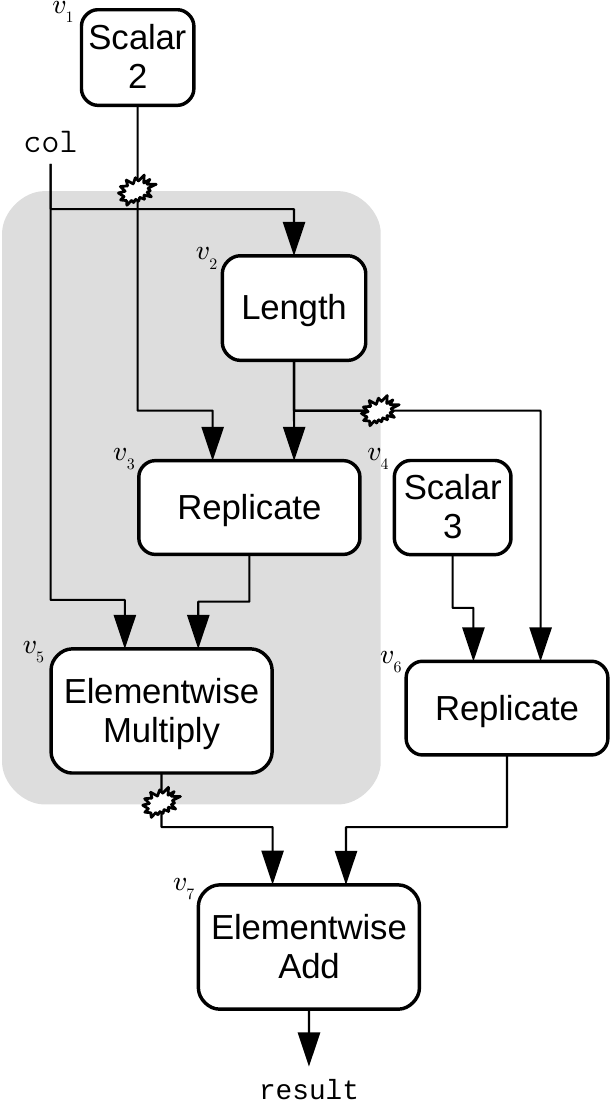}
    \indexPlanOperator{Scalar}
    \indexPlanOperator{Length}
    \indexPlanOperator{Replicate}
    \indexPlanOperator{Elementwise}
    \subcaption{A marked subset $V' \inc V$ and a snipping of its interface wires\\ \mbox{}}
    \label{subfig:columnar-circuit:induced-subcircuit:example:snip-subset}
  \end{subfigure}
  \hspace{0.25cm}
  \begin{subfigure}[b]{0.2\linewidth}
    \centering
    \includegraphics[scale=0.65]{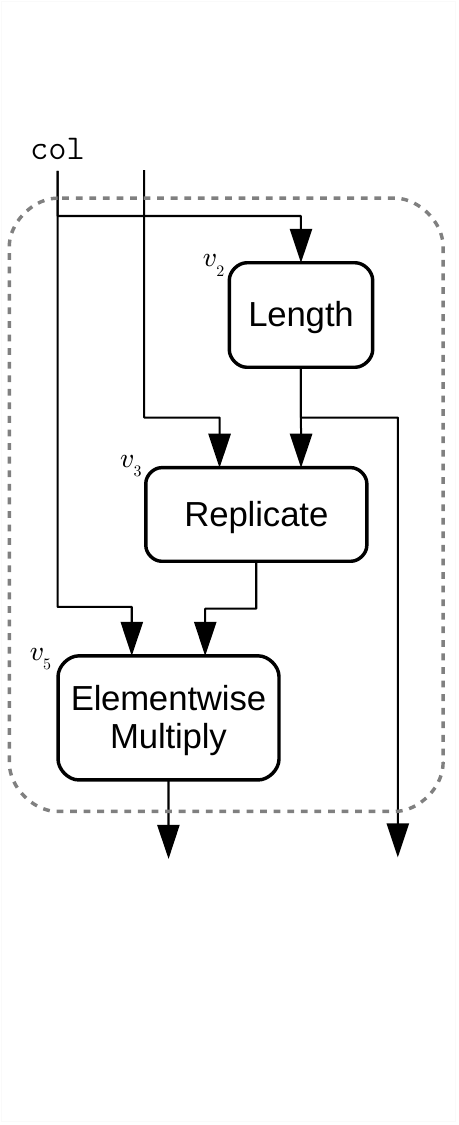}
    \indexPlanOperator{Length}
    \indexPlanOperator{Replicate}
    \indexPlanOperator{Elementwise}
    \subcaption{The interface wires extended to ports of the sibcircuit \\ \mbox{}}
    \label{subfig:columnar-circuit:induced-subcircuit:example:draw-ports-out}
  \end{subfigure}
  \hspace{0.25cm}
  \begin{subfigure}[b]{0.2\linewidth}
    \centering
    \includegraphics[scale=0.65]{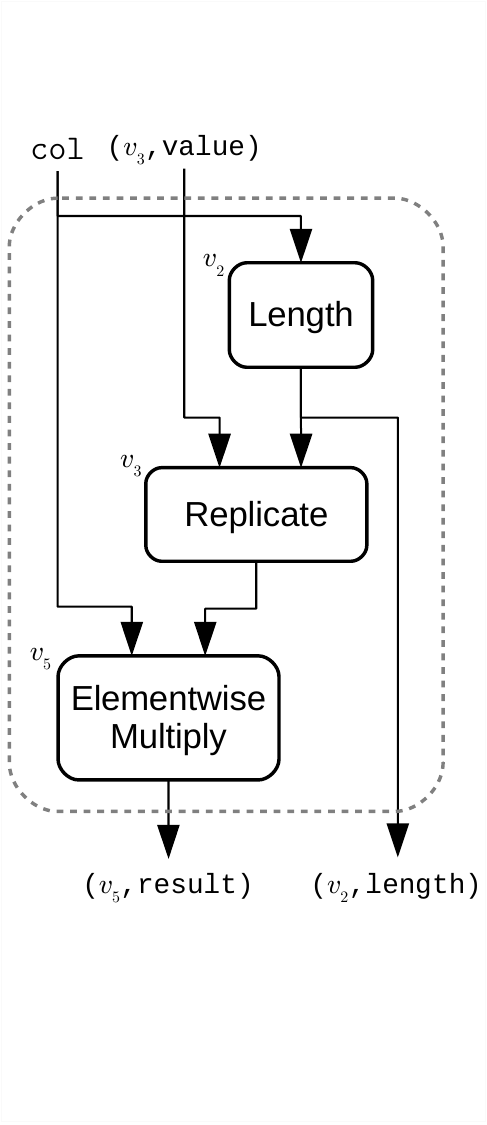}
    \indexPlanOperator{Length}
    \indexPlanOperator{Replicate}
    \indexPlanOperator{Elementwise}
    \subcaption{Subcircuit ports are named by the original-circuit ports they had connected}
    \label{subfig:columnar-circuit:induced-subcircuit:example:name-ports-and-finalize}
  \end{subfigure}
  \caption[An induced columnar subcircuit] {An induced subcircuit of the columnar circuit from \autoref{fig:columnar-circuit:function-resolution:example}}
  \label{fig:columnar-circuit:induced-subcircuit:example}
\end{figure}

\paragraph{Induced subcircuit replacement}
Let $C$ be columnar circuit, let $V' \inc \gverts{C}$ (and let $V'' = V \setminus V'$) and let $C_r$ be a replacement circuit for subcircuit $\domrestrict{C}{V'}$. The \emph{replacement of $\domrestrict{C}{V'}$ with $C_r$ in $C$} can be thought of in the same terms as replacing, say, an audio extension card in a personal computer, physically: The card has a line in  and line out ports --- part of the computer case's overall ports facing the outside world --- and the replacement card's similar ports replace those. Additionally, the card has a a sequence of specifically-positioned metal strips which fit into the computer's motherboard --- its internal connections to rest of the computer system: While a card is connected these are irrelevant, or unexposed; when we disconnect the old card these spring into effective existence, and the new card must match these connections in order to physically fit in to the system.

More formally, let $\fndef{\rho}{\circuitports[C_r] \setminus \circuitports[C]}{\circuitports[\domrestrict{C}{V'}] \setminus \circuitports[C]}$ be a bijection between the intra-$C$ ports of the replaced and replacement circuits. The circuit $\smashtilde{C}$, resulting from the replacement, is defined by
\begin{align*}
\csignature[\smashtilde{C}] &= \csignature[C] \\
\gverts{\smashtilde{C}}     &= V'' \union \gverts{C_r} = \paren*{\gverts{C} \setminus \gverts{C'}} \union \gverts{C_r} \\
\copof[\smashtilde{C}]      &= \paren*{\copof[C] \setminus \copof[C']} \union \copof[C_r] \\
\gedges{\smashtilde{C}}     &= \begin{aligned}[t]
                                 &\domrestrict{\gedges{C}}{V''} \dunion \gedges{C_r} \dunion \\
                                 &\conditionalset{ \tuple{p_r, p''} }{ \graphedgebetween{\fapply{\rho}{c_r}}{p''} \in \gedges{C}} \dunion \\
                                &\conditionalset{ \tuple{p'', p_r} }{ \graphedgebetween{p''}{\fapply{\rho}{c_r}} \in \gedges{C}} \\
                               \end{aligned} \\
\fapply{\csigmap[\smashtilde{C}]}{\ell}
                            &= \begin{cases}
                               \csigmapof{C}{\ell} &\fapply{\csigmap[C]}{\ell} \in \gportsof{C}{V''} \\
                               \fapply{\rho}{\csigmapof{C}{\ell}} & \text{otherwise} \\
                               \end{cases}
\end{align*}

We have taken some pains to formally define this transformation of columnar circuits, due to its relevance to execution plan optimization in column stores; it is a general form of the key step in a typical plan execution optimizations in a column store: Find a subplan which is amenable to improvement; then replace it with with a more useful subplan.

\medskip

\paragraph{Operator lifting}
A special case of induced subcircuit replacement with a different motivation for interest is the case of one of the circuits having just one node --- one computational operator:
Let \uplanop{Op} be a columnar operator. The tuple $C = \breakingtuple{\csignature[], \gverts{}, \gedges{}, \copof[C], \csigmap[C] }$ with
\[
\begin{aligned}
\csignature[] &= \opsig[\uplanop{Op}] \\
\gverts{}     &= \setof{v_{\uplanop{Op}}} \\
\gedges{}     &= \emptyset \\
\copof[C]     &= \setof{\tuple{v_{\uplanop{Op}}, \uplanop{Op}}} \\
\csigmap[C]   &= \condset{\graphedgebetween{\ell}{\tuple{v_{\uplanop{Op}},\ell}}}{\ell \in \circuitinlabels{C'} \dunion \circuitoutlabels{C'}}
\end{aligned}
\]
is a circuit, referred to as the \emph{lifting} of operator $\uplanop{Op}$ into a circuit $\circuitofop{C}{Op}$.

Note that any circuit with a single vertex is, in fact, a lifted operator, up to a renaming of its external ports. Now, consider a larger circuit $C$, in which we perform a subcircuit substitution of a single-node subcircuit $C'$ with another circuit $C_r$. This is essentially the replacement of the operator lifted into $C'$ with a possible implementation, $C_r$. In programming language terms, it is reminiscent of the inlining of a function at its call site (and a transformation one expects to be followed by various local optimizations such as avoiding redundant copies, operators followed by their inverse and so on).

The opposite kind of subcircuit replacement is the replacement of a (larger) subcircuit by a lifted operator; we refer to this transformation as a \emph{subcircuit fusion}. It is of key importance for efficient implementations of the columnar circuit model; see \autoref{sec:jit} for details.
%The opposite transformation --- replacing a subcircuit with a lifted operator --- is the equivalent of factoring out a new function from within an existing one. Of course, this analogy is not perfect: In programming languages, functions typically have a single definition, while a single operator can theoretically be replaced with any circuit with the same signature. Even if the semantic implications of the substitution are constrained to be equivalent, there may be numerous $C_r$ constituing an ``implementation'' of \uplanop{Op} which may be used to substitute $C'$.

\subsubsection*{Side note: ``Primitives'', operators and composition}
\label{subsubsec:primitives-vs-composite}

Many DBMSes distinguish between relational-algebraic ``operators'' and computational primitives: With this distinction, operators are what one might find in a logical SQL plan --- high-level, non-granular. They are implemented using ``primitives'', which are simpler and of higher granularity. Some primitives are atomic --- and the system is unable to reason about their innards --- while some are composites of other primitives. For a recent overview of this distnction, see \cite[\S 1,2]{GBDPS2018}. In particular, it references an (early) overview of the primitives in the columnar MonetDB \cite{BK1999} and essentially-columnar Voodoo \cite[Table 2]{PMZM2016}; one notes that they are clearly distinct from relational algebra operators, in that the former take columns (or scalars, partitioned columns etc.) --- not relations.

We do not follow this taxonomy. It does not concern itself with ``operators'' in the relational-algebraic sense (although a column store will likely acknowledge those during query plan compilation);  We only have columnar ``operators'', as defined in \autoref{def:operator}. We also have no notion of ``primitivity'' or ``atomicity''. All operators as potentially composable --- through the construction of circuits and finally their fusion into a new, composite operator; and they are all potentially decomposable (in a sense), by replacement with a multi-operator circuit computing the same function.

% ---------------------------------------
% ---------------------------------------
% ---------------------------------------
% ---------------------------------------
% ---------------------------------------

\chapter{Columnar representation}
\label{chap:columnar-representation}

A column store produces and utilizes a lot of data which isn't merely plain-vanilla columns --- whether it be meta-data for schema columns; indexing structures; lower-dimension projections; or intermediate results during the processing of a query. Having kept our data structures simplistic and uniform in the definition of the basic model of computation, this section will show how little expressive power was lost --- using simple columns as building-blocks for representing more complex structures, thus promoting them to ``first-class citizens'' within a column store.

\section{Encoding/decoding schemes for representation}
\label{sec:encoding-and-decoding-schemes-for-representation}

Fix, throughout this section, two unidirectional signatures $\sigma_e$ and $\sigma_d$, for encoded and original/not-encoded forms respectively. Also, denote $\mathcal{D} = \prod_{\tau \in \typesofsig{\sigma_d}}\colspace{\tau}$, and fix $\undecodedset \inc \mathcal{D}$ --- the set of families of (labeled) columns of which we wish to be able to decode and encode. Finally, since it will often be the case that the same structure has multiple columnar representations, we also fix $\sim$, an equivalence relation on $\undecodedset$: Intuitively, if $X \sim Y$, they must represent the same entity or structure.

\begin{definition}[Decoding scheme]
\label{def:concrete-decoding-scheme}
\label{def:representation-scheme}
Let $\tuple{C_d,C_v}$ be a pair of columnar circuits --- the \emph{decoder} and the \emph{encoded-form verifier} --- computing partial functions $\circfunc{d}$ and $\circfunc{v}$, respectively. Such a circuit pair constitutes a \emph{concrete decoding scheme} for $\undecodedset$ \wrt $\sim$ if:

\begin{enumerate}
\item The decoder has input signature $\sigma_{e}$ and output signature $\sigma_{d}$.
\item The verifier has the same input signature as the decoder ($\sigma_{e}$).
\item The verifier is a complete decision circuit (as per \autoref{def:decision-circuit}).
\item The decoder produces output on all valid encoded forms, i.e. $\fpreimageof{\circfunc{v}}{\setof{\true}} \subseteq \fpreimageof{\circfunc{d}}{\mathcal{D}}$.
\item The decoder only produces outputs in $\undecodedset$.
\item Every $X \in \undecodedset$ has a valid encoded form which decodes into $X$, or into an equivalent family $Y \sim X$ (i.e.  $\circfunc{d}$ hits every equivalence class of $\undecodedset$ under $\sim$).

\end{enumerate}
A \emph{decoding scheme} for $\undecodedset$ \wrt $\sim$ is a function $f_d$, for which there exists a concrete decoding scheme $\tuple{C_d, C_v}$ where the domain of $f_d$ are the valid encoded forms for the concrete scheme, and $\circfunc{d}$ is identical to $f_d$ on those valid encoded forms (i.e. $\fdomain{f_d} = \fpreimageof{\circfunc{v}}{\setof{\true}}$ and $f_d = \domrestrict{\circfunc{d}}{\fdomain{f_d}}$).
\end{definition}
\begin{note}The verifier circuit in \autoref{def:concrete-decoding-scheme} is an artifice for compartmentalizing the potential concerns regarding decoder input validity. It lets us describe decoding circuits which presume their inputs are valid --- significantly simplifying their structure and the operators they are made up of. Indeed, the computational complexity of deciding input validity may be higher than actually perfoming the decoding on valid inputs (as our model is circuit-based, not timed-machine based). Verifiers will not be of much significance in the remainder of this monograph, which is not concerned with invalid inputs and their handling. %Also note that, had we restricted decoder circuits to only produce output for valid input columns, we could have simply inserted the verifier circuits into the decoders and fail to produce output if the verifier rejects an input.
\end{note}

A decoding scheme in which the decoder produces a single column ($k=1$) is a \emph{column decoding scheme}.
%We refer to the inputs accepted by the verification function as \emph{encoded forms} of their decoding function output.
%For a decoding scheme with $\undecodedset$ comprising all possible tuples of column of the relevant output types, we also refer to the valid inputs as \emph{representations} of the column tuples resulting from their decoding, and to the decoding scheme as a \emph{representation scheme} for column tuples of types $\vec{\tau}$.

\begin{definition}[Encoding scheme]
A columnar circuit $C_e$, computing function $\circfunc{e}$ constitutes a A \emph{concrete encoding scheme} for $\undecodedset$ \wrt $\sim$ if:

\begin{enumerate}
\item The (encoder) circuit's input signature is $\sigma_{d}$ and its output signature is $\sigma_{e}$.
\item All families in $\undecodedset$ are encodable, i.e. $\undecodedset \inc \fdomain{\circfunc{e}}$.
\item If $X, Y \in \undecodedset$ satisfy $\fapply{\circfunc{e}}{X} = \fapply{\circfunc{e}}{Y}$ then $X \sim Y$.
\end{enumerate}
An \emph{encoding scheme} for $\undecodedset$ \wrt to $\sim$ is a function $f_e$, for which there exists a concrete encoding scheme $C_e$ with $f_e = \domrestrict{\circfunc{e}}{\undecodedset}$.
\end{definition}

\begin{definition}[Codec]
\label{def:codec}
A \emph{concrete codec} for $\undecodedset$ \wrt $\sim$ is a tuple $\tuple{C_e, C_d, C_v}$ such that $C_e$ is a concrete encoder and $\tuple{C_d,C_v}$ is a concrete decoder (\wrt $\undecodedset$, $\sim$), and the following hold:

\begin{enumerate}
\item All encoder outputs for $\undecodedset$ are valid encoded forms, i.e. $\fapply{\circfunc{e}}{\undecodedset} \included \finverse{\circfunc{v}}\setof{\true}$. (The converse statement is valid for any concrete decoder and encoder for $\undecodedset$.)
\item Decoding the encoding of a tuple of columns in $\undecodedset$ produces an equivalent tuple of columns (i.e. letting $f = \fcompose{\circfunc{d}}{\circfunc{e}}$, we have $\fapply{f}{X} \sim X$ for every $X \in \undecodedset$).
%if and only if $\fapply{f}{\vec{x}} \sim \fapply{f}{\vec{y}}$, for every $\vec{x},\vec{y} \in I$).
\end{enumerate}
A \emph{codec} is a pair $\tuple{f_d, f_e}$ of a decoding and an encoding scheme satisfying the above conditions (substituting $f_d$ for $\circfunc{d}$ and $f_e$ for $\circfunc{e}$).
\end{definition}

%\begin{definition}[Representation scheme]
%A \emph{representation scheme} for columns tuples of type $\vec{\tau}$ is a codec with $\undecodedset$ being all tuples (i.e. $\prod_i \colspace{\tau_i}$ and $\sim$ being the equality relation. If $k=1$, this is a \emph{column representation scheme}.
%\end{definition}

\begin{definition}[Column representation scheme]
A \emph{representation scheme} for columns of type $\tau$ is a codec for columns (i.e. with $k=1$), with $\undecodedset$ being all columns of this element type and with $\sim$ being the equality relation.
\end{definition}

The definitions above allow for multiple valid encoded forms (or representations) for a single uncompressed column-tuple. Occasionally, we will be interested both in such ``lax'' schemes and their ``narrowing'', by the imposition of further constraints on the encoded forms:

\begin{definition}[Sub-scheme]
\label{def:sub-scheme}
A decoding scheme $f$ constitutes a \emph{sub-decoding-scheme}, or sub-scheme, of another decoding scheme $f'$, if $\fdomain{f} \included \fdomain{f'}$ and $ \domrestrict{f'}{\fdomain{f}} = f$.
\end{definition}
We similarly define a sub-codec and a sub-representation scheme.

\section{Subcolumns}
\label{sec:subcolumns}

\subsection{Explicitly-indexed columns}
\label{subsec:index-value-pairs}

Before proceeding to define and model subcolumns, we first take a detour to introduce an additional representation scheme for (complete) columns. Its measure of redundancy will make it easy to adapt it, further below, to subcolumns.

Consider the following operators:
\begin{operators}
\opitem[Iota]
\begin{oplisting}
  Input   & \colname{n}           & $\inttype$  & $1$            & Intended length \\
  Output  & \colname{result}      & $\tau$      & $\colname{n}$  & Generated column \\
\end{oplisting}
Produces the identity column of a specified length, i.e. $\colname{result}[i] = i$.
%(Actually, \plannop{Iota} can also be obtained by composing other operators we'll be listing below.

\opitem[Permute]
\begin{oplisting}
   Input  & $\colname{permutation}$ & $\inttype$  & $n$            & Permutation of $\zeroupto{n}$  \\
   Input  & $\colname{data}$        & $\tau$      & $n$            & Column to permute \\
   Output & $\colname{permuted}$    & $\tau$      & $n$            & \\
\end{oplisting}
Produces the result of applying the permutation encoded in \colname{old_pos} to the data in \colname{data}, i.e. a column satisfying
$\colval{permuted}{i} = \colval{data}{{\colvalstar{\colname{permutation}^{-1}}{i}}}$.

\opitem[Length]
\begin{oplisting}
  Input   & \colname{col}         & $\tau$      & $n$  &  \\
  Output  & \colname{result}      & $\inttype$  & $1$  & Set to $n$, i.e. $\colsize{col}$ \\
\end{oplisting}
Produces the length of the input column \colname{col} as a scalar.
\end{operators}

If we \planop{Permute} a column, the original order of elements is lost, and the original column cannot be restored without known the permutation used. However, if we were to add a second column alongside our original one, and make it indicate positions --- the permutation would leave us with sufficient information to undo them. In fact,  the result of permuting the output of \planop{Iota} \emph{is} the permutation. We thus define:

\begin{cschemes}
 \csitem[Indexed]
  \begin{cslisting}
    $\colname{pos}$  & $\inttype$   & $n$ & element positions in the column \\
    $\colname{data}$ & $\tau$       & $n$ & element values \\
  \end{cslisting}
  The scheme induced by the concrete decoder being the lifting of \planop{Permute}, and the concrete encoder mapping a column \colname{c} to $\tuple{\planop{Iota}(\collen{c}), \colname{c}}$.% (a mapping realizable in \circuitfclass{}).
\end{cschemes}

\begin{note}This representation corresponds to the Decomposition Storage Model (DSM) \cite{CK1985}. For many years it was also the default internal representation scheme for columns in MonetDB (albeit not using the formalism in this paper), referred to as a BAT (Binary Association Table); $\tuple{\colname{pos},\colname{data}}$ here corresponds to the BAT ``head'' and ``tail'' parts \cite[\S 3]{MKB2009}. Recently, MonetDB has dropped this scheme \cite{monetdb-goes-headless} except for some vestigial aspects of the API for columns. In C\=/Store (and likely Vertica), columns are not materialized this way --- but iteration over position-and-value pairs is a fundamental part of the interface for all columns \cite[\S 2]{AMDM2007}.
%, its interfaces still includes a degenerated, `vestigial' head part.
\end{note}

\begin{figure}[H]
  % ----------------------------------------------------
  % Example of the Indexed column representation scheme
  % ----------------------------------------------------
  \centering
  \begin{subfigure}[b]{0.3\linewidth}
    \centering
    \includegraphics[height=4cm]{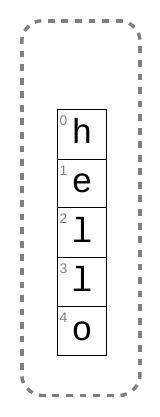}
    \subcaption{Standard {representation} \\ \mbox{}\\ \mbox{}}
    \label{subfig:scheme:column:example:indexed:decoded}
  \end{subfigure}
  \hspace{0.5cm}
  \begin{subfigure}[b]{0.3\linewidth}
    \centering
    \includegraphics[height=4cm]{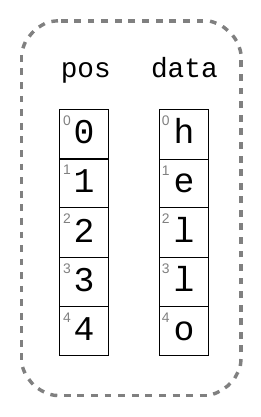}
    \subcaption{\cscheme{Indexed} representation of the same column --- result of encoding the column}
    \label{subfig:scheme:column:example:indexed:canonical}
  \end{subfigure}
  \hspace{0.5cm}
  \begin{subfigure}[b]{0.3\linewidth}
    \centering
    \includegraphics[height=4cm]{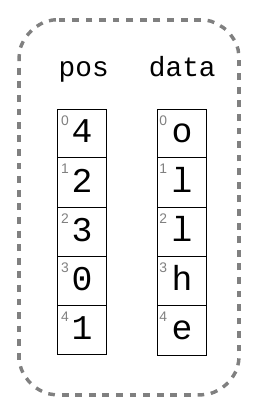}
    \subcaption{\cscheme{Indexed} representation --- a valid encoded form not obtained by applying the encoder}
    \label{subfig:scheme:column:example:indexed:any}
  \end{subfigure}
  \caption{The \capucscheme{Indexed} column representation scheme}
  \label{fig:scheme:column:example:indexed}
\end{figure}

This representation scheme is highly redundant. It does, however, preclude the need to consider the index of corresponding elements within the two column, i.e. one can make do with just the tuple set $\conditionalset{\tuple{\colval{indices}{i}, \colval{data}{i}}}{0 \leq i < n}$, decoding each tuple independently.

\subsection{A subcolumn representation}
\label{subsec:subcolumns}

\begin{definition}[Subcolumn]
\label{def:subcolumn}
A \emph{subcolumn} is a partial function $\fndef{\colname{sc}}{\naturals}{\tau}$ whose domain (of definition) is a finite subset $D \inc \naturals$.
%, or $\fdomain{\colname{sc}}$).
\end{definition}

A subcolumn \colname{sc} is said to be \emph{contiguous} if its domain is a contiguous subset of $\zeroupto{\opmax{D}}$, and \emph{incontiguous} otherwise.

\begin{observation}
A \emph{column} can be thought of as the special case of a contiguous subcolumn whose domain contains 0 (hence, with domain $D = \zeroupto{n}$). This reflects our choice to define subcolumns independently of any specific super-column, or even a specific-size supercolumn domain.
\end{observation}

%\begin{definition}[Subcolumn of a specific column]
Let \colname{sc}, $\colname{sc}'$ be two subcolumns of the same element type which agree as partial functions. \colname{sc} is said to be a \emph{subcolumn of $\colname{sc}'$} if their domains satisfy $\domainof{\colname{sc}} \inc \domainof{\colname{sc}'}$, and they agree on $\fdomain{\colname{sc}}$; this is denoted $\colname{sc} \subcolof \colname{sc}'$.
%\end{definition}
%In this case we refer to $\colname{sc}'$ as a super-subcolumn  of \colname{sc} (or a super-column of \colname{sc} if $\colname{sc}'$ is a full column).

%\begin{definition}
%The \emph{expansion set} of a subcolumn \colname{sc}, denoted $\exset{n}{sc}$ is the set of all super-subcolumns and supercolumns of \colname{sc} of length $n$.
%\end{definition}

As in the case of columns, we choose one representation scheme as the standard scheme; any other reprersentation scheme is defined by a codec relative to the standard one.

\begin{cschemes}
 \csitem*[\cscheme{Subcolumn}\text{ (Standard representation of a subcolumn)}]
  \begin{cslisting}
    $\colname{pos}$  & $\inttype$    & $n'$   & Non-negative values\\
    $\colname{data}$ & $\tau$        & $n'$   & \\
  \end{cslisting}
  The represented subcolumn $\colname{sc}$ is the partial function $x \mapsto \fapply{\colname{data}}{\fapplyinv{\colname{pos}}{x}}$, i.e. $\colval{sc}{i} = \colval{data}{\colvalstar{\colname{pos}^{-1}}{i}}$. Its domain is $\fimage{\colname{pos}}$.
\end{cschemes}

\begin{observation}
For $\fdomain{sc} = \zeroupto{n}$, the full domain, a standard representation of a subcolumn is a representation of it as a column using the \cscheme{Indexed} encoding scheme; as in that scheme, there are multiple standard representations for each subcolumn (one for every of the $\factorial{\setsize{D}}$ permutations of the domain elements).
\end{observation}

\begin{figure}[H]
  \centering
  \begin{subfigure}[b]{0.15\linewidth}
    \centering
    \includegraphics[scale=0.75]{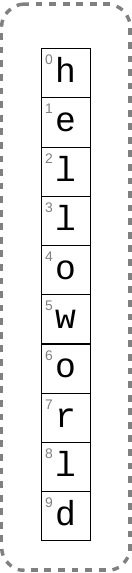}
    \subcaption{The original column \\ \mbox{} }
    \label{subfig:scheme:subcolumn:example:standard-representation:original}
  \end{subfigure}
  \hspace{0.25cm}
  \begin{subfigure}[b]{0.15\linewidth}
    \centering
    \includegraphics[scale=0.75]{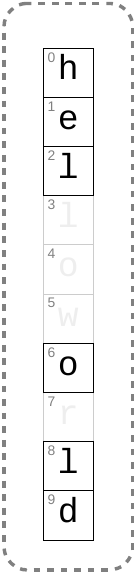}
    \subcaption{After removing some elements --- a subcolumn}
    \label{subfig:scheme:subcolumn:example:standard-representation:original-with-elements-removed}
  \end{subfigure}
  \hspace{0.25cm}
  \begin{subfigure}[b]{0.17\linewidth}
    \centering
    \includegraphics[scale=0.75]{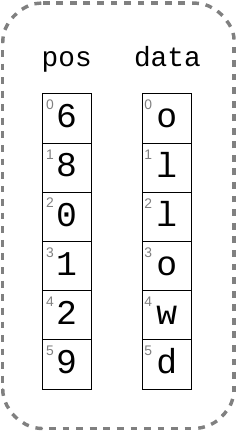}
    \subcaption{A standard representation of the subcolumn}
    \label{subfig:scheme:subcolumn:example:standard-representation:non-canonical-1}
  \end{subfigure}
  \hspace{0.25cm}
  \begin{subfigure}[b]{0.17\linewidth}
    \centering
    \includegraphics[scale=0.75]{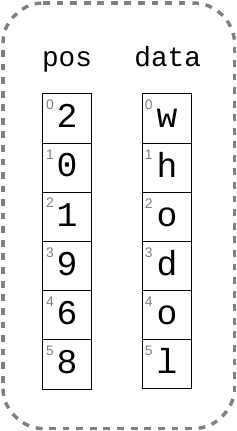}
    \subcaption{Another standard representation of the subcolumn}
    \label{subfig:scheme:subcolumn:example:standard-representation:non-canonical-2}
  \end{subfigure}
  \hspace{0.25cm}
  \begin{subfigure}[b]{0.17\linewidth}
    \centering
    \includegraphics[scale=0.75]{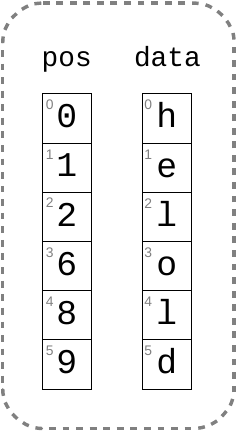}
    \subcaption{The canonical standard representation of the subcolumn}
    \label{subfig:scheme:subcolumn:example:standard-representation:canonical}
  \end{subfigure}
  \caption{The standard subcolumn representation scheme}
  \label{fig:scheme:subcolumn:examples:standard-reprersentation}
\end{figure}

\begin{definition}[Canonical representation of subcolumns] The \emph{canonical} representation of a subcolumn is its standard representation in which the elements of \colname{pos} are monotone increasing.
\end{definition}

\begin{observation}
A subcolumn \colname{sc} is contiguous if and only if the \colname{pos} column in its canonical representation is contiguous.
\end{observation}

\begin{definition}[Subcolumn decoding \& encoding schemes]
A \emph{subcolumn decoding (respectively, encoding) scheme} is a decoding (respectively, encoding) scheme (see \autoref{def:concrete-decoding-scheme} and the following text) for pairs of columns, of types $\tuple{\inttype, \tau}$ --- an integer type $\inttype$ and an arbitrary type $\tau$ ---  with $\undecodedset$ being the set of all standard subcolumn representations, and $\sim$ being the equivalence relation among standard subcolumn representations. A \emph{subcolumn codec} is a \emph{codec} for the type pair $\tuple{\inttype, \tau}$.
\end{definition}

As in the case of columns, decoding schemes for the set of all subcolumns of a type $\tau$ induce representations of type-$\tau$ subcolumns in addition to the standard one.

\subsection{Combining subcolumns}
\label{subsec:combining-subcolumns}

\begin{definition}[Subcolumn overlay]
Let $\colname{sc}_1$, $\colname{sc}_2$ be two $n$-subcolumns with domains $D_1$, $D_2$ respectively. The \emph{overlay of $\colname{sc}_1$ by $\colname{sc}_2$}, denoted $\colname{sc}_1 \leftarrow \colname{sc}_2$, is the partial function:
\[
\colname{sc}(x) =
\begin{cases}
  \colname{sc}_1(x) & x \in D_1 \setminus D_2 \\
  \colname{sc}_2(x) & x \in D_2
\end{cases}
\]
\end{definition}

\begin{figure}[H]
  \centering
  \includegraphics[height=4cm]{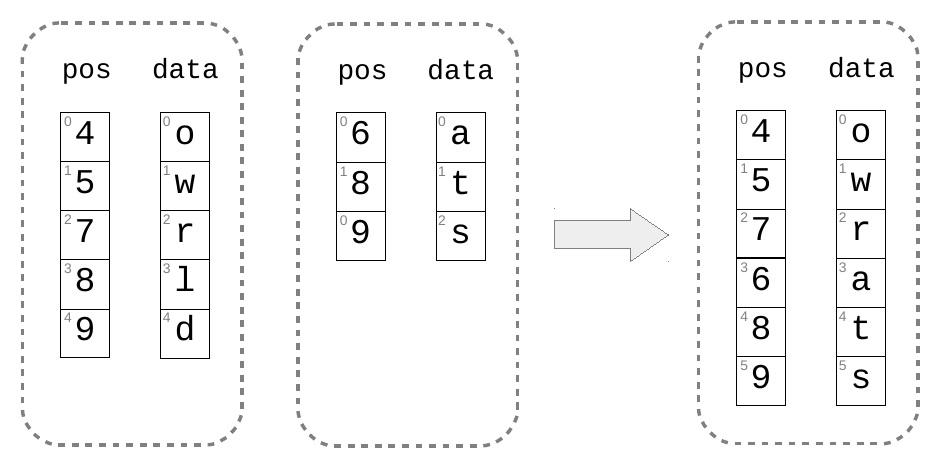} \\
  \caption[An overlay of partially-disagreeing subcolumns]{A subcolumn of ``helloworld'' (on the left) is overlaid by a subcolumn of ``hellowarts'' (middle), producing the subcolumn on the right.}
  \label{fig:scheme:subcolumn:examples:overlay}
\end{figure}

%Two subcolumns are said to be \emph{(weakly) compatible} if they agree (as functions) on all elements of the intersection of their domains; if their domains are disjoint, they are said to be \emph{strongly compatible}.
Two subcolumns are said to be \emph{compatible} if they agree (as functions) on all elements of the intersection of their domains.

\begin{observation}
The following are equivalent regarding two subcolumns $\colname{sc}_1$, $\colname{sc}_2$ of the same type:
\begin{enumerate}
 \item $\colname{sc}_1$ and $\colname{sc}_2$ are compatible.
 \item There exists a column $\colname{c}$ such that $\colname{sc}_1 \subcolumnof \colname{c} $ and $\colname{sc}_2 \subcolumnof \colname{c} $.
 \item $\colname{sc}_1 \leftarrow \colname{sc}_2$  is a super-column of both $\colname{sc}_1$ and $\colname{sc}_2$.
\end{enumerate}
\end{observation}

Two subcolumns being compatible makes for a special case of the above definition:

\begin{definition}[Subcolumn union]Let $\colname{sc}_1$, $\colname{sc}_2$ be two compatible subcolumns. The \emph{union subcolumn of $\colname{sc}_1$, $\colname{sc}_2$} is the subcolumn $\colname{sc}_\cup = \colname{sc}_1 \leftarrow \colname{sc}_2$.
\end{definition}

\begin{figure}[H]
  \centering
  \includegraphics[height=4cm]{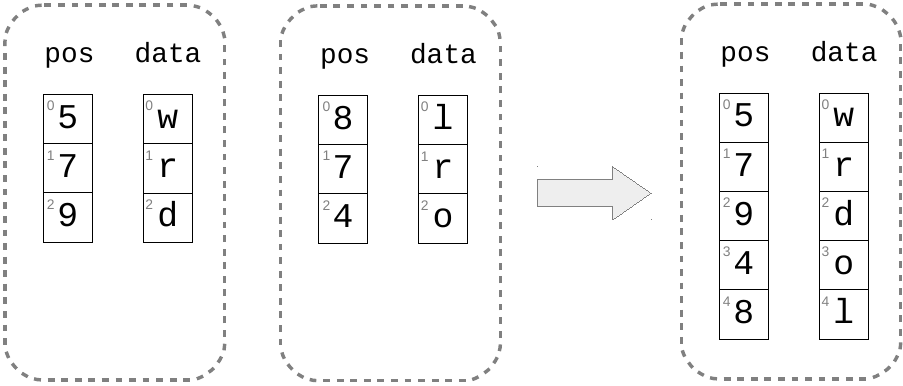}
  \caption[A union of ompatible subcolumns]{Two compatible subcolumns of ``helloworld'' (on the left), and the result of their union (on the right)}
  \label{fig:scheme:subcolumn:examples:union}
\end{figure}

\begin{observation}The relation $\subcolumnof$ is a partial order over all subcolumns of a fixed column; moreover, the set of such subcolumns and the $\subcolumnof$ partial order form a lattice \wrt intersection and union --- isomorphic to the lattice of partial sets of a fixed set.
\end{observation}

The above modes of combining columns and subcolumns are immediately useful in defining representation schemes:

\begin{cschemes}
\csitem[SubcolumnUnion]
  \begin{cslisting}
    $\colname{pos}_1$  & $\inttype$    & $n'$   & First subcolumn, element positions \\
    $\colname{data}_1$ & $\tau$        & $n'$   & First subcolumn, element data      \\
    $\colname{pos}_2$  & $\inttype$    & $n''$  & Second subcolumn, element positions \\
    $\colname{data}_2$ & $\tau$        & $n''$  & Second subcolumn, element data       \\
  \end{cslisting}
  A subcolumn representation scheme constituting standard representations of two compatible subcolumns.
\csitem[DisjointSubcolumnUnion] is a restriction of \cscheme{SubcolumnUnion} to pairs of disjoint subcolumns. This is still a representation scheme, since the restriction is of the representations (that is, encoded forms), not of the represented subcolumns.
\csitem[SubcolumnOverlay] is the subcolumn representation scheme similar to $\cscheme{SubcolumnUnion}$, except that the two subcolumns need not be compatible, and the decoding result is the overlay of the first subcolumn by the second one.
\end{cschemes}

In fact, some of these representation schemes inspire corresponding (full-)column representation schemes:

\begin{cschemes}
\csitem[ComplementingSubcolumns]
  \begin{cslisting}
    $\colname{pos}$    & $\inttype$    & $n'$     & Element positions \\
    $\colname{data}_1$ & $\tau$        & $n'$     & First subcolumn, element data      \\
    $\colname{data}_2$ & $\tau$        & $n -n'$  & Second subcolumn, element data       \\
  \end{cslisting}
  An adaptation of \cscheme{DisjointSubcolumnUnion} to the full-column case: $\colname{pos}_1$ and $\colname{pos}_2$ are each other's complement, so only one of them (say, the first) is encoded; and the decoding produces a standard representation of a column, not a subcolumn.

\csitem[ColumnByOverlay]  Same as \cscheme{ComplementingSubcolumns}, except that the two subcolumns need not have disjoint domain, and the decoding result is the overlay of the first subcolumn by the second one.
\csitem[OverlaidColumn]
  \begin{cslisting}
    $\colname{data}$           & $\tau$        & $n$   & Main column         \\
    $\colname{overlay_pos}$    & $\inttype$    & $n'$  & Subcolumn positions \\
    $\colname{overlay_data}$   & $\tau$        & $n'$  & Subcolumn data      \\
  \end{cslisting}
Similar to \cscheme{ColumnByOverlay}, except that instead of two subcolumns we now have a full column and an overlaying subcolumn. This scheme can be decoded by the lifting of the \planop{Scatter} operator (see below). (Also refered to as \cscheme{Patched} --- the individual position-datum pair are like \emph{patches} applied to the fabric of the \colname{data} column).
\end{cschemes}

These schemes can all be shown to have concrete codecs. To illustrate what some of the decoders are like, consider the following operators:

\begin{operators}
  \opitem[Concatenate]
  \begin{oplisting}
    Input  & $\colname{col}_1$  & $\tau$ & $n_1$ & \\
    Input  & $\colname{col}_2$  & $\tau$ & $n_2$ & \\
    \vdots & $\vdots$           & \vdots & \vdots \\
    Input  & $\colname{col}_k$  & $\tau$ & $n_k$ & \\
    Output & $\colname{result}$ & $\tau$ & $\sum_{j=1}^{k}{n_j}$ & \\
  \end{oplisting}
  Produces the concatenation of the two input columns, i.e. the column
  \[
  \colname{result}(x) =
  \begin{cases}
    \colname{col}_1(x) & 0 \leq x < n_1 \\
    \colname{col}_2(x - n_1) & n_1 \leq x < n_1+n_2 \\
    \vdots \\
    \colname{col}_k(x - \sum_{j=1}^{k-1}{n_j}) & \sum_{j=1}^{k-1}{n_j} \leq x < \sum_{j=1}^{k}{n_j} \\
  \end{cases}
  \]
  This operator does not concatenate individual elements into larger elements --- it's only columns that get concatenated.
  \opitem[Scatter]
  \begin{oplisting}
    Input  & $\colname{col}$    & $\tau$ & $n$  & Target to scatter onto\\
    Input  & $\colname{pos}$    & $\tau$ & $n'$ & Positions in target into which to scatter\\
    Input  & $\colname{data}$   & $\tau$ & $n'$ & Data to place into target\\
    Output & $\colname{result}$ & $\tau$ & $n$  & \makecell{A copy of $\colname{col}$ with the positions specified by \colname{pos} \\ overriden by the corresponding elements in \colname{data}.}
  \end{oplisting}
  Produces the column \[\colval{result}{i} = \begin{cases} \colval{data}{\finverseapply{\colname{pos}}{i}} & i \in \colimage{pos} \\ \colval{col}{i} & \text{otherwise} \end{cases}\]
\end{operators}

\begin{figure}[H]
  \centering
  \begin{subfigure}[b]{0.3\linewidth}
    \centering
    \includegraphics[scale=0.75]{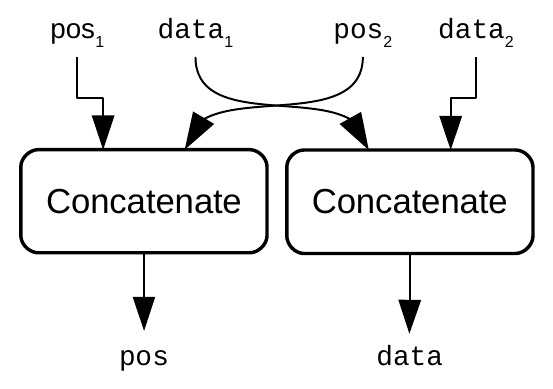}
    \subcaption{\cscheme{DisjointSubcolumnUnion} subcolumn representation decoder}
    \label{subfig:scheme:subcolumn:decoder:disjoint-subcolumn-union}
  \end{subfigure}
  \hspace{0.5cm}
  \begin{subfigure}[b]{0.3\linewidth}
    \centering
    \includegraphics[scale=0.75]{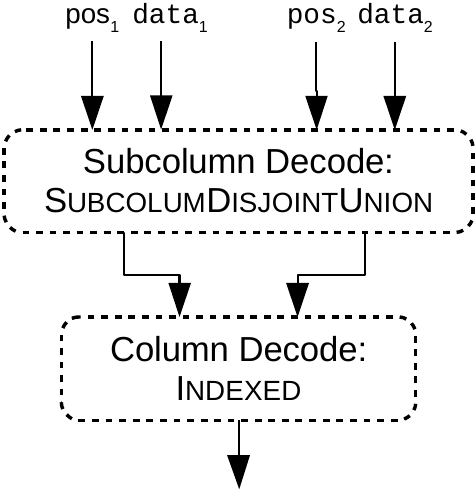}
    \subcaption{\cscheme{ComplementingSubcolumns} column representation decoder}
    \label{subfig:scheme:column:decoder:disjoint-subcolumn-union}
  \end{subfigure}
  \hspace{0.5cm}
  \begin{subfigure}[b]{0.25\linewidth}
    \centering
    \includegraphics[scale=0.75]{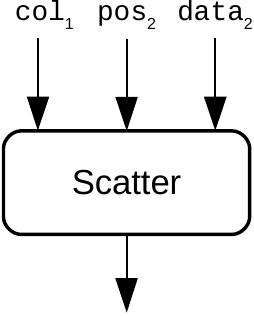}
    \subcaption{\cscheme{SubcolumnOverlay} column representation decoder}
    \label{subfig:scheme:column:decoder:subcolumn-overlay}
  \end{subfigure}
  \caption{Decoders for representation schemes involving subcolumn combinations}
  \label{fig:disjoint-subcolumn-union-decoders}
\end{figure}

Examples of the use of these three operators appear can be seen in \autoref{fig:disjoint-subcolumn-union-decoders}.

% Commenting the following subsection out - not because it's not worth putting it,
% but because it's not used anywhere earlier or later.
\begin{comment}
\subsection{Length-marked subcolumns}
\label{subsec:length-marked-subcolumns}
In defining subcolumns, we have made the choice not to have subcolumns retain information about an original column --- that is, not include an overall column size. An alternative choice would be retaining at least the original column's length, if not its values:

\begin{definition}[Length-marked subcolumn]
\label{def:length-marked-subcolumn}
A \emph{(domain-size-)marked subcolumn} is a pair $\tuple{d,\colname{sc}}$ where $d \in \naturals$ is a (full column) domain size and $\colname{sc}$ is a partial function $\fndef{\colname{sc}}{\zeroupto{d}}{\tau}$.
\end{definition}

The machinery we've introduced for subcolumns --- decoding \& encoding schemes, overlays, unions etc. --- can be readily adaptaed to marked subcolumns. Specifically, each subcolumn representation scheme has a corresponding marked-subcolumn scheme obtained by adding a length-1 column (scalar) of type $\inttype$ holding the subcolumn's $d$ value.
\end{comment}

\section{Segmented columns}
\label{sec:segmented-columns}

\paragraph{Motivation} Much of the processing in concrete column stores is applied limited-size chunks, segments or blocks: Pages on a magnetic disk drive or in DRAM; the data fitting inside one of a CPU's cache levels; a GPU multiprocessor's shared memory; an FPGA's bus width and its synthesized circuit data width; and so on. Many, if not most, existing column stores have chunks or segments as a fundamental abstraction, and partitions into chunks as a fundamental system feature (Some examples: Vectorwise \cite{ZBNH2005,ZHNB2006}, Google PowerDrill \cite[\S 2.3]{HBBGN2012}, C\=/Store/Vertica \cite{LFVTVDB2012}; a notable exception is MonetDB \cite{monetdb}). Yet, the model of computation presented in \autoref{chap:columnar-computational-model} is based on a uniform, unbroken definition of a column. We have opted for this simpler abstraction, as both the model itself and the representation and compression schemes this work presents would have been very unwieldy, had they beed saddled with segmentations of each column.

At the same time, we do intend to model existing systems, so we cannot simply ignore column segmentation. Segmentations are also inherent to several key data compression schemes, a subject which \autoref{chap:columnar-compression} will explore in more depth. We therefore presented segmented columns through the use of non-segmented ones, rather than the other way around.

\medskip
%\subsection{Index set segmentation}
%\label{subsec:index-set-segmentation}

First, suppose we merely wish to regard a column's indices --- the sequence $\zeroupto{n}$ --- in contiguous segments. These can be represented as follows:

\begin{cschemes}
  \csitem[Segmentation]
  \begin{cslisting}
    $\colname{start}$               & $\inttype$    & $m$  & \ldots of the first element of each segment\\
    $\colname{length}$              & $\inttype$    & $m$  & Non-negative number of elements in each segment
  \end{cslisting}
  There are no gaps between the segments, that is, $\colval{start}{i} = \colval{start}{i-1} + \colval{length}{i}$ for every $i \in \rngset{1}{m-1}$. It must also hold that $\colval{start}{0} = 0$ and $\colval{start}{m-1} + \colval{length}{m-1} = n$.
  \csitem[UniformSegmentation]
  \begin{cslisting}
    $\colname{segment_length}$      & $\inttype$    & $1$  & \\
    $\colname{overall_length}$      & $\inttype$    & $1$  & Length of the segmented columns \\
  \end{cslisting}
  A degenerate scheme for the special case of segment length being uniform; the number of segments is $\ceil{\colname{overall_length} / \colname{segment_length}}$.
\end{cschemes}
One may verify that these two qualify as representation schemes, with the represented columns being the identity functions of any length. Also, the $\colname{length}$ values may be 0 for the non-inform-length case; when this occurs, the segmentation is said to be \emph{degenerate}.

\medskip

Now let's apply these segmentation schemes to columns of actual data:

\begin{cschemes}
  \csitem[Segmented]
    \begin{cslisting}
      $\colname{data}$               & $\tau$       & $n$  & actual column data \\
      $\colname{segment_start_pos}$  & $\inttype$   & $m$  & of the first element of each segment\\
      $\colname{segment_length}$     & $\inttype$   & $m$  & Non-negative number of elements in each segment
    \end{cslisting}
    A standard representation of a column and an accompanying \cscheme{Segmentation}.
  \csitem[UniformlySegmented]
    \begin{cslisting}
      $\colname{data}$               & $\tau$       & $n$  & actual column data \\
      $\colname{segment_length}$     & $\inttype$   & $1$  & segment length \\
    \end{cslisting}
    A standard representation of a data column; only one of the scalars in a \cscheme{UniformSegmentation} is additionally necessary, as the data column length subsstitutes for the \colname{overall_length} of a \cscheme{UniformSegmentation}.
\end{cschemes}

\medskip

The two kinds of column segmentation --- fixed- and variable-length segments --- differ quite significantly not just in their representation, but in their motivations for use; but --- that discussion will wait until \autoref{subsec:scheme-segmentization}, below. The rest of this section will focus on the interpretation and manipulation of segmented columns (mostly with fixed-segment-length).

\subsection{Uniform segmentation and nearly-matrices}
\label{subsec:segmented-representation-scheme}

\begin{definition}[Segmented view]
Let \colname{c} be a column of length $n$ and $\ell \in \naturals$. The \emph{$\ell$-segmented view} of $\colname{c}$ is the function $\tuple{i,j} \mapsto \colval{c}{j \* \ell + i}$, defined whenever $j\*\ell + i < n$.
\end{definition}
With this function in mind, and when $\ell$ is obvious from the context, we abbreviate $\colval{c}{j \* \ell + i}$ as $\colval{c}{i,j}$ (the $i$\xth element of the $j$\xth segment).

The segmented view is the conceptualization of the break-up into segments defined above: Consecutive, aligned, length-$\ell$ segments are perceived as short columns in a $\ceil{\hfrac{n}{\ell}}$-column \emph{nearly-matrix}. This view is (row-major) matrix-like; but it cannot generally correspond to a proper matrix, since the last shorter column --- the last segment --- will be missing some elements when $\ndivides{\ell}{n}$. We also refer to this last segment as the \emph{slack} segment.
%; and this has many implications on how such segmented views can be handled.

\begin{figure}[H]
  \centering
  \begin{subfigure}[b]{0.125\linewidth}
    \centering
    \includegraphics[height=4cm]{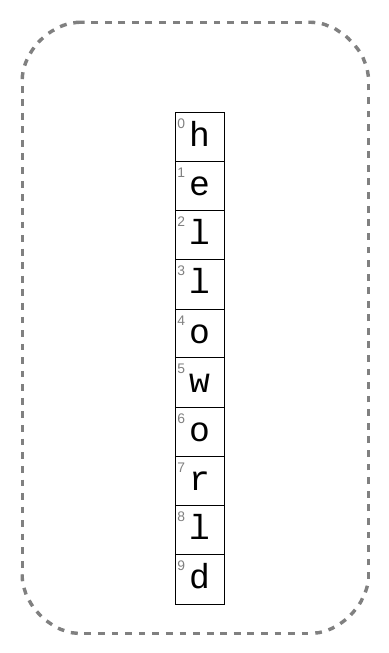}
    \subcaption{Standard \\ representation \\ \mbox{} \\ \mbox{}}
    \label{subfig:scheme:column:example:segmented:decoded}
  \end{subfigure}
  \hspace{0.05cm}
  \begin{subfigure}[b]{0.19\linewidth}
    \centering
    \includegraphics[height=4cm]{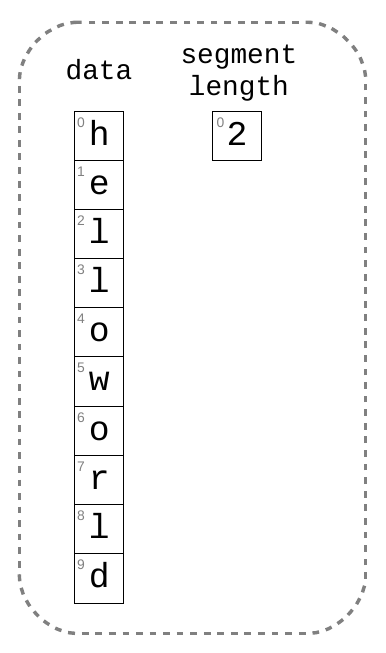}
    \subcaption{\cscheme{UniformlySegmented} with segment length 2, plain-column view}
    \label{subfig:scheme:column:example:segmented:2-plain}
  \end{subfigure}
  \hspace{0.05cm}
  \begin{subfigure}[b]{0.19\linewidth}
    \centering
    \includegraphics[height=4cm]{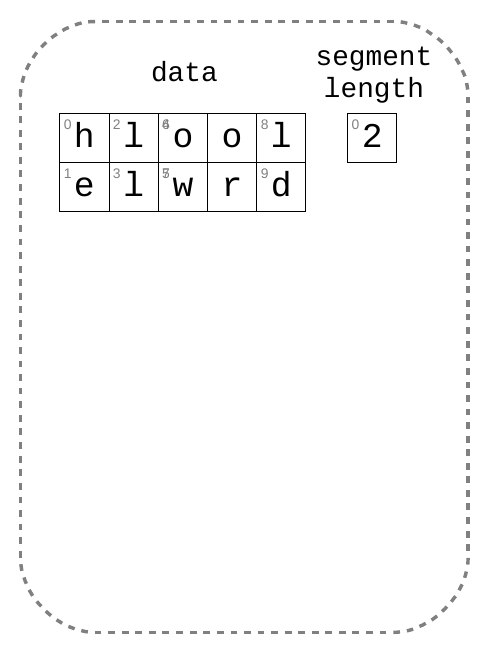}
    \subcaption{\cscheme{UniformlySegmented} with segment length 2, 2-segmented view}
    \label{subfig:scheme:column:example:segmented:2-segmented}
  \end{subfigure}
  \hspace{0.05cm}
  \begin{subfigure}[b]{0.19\linewidth}
    \centering
    \includegraphics[height=4cm]{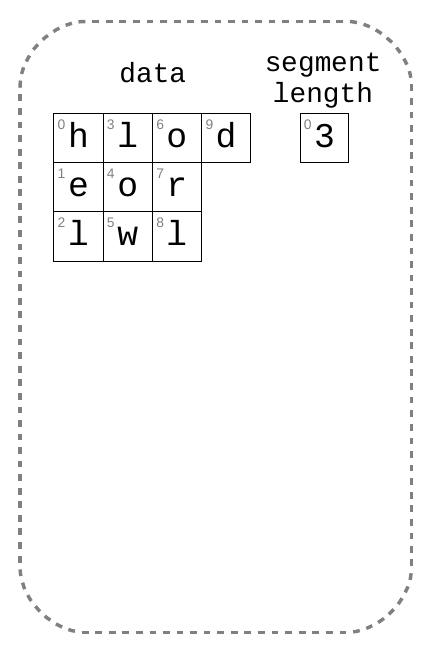}
    \subcaption{\cscheme{UniformlySegmented} with segment length 3, 3-segmented view}
    \label{subfig:scheme:column:example:segmented:3}
  \end{subfigure}
  \hspace{0.05cm}
  \begin{subfigure}[b]{0.19\linewidth}
    \centering
    \includegraphics[height=4cm]{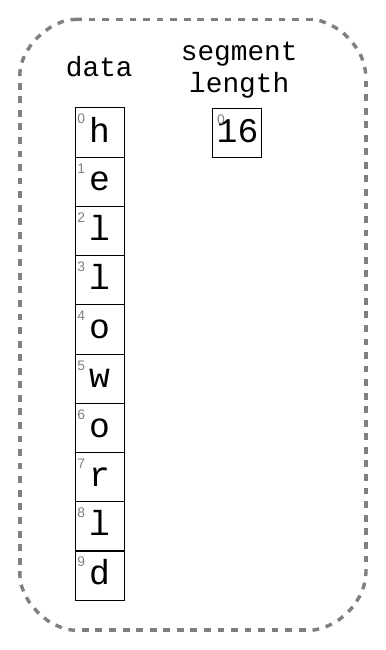}
    \subcaption{\cscheme{UniformlySegmented} with segment length 16 (plain-column view)}
    \label{subfig:scheme:column:example:segmented:16}
  \end{subfigure}
  \caption{The {\normalfont\ucscheme{UniformlySegmented}} representation scheme, $\ell$-segmented views and nearly-matrices}
  \label{fig:scheme:column:examples:segmented}
\end{figure}

Still adhering to the matricial interpretation, a $1$-segmented view of a column can be thought of as its transposition --- a single `row' of length $n$ (but bear in mind it is merely the replacement of $i \mapsto \colname{c}\sparen{i}$ with $\tuple{0,1} \mapsto \colname{c}\sparen{i}$). More generally, transposition is defined as follows:
\begin{operators}
  \opitem[Transpose]
  \begin{oplisting}
    Input  & $\colname{segment_length}$            & $\inttype$  & 1 & \\
    Input  & $\colname{col}$                       & $\tau$      & $\colname{segment_length} \* k$ \\
    Output & $\colname{transposed}$                & $\tau$      & $\colname{segment_length} \* k$ \\
    Output & $\colname{transposed_segment_length}$ & $\inttype$  & 1 & \\
  \end{oplisting}
  Produces a permutation of \colname{col} such that $\colval{transposed}{i,j} = \colval{c}{j,i}$ for every element offset $j$ within segment $i$. Consequently, the segment length is exchanged with the number of segments.
\end{operators}
Transposition requires \colname{col} to be a complete matrix (i.e. have a full-lnegth last segment), as otherwise its result does not form a nearly-matrix, i.e. there would be gaps in the resulting `column' due to their origin elements under transposition missing. This contrasts with a a segment length change (a re-segmentation if you will), which is possible for any column length.

\begin{figure}[H]
  \centering
  \begin{subfigure}[b]{0.25\linewidth}
    \centering
    \includegraphics[height=3.5cm]{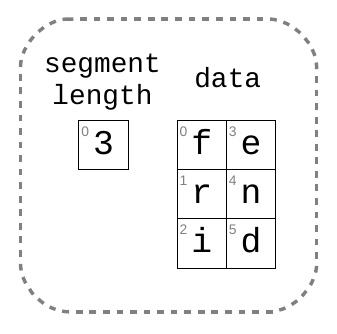}
    \subcaption{Original column segmented view}
    \label{subfig:operator:transpose:example:original}
  \end{subfigure}
  \hspace{0.25cm}
  \begin{subfigure}[b]{0.25\linewidth}
    \centering
    \includegraphics[height=3.5cm]{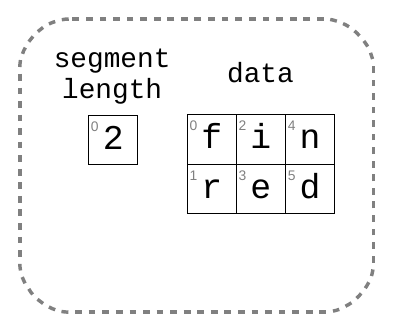}
    \subcaption{Change of the original's segment length from 3 to 2}
    \label{subfig:operator:transpose:example:segment-length-change}
  \end{subfigure}
  \hspace{0.25cm}
  \begin{subfigure}[b]{0.25\linewidth}
    \centering
    \includegraphics[height=3.5cm]{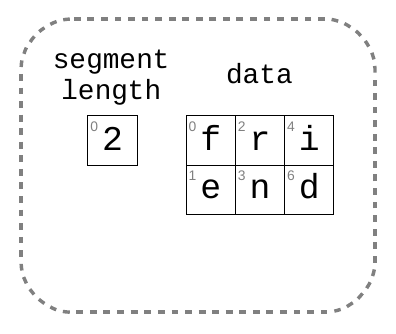}
    \subcaption{\planop{Transpose} applied to the original column}
    \label{subfig:operator:transpose:example:transposed}
  \end{subfigure}
  \hspace{0.25cm}
  \caption{Rearranging data in a segmented view: Segment length change vs. transposition}
  \label{fig:operator:transpose:example}
\end{figure}

\subsection{Data replication within and across segments}
\label{subsec:segmentization-and-data-replication}

Columnar circuits for decoding and encoding schemes occasionally involve replicating the values of short columns or scalars into longer columns; and this is the case also for column segments (as will become apparent in \autoref{subsec:segmentizing-operators}). Additionally, the matrix-like segmented views lend themselves to thinking of two kinds of replication of data, corresponding to the two dimensions of the segmented-view `matrix': Column-wise and row-wise replication:

\begin{operators}
  \opitem[ReplicateSegments]
  \begin{oplisting}
    Input  & $\colname{col}$            & $\tau$        & $n = \colname{segment_length} \* k$ & \\
    Input  & $\colname{segment_length}$ & $\inttype$    & 1                                   & \\
    Input  & $\colname{factor}$         & $\inttype$    & 1                                   & \\
    Output & $\colname{replicated}$     & $\tau$        & $\colname{factor} \* n$             & \\
    Output & $\colname{segment_length}$ & $\inttype$    & 1                                   & \makecell{Equal to the input of the same \\ name.}
  \end{oplisting}
  Produces a \cscheme{UniformlySegmented} column with the \emph{same} segment length as the input but with $\colname{factor} \* k = \colname{factor} \* \paren{n / \colname{segment_length}}$ segments, with column $i$ in the segmented view of \colname{replicated} being equal to column $i / \colname{factor}$ in the segmented view of $\colname{col}$. In other words, each segment is replicated into $\colname{factor}$ consecutive copies of itself in the output.
  \opitem[Replicate]
  \begin{oplisting}
    Input  & $\colname{col}$             & $\tau$      & $n = \colname{segment_length} \* k$ & \\
    Input  & $\colname{segment_length}$  & $\inttype$  & 1                                   & \\
    Input  & $\colname{factor}$          & $\inttype$  & 1                                   & \\
    Output & $\colname{replicated}$      & $\tau$      & $\colname{factor} \* n$             & \\
    Output & $\colname{segment_length}'$ & $\inttype$  & 1                                   & $= \colname{factor} \* \colname{segment_length}$ \\
  \end{oplisting}
  Produces a \cscheme{UniformlySegmented} column with \emph{longer segments} than the the input, but the \emph{number} of segments remains \emph{the same}. The new segment length is $\colname{factor}$ times the old length; each element in each of these segments is replicated $\colname{factor}$ times, consecutively, within its segment, so that the $j$\xth element in the result segment equals the $(j / \colname{factor})$\xth element of the original segment.
\end{operators}
Note that the \planop{Replicate} operator, defined in \autoref{subsec:operators} for unsegmented columns of length $1$, agrees with this definition --- it is the special case of this definition where the input has just a single segment --- the entire column. See also \autoref{subsec:segmentizing-operators} below regarding similar generalizations.

\begin{figure}[H]
  \centering
  \begin{subfigure}[b]{0.45\linewidth}
    \centering
    \includegraphics[scale=0.67]{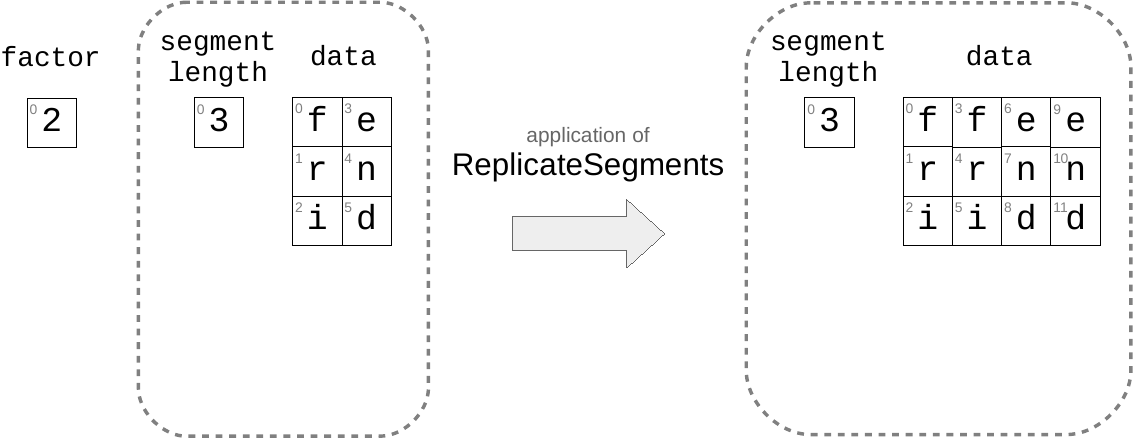}
    \subcaption{\planop{ReplicateSegments}}
    \label{subfig:scheme:operator:example:replicate-segments}
  \end{subfigure}
  \hspace{0.5cm}
  \begin{subfigure}[b]{0.45\linewidth}
    \centering
    \includegraphics[scale=0.67]{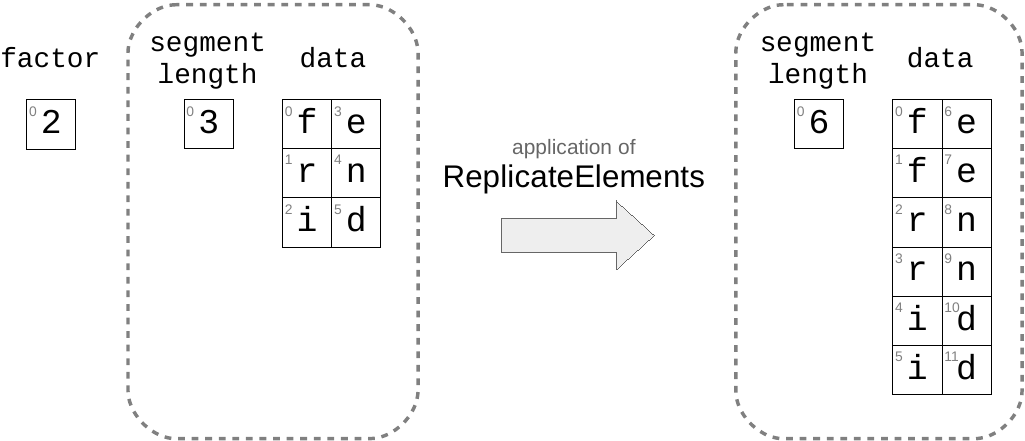}
    \subcaption{\planop{Replicate}}
    \label{subfig:scheme:operator:example:replicate-elements}
  \end{subfigure}
  \caption[Replication in the two dimenions of a segmented view ]{Replication in the two dimenions of a segmented view --- columns and rows respectively (note the progressing of indices within the \colname{data} column)}
  \label{fig:operators:examples:replication}
%  {\footnotesize The progression of indices along the element boxes indicates actual order in the layout in memory.}
\end{figure}

\subsubsection{Uniform segmentation and subcolumns}
\label{subsubsec:uniform-segmentation-and-subcolumns}

We would like to have a meaningful concept of ``taking a subcolumn'' of an $\ell$-segmented view, rather than a plain column; naturally, this means respecting the segmentation when discarding some of the data. To be more explicit: A permutation of $\zeroupto{n}$ is said to \emph{respect $\ell$-segmentation} if the image of every $\ell$-segment of the domain is an $\ell$-segment as well (e.g.  $\itemrng{\ell \* i}{\ell \* (i+1) - 1}$ becomes $\itemrng{\ell \* i'}{\ell \* (i'+1) - 1}$); and the final sub-$\ell$-length segment, if it exists, remains in place. A subcolumn \colname{sc} is said to \emph{respect the $\ell$-segmentation} of a supercolumn column \colname{c} if each (short) column in \colname{c}'s $\ell$-segmented view is either entirely within \coldom{sc} or does not intersect $\coldom{sc}$ at all.

\begin{observation}An $\ell$-segmentation-respecting permutation of the \colname{pos} column of an $\ell$-segmentation-respecting subcolumn can be uniquely determined with just the first of every $\ell$ contiguous elements.
\end{observation}
For our well-behaved subcolumns and permutations, this observation leads to a more space-efficient representation:

%Now consider the canonical representation of $\ell$-segmentation-respecting subcolumn. We know for certain that its \colname{pos} column will consist of internally-contiguous length-$\ell$ sequences (and possibly the last, shorter sequence): Some of these will be missing, some present. While not every one of the column's standard representations exhibits this feature, all representations in which consecutive subcolumn elements within the same segment appear consecutively, will. Conditioning on a representation having this feature, $\ell-1$ of every $\ell$ position indices are in fact redundant; and it is sufficient to simply indicate the positions of the beginning of an $\ell$-segments. Indeed, doing so  results in another subcolumn representation:

\begin{cschemes}
  \csitem[SegmentedSubcolumn]\nopagebreak
  \begin{cslisting}\nopagebreak
    $\colname{segment_length}$ & $\inttype$  & $1$                             & \\
    $\colname{segment_pos}$    & $\inttype$  & $n / \colname{segment_length}$  & in units of $\ell$ elements \\
    $\colname{data}$           & $\tau$      & $n$                             & \\
  \end{cslisting}
  Instead of a staight-up permutation of the entire index set, this scheme describes a permutation of $n'$ elements
  The positions are in units of $\ell$ elements, i.e. if $k \in \colimage{segment_pos}$ then the decoded subcolumn's domain contains $k \ell, k\ell+1 , \ldots, k \ell - 1$.
\end{cschemes}

\begin{figure}[H]
   % --------------------------------------------
   % Example of the 'Segmented' encoding scheme
   % --------------------------------------------
  \centering
  \begin{subfigure}[b]{0.35\linewidth}
    \centering
    \includegraphics[height=4cm]{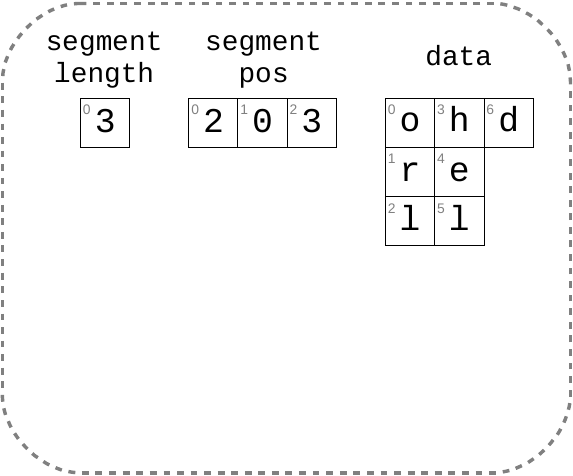}
    \subcaption{\cscheme{SegmentedSubcolumn} encoded form (segmented view)}
    \label{fig:scheme:subcolumn:example:segmented:matrix}
  \end{subfigure}
  \hspace{0.1cm}
  \begin{subfigure}[b]{0.3\linewidth}
    \centering
    \includegraphics[height=4cm]{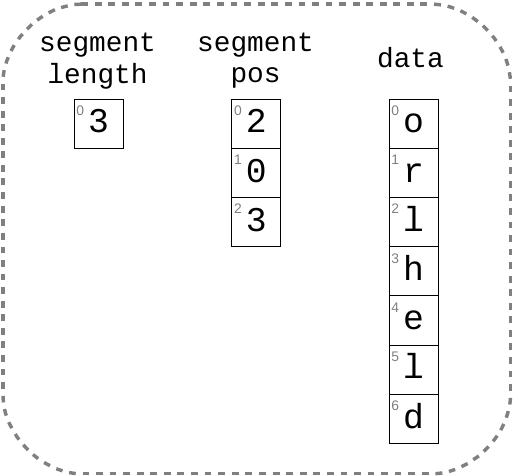}
    \subcaption{\cscheme{SegmentedSubcolumn} encoded form (plain column view)}
    \label{fig:scheme:subcolumn:example:segmented:plain}
  \end{subfigure}
  \hspace{0.1cm}
  \begin{subfigure}[b]{0.225\linewidth}
    \centering
    \includegraphics[height=4cm]{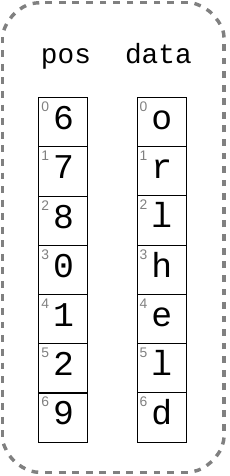}
    \subcaption{Standard \\ representation}
    \label{fig:scheme:subcolumn:example:segmented:decoded}
  \end{subfigure}
  \caption[A subcolumn encoded with the \capucscheme{SegmentedSubcolumn} scheme]{The subcolumn from \autoref{fig:scheme:column:examples:segmented} (``helloworld'' without the ``low'') encoded with the \capucscheme{SegmentedSubcolumn} scheme.}
  \label{fig:scheme:subcolumn}
\end{figure}

\begin{figure}[H]
  \centering
  \begin{subfigure}[b]{0.25\linewidth}
    \centering
    \includegraphics[scale=0.8]{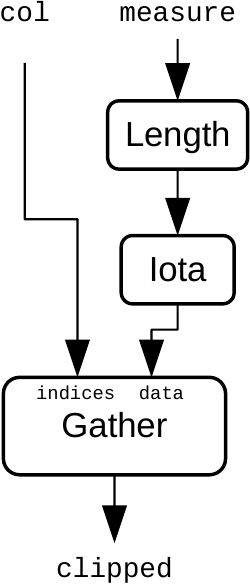}
    \subcaption{Implementation of a \uplanop{ClipBy} operator}
    \label{fig:scheme:subcolumn:decoder:segmented:clip-by}
  \end{subfigure}
  \hspace{0.1cm}
  \begin{subfigure}[b]{0.25\linewidth}
    \centering
    \includegraphics[scale=0.8]{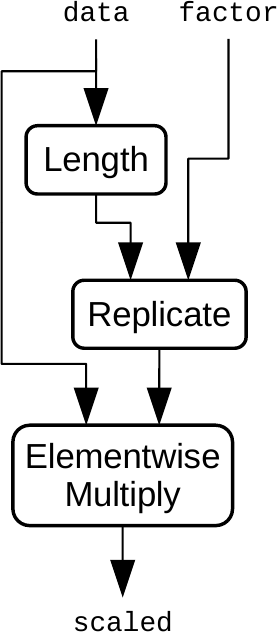}
    \subcaption{Implementation of a \uplanop{Scale} operator}
    \label{fig:scheme:subcolumn:decoder:segmented:scale}
  \end{subfigure}
  \hspace{0.1cm}
  \begin{subfigure}[b]{0.4\linewidth}
    \centering
    \includegraphics[scale=0.8]{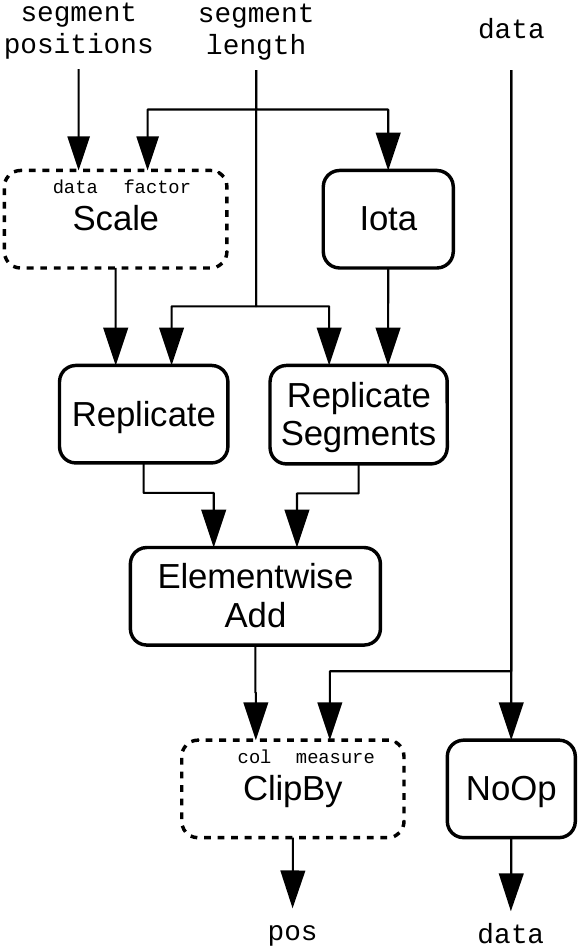}
    \subcaption{The actual decoder, using the two helper operators}
    \label{fig:scheme:subcolumn:decoder:segmented:actual-decoder}
  \end{subfigure}
  \caption{A decoder for the \capucscheme{SegmentedSubcolumn} representation scheme}
  \label{fig:scheme:subcolumn:decoder:segmented}
\end{figure}

\subsection{Adapting operators to column segmentation}
\label{subsec:segmentizing-operators}

To actually utilize segmented columns --- to apply columnar circuits to them --- we require operators which respect such segmentation (either fixed- or variable-length). Specifically, we would want to adapt operators we have already defined for non-segmented columns to take segmented column inputs, and produce outputs for each of the input column segments.

\begin{definition}[Simple, fixed-length operator segmentization]
\label{def:segmentization-of-a-simple-op}
Let \uplanop{Op} be a columnar operator taking an input column labeled $\colname{c}$ of type $\tau$ and producing a single output labeled $\colname{c}'$, whose length is a function of $\collength{c}$. The \emph{segmentization of \uplanop{Op}} is an operator $\uplanop{Segmentize(Op)}$, taking a \cscheme{UniformlySegmented} input and producing a \cscheme{UniformlySegmented} output. $\uplanop{Segmentize(Op)}$ applies \uplanop{Op} to every one of the $m = \ceil{\hfrac{\colsize{c}}{\colname{segment_length}}}$ segments of the segmented view of $\colname{c}$. The resulting  would-be output columns (denote them  $\colname{c}'_1$ through $\colname{c}'_m$) are concatenated together into the \colname{data} component of $\colname{c}'$. By our constraint on $\uplanop{Op}$, all segment outputs are of the same length, which is used as the $\colname{segment_length}$ component of $\colname{c}'$.
\end{definition}

This definition can be generalized to the case of multiple input and output columns:

\begin{definition}[General operator segmentization]
\label{def:segmentization-of-an-op}
Consider a signature $\opsig = \tuple{\opsigin, \opsigout}$, let $L_\text{shared} \included \labelsofsig{\sigma_\text{in}}$ be a subset of inputs to be shared among all segments, let $L_\text{segmented} = \inlabelsofsig{\opsig} \setminus L_\text{shared}$, and let \uplanop{Op} be a columnar operator with signature $\opsig$.
% Do we need the following constraint? I hope not
%monotone-nondecreasing \wrt $L_\text{segmented}$ in a weak sense: If all lengths of its inputs in $L_\text{segmented}$ increase, $\uplanop{Op}$'s output length does not decrease.
The \emph{segmentization of \uplanop{Op}} is an operator $\uplanop{Segmentize(Op)}$ with the following characterization.
$\uplanop{Segmentize(Op)}$ has the input labels in $L_\text{shared}$, and additionally, for every input label in $L_\text{segmented}$, $\uplanop{Segmentize(Op)}$ takes a segmented column (in the \cscheme{Segmented}  or \cscheme{UniformlySegmented} encoding scheme) of the same data element type. $\uplanop{Segmentize(Op)}$ produces output only if the its inputs from $L_\text{segmented}$ all have the same number of segments. When that is the case, for each label $\ell_\text{out} \in  \opsigout$, $\uplanop{Segmentized(Op)}$ produces a \cscheme{Segmented} column, with the data being concatenatation of $\uplanop{Op}$'s $\ell_\text{out}$ outputs for the input families for the each of the segments, and a segmentation corresponding to these outputs' lengths.

\end{definition}

In formulating this definitions, one notices how fixed-length segmentation in columns lacks robustness: If an operator is applied whose output length depends on input \emph{data} rather than just input \emph{lengths} (say, a range filter) --- different output column segments will have different lengths; thus the outputs must be defined as variable-length segmented. For this reason the definition above does not use \cscheme{UniformlySegmented} outputs; for those to be possible, we would need a constraint on the operator involved to conserve length uniformity.

\begin{figure}[H]
   % --------------------------------------------
   % Example of (non-simply) segmentation
   % --------------------------------------------
  \indexPlanOperator{Permute}
  \centering
  \begin{subfigure}[b]{0.375\linewidth}
    \centering
    \includegraphics[scale=0.7]{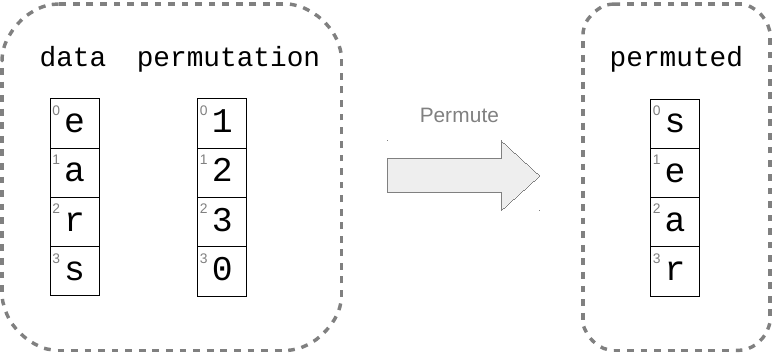}
    \subcaption{An application of the unsegmentized operator}
    \label{fig:operator-segmentization:example:before}
  \end{subfigure}
  \hspace{0.5cm}
  \begin{subfigure}[b]{0.575\linewidth}
    \centering
    \includegraphics[scale=0.7]{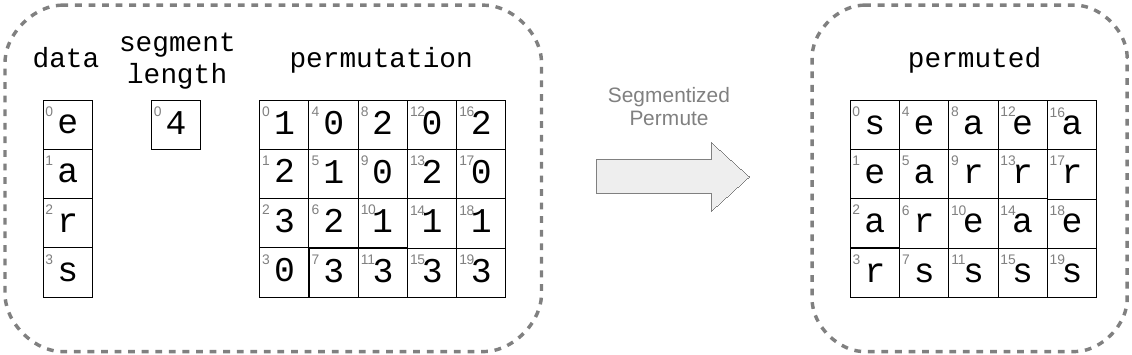}
    \subcaption{An application of the segmentized operator --- multiple permutations of the same data; the input is presented in 4-segmented view.}
    \label{fig:operator-segmentization:example:after}
  \end{subfigure}
  \caption[Operator segmentization example]{A segmentization of the {\rm \uplanop{Permute}} operator with {\rm $L_\text{shared} = \setof{\colname{data}}$} and {\rm $L_\text{segmented} = \setof{\colname{permutation}}$}.}
  \label{fig:operator-segmentization:example}
\end{figure}

\section{Representing additional constructs}
\label{sec:representing-additional-constructs}

\subsection{Index subsets}
\label{subsec:index-subsets}

%While the column formalism presented above is strictly ordered, and the computational circuit model inherits this feature regarding input data, sets of indices do tend to pop up when perfoming columnar computations. First, in the context of relational databases, relations \emph{are} sets, and must be represented somhow; but additionally, sets of indices tend to pop-up even in computations on originally fully-ordered data.

Recall that our standard representation of a subcolumn contains two columns: \colname{positions} and \colname{values}. Now consider the result of dropping the \colname{values} column: Doing so will leave us with the information regarding which indices are the subcolumn's domain; and since the order of indices in this columns is incosequential
%(or if you will, we may consider the \colname{indices} columns of an entire equivalence class of standard subcolumn representations)
--- what we are left with is the specification of an \emph{index subset}. As in the definition of subcolumns, we must choose whether to be ignorant of the overall index set size (resp. subcolumn domain); and for index sets we make the opposite choice, marking the domain size by default:

\begin{cschemes}
 \csitem*[\cscheme{SparseIndexSet}\text{ (Standard representation of an index subset)}]
  \begin{cslisting}
    $\colname{full_length}$  & $\inttype$  & 1                             & size of the full set of indices \\
    $\colname{elements}$     & $\inttype$  & up to $\colname{full_length}$ & values of elements in the index subset, in any order
  \end{cslisting}
\end{cschemes}
If the $\colname{elements}$ column is also sorted, the subset representation is said to be \emph{canonical}.

\medskip

\begin{figure}[H]
   % --------------------------------------------
   % Encoder and Decoder for SprarseIndexSet
   % --------------------------------------------
  \centering
  \begin{subfigure}[b]{0.4\linewidth}
    \centering
    \includegraphics[scale=0.7]{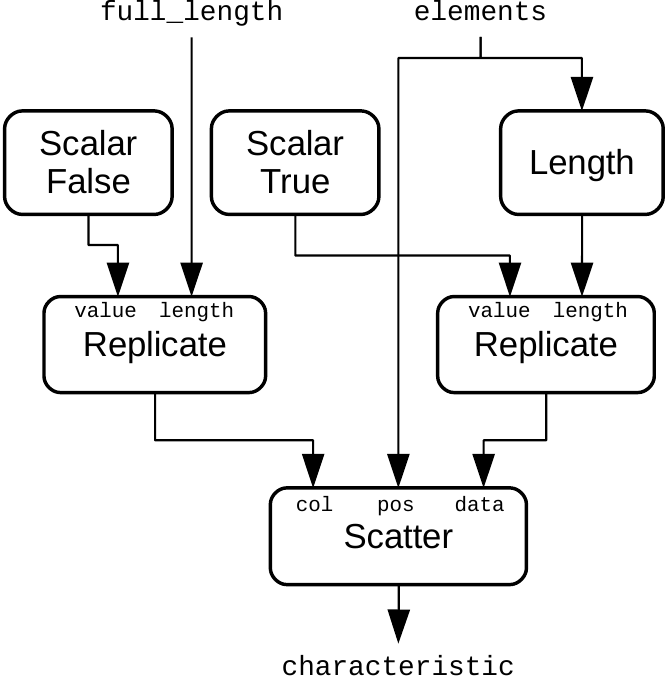}
    \indexPlanOperator{Scalar}
    \indexPlanOperator{Replicate}
    \indexPlanOperator{Scatter}
    \subcaption{Encoder (from \ucscheme{SparseIndexSet})}
    \label{fig:scheme:index-set:dense:example:encoder}
  \end{subfigure}
  \hspace{0.5cm}
  \begin{subfigure}[b]{0.3\linewidth}
    \centering
    \includegraphics[scale=0.7]{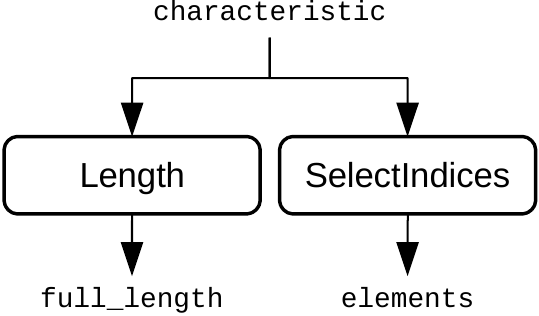}
    \indexPlanOperator{Length}
    \indexPlanOperator{SelectIndices}
    \subcaption{Decoder (to \ucscheme{SparseIndexSet})}
    \label{fig:scheme:index-set:dense:example:decoder}
  \end{subfigure}
  \caption{The \capucscheme{DenseIndexSet} representation scheme}
  \label{fig:scheme:index-set:dense:codec}
\end{figure}

The standard representation is sparse in the sense of the underlying assumption that elements not specified in it are missing from the set. Alternatively, we can choose to have an explicit indication of presence for every potential element --- a \emph{dense} representation --- using the following:

\begin{definition}[Set characteristic column]
Let $S \inc \zeroupto{x}$. The \emph{characteristic column} of $S$ (a.k.a. the \emph{characteristic function} of $S$) is the function $\fndef{\chi_S}{\zeroupto{n}}{\setof{\true,\false}}$ with
\[
\chi_S(x) =
\begin{cases}
 \true  & x \in S \\
 \false & \text{otherwise}
\end{cases}
\]
\end{definition}
The set of characteristic columns for all $S \inc \zeroupto{n}$ is in bijection with the power set $\powerset{\zeroupto{n}}$ itself; and thus we choose the characteristic columns --- the evaluated characteristic functions --- as the \cscheme{DenseIndexSet} reprersentation scheme for subsets of $\zeroupto{n}$. A concrete encoder and decoder for this scheme appear in \autoref{fig:scheme:index-set:dense:codec}.

The decoding and encoding circuits we've presented for \cscheme{DenseIndexSet} use two yet-undefined operators. These are:

\begin{operators}
  \opitem[SelectIndices]
  \begin{oplisting}
    Input  & $\colname{characteristic}$  & $\bittype$ or $\setof{\true, \false}$ & $n$       & \\
    Output & $\colname{indices}$         & $\inttype$                            & up to $n$ & \\
  \end{oplisting}
  Produces a column containing the indices of all elements of $\colname{dense}$ with value \true{} (or $1$ if we're using the numeric type). The output is not necessarily sorted in increasing order.

  \opitem[Length]Given a column of type $\tau$, produces its length as a scalar (i.e. a column of length 1).
\end{operators}

Columns of bits corresponding to a subset a column's (or rather, DBMS table's) indices have long been in use in analytic query processing, well before the advent of column stores (an early example is the Model 204 DBMS by Compute Corporation of America \cite{O1989}, which predates even the era of relational DBMSes and SQL).

\medskip

% Perhaps bring this commented-out scheme as a ``lifting'' of an operator into a compression scheme.
%
% The characteristic function of a subset is often computed in practice using some criterion or predicate on a column's values. This motivates the following scheme, in which the application of the criterion is ``lazy'' (i.e. only applied when decoding):
% 
% \begin{cschemes}
%  \csitem*[\cscheme{ImplicitDenseIndexSet}_p]
%   \begin{cslisting}
%     $\colname{predication_object}$ & $\tau$      & $n$ &  data to which the predicate $p$ is applied \\
%   \end{cslisting}
%   $p$ must be a function which is not uniform on the domain of $\tau$. Specifically, let $a,b$ be values of type $\tau$ with $p(a) = \false$ and $p(b) = \true)$.
%   This scheme is decoded by applying $\planop{Elementwise}_p$, then applying the decoder for \ucscheme{DenseIndexSet}. Encoding begins with a \ucscheme{DenseIndexSet} encoder, followed by the elementwise mapping of $\false \mapsto a$ and $\true \mapsto b$.
% \end{cschemes}

\medskip

Finally, we define highly concise representations for subsets of $\zeroupto{n}$ which are either contiguous, or are a union of contiguous ranges.

\begin{cschemes}
 \csitem[ContiguousIndexSet]
  \begin{cslisting}
    $\colname{start}$  & $\inttype$  & 1 &  Index of the first element in the set \\
    $\colname{length}$ & $\inttype$  & 1 &  Number of elements in the set
  \end{cslisting}
  The represented set is $\rngset{\colval{start}{0}}{\colval{start}{0} + \colval{length}{0}}$.
%  \csitem[ContiguousRangesIndexSet]
%   \begin{cslisting}
%     $\colname{start}$  & $\inttype$  & m &  of the first element of each contiguous range\\
%     $\colname{length}$ & $\inttype$  & m &  Number of elements in each contiguous range
%   \end{cslisting}
%   The represented set is $\Union_{i=1}^m{\rngset{\colval{start}{i}}{\colval{start}{i} + \colval{length}{i} - 1}}$. For each $i$, it must hold that $\colval{start}{i} + \colval{length}{i} \leq n$.
\end{cschemes}

% As with some other representations, we have opted for a definition preferring explicitness over terseness. Also, we have not constrained the ranges to be sorted, nor disjoint, nor even distinct. If all these conditions hold for an encoded form of \cscheme{ContiguousRangesIndexSet}, we call that form \emph{canonical}.

\subsection{Partitions of columns and index sets}
\label{subsec:partitions}

In \autoref{sec:segmented-columns} we presented representation schemes for the break-up of colums into contiguous segments; let us now generalize this to arbitrary partitions --- in which the parts are not necessarily contiguous.

If the number of parts is fixed (say, $k$), we could simply adjoin the (standard) representations of $k$ subcolumns:

\begin{cschemes}
\csitem*[\cscheme{Partitioned}_k]
  \begin{cslisting}
    $\colname{pos}_1$  & $\inttype$    & $n_1$  & Element positions of the first subcolumn   \\
    $\colname{data}_1$ & $\tau$        & $n_1$  & Element data of the first subcolumn        \\
    \vdots             & $\vdots$      & \vdots & \vdots                                     \\
    $\colname{pos}_k$  & $\inttype$    & $n_k$  & Element positions of the $k$\xth subcolumn \\
    $\colname{data}_k$ & $\tau$        & $n_k$  & Second subcolumn, element data             \\
  \end{cslisting}
  The position columns are distinct, and the union of their support sets is the entire index set, i.e. $\Union_{i} \fsupport{\colname{pos}_i} = \zeroupto{n}$.
\end{cschemes}

And if we only consider the index sets in this scheme, we have an index set partition representation scheme: $\cscheme{PartitionedIndexSet}_k$.

\begin{note}Given our constraints, only $k-1$ of the position columns are actually necessary. For the case of $k=2$, dropping the redundant complementing position column results in the \cscheme{ComplementingSubcolumns} scheme.
\end{note}

\medskip

Having separate columns for each part in the partition becomes rather unwieldy with many parts, and essentially irrelevant when wishing to represent a number of parts proportional to the column length. We can do with something simpler than $k$ subcolumns, when we consider set-theoretic definition of a partition: A function from the set of indices to the set of parts. And a column, after all, is merely an evaluated function from indices to the domain of the element type; hence:

\begin{cschemes}
  \csitem*[\cscheme{Partition}\text{ (Standard representation of a column index set partition)}]
  An alias for the standard representation of a column, with the domain being a set of part identifiers.
\end{cschemes}

A representation of a partition in the \cscheme{Partition} scheme is said to be \emph{canonical} if its image (as a function) is $\zeroupto{\setsize{\frange{k}}}$ for some $k \in \naturals$. With a canonical representation, there are no implicitly-empty parts.

\begin{figure}[H]
  % --------------------------------------------------------
  % Materializing a partition
  % --------------------------------------------------------
  \centering
  \begin{subfigure}[b]{0.4\linewidth}
    \centering
    \includegraphics[scale=0.7]{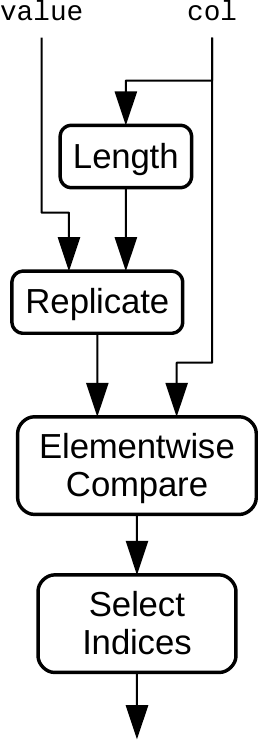}
    \indexPlanOperator{Length}
    \indexPlanOperator{Replicate}
    \indexPlanOperator{Elementwise}
    \indexPlanOperator{SelectIndices}
    \subcaption{An implementation of a \uplanop{MatchScalar} operation, producing the set of indices on which a column matches a scalar value.}
    \label{subfig:scheme:partition:materialize:match-scalar}
  \end{subfigure}
  \hspace{0.05\linewidth}
  \begin{subfigure}[b]{0.4\linewidth}
    \centering
    \includegraphics[scale=0.7]{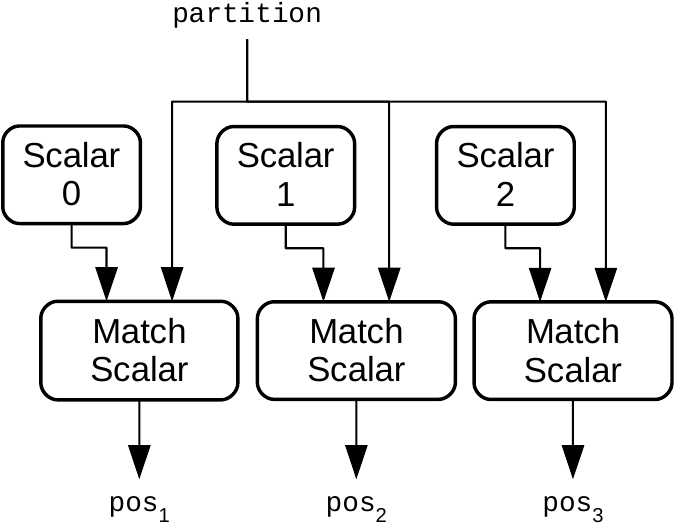}
    \indexPlanOperator{Scalar}
    \subcaption{The full materialization circuit for $k = 3$, resulting in a $\capucscheme{Partitioned}_3$ instance. \\ \mbox{}}
    \label{subfig:scheme:partition:materialize:full-circuit}
  \end{subfigure}
  \caption[Materialization of an index set partition into the $\capucscheme{Partitioned}_k$ scheme]{Materialization of an index set partition, from the standard scheme (\capucscheme{Partition}) into the $\capucscheme{Partitioned}_k$ scheme.}
  \label{fig:scheme:partition:materialization}
\end{figure}

\begin{note} The circuit in \autoref{fig:scheme:partition:materialization}, when adapted for $k=2$, transforms a \cscheme{DenseIndexSet} instance (after a $\bittype$ to $\inttype$ upcast) into an index subset in the default representation, along with its complement.
\end{note}

\subsection{Element de/composition, interleaving and type-punning}
\label{subsec:interleave-decompose-pun}

Let $\vec{\tau} = \idxrngtuple{\tau}{1}{k}$, and consider the following operator:

\begin{operators}
  \opitem*[\planop{Zip}_k]
  \begin{oplisting}
     Input   & $\colname{component}_1$              & $\tau_1$                           & $n$ & \\
     Input   & $\colname{component}_2$              & $\tau_2$                           & $n$ & \\
     \vdots  & $\vdots$                             & \vdots                             & \vdots \\
     Input   & $\colname{component}_k$              & $\tau_k$                           & $n$ &  \\
     Output  & $\colname{zipped}$                   & $\prod_{1 \leq i \leq k}{\tau_i} $ & $n$ &  \\
  \end{oplisting}
  The $\uplanop{Elementwise}_f$ operator, with $f$ being the $k$-tuple constructing function $\idxrng{x}{1}{k} \mapsto \idxrngtuple{x}{1}{k}$. (a.k.a. \IntroducePlanOperatorAlias{Compose}{Zip}; see \cite[\S 3.1]{GBDPS2018}).
\end{operators}

this operator induces the following representation scheme for $\vec{\tau}$ elements:

\begin{cschemes}
\csitem[Components] Each multi-type element $\vec{x} = \idxrngtuple{x}{1}{k}$ is represented by $k$ elements in disjoint columns, each with a single, simpler, uniform type. It is decoded by applying $\planop{Zip}_k$, and as that operation is invertible, it is encoded by applying $\finverse{\planop{Zip}_k}$.
\end{cschemes}

The \cscheme{Components} scheme 
%for a product type $\tau_c$ 
can also be thought of as a Structure-of-Arrays representation of a product-type column, as opposed to the standard representation, which is an Array-of-Structs. Also, if a relational DBMS table only has types of fixed width, and $\vec{\tau}$ is the tuple of the table column element types --- \cscheme{Components} constitutes a representation of the entire table as a single column.

% Moving away from the SoA-vs-AoS metaphor, let us generalize \cscheme{Components} from use of the $f_\text{cons}$ function to any injective partial function $\fndef{f}{\vec{\tau}}{\tau_c}$: Such a function can be perceived as an alternative, non-trivial method of losslessly \emph{composing} elements of types $\idxitemrange{\tau}{1}{k}$ into an element of type $\tau_c$, yielding:
% 
% 
% \begin{cschemes}
% \csitem*[\cscheme{Components}_f] A generalization of $\cscheme{Components}$, replacing $f_{\text{cons}}$ with an arbitrary function.
% The encoded form of a column \colname{col} is its image under $f^{-1}$, i.e. $k$ columns of types $\idxrngtuple{\tau}{1}{k}$, each of length $\colsize{col}$. Decoding is performed by applying $\planop{Elementwise}_f$.
% \end{cschemes}

Now suppose that all types in $\vec{\tau}$ are identical, i.e. the case of $\tau_i = \tau_d$ for all $i \in \rngset{1}{k}$. For the distinct-type case we needed a separate column for each component; but for the uniform-type case, we could just have all of them within a single column. Below are two alternative formalizations of such a scheme, differing by the placement of component data within the single column:

\begin{operators}
\opitem*[\planop{ComposeSegments}]
  \begin{oplisting}
     Input  & $\colname{segment_length}$  & $\inttype$        & $1$ & \\
     Input  & $\colname{components}$      & $\tau_d$          & $n$ & $n$ is divisible by \colname{segment_length} \\
     Output & $\colname{composed}$        & $\tau_c$          & $\colname{segment_length}$  \\
  \end{oplisting}
%  For an input $\colname{segment_length}$ of $\ell$, an $\ell$-segmented view of \colname{components} would yield there are $\colname{segment_length}$ columns; and 
  The output of this operation is the equivalent of applying $\planop{Zip}_{n'}$ to the different segments of a segmented view of your input, each considered as one of the component columns.
columns, so that the output satisfies $\colval{composed}{i} = \breakingtuple{\itemrng{\colval{components}{i}, \colval{components}{n' + i}}{\colval{components}{(k-1)n' + i}}}$. Note that a different $\planop{Zip}$ operator will be applied for different lengths of the \colname{components} column - so that this operator and $\planop{Zip}$ are quite distinct, despite the similarity in practice.
\opitem*[\planop{Assemble}_k]
  \begin{oplisting}
     Input  & $\colname{segment_length}$  & $\inttype$        & $1$ & Has value $k$\\
     Input  & $\colname{components}$      & $\tau_d$          & $n' \* k$ & \\
     Output & $\colname{composed}$        & $\tau_c$          & $n'$     &  \\
  \end{oplisting}
  While the \colname{segment_length} scalar is redundant, its presence makes this a uniformly-segmented scheme. In a segmented view of \colname{components}, the output is obtained by a applying the elementwise function of $\planop{Zip}_\colname{segment_length}$ to each of the segments. Thus the output satisfies $\colval{composed}{i} = \breakingtuple{\itemrng{\colval{components}{k\*i}}{\colval{components}{k\* i + k - 1}}}$.
\end{operators}
%These operators both have an effectively two-dimensional input; but only one of them constitutes a segmentization
%
%but only one of them applies constitutes a segmentization$\planop{ComposeSegments}_f$ does not apply a segmentized operator segments are of varying l does not observe a fixed-length segmentation: It applies the composing function $f$ to elements at offsets of $n'$ apart (i.e. it expects a concatenation of the columns); $\planop{Assemble}$ applies $f$ to consecutive elements of the \colname{components} column. 
Each of these operators can be obtained from the other by applying it after a \planop{Transpose}.

\begin{figure}[H]
  % --------------------------------------------------------
  % Examples of 'Assemble' and 'ComposeSegments'
  % --------------------------------------------------------
  \centering
  \includegraphics[height=3.5cm]{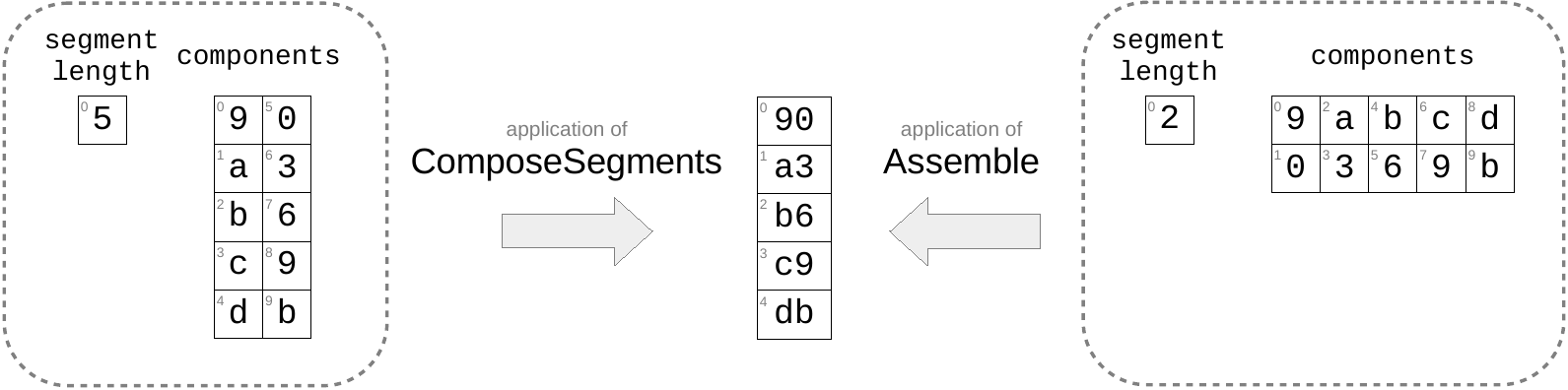}
  \indexPlanOperator{Assemble}
  \indexPlanOperator{ComposeSegments}
  \caption[The {\normalfont\uplanop{Assemble}} and {\normalfont\uplanop{ComposeSegments}} operators]{The {\normalfont\uplanop{Assemble}} the {\normalfont\uplanop{ComposeSegments}} operators.}
  \label{fig:operator:assemble-and-compose}
\end{figure}

The two operators translate into two compression schemes --- two specializations of \cscheme{Components} for the case of a uniform elementwise function argument type:

\begin{cschemes}
\csitem[ConcatenatedComponents] A segmented column decompressible by \planop{ComposeSegments}; the input column length must be divisible by the segment length, i.e. the segmented view of the must be a full matrix.
\csitem[Shattered] A segmented column decompressible by \planop{Assemble}; the input column length must be divisible by the segment length.
\end{cschemes}

\subsubsection{Bitwise decomposition}
\label{subsubsec:bitwise-decomposition}

The deepest one can go in applying \cscheme{ConcatenatedComponents} is breaking elements up into individual bits: A $w$-bit-wide column becomes $w$ 1-bit-wide component columns, concatenated into a single, $w$-times-longer, $1$-bit wide column. Such decomposition has seen some research attention under the name of ``bitwise-decomposed storage''. Specifically, when I/O is very expensive (or slow) relative to computation (as in the case of GPU accelerators obtaining data from main system memory over a PCIe bus) --- it may be worthwhile to send over just the highest bit or several bits of a wider column, and begin some computation 
of values (e.g. for CPU-GPU co-execution of queries, as explored in \cite{PMK2014}).

%of columns been studied as an avenue for splitting column data between faster GPUs and slower CPUs with access to the much larger system memory, in  (with the idea of performing approximate initial computation, using some prefix bits, on the GPU, so that the CPU only had to do work for a much smaller set of column elements).
\medskip

Another kind of decomposition into bits --- less orthogonal and space-frugal --- is per-value indicator bit columns. For example, suppose $\tau_c$ is a type with domain $\setof{a,b,c,d,e}$\; instead of using 3 one-bit component columns (with which values will correspond to \texttt{000},\texttt{001},\texttt{010},\texttt{011},\texttt{100}), we would use 5 one-bit indicator columns, such that for every index $i$, exactly one of them is set (i.e. \texttt{00001},\texttt{00010},\texttt{00100},\texttt{01000},\texttt{10000}).

\begin{cschemes}
\csitem[ValueIndicators]
  \begin{cslisting}
     $\colname{domain_size}$     & $\inttype$     & $1$ & \\
     $\colname{bitmaps}$         & $\bittype$     & $n \* \colname{domain_size}$ & \\
  \end{cslisting}
  $\colname{num_bits}$ concatenated columns (via a segmented view); the $j$\xth $n$-bit column is a dense subset representation of those indices within $\zeroupto{n}$ having value $n-1$. (Inaccurate but common shorthand name: \IntroduceCompressionSchemeShorthand{Bitmap}{ValueIndicators}; a.k.a.``bit-vector encoding'' \cite[422]{ABHIM2013}.)
\end{cschemes}
While this representation scheme is exponentially inefficient, it has the advantage of offering what are essentially pre-computed results for constant-column equality operator applications, and requiring access to only a small part of the column when only a small subset of values are of interest.

\subsubsection{Type punning and member access}
\label{subsubsec:type-punning}

The bitwise decomposition described in \autoref{subsubsec:bitwise-decomposition} is a specific example of using the \cscheme{Shattered} for reinterpreting, rather than transforming, data. In our model of computation, there is no ``reinterpretation'', nor access to a part-of-an-element --- a column is never replaced, reconceived, or trimmed of uninteresting data; but in actual programming, A single value of an aggregate or product type may be perceived as the sequence of its constituent types, in order of their appearance in physical memory. For example: a complex number could be represented by its real component followd by its imaginary components, separately. Descending a level further, non-aggregate numeric types may be perceived as sequences of bytes, or bits, in accordance to the idiosyncracies of their representation in computer memory. This practice is known as \emph{type punning}; applying the \cscheme{Shattered} scheme with an appropriate function on an input column produces the equivalent of such punning (and will likely result in actual punning in an implementation of the computationa model).

\subsection{Variable-width columns and string representation}
\label{subsec:variable-width-columns}

Our basic definition of a column (\autoref{def:column}) was chosen for its simplicity, which in turn simplifies the formalism for our model of computation. An important aspect of this choice is the limitation of columns to have a fixed width, i.e. element types with a fixed-size representation. This ensures the memory representation of a column is perfectly straightforward: The location in memory of the representation of each element can be determined without having to pre-read any other information from the column (no ``read-after-read'' dependencies). A column store must, however, also support variable-width data --- so as to hold columns of strings, arbitrary-precision numbers or just opaque sequences of octets. Variable-width data also features extensively in many CPU-targeted lightweight compression schemes (see, e.g. \cite{DHHL2017}); most of which are usable or in-use in column stores, further motivating consideration. To do so, we will not amend the definition of a column, but rather make a separate one:

\begin{definition}
A \emph{variable-width column} is a function $\fndef{\colname{vwc}}{\rngset{0}{n-1}}{\fdom{\Dunion_{k \in \naturals} \prod_{j=0}^{k} {\tau}}}$ (where the empty product of $\tau$'s is $\tau_\text{unit}$)
for some fixed-representation-size element type $\tau$ and non-negative integer $n$. $n$ is the column's \emph{length}, and $\tau$ is its \emph{base element type}. The length (in units of $\tau$) of the $i$\xth element of $\colname{vwc}$, also referred to as the \emph{width of $\colname{vwc}$ at index $i$}, is the value $k$ such that $f(i) \in \fdom{\prod_{j=0}^{k} \colspace{\tau}}$.%\footnote{$\tau_\text{empty}$ is used in other contexts to express a null value, but here it signifies the single sequence of $\tau$'s of length $0$}.
\end{definition}
A variable-width column is \emph{not a column} by our definition (\autoref{def:column}); it may only accidentally constitute a column if it so happens that all of its elements have the same length, it will therefore not be decodable into the standard column representation, nor could we encode it in other column representation schemes. The concatenation of the $\tau$-units of all column elements, however, is a plain column; and the additional information it needs is the association of contiguous subsequences of $\tau$ elements with their corresponding variable-width element indices. That is exactly a partition into segments, discussed in \autoref{subsec:partitions}; and our representation scheme will combine a \cscheme{Segmentation} form with the raw data:

\begin{cschemes}
\item[(Standard representation of a variable-width column)]
  \begin{cslisting}
     $\colname{start_position}$ & $\inttype$  & $n$ & offsets into \colname{data} \\
     $\colname{length}$         & $\inttype$  & $n$ & \makecell{ ... of the sequences of $\tau$ elements for each                \\
                                                                  element of the variable-width column; of non-negative value. } \\
     $\colname{data}$           & $\tau$      & ?   & \makecell{ individual $\tau$ values whose concatenations make up           \\
                                                                 the different elements of the variable-width columns          } \\
  \end{cslisting}
  $\colname{start_position}$ and $\colname{length}$ correspond to $\colname{start}$ and $\colname{length}$, respectively, in scheme \cscheme{Segmentation}.
  For an index $i$, the $i$\xth element of the variable-width column occupies the range of $\tau$ elements $\itemrange{\colval{data}{i}}{\colval{data}{i + \colval{lengths}{i} - 1}}$ (and specifically, occupies nothing if $\colval{lengths}{i} = 0$).
\end{cschemes}

This ``verbose'' choice of standard representation has more to it than meets the eye, as we have not constrained \colname{start_positions} to even be monotone increasing. When it also holds that $\colval{start_positions}{i} = \sum_{j < i}\colval{lengths}{j}$, things are much simpler, and we refer to this case as a \emph{canonical representation of a	 variable-width column}.

\medskip

A special case of variable-width columns are those columns whose width is not \emph{fixed}, but is at least \emph{bounded}. In the SQL standard, \sqlcode{VARCHAR(k)} columns are of this kind: strings with a per-column maximum length. These columns have a more straightforward, though wasteful, representation if we expand all elements to actually take up that maximum width:
\begin{cschemes}
\csitem[CappedWidth]
  \begin{cslisting}
     $\colname{max_length}$      & $\inttype$        & $1$                         & Cap on single element length in units of $\tau$ \\
     $\colname{lengths}$         & $\inttype$        & $n$                         & \makecell{ ... of the sequence of $\tau$ elements for each element of the \\
                                                                                                variable-width column} \\
     $\colname{data}$            & $\tau$            & $n \* \colname{max_length}$ & \makecell{ Sequences of $\tau$ elements for each element of \\
                                                                                                the variable-width column, and arbitrary filler values. }
  \end{cslisting}
  Considering a \colname{max_length}-segmented view of \colname{data}, each column is a small buffer holding a single variable-size element, and \colname{lengths} indicates how much of the buffer is actually taken up by the element; the few remaining type-$\tau$ elements may have any value and are ignored.
\end{cschemes}

\begin{note}
The above is insufficient to fully model the representation of strings, as it does not go into the character set encoding --- which often intorudce another layer of variability-of-length: Each single character within a string can have variable length (e.g. when the popular UTF-8 encoding is used). While this work covers composite compression schemes in \autoref{chap:columnar-compression}, String representation is not explored further
\end{note}

%More compact and less wasteful representations of capped-length data are explored as compression schemes --- in sections \Autoref{sec:compression-scheme-composition-patterns,sec:composite-compression-schemes}.

\subsection{Nullable and union-type columns}
\label{subsec:nullable-and-union-type-columns}

In SQL tables, a column may be allowed to have \sqlnull values at some positions rather than a value of the actual type. This can be interpreted as expanding the column type $\tau$ by a single value, i.e. the replacement of $\tau$ by the union type $\tau \dunion \unittype$ (with $\unittype$ having a single possible value); and this interpretation motivates considering the case of \sqlnull{}'s as a special-case of disjoint-union types: $\tau_1 \dunion \tau_2$. How would we represent a column with such an element type?

Three approaches come to mind:

\begin{enumerate}
 \item Actually expand the representation of $\tau_1$ values to accomodate all $\tau_2$ values as well; this often means increasing column elements' width in bits.
 \item Hold two complementary subcolumns, one of $\tau_1$'s and the other of $\tau_2$'s.
 \item Hold a column of just one of the types (\wloss, $\tau_1$), and an overlay subcolumn of elements of the other type, which are to be \emph{patched} into the base column.
\end{enumerate}

The first approach is typically the most burdensome: In addition to the width increase, one must also pay up-front the price of extra condition checks and reinterpretation work to obtain a usable $\tau_1$ or $\tau_2$ from the combined domain. However, by definition, this is our standard approach: Ignoring how the type was formed and yielding a standard representation (in the sense of \autoref{subsec:columns}). The second and third approaches correspond, repsectively, to the \cscheme{ComplementingSubcolumns} and to \cscheme{Patched}.

\smallskip

Now let us consider these approaches for the specific case of \sqlnull{}'s. The \emph{first} approach isn't often used in DBMSes in practice (in part due to the shortcomings listed above). Some DBMSes do, however, artificially restrict many of their data types to ensure at least one ``invalid'' value, which is then used to represent \sqlnull{}. Two concrete examples from MonetDB, which takes this approach \cite{MonetDBTypeSystem}: In the case of fixed-width integer types, the lowest or highest representable value is sacrificed to serve as the encoding of \sqlnull; and in a UTF-8 encoded string, one of the invalid sequence of octets is used (as not any sequence of octets is a valid encoded UTF-8 string).

Using the second approach for \sqlnull{}s, we would compose \cscheme{ComplementingSubcolumns} with the \cscheme{Constant} scheme used for the $\tau_2 = \unittype$ subcolumn, i.e. out of \colname{pos}, $\colname{data}_1$ and $\colname{data}_2$, we would only need the first two constituent columns --- the non-null positions and the non-null values --- alongside the overall column length and the \sqlnull scalar. 

Finally, the third option would also require just two (non-scalar) columns due to the use of \cscheme{Constant}: A full-length column with junk values at the overlaid positions, and the column of overlay positions. In this case, and for a column of basic type $\tau$, of length $n$ and with $m$ \sqlnull values, the overhead would be $m \* \setsize{\inttype}$ bits, or $m / n$ relatively to a non-nullable column of type $\tau$ and length $n$.

% ---------------------------------------
% ---------------------------------------
% ---------------------------------------
% ---------------------------------------
% ---------------------------------------

\chapter{Column compression}
\label{chap:columnar-compression}

We have so far modeled columnar computations; and have used them to represent several fundamental combinatorial structures in columnar schemes. This is sufficient machinery for us to turn our attention to encoding schemes effecting data \emph{compression}. The following sections are not intended to be an exhaustive re-formulation of existing academic work on compression schemes for DB columns; they do, however, cover the lightweight compression schemes most commonly studied academically and used in practice (see \cite[\S 4.2]{ABHIM2013} for a brief survey), formulating them in columnar terms. Other schemes developed below result from the application of the same principles and approach as in previous chapters, as well as the composition of existing schemes --- and have mostly not been formulated in previously published research.

\begin{definition}[Compression scheme]
A \emph{column compression scheme} is a codec $\tuple{f_e,f_d}$ for single columns, as per \autoref{def:codec}. Respectively, a \emph{concrete column compression scheme} is a concrete codec $\tuple{C_e,C_d,C_v}$ for single columns.
\end{definition}

Column compression schemes are often not representation schemes for columns, in that only a subset of all possible columns is representable by a compressed form in a given scheme (other columns are ``not compressible'' by the scheme).

\begin{definition}[Representation size]
The \emph{size (in bytes)} of a columnar representation of a column or other structure is the sum of the constituent columns' sizes in bytes, which in turn is their element type's fixed width in bytes, times the column length.
\end{definition}

\begin{definition}[Compression ratio]
The \emph{compression ratio} of a compressed form $\colname{col}'$ under a concrete compression scheme $\tuple{C_c,C_d}$ is the ratio of sizes of $\circfunc{d}(\colname{col}')$ and ${\colname{col}'}$ in bytes.
The compression ratio of an uncompressed column \colname{col} under scheme $\tuple{C_c,C_d}$ is the compression ratio of its compressed form  $\circfunc{c}(\colname{col})$ (which is the ratio of sizes of ${\colname{col}}$ and $\circfunc{c}(\colname{col})$ in bytes).
\end{definition}

\section{Basic schemes}
\label{sec:basic-schemes}

We begin with a formulation of relatively simple compression schemes, each exploiting a single aspect of redundancy or excess in the representation of the input: Predictability of values using the element index; or a support set much smaller than $\fdomain{\tau}$. These schemes are also characterized by decompressed elements directly depending on a single element of the compressed column, at the most --- so that the basic schemes are also the ``most parallelizable''.

\subsection{Output-index-based compression}
\label{subsec:output-index-based-schemes}

Our first compression scheme applies in cases where the entire input column is simply an evaluation of a known function:

\begin{cschemes}
\csitem*[\cscheme{Generated}_f]
  \begin{cslisting}
    $\colname{length}$ & $\inttype$ & $1$ & ... of the represented column \\
  \end{cslisting}
  The decompressed column is the evaluation of $f$ over domain $\zeroupto{\colval{length}{0}}$ (i.e the result of applying the composition of $\planop{Elementwise}_f$ after \planop{Iota} to the compressed input).
\end{cschemes}

Of course, there exists only one such column for each choise of $f$ at each length, which means such a scheme will not often be directly used in practice; still, auto-generated columns do occur in real-life databases (e.g. matrix row and column indices based on the overall record index). To slightly expand the expressivity beyond this scheme, we introduce some parameters to choose $f$ within a larger (though fixed-dimensional) space. For example, let it be just a single parameter for selecting among the space of uniform-valued functions from $\naturals$ to $\fdomain{\tau}$:

\begin{cschemes}
\csitem[Constant]
  \begin{cslisting}
    $\colname{value}$  & $\tau$      & $1$ & the single uniform value of all column elements\\
    $\colname{length}$ & $\inttype$  & $1$ & ... of the represented column \\
  \end{cslisting}
  The decompressed column has \colname{length} elements, all equal to \colname{value}.
\end{cschemes}

This is, in fact, more than a toy example; it is quite the useful plug-in for more complex compression schemes --- where it might stand in for a full-blown component column. Examples of such schemes can be found in \autoref{subsec:nullable-and-union-type-columns} and \autoref{subsec:run-encoding} below. It may also be used as the default representation scheme for newly-created non-\sqlnull fixed-width columns, so as to avoid the work of actually initialize them.

For columns with a ring-like element type, admitting addition and multiplication (e.g. the integral types), the \cscheme{Constant} can be seen as a semi-degenerate of larger spaces of functions: The single value $\colname{value}$ is a scalar coefficient, for a single basis function $x \mapsto 1$. If we allow $k$ functions, rather than just one, as part of the definition of the scheme, we have:

\begin{cschemes}
\csitem*[\cscheme{Generated}_{\vec{f}}]
  \begin{cslisting}
    $\colname{length}$       & $\inttype$    & $1$  & ... of the intended column \\
    $\colname{coefficients}$ & $\tau$        & $k$  & \makecell{ coefficients of the linear combination of $\vec{f}$ \\
                                                                 forming the generating function }
  \end{cslisting}
  The scheme is parameterized on a sequence of $k$ univariate basis functions, $\vec{f} = \idxrngtuple{f}{1}{k}$, so that the decompressed column is an evaluation of $\tilde{f} = \sum_{i=1}^{k} \colval{coefficients}{i-1} \* f_i$  over the domain $\zeroupto{n}$.
\csitem[GeneratedPolynomial] The $\cscheme{Generated}_{\vec{f}}$ scheme, with the choice of increasing-power monomials for $\vec{f}$, i.e. $f_i(x) = x^{i-1}$.
\end{cschemes}

Of course, real-life data does not match a generated function perfectly --- as there are almost always some discrepancies (see \autoref{subsec:patching}) or what could be described as noise (see \autoref{subsec:elementwise-composition}).

\subsection{Column support compaction}
\label{subsec:column-support-set-compaction-1}

The support of a column may only constitute a small subset of the domain of its element type, motivating compression schemes which use less space to indicate which element of that support set is used. We consider an initial, simplistic taxonomy of small support sets, by order of increasing complexity.

%Unfortunately, we cannot yet present all schemes relevant to small-support-set cases, so we begin with a limited taxonomy of just three cases:

\begin{enumerate}
 \item Singleton set
  \tabto{4.0cm} $\rightarrow$ \cscheme{Constant}
 \item Set having a low maximum within $\tau$ with respect to its physical (binary) representation, containing most values below that maximum
  \tabto{4.0cm} $\rightarrow$ \cscheme{DiscardHighBits}
 \item Set with none of the above features (scattered in a non-trivial fashion within $\tau$)
  \tabto{4.0cm} $\rightarrow$ \cscheme{Dictionary}
\end{enumerate}
%the taxonomy will be expanded in \autoref{subsubsec:support-set-compaction-2} further below with small support sets which require more complex schemes to exploit.

\paragraph{Singleton support sets} Single-value columns admit the \cscheme{Constant} compression scheme from \autoref{subsec:output-index-based-schemes} above. \cscheme{Constant} is a degenerate case of a small support set, requiring no per-element compressed data. %; see also \autoref{subsec:differentiation}.

\paragraph{Dense low-bounded support sets} A database column will often be declared to hold integers of some default (larger) width, while in fact its data could be represented with a lower width. This makes the top bits in the representation always-zero (or in the case of signed integers, a sign-extension). If it also happens that within the non-redundant bits, the column's support set is `dense` --- containing more than half of the possible values --- it does not make sense to try to represent an element of the support set other than by its lower bits. Generalizing from the case of unsigned integers, we define:

\begin{cschemes}
\csitem[DiscardHighBits] Let $\tau, \tau'$ be types, $\preceq$ be the natural well-ordering of $\fdomain{\tau}$ and $b$ a positive integer, such that $\fdomain{\tau'} \pincluded \fdomain{\tau}$ and that $\setmax{\fdomain{\tau'}} $ is lower by $\preceq$ than the $2^b$-th lowest element in $\fdomain{\tau}$. The compression by \cscheme{DiscardHighBits} is the mapping of each element of type $\tau$ to itself, but as an element of type $\tau'$; and the decompression is the opposite. In ``physical'' terms, this is typically the dropping the high bits in the representation of each element%, or extending the element with 0's or with 1's.
Abusing the above definition somewhat, we also refer to the discarding of high bits of signed-integer types as encoding with this scheme.

\item[\IntroduceCompressionSchemeShorthand{NullSuppression}{DiscardHighBits}] A commonly used alternative name for \cscheme{DiscardHighBits}; the scheme does \emph{not} relate to the SQL \sqlnull value. (Shorthand name: \IntroduceCompressionSchemeShorthand{NS}{DiscardHighBits})
\end{cschemes}

\paragraph{Explicitly-enumerated support sets}
The support set may sometimes be small, but not limited mostly to the lower elements of an ordered type. While it may still fit some pattern or rule, it is often simple and effective to specify all elements of the set, using an evaluated function whose image covers the entire set. This is (inaccurately) referred to as a \emph{dictionary} for the column, used in an eponymous compression scheme:

\begin{cschemes}
\csitem[Dictionary]
  \begin{cslisting}
    $\colname{dictionary}$ & $\tau$      & $d$ & \\
    $\colname{indices}$    & $\inttype$  & $n$ & ... into \colname{dictionary} \\
  \end{cslisting}
  Letting $\colname{col}$ denote the uncompressed column, $\colimage{dictionary}$ is a superset of $\colimage{col}$. The compressed form is the index in the evaluated $\colname{dictionary}$ instead of the original value, so that $\colname{dictionary} \circ \colname{indices} = \colname{col}$ (with $\circ$ denoting function composition), and elementwise we have $\colname{col}\sparen{i} = \colname{dictionary}\sparen{\colname{indices}\sparen{i}}$.
\end{cschemes}

Decompressing the \cscheme{Dictionary} scheme involves performing essentially the opposite of the \planop{Scatter} operator (see \autoref{subsec:combining-subcolumns}):

\begin{operators}
  \opitem[Gather]
  \begin{oplisting}
    Input  & $\colname{pos}$    & $\inttype$  & $n_2$ & positions in \colname{data}\\
    Input  & $\colname{data}$   & $\tau$      & $n_2$ & \\
    Output & $\colname{result}$ & $\tau$      & $n_1$ & \\
  \end{oplisting}
  Performs an indirect lookup into $\colname{data}$, i.e. gathers various elements of $\colname{data}$, so that $\colval{result}{i} = \colval{data}{\colval{pos}{i}}$.
\end{operators}

\subsubsection{De-duplication and use of dictionary indices as surrogates}
\label{subsubsec:de-duplication-and-use-of-surrogates}

With a compression scheme focused only on the support set, it is natural to wish to use the compressed-form elements (the representations of individual uncompressed elements) as \emph{surrogates} for the original column elements: To avoid decompression, and apply columnar operators to the compressed form of the data. This is possible if the compression function is injective, or better still, when the order relation within the compressed domain commutes with the decompression function. Key operators we can apply to the compressed surrogates are:

\begin{itemize}
 \item Equality and range filters, using a compression of a single match value or a pair of extrema to compare against;
 \item \planop{Sort}ing or partitioning by relative order;
 \item Obtaining the column's support set;
 \item \planop{Join} of a pair of columns which share the same compression scheme.
\end{itemize}
This potential for operator ``push-down'' via surrogates was noticed already in early research into the use of compression in DBMSes \cite[Page 4]{GS1991}.

\medskip

Use of surrogates is clearly possible with \cscheme{DiscardHighBits}; but the \cscheme{Dictionary} scheme does not guarantee, by definition, usable surrogates. To secure this guarantee we must impose additional constraints on the \colname{dictionary} column of the scheme:

\begin{enumerate}
 \item \emph{Injectivity}: For surrogate use of the compressed elements in \emph{exact-match comparisons}, $\finverse{\colname{dictionary}}$ must be a proper function, i.e. all dictionary entries must be distinct.
 \item \emph{(Strong) monotonicity}:  For surrogate use of the compressed elements in \emph{order relation comparisons}, $\colname{dictionary}$ must strictly preserve that order, i.e. $i < j$ iff $\colval{dictionary}{i} < \colval{dictionary}{j}$. (Note that if $\colname{dictionary}$ is known to be injective, this is equivalent to the additional constraint of weak monotonicity.)
 \end{enumerate}

By adding one or both of these constraints to the definition of \cscheme{Dictionary}, we obtain a different --- more restricted --- compression scheme:

\begin{cschemes}
\csitem[UniqueDictionary] A sub-scheme of \cscheme{Dictionary}, where compressed forms have no duplicate elements in the $\colname{dictionary}$ column. (Shorthand name: \IntroduceCompressionSchemeShorthand{UDICT}{UniqueDictionary}.) With this scheme, $\colname{indices} = \colname{dictionary}^{-1} \circ \colname{col}$ (with $\circ$ denoting the composition of functions).
\csitem*[\cscheme{MonotoneDictionary}_{\preceq}] A sub-scheme of \cscheme{UniqueDictionary}, where the \colname{dictionary} column is also (strong-)monotone-increasing with respect an order relation $\preceq$ over $\fsupport{\colname{dictionary}}$; we omit $\preceq$ if it is the default order relation for the uncompressed column's element type. (Shorthand name: \IntroduceCompressionSchemeShorthand{MDICT}{MonotoneDictionary}.)
\end{cschemes}

%the $\colname{dictionary}$ column is an enumeration of the domain of the uncompressed column (denote it $\colname{col}$), i.e.

De-duplication and sorting change the cost-benefit tradeoff of the unconstrained \cscheme{Dictionary} scheme: We gain the use of surrogates, described above --- at the cost of additional, less-parallelizable work in performing the compression. But there may be additional, secondary gains in using \cssh{UDICT}{UniqueDictionary} or \cssh{MDICT}{MonotoneDictionary}: When a unconstrained dictionary contains many duplicate entries, their removal reduces its size. This, in turn, may allow for a smaller width of the dictionary indices as well. The compression ratio may thus improve somewhat; and a smaller dictionary is also likely to fit more of itself into a processor's cache or local memory space. Also, a sorted dictionary is more likely to admit additional compression utilizing the similarity of consecutive elements (see \Autoref{sec:compression-scheme-composition-patterns, subsec:encoding-patch-positions-with-less-redundancy} below).

Finally, note that the choice of indices in a dictionary can be used to express an order relation among values of the encoded column, other than the default order on the uncompressed column's element type. For example, the dictionary entries could be ordered by descending frequency within the uncompressed column, or even so that the dictionary index is a hash of the value (see also \autoref{sec:variable-width-dicts-and-surrogacy}).

\section{Patterns of scheme composition and transformation}
\label{sec:compression-scheme-composition-patterns}

A reader familiar with existing analytic DBMSes (columnar or otherwise), may be puzzled by the previous section: Very few schemes were presented as ``basic''; most of them cannot even encode an arbitrary input column (regardless of compression ratio); and several schemes which, in other systems, are not considered `composite', were not mentioned (such as \ucscheme{RLE} and \ucscheme{FOR}). The reason for this is that our model of computation, together with the machinery for columnar representations of data, allow us to formulate all those additional schemes by composing the basic ones, or transforming them using a few additional operations in the encoding and decoding circuits.

Each such ``composite compression scheme'' is simply a codec, with the encoding and decoding partial-functions being compositions of several functions. A composite scheme may be concretized by composing the decoding and encoding circuits of its constituent schemes: Assigning outputs of one decoder to inputs of the other, and assigning encoder outputs in the opposite direction; or, in some cases, taking the union of these circuits; both of these transformations were described in \autoref{sec:circuit-composition-and-transformation} above.

One can compose compression schemes almost to no end, choosing schemes using knowledge of the intended input distributions. This section will describe some patterns for composing schemes --- which will be put to use in concrete composite schemes of particular interest, throughout the rest of this \namecref{chap:columnar-representation}. Each of the patterns presented here exemplifies some concept worthy of explicit description --- and are commonly useful in terms of compression ratio gains.
%; \autoref{sec:composite-compression-schemes} will present several specific compositions of particular interest.

\subsection{Elementwise composition and ``noise-signal''-like separation}
\label{subsec:elementwise-composition}

In signal processing, a fundamental task is restoring or isolating an original transmitted signal from (literal) noise it picks up as it is transmitted over physical media. The signal carries limited and well-formed information; the noise is unconstrained with the encoding or modulation of the signal, but is of limited power. In our setting, there is `noise' per se --- information can discard; but the analogy still inspires a decomposition of an unencoded column: $\colval{col}{i} = \colval{signal}{i} + \colval{noise}{i}$, with \colname{signal} admitting a more effective compression (typically better than, say, \ucscheme{DiscardHighBits}), and \colname{noise} being of limited magnitude. These can be compressed separately, with \colname{noise} using \cscheme{DiscardHighBits}.%; and with the length of $\colname{noise}$ replacing a scalar specifying the length of \colname{signal} (as necessary).

The trivial --- though useful --- example of this kind of scheme would follow through with the analogy to signal processing:

\begin{cschemes}
\csitem*[\cscheme{NoisyGenerated}_{\vec{f}}]
  \begin{cslisting}
    $\colname{coefficients}$ & $\tau$        & $k$  & \makecell{ coefficients of the linear combination of $\vec{f}$ \\
                                                                 forming the generating function } \\
    $\colname{noise}$ & $\tau_\text{noise}$ & $n$ & elementwise additive noise
  \end{cslisting}

  The type $\tau_\text{noise}$ is a subtype of $\tau$, typically with fewer bits in its representation.
  The decompressed column is the addition of the noise, up-cast from $\tau_\text{noise}$ to $\tau$, to the generated function.
  %(In practice, one could drop the \colname{length} scalar as it is easily derivable from \colname{noise}; but that would be idiosyncratic to this particular scheme.)
% \csitem[NoisyAffine]
%   The specific case of \cscheme{NoisyGenerated} with $k=2$, $f_0(x) = 1$ and $f_1(x) = x$, i.e. where the generated function is a line.
\end{cschemes}
The more general scheme above scheme may well be used in practice to store samples of some real-world signal, with signal-processing techniques employed to determine the basis coefficients. %(e.g. a discrete Fourier transform for a base of sinusoids).
%The more specific case, despite its weaker model expressivity, should also be rather useful considering its simplicity; see also \autoref{par:segmentized-frame-of-reference}.
\medskip

Generalizing the `signal-noise' decomposition, elementwise addition may combine arbitrary schemes. Specifically, \cscheme{Generated} with $k$ base functions can be perceived as the application of $k-1$ compositions-by-addition of $\cscheme{Generated}_f$ schemes for the different basis functions.

\begin{note}
While lossy compression of information is out of scope for this monograph --- $\cscheme{NoisyGenerated}_{\vec{f}}$ schemes could be used for lossy compression floating-point data, which is of inherently limited accuracy: The \colname{noise} values might be encoded at lower precision, seeing how after their addition to \colname{signal}, those less-significant bits may fall below the floating point precision threshold.
\end{note}

\subsection{Discrete differentiation and integration}
\label{subsec:differentiation}

Column stores processing queries, and performing decompression, frequently need to consider both the \emph{positions} (or indices into a column) of elements and \emph{lengths} or \emph{distances} between elements  (within a column), or to switch between these two kinds of information.
%This transformation can be avoided in representations which exhibit some redundancy, such as the \ucscheme{VariableLengthSegmented} scheme we've defined in \autoref{subsubsec:uniform-segmentation-and-subcolumns} above %(whose verifier must ensure the encoded form position differences are exactly the encoded lengths).
When compressing, we can make do with only one kind: We may store the positions and take their derivative as necessary, or store only lengths, and integrating over them. To be more concrete, circuits involving such compressed data will use one of the following operators:

\begin{operators}
 \opitem[Derivative]
  \begin{oplisting}
    Input  & $\colname{col}$          & $\tau$                & $n$                      & \\
    Output & $\colname{differences}$  & $\tau$                & $n - 1$                  & \\
  \end{oplisting}
  Sets $\colval{differences}{i}$ to $\colval{col}{i+1} - \colval{col}{i}$ --- the discrete derivative.

 \opitem*[\IntroducePlanOperatorAlias{PrefixSum}{PrefixAggregate}]
  \begin{oplisting}
    Input  & $\colname{data}$        & $\tau$                 & $n$                      & \\
    Output & $\colname{aggregates}$  & $\tau$                 & $n$                      & \\
  \end{oplisting}
  Sets $\colval{aggregates}{i}$ to $\sum_{j=0}^{i} \colval{data}{j}$. (a.k.a. $\planopalias{InclusivePrefixSum}{PrefixSum}$, $\planopalias{Scan}{PrefixSum}$ --- but this is not a relational``table scan''; see also \cite[\S 3.2]{GBDPS2018}).

 \opitem*[\IntroducePlanOperatorAlias{ExclusivePrefixSum}{ExclusivePrefixAggregate}]
  \begin{oplisting}
    Input  & $\colname{data}$        & $\tau$                 & $n$                      & \\
    Output & $\colname{aggregates}$  & $\tau$                 & $n$                      & \\
  \end{oplisting}
  Sets $\colval{aggregates}{i}$ to $\paren{0 + \sum_{j=0}^{i-1} \colval{data}{j}}$. Similar to \uplanop{PrefixSum}, except that the $i$\xth aggregate does not involve the $i$\xth element.

 \opitem*[\planop{PrefixAggregate}_{\oplus}]
   The generalization of $\planopalias{PrefixSum}{PrefixAggregate}$ to an arbitrary (commutative) binary operator $\oplus$.
 \opitem*[\planop{ExclusivePrefixAggregate}_{\oplus}]
   The generalization of $\planopalias{PrefixSum}{PrefixAggregate}$ to an arbitrary (commutative) binary operator $\oplus$. There must be a neutral element in $\fdomain{\tau}$ w.r.t. $\oplus$, for use as the aggregate of no elements.
\end{operators}

% TODO: Consider adding here:
%
% A definition of a Reduce operator (no partial sums, just the full one)
% A Fold alias for Reduction, or for Prefix Sum; perhaps the latter ?

\subsection{Segmentization}
\label{subsec:scheme-segmentization}

In \autoref{sec:segmented-columns} we formalized representations for segmented columns, and the process of segmantization of operators. Now, given some compression scheme for columns --- it can be applied separately to consecutive segments of an (uncompressed) column, with the compressed forms' constituent columns for every segment being concatenated, to form a compressed form for the column overall. Doing so is applying a segmentized version of the original scheme. Segmentized schemes are of two varieties, depeding on the nature of the segmentation:

\paragraph{Fixed-length segments} As mentioned in \autoref{sec:segmented-columns}, segmented columns with fixed segment length are prevalent in many column stores regardless of compression; and they therefore often choose to compress distinct segments separately (possibly with some column-level data) --- to avoid inter-segment dependencies during decompression. Provided segments are not too short, this may help compression ratio: The overhead on per-segment data is often lower than the savings due to localization: A smaller range of variation to cover; and no need to tolerate or account for anomalies throughout the entire column.

%The interest in considering compressed data in separate segments can be either that of better compression, or alternatively, more efficient decompression.

%

\paragraph{Variable-length segments} Letting the number of elements per segment vary is a costly choice, regardless of the compression scheme. First, there is the space required for specifying segment lengths. There's also a cost in parallel decompression time, due to read-after-read dependencies: A decompressed column element's segment index cannot be determined in advance (and, in fact, often requires multiple reads into a \cscheme{Segmentation}); and only when it is known can the per-segment data be read. Such a choice would therefore be the result of necessity rather than convenience:

\begin{itemize}
\item  The \emph{column data} itself may require adapting segment length to capture abrupt local changes within a column. % Eexample: An abrupt change of pattern in the data, between stretches of well-behaving elements).
 Alternatively, the data may exhibit gradual, but non-uniformly-progressing, changes in behavior; in this case the break-up into segments would be rather artifical, but required for limiting divergence between the actual data and some localized model. As an example, think of columns obtained by the following random process: Elements are uniformly and independently sampled from a small-support local distribution, and this distribution changes at a non-uniform pace. %While gradual changes cam typically also be captured by fixed-length segments, but since the change itself is non-uniform, one would either have many short segments, or segments where data changes too much --- hurting the compression ratio either way. In this context, the non-uniformity is a choice of better compression ratio over better compression speed.
 \item  In some cases, it is the  \emph{decompression process} which necessitates variable-length segmentation. We observed in \autoref{subsec:segmentizing-operators} that segmentized operators do not preserve uniformity: The dependence of operator output length on the input data causes uniformly-segmented data to be transformed into non-uniform-length output segments. Thus, even if a variable-length segmentation is not immediately apparent in a description of the compressed form, a decoder may have to introduce it within a decoding circuit.
\end{itemize}

%Finally, the existence of multiple, different motivations for using segmentation in compression scheme results in there occasionally being cause for \emph{composite} uses of it, e.g. variable-length segmentation fundamentally, then fixed-length segmentation to localize searches.

%The choice between fixed and variable-length segments is not trivial. Consider  the motivating case of a ``column'' of variable-width elements. If the per-element data is stored using plain concatenation, with no auxiliary information --- element lookup then necessitates a sequential search over large swaths of this data. A fixed-segmentation mitigate this issue by delimiting the search area for each given index using the segmentation: Element $i$ lies in segment $j = \floor{i / \colname{segment_length}}$, and thus between the positions $\colval{segment_start_pos}{j}$ and $\colval{segment_start_pos}{j} + \colname{segment_length}$. Variable-length segmentation may narrow down the search space even further, due to adaptivity --- but then the segmentation itself requires some searching to locate the approrpiate segment.

\subsection{Patching and outlier removal}
\label{subsec:patching}

\begin{definition}[Hamming distance \cite{Hamming1950}] Let $\colname{col}_1$, $\colname{col}_2$ be columns of type $\tau$ and length $n$. The \emph{Hamming distance between $\colname{col}_1$ and $\colname{col}_2$} is the number of indices on which $\colname{col}_1$ and $\colname{col}_2$ disagree, i.e. $\setsize{{\Large{\mathstrut}} \condset{i \in \zeroupto{n}}{\colname{col}_1\paren{i} \neq \colname{col}_2\paren{i}}}$. The \emph{Hamming distance between a column $\colname{col}$ and a set of columns $S$}, all of the same length as $\colname{col}$, is the infimum of distances between $\colname{col}$ and every column in $S$.
\end{definition}

Suppose some compression scheme \ucscheme{MainScheme} (that is, some codec $\tuple{f_d, f_e}$) has certain desirable features, such as high compression ratio or well-parallelizable decompression. Now suppose we wish to compress a column $\colname{col}$ which is not in $\fdomain{f_e}$; rather, it is within a small Hamming distance from it: If not for a few outlier values, the column could be compressed using \ucscheme{MainScheme}. Let $\colname{col}'$ denote a column in $\fdomain{f_e}$, closest to $\colname{col}$; instead of giving up on the efficient compression of $\colname{col}$, it is intuitive to consider using the compressed representation of $\colname{col}'$ instead, with some information allowing for post-decompression correction of $\colname{col}'$ into $\colname{col}$. The necessary information is in fact simply the subcolumn of  $\colname{col}$ on $\breakingcondset{i \in \zeroupto{n}}{\colname{col}\paren{i} \neq \colname{col}'\paren{i}}$, which we adjoin in the standard representation (see \autoref{def:subcolumn}). The application of this subcolumn --- the \emph{patching} of $\colname{col}'$ with $\colname{col}$ values --- is an application of the \planop{SubcolumnOverlay} operation; and we have thus obtained a new compression scheme, being the composition of \cscheme{Patched} (defined in \autoref{subsec:combining-subcolumns}) after \ucscheme{MainScheme} on the main column.

\begin{figure}[H]
  \centering
  \includegraphics[height=5cm]{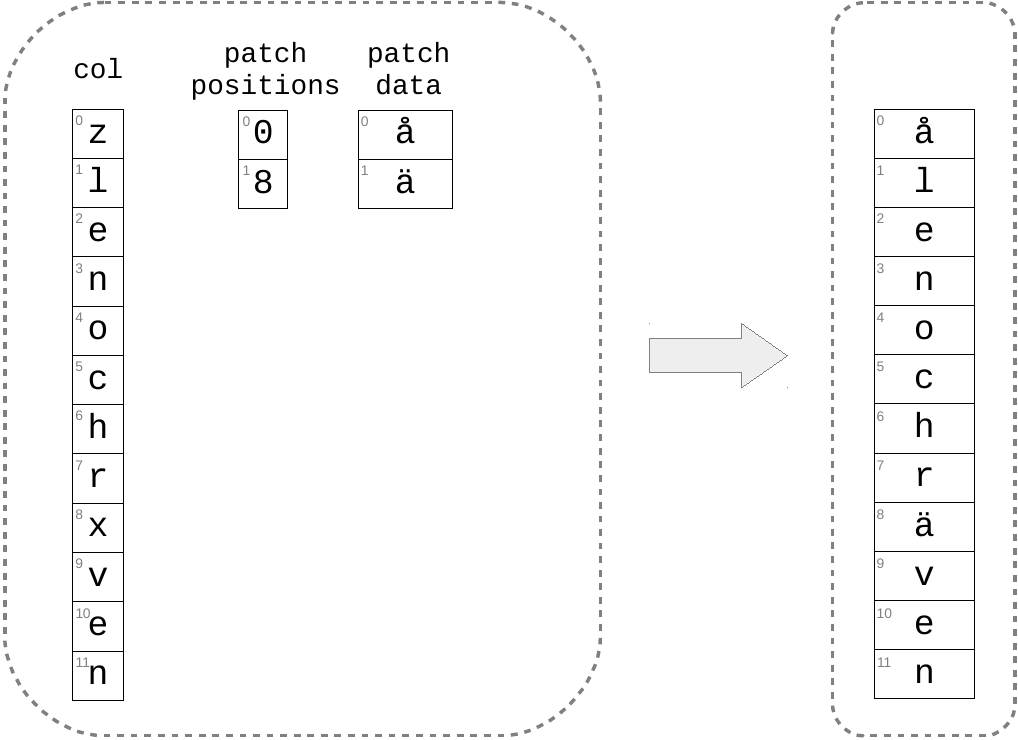}
  \caption[An example of patching with simple inputs]{An example of patching with standard-representation-coded inputs. The column contents is the phrase ``the eels and the foxes'' in Swedish, without spaces; most character UCS-2 codes fit into single octets, while the patches use the full 16-bit width. The choice of 8-bit characters later overridden by patch values is arbitrary.}
  \label{fig:scheme:column:example:patching}
\end{figure}

\begin{figure}[H]
  \centering
  \includegraphics[height=5cm]{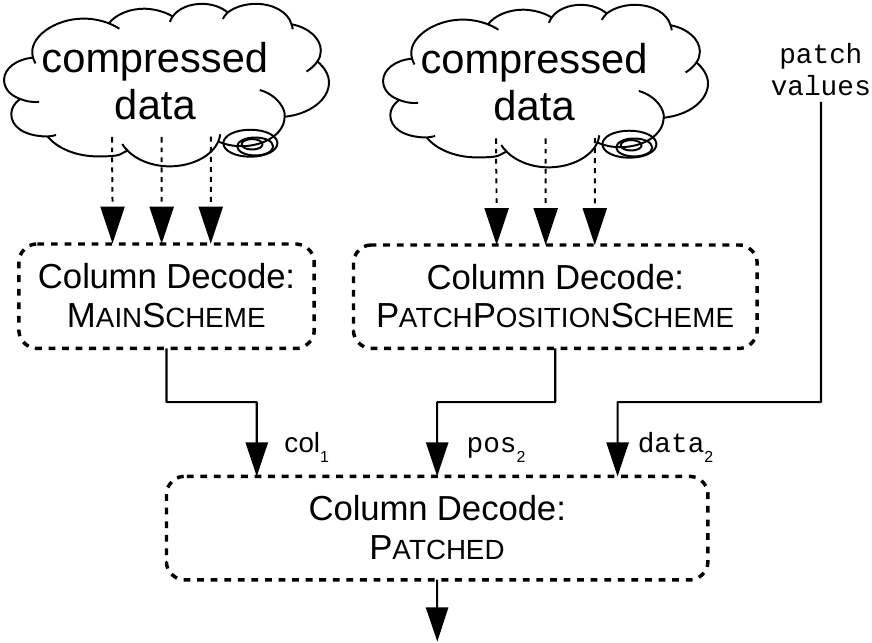}
  \indexCompressionScheme{Patched}
  \caption[A decoder for patched columns using \capucscheme{PatchPositionsScheme} for patches]{A decoder for patched columns, where most of the data is encoded with scheme \capucscheme{MainScheme} and the positions of outliers encoded with \capucscheme{PatchPositionsScheme} respectively.}
    \label{fig:scheme:column:decoder:patching}
\end{figure}

% Talk about the metric relation
%The composition with \cscheme{Patched} (which, in itself, is not a compression scheme) is therefore a useful transformation for what could be described as an expansion of the domain, or the expressivity, of a compression scheme. It is closely-related to composition-by-addition of \cscheme{DiscardHighBits} ``noise'': Hamming distance corresponds to the $L^0$ metric over the space of length-$n$ columns, while

The decoding example is immediately reminiscent the use of the \cscheme{Patched} scheme for representing \sqlnull values, in \autoref{subsec:nullable-and-union-type-columns}
Indeed, if the column element type is $\tau = \tau_1 \dunion \tau_2$, and there are only a few elements from $\fdomain{\tau_2}$, the patches can have type $\tau_2$ and the column overall can be compressed as a purely $\tau_1$-type column.

The use of patched compression schemes in column stores was pioneered by Vectorwise, as presented in \cite{ZHNB2006}. However, work published so far on this subject (as surveyed\cite{DHHL2017}) does not present patching as a meta-scheme, or a scheme transformation, applicable generically to arbitrary base schemes; rather, very specific patched schemes are presented. This stems from the orientation of existing work towards less-parallelized CPU-based decompression: These schemes ``package'' data from both the main column and the patches subcolumn into consecutive chunks of the (not immediately columnar) compressed form --- so that it fits well into a few consecutive CPU cache lines. This is done with some finesse in carefully designing compressed block layout --- which makes for idiosyncracy rather than generality. See a more detailed discussion of this point in \cite{RB2017}.

%But the flexibility of the columnar formulation has significant potential benefits, as will be demonstrated in \autoref{subsec:encoding-patch-positions-with-less-redundancy} below.

\medskip

Patching can increase compression efficiency not only by reaching the domain of a compression scheme: It also allows selection of better-compressible column within a compression scheme's domain. For example, suppose the \cssh{RLE}{RunLengthEncoding} scheme is applied to a column with very long stretches of similar lengths. But suppose also that at a low frequency within those stretches, an interrupting value appears once or twice. The distance between these interruptions has high variance, making it difficult to estimate the index of the run covering a given position in the output; also, the interruptors take up close to half of the \cssh{RLE}{RunLengthEncoding} compressed form in terms of bytes used. Patching away a single one of these interrupting values saves the space of two values and two run lengths in the main \cssh{RLE}{RunLengthEncoding} compressed form, but costs one-or-two values and one-or-two position indices in patch information; this is not much of a saving, if at all (especially if the type for position indices is wider than the type of run lengths). However, the overall length of the compressed main column has now dropped significantly --- as most runs had only existed separately because of the interruptors --- and the variances in run lengths have also decreased. This may allow for \emph{faster decompression} of the column overall, as well as \emph{faster processing of the compressed form}: Computing, say, a sum of all column values may become faster due to the smaller number of runs, while the re-integration of the sum of patch values and the sum of the main column will have very low overhead.

%\begin{note}
%The kind of patching described in this \namecref{subsec:patching} is applied \emph{aposteriori} --- after the decompression of the main column. Not every compression scheme is amenable to this approach: It is mostly the elementwise schemes, where the value of single elements has no direct effect on the decompression of surrounding or other elements. When decompression does involve multiple elements, it is only possible to apply patches \emph{earlier} in the decompression process --- if at all. An example of doing so can be found in \autoref{subsec:segmentized-differences} below.
%\end{note}

\medskip

\begin{note}
As part of a later discussion of composite compression schemes, a converse-of-sorts of column patching will be presented in \autoref{subsec:small-dictionary-fitting}.
\end{note}

\subsection{Alternating schemes within the same column}
\label{subsec:alternating-schemes}

This section is concerned with the variation in the distribution and patterns of data within a single column to be compressed. We have already touched upon several approaches to handle such variation, under certain constraints: The use of (low-complexity) models; separate handling of outliers by patching; and the localization of the compression scheme's parameters, by segmentizing the scheme. But what if significant fractions of the column behave so differently, that forcing the same compression scheme on the entire column is extremely costly? Alternatively, what if two possible schemes are each better fit at different parts of the column, so that while none of them does poorly overall, each of them could do really well if it could just be limited to that part? Indeed, it is intuitively useful to be able to alternate between two (or more) different schemes in different parts of the column.

In existing column stores supporting compression, this is achieved as a side-effect of column segmentation. In SybaseIQ, for example, each page in magnetic disk storage may have data in a different compression scheme \cite[\S 2.1]{ZHNB2006}. In Vectorwise, there are (at least) two levels of segmentation --- one spanning a certain number of secondary storage pages (``chunks''; a compression scheme is chosen at this resolution) and within each chunks, multiple segments which may either be compressed or uncompressed \cite[\S 3.1.1]{ZHNB2006}.

It so happens, that this capability is inherent already in the machinery developed so far in this and the previous chapters, for representation and compression schemes: For each of the compression schemes to be alternated, the elements of the column compressed using this scheme constitute a subcolumn; and these subcolumns constitute a partition of the full column. We have already discussed the representation of partitions, in \autoref{subsec:partitions}; one simply has to adjoin the \cscheme{Partition} representation scheme with the compressed forms for the concatenated elements of each of the compressed subcolumns.

Of course, the column representing the partition (holding compression scheme indices) may itself be compressible, reducing the overhead of scheme alternation. A specific case of interest is alternation aligned with transitions between equi-sized pages or chunks on disk or other storage: In this common case, we call the partitions are $\ell$-segmentation-respecting, for some appropriate $\ell$. Given $\ell$, and with $k$ alternating compression schemes --- instead of keeping a \colname{partition} with $\ceil{\log{k}}$ bits per each compressed column element, we would only require $\ceil{\ceil{\log{k}}/\ell}$ bits per element to indicate which scheme is in use. For typical values of $\ell$ in practice, this is rather negligible.

\medskip

Using a single \colname{partition} column is not always the best choice, especially for an $\ell$-segmentation-respecting partition. In the latter case, we may prefer to represent each of the subcolumns using \cscheme{SegmentedSubcolumn} before each subcolumn's specific compression scheme. While the space overhead is somewhat higher --- up to $\ceil{\ceil{\log{n/\ell}}/\ell}$ per element for a column of length $n$ --- this representation affords us immediate use of each subcolumn separately. Also, a concatenation of the different \colname{segment_pos} columns will constitute a valid \cscheme{Segmentation} scheme instance.

%The above works for any partition, not just partitions respecting $\ell$-uniform segmentations. If the partition \emph{does} respect it, i.e. in the case of length-$\ell$ ``compression blocks'', we could alternatively use a composition of each of the compression schemes with \cscheme{SegmentedSubcolumn}; data thus compressed would be decompressed into the appropriate subcolumn for each scheme, and the entire uncompressed column is obtained by taking their (disjoint) union. The decoding process for this alternative is illustrated in \autoref{sec:circuit-composition-and-transformation} (for the case of $k = 2$ for simplicity):

%Now, the ``compressed blocks'' described earlier in existing systems are simply the special case of the above of the alteration potentially occuring at fixed index intervals into the column --- to use our own terminology, at segment boundaries (with segment length $\ell$). For this case, we can use the \cscheme{UniformlySegmented} representation scheme to fully represent one subcolumn; and for the other one, we may use the complemenet of the index set (not storing anytinhg), keeping only the compressed \colname{data} component. This --- as well as the entire decoding process --- is illustrated in \autoref{fig:scheme:subcolumn:decoder:segmented-combination-of-two-schemes}.

\begin{figure}[H]
   % --------------------------------------------
   % Encoder and Decoder for SprarseIndexSet
   % --------------------------------------------
  \centering
  \begin{subfigure}[b]{0.45\linewidth}
    \centering
    \includegraphics[scale=0.7]{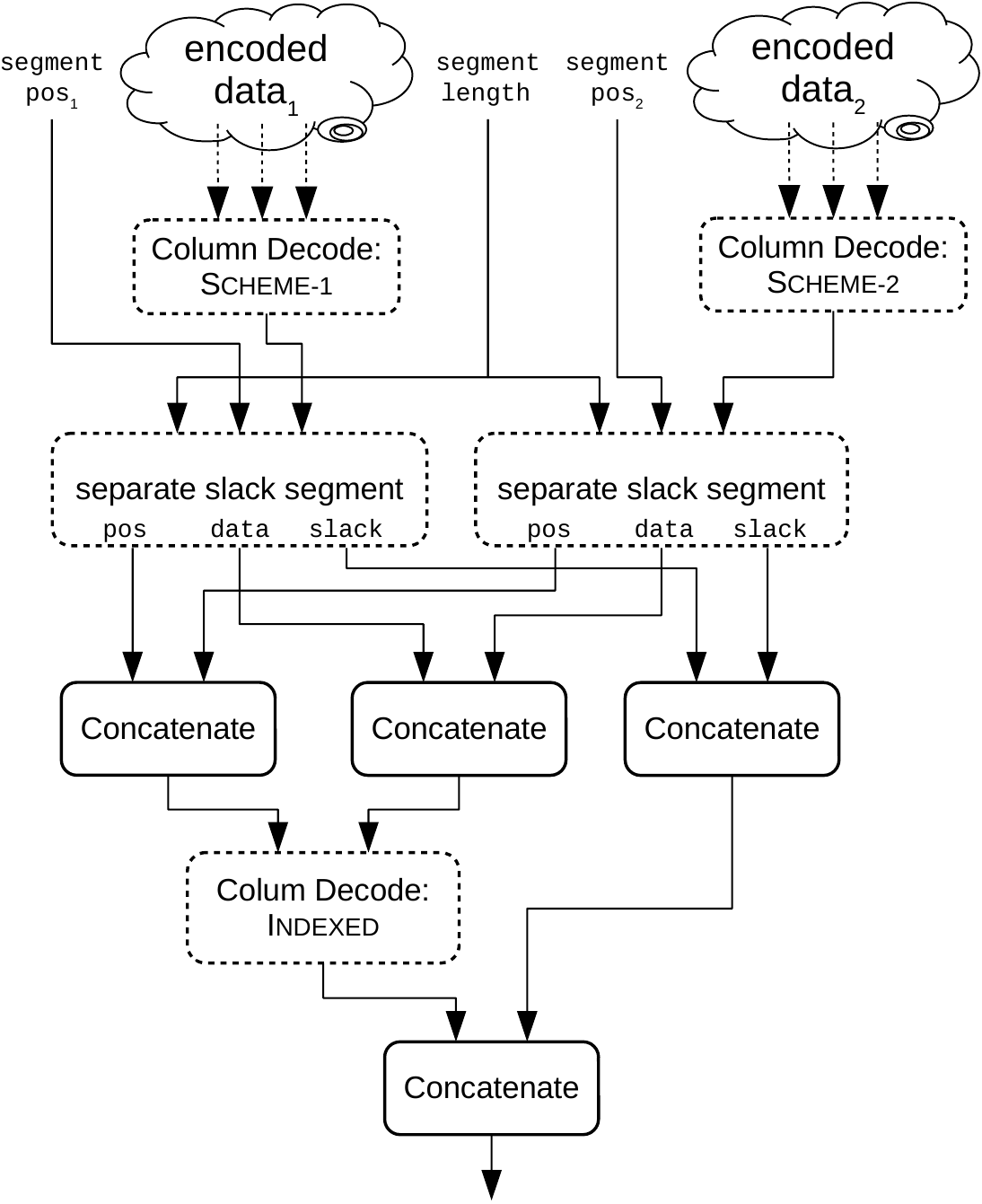}
    \subcaption{Alternation using a \cscheme{SegmentedSubcolumn}-based representation. Note at most one of the subcolumns has a non-empty slack segment.}
    \label{subfig:scheme:column:decoder:alternating-schemes:via-segmented-subcolumn}
  \end{subfigure}
  \hspace{0.05\linewidth}
  \begin{subfigure}[b]{0.45\linewidth}
    \centering
    \includegraphics[scale=0.7]{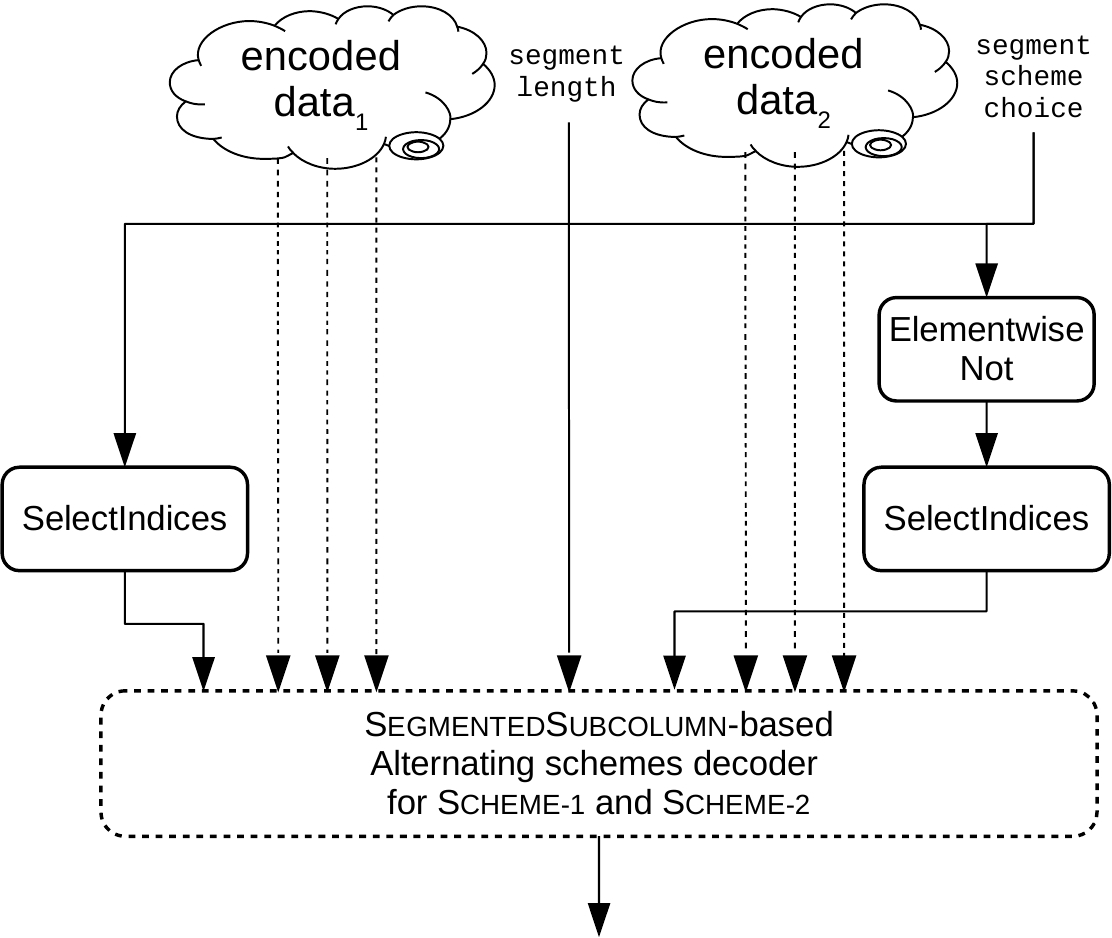}
    \subcaption{Alternation using a \cscheme{Partition} representation for segment scheme selection --- decoded by reduction to the \cscheme{SegmentedSubcolumn}-based representation}
    \label{subfig:scheme:column:decoder:alternating-schemes:via-partition}
  \end{subfigure}
  \caption[Decoders for columns compressed with two alternative schemes]{Decoders for columns compressed with two schemes, \capucscheme{Scheme-1} and \capucscheme{Scheme-2}, alternating in an $\ell$-segmentation-respecting fashion.}
  \label{fig:scheme:column:decoder:alternating-schemes}
\end{figure}

\bigskip
Having presented several patterns for composing compression or encoding schemes, in subsequent sections we will use them to build up a rising pyramid of actual composite compression schemes, exploring their uses, strengths and weaknesses.

%\section{Composite compression schemes}
%\label{sec:composite-compression-schemes}

\section{Variable-width columns: Dictionaries and surrogates}
\label{sec:variable-width-dicts-and-surrogacy}

In \autoref{subsec:column-support-set-compaction-1} above we explored the use of dictionaries for compressing individual elements independently, and how their use may be adapted so that dictionary indices are usable as surrogates for the entire value (\autoref{subsubsec:de-duplication-and-use-of-surrogates}). The benefits of achieving the same for variable-width columns are potentially even more dramatic: It is typical, for example, for a database to have long columns of strings, each of which is a word or a phrase taken from a small set; and users are notorious for performing \planop{Join}'s on these columns.

Recall, that the standard representation of variable-width columns (defined in \autoref{subsec:variable-width-columns}) already has a dictionary-like aspect: Access to element data for a given index in the column is indirect, going through a column of pointers / position indices into the raw data. If all elements of a variable-width column happen to have length 1, then its \colname{lengths} column is redundant, and we in fact have a \cscheme{Dictionary}-encoded regular column (with \colname{start_positions} assuming the role of \colname{indices} and \colname{data} assuming the role of \colname{dictionary}). It can thus be said that the standard representation variable-width columns \emph{already incorporates} a dictionary, except that dictionary keys are (position, length) pairs rather than merely a position; and that no additional compression scheme need be defined.

If the support set is small enough that indicating the length of each element is too wasteful, the pair of columns \colname{start_positions} and \colname{lengths} may be compressed using the \cscheme{Dictionary} scheme --- with identical indices. The resulting compression scheme merits a proper named definition:
%
%i.e. for each variable-width column element only one dictionary index will be held, but rather than indexing the \colname{data} column, it will index the \colna
%
%
%In this ``reeduction'' of fixed-width-column \cscheme{Dictionary} to a degenerate variable-width column representation, pairs of column elements have either dijoint or identical element data in \colname{data}; this if data is shared, the lengths of two elements are identical, and consequently --- it's sufficient to keep a single length value for every stretch of \colname{data} corresponding to a single variable-width element. We thus define a first, simpler, extension of \cscheme{Dictionary} to variable-width columns as follows:

\begin{cschemes}
\csitem[VariableWidthDictionary]
  \begin{cslisting}
    $\colname{indices}$               & $\inttype$  & $n$                                    & enumeration indices of the original elements \\
    $\colname{entry_start_positions}$ & $\inttype$  & $m$                                    & offsets into \colname{data} \\
    $\colname{entry_lengths}$         & $\inttype$  & $m$                                    & variable-size ``element'' lengths in units of $\tau$\\
    $\colname{entry_data}$            & $\tau$      & $\sum_{i=0}^{m-1} \colval{lengths}{i}$ & the concatenated dictionary entries \\
  \end{cslisting}
  The $i$\xth element of the uncompressed (variable-width) column is the concatenation of the elements of $\colname{entry_data}$ at indices $\colval{entry_start_positions}{i}$ through $\colval{entry_start_positions}{i} + \colval{entry_lengths}{i} - 1$.
  (Shorthand name: \IntroduceCompressionSchemeShorthand{VWDICT}{VariableWidthDictionary}.)
\end{cschemes}

\begin{figure}[H]
  % ----------------------------------------------------
  % Example of the VariableWidthEntryDictionary scheme
  % ----------------------------------------------------
  \centering
  \begin{subfigure}[b]{0.275\linewidth}
    \centering
    \includegraphics[scale=0.75]{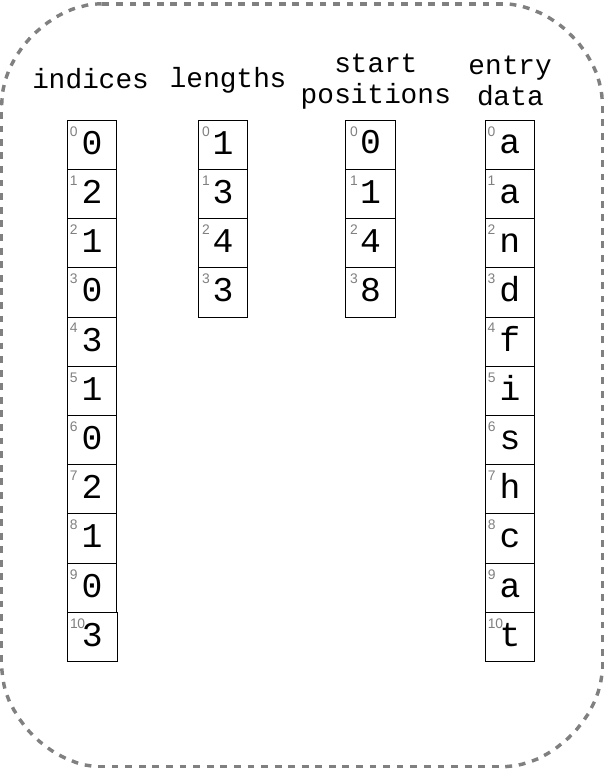}
    \subcaption{\ucscheme{VWDICT}-compressed form \\ \mbox{} \\ \mbox{}}
    \label{subfig:scheme:column:example:vldict:compressed-form}
  \end{subfigure}
  \hspace{0.3cm}
  \begin{subfigure}[b]{0.25\linewidth}
    \centering
    \includegraphics[scale=0.75]{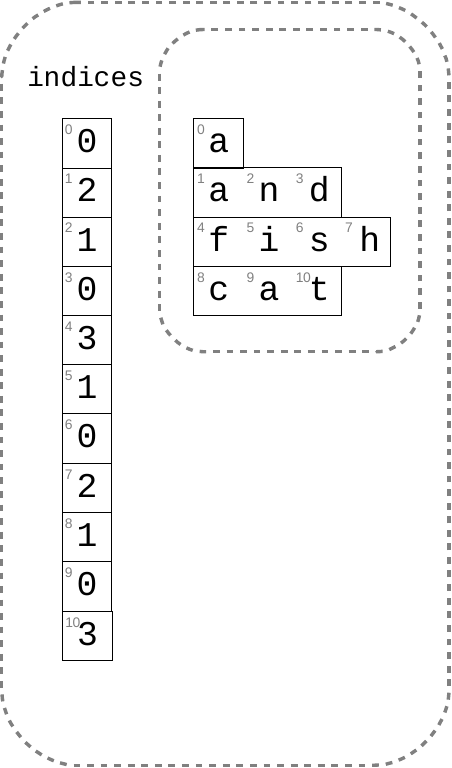}
    \subcaption{Compressed form, variable-width view of the dictionary \\}
    \label{subfig:scheme:column:example:vldict:compressed-with-vl-view-of-dictionary}
  \end{subfigure}
  \hspace{0.3cm}
  \begin{subfigure}[b]{0.25\linewidth}
    \centering
    \includegraphics[scale=0.75]{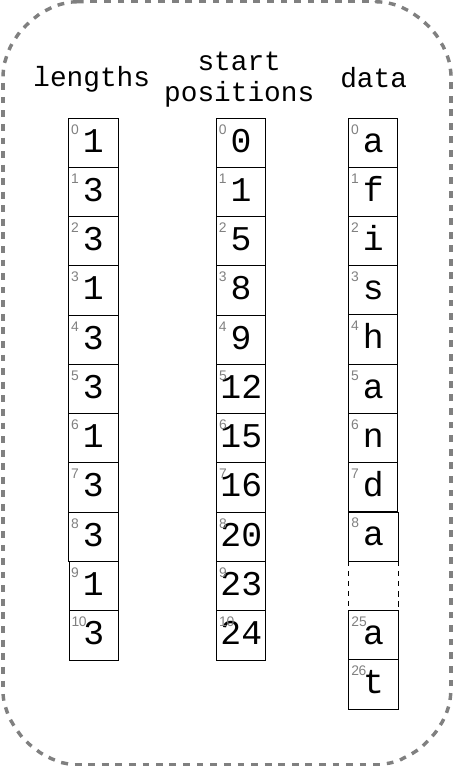}
    \subcaption{Uncompressed form (canonical representation; data column clipped to fit)}
    \label{subfig:scheme:column:example:vldict:uncompressed-form}
  \end{subfigure}
  \hspace{0.3cm}
  \begin{subfigure}[b]{0.125\linewidth}
    \centering
    \includegraphics[scale=0.75]{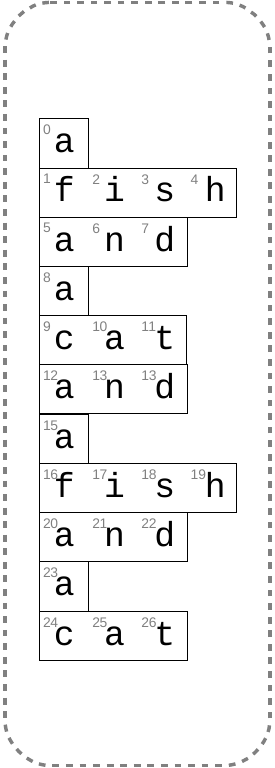}
    \subcaption{Uncompressed form (variable-width view)}
    \label{subfig:scheme:column:example:vldict:uncompressed-form-vl-view}
  \end{subfigure}
  \caption[The \capucscheme{VariablLengthDictionary} compression scheme ]{The \capucscheme{VariableWidthDictionary} compression scheme applied to the sequence of words \texttt{"a", "fish", "and", "a", "cat", "and", "a", "fish", "and", "a", "cat"}. Note that in a realistic scenario, the \colname{indices} column would be much longer than the dictionary columns.}
  \label{fig:scheme:column:examples:indexed}
\end{figure}

With \cschemesh{VWDICT}{VariableWidthDictionary} defined, we now have a second alternative for a surrogate scheme for variable-width data: Either use the default representation's (position, length) pairs, or just use \cschemesh{VWDICT}{VariableWidthDictionary} positions. And for each of these alternatives we can define the two sub-schemes in which the dictionary entries are unique and strictly-monotone respectively (see \autoref{subsubsec:de-duplication-and-use-of-surrogates}).

\subsection{Creatively positioning variable-length element data}
\label{subsec:creatively-positioning-variable-length-element-data}

Two final potential benefits of using dictionaries with variable-length data involve either the \emph{overlap} or the \emph{spacing-out} of element data within the \colname{data} array.

Recall that there are no constraints --- in either the standard representation or in \cschemesh{VWDICT}{VariableWidthDictionary} --- on the positions within the raw data column of elements' data ; this is unlike in the canonical column representation, where \colname{data} is a concatenation of the elements in order. Specifically, there is nothing preventing the stretches of individual $\tau$ values for different elements of the variable-width column from \emph{overlapping} each other, or even beginning \emph{at the same position}: A variable-length element is determined also by its length. Thus with base data ``abc'', we could have the dictionary consisting off ``a'', ``b'', ``c'', ``ab'', ``bc'', ``abc'' and the empty string --- 7 different elements. Overlap use still has a small price with \cschemesh{VWDICT}{VariableWidthDictionary}, where each such element costs another length value, but in the standard representation, it costs nothing.

Alternatively, we may want to use the dictionary indices to make searching for element more efficient. Suppose we wish to determine whether a column contains a certain variable-length value. If the column values are sorted (regardless of the compression or representation scheme), a binary search would work. if we were guaranteed the dictionary only holds extant values, and we were using the sorted sub-schemes of the default reprersentation or of \cschemesh{VWDICT}{VariableWidthDictionary}, we could apply a binary search to the dictionary --- which is likely much faster to do. But finally, suppose we have a reasonable hash function $\fndef{h}{\Union_{k \in \naturals} \prod_{j=1}^{k} {\tau}}{\integers}$; in this case, we could have the data for element $\colval{col}{i}$ appear in \colname{data} starting at index $\fapply{h}{\colval{col}{i}}$ (ignoring collisions). If we then search for our arbitrary element at that exact position, the search would only require constant time (independently of the column length) on average \cite[Chapter 11]{CLRS2009}.

Of course, collisions can and do occur. One can construct a proper hash table to handle them (e.g. of the linear probing \cite{Litwin1980} variety); but another option is to give up on hashing colliding elements, so that \colname{data} is split into \colname{hashed_data} and \colname{unhashed_data}. MonetDB applies this approach to some extent for representing string columns: For string columns with small support size, the strings are placed inside a hash, Once the hash exceeds some size, storage reverts to a more straightword manner, which is not even a dictionary, so that later-inserted strings may appear multiple times. (This feature has not been described in published scholarly work, and would require examining the source code \cite{MonetDBSources}).

Hash-based placement of element data also has uses when \planop{Join}'ing columns. Of particular interrest is the case of a pair of columns to be \planop{Join}'ed, both of which share such a compression scheme: A hash-table layout of the dictionary can serve as an imperfect surrogate of a hash table of the entire column --- saving most of the work of a hash-based \planop{Join} implementation, except for those elements with hash collisions.

\section{Segmentized schemes}
\label{sec:composite-segmentized-schemes}

\subsection{Uniform-value-run encoding}
\label{subsec:run-encoding}

Database table columns may exhibit long sequences of identical values --- a \emph{run} --- without being constant overall. The typical example are the columns of a multi-column key by which a table is sorted, with the exclusion of the last column; and these are quite prevalent in practice (as the authors of C\=/Store report, for example, in \cite[\S 4.3]{AMF2006}). With our extensive previous discussion in \autoref{subsec:scheme-segmentization}, and recalling the basic schemes in \autoref{sec:basic-schemes}, we can capture these runs by applying the \cscheme{Constant} compression scheme to each of them, and treating the concatenation of these stretches as a segmented column, i.e. segmentizing the \cscheme{Constant} scheme. The resulting scheme is:

\begin{cschemes}
\csitem[RunEncoding]
  \begin{cslisting}
    $\colname{start_position}$  & $\inttype$     & $n'$ & ... of each run \\
    $\colname{length}$          & $\inttype$     & $n'$ & ... of each run \\
    $\colname{value}$           & $\tau$         & $n'$ & The repeating value of each run \\
  \end{cslisting}
  The uncompressed column has the value $\colval{value}{i}$ at all positions $\colval{position}{i} \ldots \colval{position}{i} + \colval{length}{i} - 1$.
\end{cschemes}
This scheme achieves effective compression only if the run lengths are not too short on average. To quantify ``not too short'', consider the specific case of $\tau = \inttype$: The representation of each run requires one extra element in each column, which in our limited case is 3 $\inttype$ elements; thus, for a compression rate of 2, the average run length would have to be at leeast 6. This begs for some further space savings, so as to be useful with more modest average-length of runs. Reducing the overhead is indeed possible, since $\colname{start_position}$ and $\colname{length}$ columns are almost entirely redundant with each other: The former is a discrete integration of the latter. Dropping any one of these columns results in a different scheme:

%If the column has \emph{very} long runs --- perhaps this is inconsequential; but using run lengths for compression can be relevant even for relatively short runs --- of merely a few elements on average, depending on the type widths specifics. Naturally, we should consider dropping one of the two mutually-redundant columns; and our choice results in different compression schemes:

\begin{cschemes}
\csitem[RunPositionEncoding] \cscheme{RunEncoding} without the $\colname{length}$ column, but with an additional scalar $\colname{overall_length}$. Or, a standard sucolumn representation of the first elements of each run, plus the overall length necessary to bound off the last run. (Shorthand name: \IntroduceCompressionSchemeShorthand{RPE}{RunPositionEncoding}.)
\csitem[RunLengthEncoding] \cscheme{RunEncoding} without the $\colname{start_positions}$ column. (Shorthand name: \IntroduceCompressionSchemeShorthand{RLE}{RunLengthEncoding}.)
\end{cschemes}

There is a trade-off in the choice of which of the redundant columns remains and which is discarded: If we keep the $\colname{start_position}$ column, we get simplified decompression --- as the value of any element of the output is determined by the triplet of elements for a single run, without influence by any of the other runs; decompressing using the $\colname{length}$ requires discrete integration --- a much more costly operation with deep dependency chains among elements. On the other hand, run lengths are typically much shorter than the entire column length, in which case we can further compress them with \cscheme{DiscardHighBits} --- with minimal decompression overhead; this motivates keeping $\colname{length}$ over  $\colname{start_position}$. In fact, we can ensure that this typical occurs always, with a slightly modification of the \cschemesh{RLE}{RunLengthEncoding} scheme:

\begin{cschemes}
\csitem[CappedRunLengthEncoding] A scheme similar to \cschemesh{RLE}{RunLengthEncoding}, but with $\fdomain{\colname{length}} \inc \zeroupto{r}$ for some $\colname{r}$ provided as an extra scalar. When encoding, runs longer than $r$ elements are broken up into consecutive runs of length $r$, and a final run of up to $r$ elements --- all with the same run value. (Shothand name: \IntroduceCompressionSchemeShorthand{CappedRLE}{CappedRunLengthEncoding}.)
\end{cschemes}

\begin{note}
The \cschemesh{RLE}{RunLengthEncoding} scheme is very well-known --- albeit not in a columnar formulation --- and in extensive practical use in a variety of applicative fields, dating back as far as analog compression of US Television broadcast signals in the 1960s \cite{Cherry1963, RC1967}. The capping of run lengths, so as to limit the amount of information required to encode it, and the consequent introduction of additional runs of the same value, was discussed and put into practice already in those early days of digital telecommunications \cite[\S F]{Cherry1963}. The \cschemesh{RPE}{RunPositionEncoding} scheme is not as frequently used; this is likely because it is less suited for communication streams, being anchored to some starting position. Also, the decompression penalty when using \cschemesh{RLE}{RunLengthEncoding} only occurs in a parallelized setting; with serial decompression, integration is just as easy as differentiation (neglecting numeric stability issues), slanting the trade-off mentioned above in favor of \cschemesh{RLE}{RunLengthEncoding}. The \cschemesh{RPE}{RunPositionEncoding} scheme has, however, recently proved quite useful for linear algebra computations on compressed data (see \cite[``Offset List Encoding'' in \S 4]{EBHRR2016}).
\end{note}

\begin{figure}[H]
   % ---------------------------------------------------
   % Examples for the RE, RLE, RPE compression schemes
   % ---------------------------------------------------

  \centering
  \begin{subfigure}[b]{0.15\linewidth}
    \centering
    \includegraphics[scale=1.0]{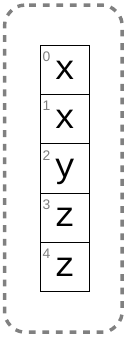}
    \subcaption{\centering Uncompressed column}
    \label{subfig:scheme:column:example:rlpe:decoded}
  \end{subfigure}
  %\hspace{0.05cm}
  \begin{subfigure}[b]{0.275\linewidth}
    \centering
    \includegraphics[scale=1.0]{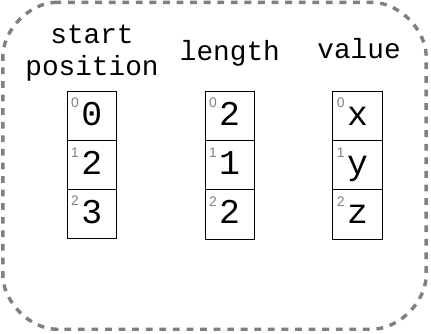}
    \indexCompressionScheme{RunEncoding}
    \subcaption{\capucscheme{RunEncoding}\\ \mbox{}}
    \label{subfig:scheme:column:example:rlpe:rlpe}
  \end{subfigure}
  %\hspace{0.05cm}
  \begin{subfigure}[b]{0.225\linewidth}
    \centering
    \includegraphics[scale=1.0]{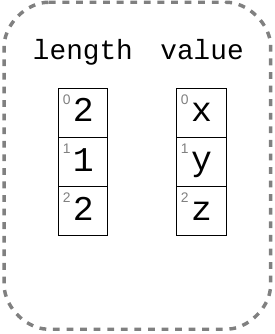}
    \indexCompressionScheme{RunLengthEncoding}
    \subcaption{\capucscheme{RunLengthEncoding}\\ \mbox{}}
    \label{subfig:scheme:column:example:rlpe:rle}
  \end{subfigure}
  %\hspace{0.05cm}
  \begin{subfigure}[b]{0.3\linewidth}
    \centering
    \includegraphics[scale=1.0]{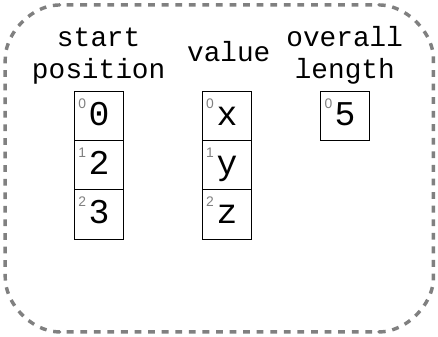}
    \indexCompressionScheme{RunPositionEncoding}
    \subcaption{\capucscheme{RunPositionEncoding}\\ \mbox{}}
    \label{subfig:scheme:column:example:rlpe:rpe}
  \end{subfigure}
  \caption{A column and its compression with \capucscheme{RunEncoding} and related schemes.}
  \label{fig:scheme:column:example:rpe}
\end{figure}

\begin{figure}[H]
   % ---------------------------------------------------
   % Decoder & Encoder for the RPE compression scheme
   % ---------------------------------------------------
  \centering
  \begin{subfigure}[b]{0.4\linewidth}
    \centering
    \includegraphics[height=7.5cm]{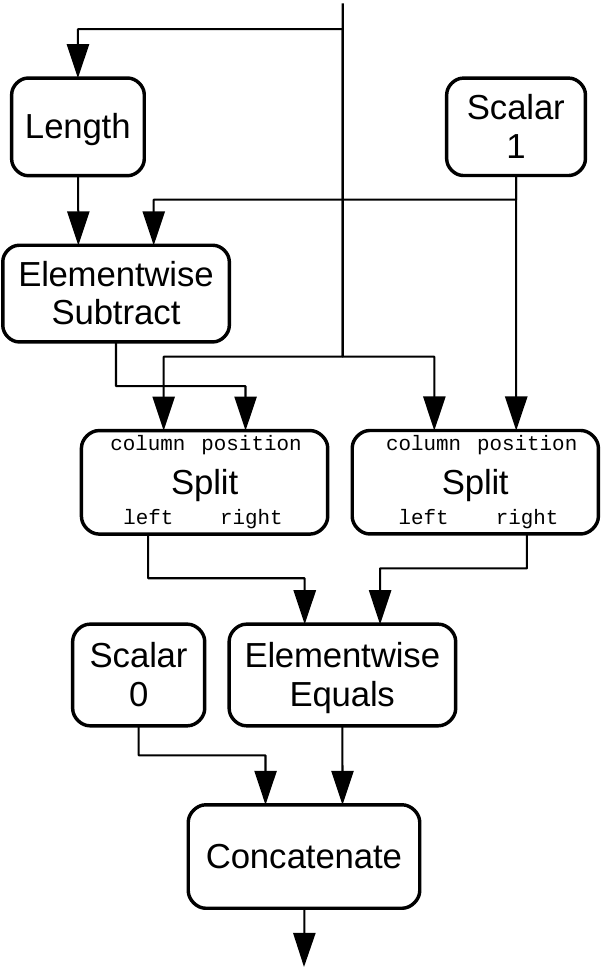}
    \indexPlanOperator{Split}
    \indexPlanOperator{Length}
    \indexPlanOperator{Elementwise}
    \indexPlanOperator{Concatenate}
    \subcaption{An implementation of the \mbox{\uplanop{IsSameAsPrevious}} operator}
    \label{subfig:operator:is-same-as-previous:circuit}
  \end{subfigure}
  \hspace{0.25cm}
  \begin{subfigure}[b]{0.3\linewidth}
    \centering
    \includegraphics[height=7cm]{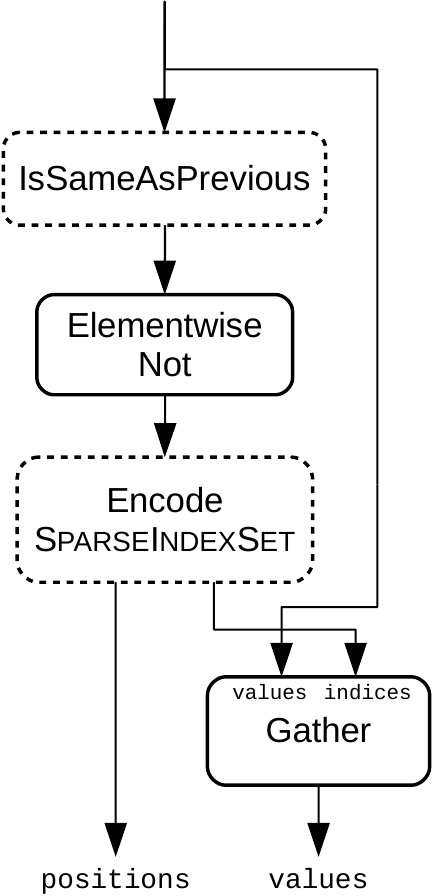}
    \indexPlanOperator{Gather}
    \indexPlanOperator{Elementwise} % Elementwise Not
    \subcaption{\ucscheme{RPE} encoder \\ \mbox{}}
    \label{subfig:scheme:column:encoder:rpe}
  \end{subfigure}
  \hspace{0.25cm}
  \begin{subfigure}[b]{0.2\linewidth}
    \centering
    \includegraphics[height=6cm]{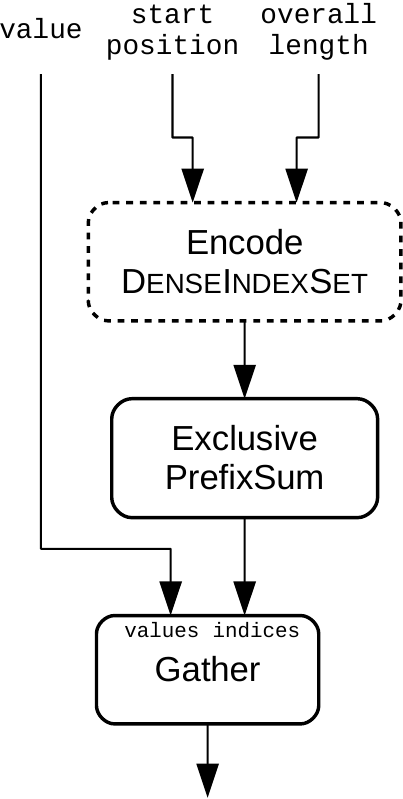}
    \indexPlanOperator{PrefixAggregate}
    \indexPlanOperator{Gather}
    \subcaption{\ucscheme{RPE} Decoder \\ \mbox{}}
    \label{subfig:scheme:column:decoder:rpe}
  \end{subfigure}
  \caption[Compressor and decompressor for the {\normalfont\ucscheme{RunPositionEncoding}} scheme]{Compressor and decompressor for the {\normalfont\ucscheme{RunPositionEncoding}} compression scheme}
  \label{fig:scheme:column:rpe}
\end{figure}

\medskip

Finally, we consider two variants of \cscheme{RunEncoding} and its related schemes, involving further segmentation:

\paragraph{Fixed-width pre-segmentization} A significant difference between the columnar formulation of \cschemesh{RLE}{RunLengthEncoding} and its use in practice has to do with the common practice of holding columnar data in segments, ``blocks'' or ``chunks'' (see \autoref{sec:segmented-columns}). Despite \cscheme{RunEncoding}, \cschemesh{RPE}{RunPositionEncoding} or \cschemesh{RLE}{RunLengthEncoding} being inherently segmented, this is a \emph{variable-length} and \emph{data-dependent} segmentation; to better reflect existing systems, we need to consider the application of an artifical fixed-width segmentation to a column before applying one of the run-based schemes. The price of this pre-segmentatization; Non-negligible decompression overhead, and no chance for exceedingly-high compression ratios with extremely-long runs. But there is also a significant gain: In addition to the bounding of search lengths, the capping of run length and the fact that position offsets are relative to the segment, so both \cschemesh{RLE}{RunLengthEncoding} and \cscheme{RPE}{RunPositionEncoding} now require less bits per run --- which changes the trade-off between them somewhat in favor of \cschemesh{RPE}{RunPositionEncoding}.

\paragraph{Implicit sub-segmentation} In this variant we stick to the single segmentation inherent in \cscheme{RunEncoding}\relax-like schemes: The runs. Instead of adding a second segmentation, we  break up existing runs into smaller runs --- so that it becomes also a sub-segmentation of a uniform $\ell$-segmentation; that is, whenever a run contains both the $\paren{i \* \ell - 1}$\xth and the $i \* \ell$\xth element - we break it into two runs on the length-$\ell$ ``segment'' boundaries. The decompression circuits for \cscheme{RunEncoding}/\cscheme{RLE}/\cscheme{RPE} can be used as-is for these schemes. Also note that applying this sub-segmentation to \cschemesh{RLE}{RunLengthEncoding} column after this refinment, and adjoining $\ell$ as a scalar, results in an instance of the \cschemesh{CappedRLE}{CappedRunLengthEncoding} scheme mentioned above.

\subsection{Splines: segmentized generated functions}
\label{subsec:splines}

We've just explored compression schemes using the segmentization of \cscheme{Constant} --- \ucscheme{RunEncoding}, \ucscheme{RPE} and \ucscheme{RLE} --- representing columns as consecutive uniform-value runs. Recalling that columns are simply functions, we've in fact been discussiong representations of \emph{step functions}. In \autoref{subsec:splines}, we generalized from constant functions to elements of linear spaces of basis functions, such as low-degree polynomials; and we now do the same in the segmentized context. Indeed, we will use the well-studied generalization of step functions (to low-degree polynomial functions): The \emph{spline}.

In fields such as computer game graphics, industrial design and typography, one often wishes to model curves (or surfaces in 3-dimensional space) using polynomial functions. However, high-degree polynomials are harder to reproduce physically and slower to compute; also, they typically exhibit high-amplitude oscillation between their points of constraint, particularly near edges of the intervals of definition (Runge's phenomenon \cite[\S 4.3.4]{DB1974}). It is thus usually more convenient to subdivide the domain, and use multiple low-degree polynomial pieces, one per segment, rather than a single high-degree polynomial. Historically, their use arose from physical splines: pieces of wood used in craft design which would bend into the minium-energy form, while being bound by physical constraints of position and angle at their endpoints. For an exposition of the historical development and varigated uses of splines, see \cite{DeBoor2001}.

A similar principle is relevant for data compression also: If the data has a certain shape locally, be it a gradually-changing shape or with points of discontinuity --- this overall tendency or shape would be captured relatively well by a spline, at a certain resolution of segmentation (whether fixed or variable, and with or without continuity and smoothness constraints).

The proper segmentization of $\cscheme{Generated}_{\vec{f}}$ yields the following scheme:

\begin{cschemes}
 \csitem*[\cscheme{GeneralizedSpline}_{\vec{f}}]
  \begin{cslisting}
    $\colname{segment_start_pos}$  & $\inttype$   & $m$      & of the first element of each segment\\
    $\colname{segment_length}$     & $\inttype$   & $m$      & Non-negative number of elements in each segment \\
    $\colname{coefficients}$       & $\tau$       & $m \*k$  & \makecell{ a concatenation of $m$ sequences of $k$ coefficients for the \\
                                                                          basis functions in $\vec{f}$, forming a different function for each \\
                                                                          segment. }
  \end{cslisting}
 \vbox{
 \csitem[Spline] The case of $\cscheme{GeneralizedSpline}_{\vec{f}}$ where the basis functions are monomials, i.e. $f_i(x) = x^i$.
  }
\end{cschemes}

These schemes have a somewhat redundant representation --- similarly to the case of \ucscheme{RunEncoding}. We may compress it further by dropping one of the two columns of the segmentation.  Here we spare ourselves the dilemma by following the traditional formulation of splines, which focuses on \emph{knots}, the pointes delimiting the segments:

\begin{cschemes}
  \csitem*[\cscheme{KnottedSpline}_k]
  \begin{cslisting}
    $\colname{knots}$              & $\inttype$   & $m+1$    & start/endpoints of segments \\
    $\colname{coefficients}$       & $\tau$       & $m \*k$  & \makecell{ a concatenation of $m$ sequences of $k$ coefficients for \\
                                                                          the polynomials $\itemrng{x^0}{x^{k-1}}$ }
  \end{cslisting}
  The first knot is always $0$ and the last is $n-1$, with $n$ being the uncompressed column length.
  \csitem*[\cscheme{EquiknottedSpline}_k]
  \begin{cslisting}
    $\colname{interval_length}$    & $\inttype$   & $1$      & the distance between consecutive knots \\
    $\colname{coefficients}$       & $\tau$       & $m \*k$  & \makecell{ a concatenation of $m$ sequences of $k$ coefficients for \\
                                                                          the polynomials $\itemrng{x^0}{x^{k-1}}$ }
  \end{cslisting}
  The uniform-segmentation equivalent of $\cscheme{KnottedSpline}_k$. The last interval may be shorter than the uniform length.
\end{cschemes}
and we could similarly define knotted versions of $\cscheme{GeneralizedSpline}$.

Spline functions are technically a generalization of step functions --- which are captured by the segmentized \cscheme{Constant} scheme. And yet, in practical use, it is in fact the step-function-describing \ucscheme{RPE} and \ucscheme{RLE} schemes that are effectively the more general, or rather more widely applicable directly. This is due to their being much more ``forgiving'' towards diversions from the modeled function: If some elements do not fit the current (constant) step/run value - we may simply add a new run for it, with only local overhead --- another run, and possibly breaking up an existing run in the middle. With splines, one needs quite a few coefficients. Also, in practice, one does indeed encouter long stretches of identical values; and possibly some long arithmetic sequences (degree-1 splines); but column data being perfectly quadratic or cubic over some stretch are rare occurrence. Thus, paradoxically, one would tend not to use the generalized scheme as-is, but rather combine it elementwise with small additive ``noise'' column (as discussed in \autoref{subsec:elementwise-composition}).

\paragraph{Smooth splines} Spline functions are often constrained to be continuous curves, i.e. constrained to be continuous at the internal knots. Stronger smoothness constraints are also sometimes introduced, so that the polynomial ``pieces'' have the same direction at the point of joining (e.g. to avoid jagged edges on the surface of a car or side of a ship, whose contour is set to a spline); and further constraints are sometimes placed on the higher derivatives. These constraints create linear dependencies between the coefficients of the polynomials (or outright equality constraints if we choose another polynomial base, as in B-splines \cite[Chapter 10]{DeBoor2001}) - reducing the number of parameters necessary per segment. One would typically still opt to keep the full sequence of redundant coefficients for each knot, as failing to do so requires reconstructing them with a pass over previous segments.

\paragraph{Splines with non-integral knots} Monomials --- the basis functions for the \cscheme{Spline} scheme --- can be applied to any positive real value rather than just to the (integral) column indices. The same would often be true for other choices of $\vec{f}$ base functions. With this in mind, one could specify \emph{non-integral} spline knots --- using some fixed-width floating-point type for the knot positions --- in the schemes above, or a non-integral interval length for equi-knotted splines and generalized splines. The column values are the evaluations at offsets which are integral relative to each other, but non-integral relative to the knot. This may reduce the complexity of fitting optimal splines to data.

\paragraph{Frame-of-Reference}
\label{par:segmentized-frame-of-reference}
The composition of $\cscheme{EquiknottedSpline}_1$ with elementwise additive differences (see ``noise-signal'' separation in \autoref{subsec:elementwise-composition}). At every segment, the additive differences can be thought of as offsets from a \emph{reference value} or a \emph{baseline value} for that segment --- the uniform value of the step function. This special case is commonly known as:

\begin{cschemes}
\csitem[FrameOfReference]
  \begin{cslisting}
    $\colname{segment_length}$  & $\inttype$           & $1$ & \\
    $\colname{reference}$       & $\tau$               & $1$ & \\
    $\colname{offsets}$         & $\tau_\text{offset}$ & $n$ & \\
  \end{cslisting}
  The $i$\xth element of the uncomrpessed column is $\colval{offsets}{i} + \colval{references}{i / \colname{segment_length}}$.
  (Shorthand name: \IntroduceCompressionSchemeShorthand{FOR}{FrameOfReference}.)
\end{cschemes}
The \cschemesh{FOR}{FrameOfReference} has been in use by several analytic DBMSes (e.g. Teradata \cite{Morri2002} and Vectorwise \cite{ZHNB2006} composed with aposteriori patching).

\subsection{Compressed Differences}
\label{subsec:segmentized-differences}

Consecutive elements in real-world database table columns are often close to their predecessors: Timestamps; samples from real-world data sources with low energy in high frequencies; arbitrary data after having undergone a sort; and so on. In such cases, each element's value is a fairly decent estimate for the value of the next one --- even if a model for the entire column is not available or is complex. When these element-to-element differences are all integral (or of fixed precision), they may fit a smaller-size type, and the following scheme achieves meaningful compression:

\begin{cschemes}
\csitem[NaiveDelta]
  \begin{cslisting}
    $\colname{base}$  & $\tau$               & $1$ & Arbitrary value \\
    $\colname{delta}$ & $\tau_\text{diff}$  & $n$ & \makecell{Differences between consecutive uncompressed elements, with $\setsize{\tau_\text{diff}} < \setsize{\tau}$ }
  \end{cslisting}
  The first element of the $\colname{delta}$ column holds the difference between the first uncompressed element and the \colname{base} value (so that the $\colname{base}$ value is not uniquely determined by the uncompressed column). 
\end{cschemes}

\begin{figure}[H]
   % --------------------------------------------
   % Examples for the 'Delta' encoding scheme
   % --------------------------------------------
  \centering
  \begin{subfigure}[b]{0.275\linewidth}
    \centering
    \includegraphics[scale=0.8]{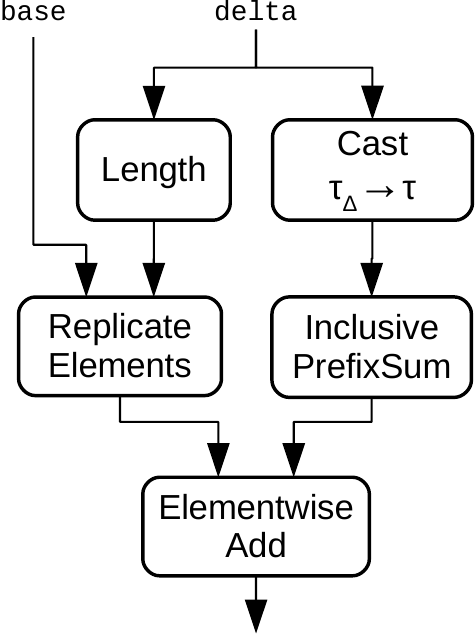}
    \indexPlanOperator{PrefixAggregate}
    \indexPlanOperator{Elementwise}
    \indexPlanOperator{Length}
    \indexPlanOperator{Replicate}
    \subcaption{\cscheme{NaiveDelta} scheme decoder \\ \mbox{}}
    \label{fig:scheme:column:decoder:naive-delta}
  \end{subfigure}
  \hspace{0.0cm}
  \begin{subfigure}[b]{0.275\linewidth}
    \centering
    \includegraphics[scale=0.8]{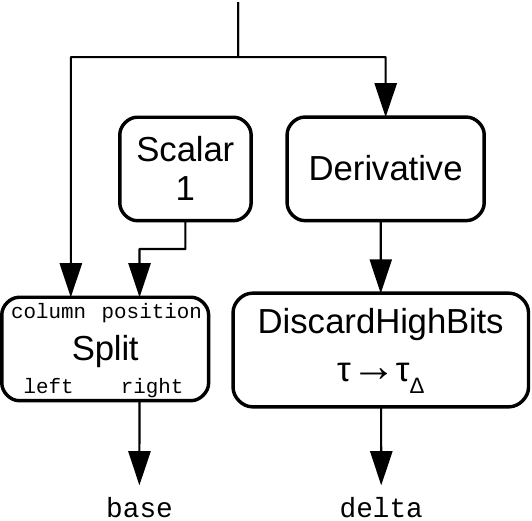}
    \indexPlanOperator{Scalar}
    \indexPlanOperator{Split}
    \indexPlanOperator{Elementwise}
    \indexPlanOperator{Concatenate}
    \subcaption{\cscheme{NaiveDelta} scheme encoder \\ \mbox{}}
    \label{fig:scheme:column:encoder:naive-delta}
  \end{subfigure}
  \hspace{0.1cm}
  \begin{subfigure}[b]{0.2\linewidth}
    \centering
    \includegraphics[scale=1]{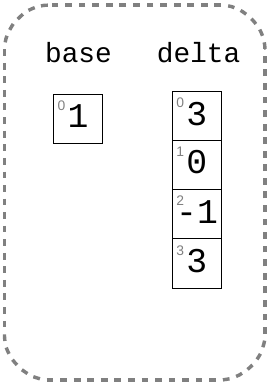}
    \subcaption{A \cscheme{NaiveDelta}-compressed column}
    \label{fig:scheme:column:example:delta:compressed}
  \end{subfigure}
  \hspace{0.15cm}
  \begin{subfigure}[b]{0.15\linewidth}
    \centering
    \includegraphics[scale=1]{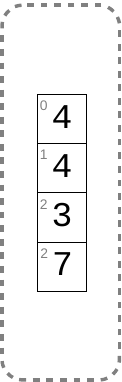}
    \subcaption{The uncompressed column}
    \label{fig:scheme:column:example:delta:uncompressed}
  \end{subfigure}
  \caption{The \capucscheme{NaiveDelta} compression scheme}
  \label{fig:scheme:column:naive-delta}
\end{figure}

As defined, \cscheme{NaiveDelta} is not a very practicable scheme: The computation of each decompressed element depends on all previous input elements --- possibly millions or billions of them. It is at this point that segmentation comes into play: Instead of a change in any compression scheme parameters per se, we apply it to \cscheme{NaiveDelta} to break the chain of dependencies; dependencies will now extend only to the segment boundary rather than all the way to the beginning of the column.

\begin{cschemes}
\csitem[Delta]
  \begin{cslisting}
    $\colname{segment_length}$  & $\tau$         & $1$                                      & \\
    $\colname{base}$            & $\tau$         & $\ceiling{n / \colname{segment_length}}$ & one value per segment\\
    $\colname{delta}$           & $\tau_\Delta$  & $n$                                      & \makecell{ Differences between consecutive uncompressed \\
                                                                                                         elements in a segment }
  \end{cslisting}
  The first element of each segment, i.e. $\colval{delta}{\colname{segment_length} \* i}$ for the ${i}\xth$ segment, holds the difference between the segment's first uncompressed element and that segment's \colname{base} value, $\colval{base}{i}$.
\end{cschemes}

\medskip

An alternative to segmentizing \cscheme{NaiveDelta}, is characterizing the behavior of the differences \emph{globally} rather than \emph{locally} --- with model functions as in \autoref{subsec:splines} above. But that is not, in fact, a different compression scheme --- as a characterization of the derivative as a span of basis functions is the same as a characterization of the underived function, with the anti-derivative of those basis functions (up to a global constant). Thus, \cscheme{NoisyGenerated} may be a reasonable alternative to \cscheme{NaiveDelta}; and while both can be segmented, it will be less essentiall to do so for the latter scheme; \cscheme{NoisyGenerated} would also be much easier to decompress, with its work being elementwise.

\paragraph{Patching segmentized differences} The long dependency chains created by representing differences rather values remains somewhat problematic even with segmentatization. In addition to the limits of the parallelization of decompression, the use of (aposteriori) patching also becomes difficult. To illustrate, suppose $\tau, \tau_\Delta$ are integeral types and that  $\opsup{\fdomain{\tau_\Delta}} < M$. Now suppose we wish to compress a column made up of $k$ occurrences of $0$, followed by at least $k$ occurrences of $(k+1)M$. One can show that at least $k$ patches are necessary to bring all consecutive pairs of elements to within $M$ of each other; consequently, at least $k$ aposteriori patches must be made to a \cscheme{Delta}-compressed form of this column. If, instead, we were to patch the \emph{compressed} column --- after casting it into $\tau$ rather than $\tau_\Delta$ values --- this would require only a single patch at index $k$.

The adjoining a subcolumn in standard representation to the compressed form of \cscheme{Delta}, of patches to be applied as we've just described, defines another compression scheme --- \cscheme{PatchedDelta} scheme or \ucscheme{PDELTA} for short. A non-columnar equivalent of this scheme is defined in \cite[\S 3.1]{ZHNB2006} as \ucscheme{PFOR-Delta}.

% \begin{cschemes}
% \csitem[PatchedDelta]
%   \begin{cslisting}
%     $\colname{segment_length}$   & $\tau$         & $1$                                      & \\
%     $\colname{base}$             & $\tau$         & $\ceiling{n / \colname{segment_length}}$ & one value per segment\\
%     $\colname{delta}$            & $\tau_\Delta$  & $n$                                      & \makecell{ difference between consecutive uncompressed \\
%                                                                                                          elements }
%     $\colname{patch_positions}$  & $\tau$         & $m$                                      & \\
%     $\colname{patch_values}$     & $\tau$         & $m$                                      & \\
%   \end{cslisting}
%   The $\delta$ column contains the difference between the corresponding element \emph{in the corresponding segment of the uncompressed column} and its predecessor in that segment --- except $\colval{delta}{\colname{segment_length} \* i}$ for $i \in \naturals$, which hold the difference between the first uncompressed element in a segment of the uncompressed and that segment's \colname{base} value (i.e. $\colval{base}{i}$.
% \end{cschemes}

\subsubsection{Segment-local dictionaries}
\label{subsec:segment-dictionaries}

Segmentation allows mid-column changes in a compression scheme's parameters. In \namecref{subsec:segment-dictionaries} above we described the localized modeling of a column in a space of low-degree functions --- fitting a different model for each segment. We now turn to a different assumption of local ``well-behavedness'': An assumption regarding the column \emph{support set}.

For example, consider a column of type $\tau$ with a poor (discrete) modulus of continuity, for which one cannot provide a well-fitting model using typical base functions (e.g. polynomials), even if one were to allow for a narrow noise column. But suppose also, that despite its discontinuous nature, the column's support set \emph{does} exhibit locality: There exists a neighborhood width $N < \setsize{\typedom{\tau}}$ around each position in the column, such that the local support size in each of these neighborhoods is significantly less than $N$. %Suppose that the overall support size is large (otherwise we could simply apply \cscheme{Dictionary} first, see below).

To exploit this locality, we break up the column into contiguous segments, each of which with its own hopefully-small local dictionary: Not too large, so as to limit the size of the support set; but not too small, so as not to incur excessive storage overhead for the dictionaries, nor hurt the decompression speed overmuch by having to read them. Keeping one dictionary for each of these segments results in the following compression scheme:

\begin{cschemes}
\csitem[UniformSegmentDictionaries]
  \begin{cslisting}
    $\colname{segment_length}$          & $\inttype$   & $1$      & with positive elements \\
    $\colname{dictionary_entries}$      & $\tau$       & $d \* \colname{segment_length}$
                                                                  & Entries for \emph{all} dictionaries, concatenated\\
    $\colname{indices}$                 & $\inttype$   & $n$      & ... into the respective segment's dictionary \\
  \end{cslisting}
  Decoded similarly to \cscheme{Dictionary}, except that the dictionary entry index is not $\colval{indices}{i}$, but rather $\colval{indices}{i} + \floor{i / \colname{segment_length}} \* d$.
\end{cschemes}

This local-dictionary scheme is used essentially as-is, albeit not in columnar formulation, in existing systems (e.g. Oracle \cite{PP2003} and Vertica \cite{LFVTVDB2012}). It also has quite a few possible variations, or further developments:

\paragraph{De-duplication} We can constrain the per-segment dictionaries to have no duplicate values and/or to be sorted --- allowing for the local-dictionary indices to serve as surrogates; these would be the same constraints described in \autoref{subsubsec:de-duplication-and-use-of-surrogates} for the case of a single all-column dictionary. Note, though, that the surrogacy in the segment-dictionaries case would be limited to the relevant segment: The comparison of indices into dictionaries of different segments does not yield a meaningful result. %This cann ; nor can this be rectified by replacing $\colval{indices}{i}$ with $\colval{indices}{i} + \paren{i / \colname{segment_length}} * d$ --- as there is definitely duplication within the concatenated dictionaries. To address this issue a proper decompression (i.e. a lookup in the dictionary) is necessary.

\paragraph{Patching} The segmented dictionaries scheme relies on complete disappearances of elements from support sets from segment to segment. In some real-life cases, this does happen; for example, when batches of data in a column are inserted in large batches from different ``categorical sources'', where the support for different categories is distinct (e.g. records involving activities in one country and records involving another country). But if the column's support has a more continuous behavior, disappearance will not be complete; Instead, we're more likely to see supports with the vast majority of weight on a small set of values, and some weight on a large set of very-infrequent values. In other words, the segment support sets will not be small, but the frequency functions of values in the column will have almost all of their weights on a small set each. To be able to ignore the distributions' ``tails'', we may use patching. The patches can theoretically be themselves segmentized rather than being column-global , but it is not clear whether this segmentation is particularly useful, so all-column patching seems more attractive.

\paragraph{Variable-length segments} We've made an off-hand choice of using a \emph{uniform} segmentation with the local dictionary, but we can also opt for a non-uniform segment length (see \autoref{sec:segmented-columns}). Such a choice may be relevant if the changes in the local support set are themselves non-uniform, i.e. if there are long stretchs where the support set hardly changes, and others where it changes at a faster pace. Remember, however, that instead of adding another level of indirection, one can also consider \emph{patching} for some problematic parts of a column; increasing the dictionary size, effectively obtaining a slower change; or using a two-level dictionary (see below).

A particularly interesting scenario where the locality of support breaks is that of a discontinuity in the occurrence frequency function: After a certain element, the support set changes abruptly, perhaps even completely (and typically, this may not be aligned with a segment end). Variable-length segments seemlessly addresses this case --- but it can also be handled by a doubling of the dictionary size: If the discontinuities are at least $\colname{segment_length}$ elements apart, a double-sized dictionary could have both the current segment and the next segment's support set covered, disappearing the discountinuity. Again, this is a trade-off between read-after-read dependencies, bit width, alignment of values (the doubling means an extra bit) and other compression scheme parameters.

\paragraph{Two-level dictionaries} A drawback of \cscheme{UniformSegmentDictionaries}, as defined, is the amount of space necessary to store all of the dictionaries, especially when the segment length is not very high relative to the dictionary size (due to relatively weak support set locality). Also, a global support set which is itself compressible, or when the original column has very large elements or variable width --- adversely affect the storage of segment dictionaries in terms of complexity, size or both. Both these issues may be addressed by composing \cscheme{UniformSegmentDictionaries} after an application of \cscheme{Dictionary} to the uncompressed column. The first level of dictionary use --- the global dictionary --- results in fixed-width integers, regardless of the original element type; and pre-reduces their sizes in the second-level dictionaries -- the segment dictionaries.

Such a two-level scheme (with de-duplication and sorting) is used in the Google PowerDrill column store \cite[\S 2.3]{HBBGN2012} (again, in a non-columnar formulation). Specifically, schema columns are physically partitioned into ``chunks'', each one being much more complex than a plain column.

%Even if the global support set covers the full domain of the column's element type ,

\paragraph{Compression of the segment dictionaries}
The overhead of storing a dictionary per each segment poses a dilemma: If our segments are short, we may need to store so many dictionaries that the space taken up by $\colname{dictionary_entries}$ is of the same order of magnitude as the space taken up by the $\colname{indices}$ column, compromising the compression benefit of the segment dictionaries in the first places. On the other hand, if segments are long, the dictionary for each segment must be larger, as we lose some of the effect of locality; and while the dictionary's size is offset by the lower number of dictionaries --- the dictionary index width will increase as well, dragging the overall compression ratio down.

One can conceive of several ways of circumventing, or resolving, this dilemma:

\begin{itemize}
 \item Expand the dictionaries slightly (hopefully without increasing the dictionary index size) beyond the size necessary for our fixed choice of segment length. Now, construct `union dictionaries' of the dictionaries of consecutive segments, which can be used for stretches of several consecutive segments. We could then apply \cscheme{RunEncoding} or \cschemesh{RLE}{RunLengthEncoding} to avoid duplicating these dictionaries.
 \item
 %Expand the dictionaries slightly, as in the previous item, so that hopefully each of them may accommodate several
 %and
 Make some effort to reduce the number of distinct dictionaries (e.g. by creating ``segment union dictionaries'' as in the item above).
 Avoid the repeated use of the same dictionary \emph{using a dictionary-of-dictionaries}, i.e. adjoin a column with one $\inttype$ element per each segment of $\colname{indices}$, indicating the index of the dictionary to use for that segment; the distinct dictionaries, having some fixed size, are concatenated in $\colname{dictionary_entries}$.
 \item We've been treating the different dictionaries as perfectly distinct stretches of data, despite them being placed together in the same column. We could remove this constraint, and \emph{allow a dictionary to start at an arbitrary position} within the $\colname{dictionary_entries}$ column. Thus less-frequent elements at the beginning of a dictionary may be discarded by moving its starting index up, but not all the way, so that many or most elements remain. The dictionary length will also be set so as to introduce the appropriate amount of new elements, and possibly also to re-introduce elements which were lost at the beginning but are still necessary; at the same time, we'll need to avoid enlarging it too much to avoid widening the indices into it.

 This approach complicates the compression somewhat, as one must tackle the combinatorial challenge of choosing a dictionary index width for which the exist a satisfying placement of elements: Such a placement must include enough redundant appearances of values to capture segments' support, but not too many appearances so as to spoil the overall effect of dictionary overlap.

 \item Instead of storing entire dictionaries, \emph{store the differences} between consecutive dictionaries (i.e. which elements were added and which removed);  due to the locality property, and the choice of smaller segment length, consecutive dictionaries mostly overlap (if we ignore placement within the dictionary), on the average. Thus, on average, their differences should take up little space. For more on the use of differences in compression, see \autoref{subsec:segmentized-differences} above.

 ... and if this option reminds the astute reader of Intra-frames and P-frames in MPEG video compression \cite{DL1996} --- the analogy is not without merit.
\end{itemize}

%This note is useful, but it doesn't fit well with the narrative flow
%\begin{note}
%A dictionary-of-segment-dictionaries is, in essence, a sub-scheme of the composition of \cscheme{FrameOfReference} after \cscheme{Dictionary}: Instead of using the actual global dictionary entries, we have a per-segment ``baseline value'' --- the offset within the concatenated dictionary entries column to apply to the indices actually stored. The slight difference is that the reference value is not stored as-is, but is divided by $d$, the dictionary size.
%\end{note}

\section{Exploiting non-uniformity in frequency distributions}
\label{sec:exploiting-non-uniform-frequencies}

In seeking features to exploit for compression, and particularly in considering a column's support set, we have so far mostly ignored the frequencies of values appearing in the column: We've ve treated each of them as equally worthy of compression, with an equal potential benefit from compression; and the only allowance for infrequency we've presented is patching, which is useful removing extremely-infrequent elements entirely. In other words, we've mostly ignored the non-uniformity of columns' \emph{frequency distribution} (as per \autoref{def:column-frequency-distribution}).

We begin with a classic (non-columnar) approach to exploiting such non-uniformity for coding --- using elementwise variable-length coding of the originally fixed-length data, so that frequent elements use less bits. Then, in \autoref{subsec:cascading-dictionaries}, we consider a conceptual of ``transposition'' of the above, where instead of having one variable-width column, we have a multiple, uniform-width columns, but of decreasing lengths, i.e. where some elements ``end'' in an earlier column. Finally, we focus on a single aspect, or step, of the multiple-column approach, involving a two-way partition of the frequency distribution function, that is useful in itself as a building block for compression schemes, and is in a sense the complement of patch removal; this is  \autoref{subsec:small-dictionary-fitting}.

\begin{note}
In information-theoretic terms, in this section we exploit low \emph{(Shannon) Entropy} \cite[\S 2.1]{CT2012} of columns' element distributions.
\end{note}

\subsection{Huffman-like encoding of variable-width columns}
\label{subsec:huffman-and-hu-tucker}

Given a column with highly non-uniform element frequencies, it may be worthwhile to apply an elementwise approach of exploiting this non-uniformity: Having frequent elements be represented in less space and infrequent elements taking up more space than in the fixed-width representation. This is an extremely common technique in communication and (serial, non-columnar) compression, particularly using the Huffman code \cite{Huffman1952} for representing elements; this code is optimal for symbol-for-symbol prefix codes (see \cite[\S 5.8]{CT2012}. A similar choice, for the additional constraint of preserving the relative order of the unencoded values in the encoded form, is the Hu-Tucker code \cite{HT1971, Hu1982}.

But what is a straightforward in a serial context is not to be considered lightly in parallelism-focused computational computation: Variable-size rather fixed-size representation of data comes at a cost, of either extra \emph{space} or extremely poor amenability to parallel decompression. With our standard representation of variable-width column, see \autoref{subsec:variable-width-columns}) we pay in a significant amount of extra space, to indicate element positions or lengths. Without those indications, however, it would generally not be possible to start inspecting the variable-length data at some mid-point and determine where an uncompressed element starts or ends; decompression would therefore have very long dependency chains to follow over numerous compressed-form column elements.

Two approaches for achieving a middle ground between minimum size and more shallow data dependencies are:
\begin{itemize}
 \item  Use of periodic, rather than per-element, position/length indication (a-la \cscheme{FrameOfReference} or \cscheme{Delta}); between these indicators, decompression is almost impossible to parallelize, in that one cannot determine where elements start or end by examining them from some midpoint between the indicators.
 \item  Interspersing \emph{synchronization elements} between actual compressed column elements, which can be identified with certainty even when the input is read starting at an arbitrary position. Without delving into the details, applying this code-theoretic concept would be based on the exploration of the subject in \cite{FR1984, Titchener1997}; but would differ due to our ability to pre-transform the column to enforce certain features or remove certain artifacts, using patching; this may allow for simpler decompression, or for an improved upper bound on the compression ratio relative to the classic case.
\end{itemize}

Finally, note that compression using variable-width columns can exploit more than the mere \emph{element} frequency distribution, but rather than frequency distribution of pairs, or longer sequences, in the uncompressed column. This would bring it somewhat closer to Lempel-Ziv-style compression schemes \cite{LZ1978}.

\subsection{Cascading dictionaries}
\label{subsec:cascading-dictionaries}

We wish to benefit from using a shorter representation for more frequent elements --- but perhaps we are loath to sacrifice so much of the uniformity and regularity of fixed-width columns, which is required for Huffman-like coding. In this case, we may adopt a compromise: Fixed-width columns, but several of them, or rather, several complementary subcolumns --- each with a different width. Of course, the representation in each width would be a fixed-width number, or more specifically: An index into a Dictionary. Elements will be assigned indices to these subcolumns (and given a dictionary entry) based on the frequencies in the uncompressed column.

Consider the first, narrowest dictionary, $D_1$; its index width is $b_1$ bits, and it covers an $\eps_1$ fraction of the column's values. One of its entries, however --- we'll choose entry 0 for simplicity --- is reserved for indicating that an element is not covered by the dictionary. We may think of the \colname{indices} column of the \cscheme{Dictionary} scheme also as a dense subcolumn representation (if we were to check it for equality with 0, elementwise). After these elements have been added to the dictionary and the problem resolved, we remain with the task of compressing merely the $1-\eps_1$ fraction of the values which do not appear in $D_1$, as a contiguous column. If the residual subcolumn still meets the criteria of this section (non-compactable support set, roughly obeys a power law) --- the process may be repeated: Dictionary $D_2$, with $b_2$ bits covering $\eps_2$ of the subcolumn; and again, a new residual subcolumn is formed, to be compressed further.

\begin{cschemes}
 \csitem*[\cscheme{CascadedDictionaries}_k]
  \begin{cslisting}
     $\colname{dictionary}_1$          & $\tau$             & upto $2^{b_1}$  & Entries of the $1\xth$ dictionary \\
     ...                               & ...                & ...             & ... \\
     $\colname{dictionary}_k$          & $\tau$             & upto $2^{b_k}$  & Entries of the $k\xth$ dictionary \\
     $\colname{indices}_1$             & $b_1$-bit integers & $n$             & \makecell{into $\colname{dictionary}_1$; an index of 0 indicates \\
                                                                                          elements to be set in later phases.}\\
     ...                               & ...                & ...             & ... \\
     $\colname{indices}_k$             & $b_k$-bit integers & $n_k$           & into $\colname{dictionary}_k$.  \\
  \end{cslisting}
  Decompression occurs in $k$ phases, each using two circuit inputs columns, \idxcolname{dictionary}{i} and \idxcolname{indices}{i}, and an additional, intra-circuit, input \idxcolname{residual}{i-1} --- one of the outputs of the previous phases. The phase produces two outputs: \idxcolname{residual}{i}, for use in the next phase, and a sparse-representation subcolumn $\idxcolname{sc}{i} = \tuple{\idxcolname{pos}{i}, \idxcolname{data}{i}}$. The phases can be interpreted as successive application of \cscheme{SubcolumnOverlay}. The final phase' $\idxcolname{residual}{k}$ output is empty, and is not used.

  The $i$\xth phase gets the positions of those output column elements which have not yet finalized, from \idxcolname{residual}{i-1}. It partitions this set of indices into two subsets: Indices to be finalized in this phase --- \idxcolname{pos}{i} --- and indices left for later phases --- \idxcolname{residual}{i}. The partition uses \idxcolname{indices}{i}: An index \idxcolval{residual}{i-1}{j} is placed in \idxcolname{residual}{i} if $\idxcolval{indices}{i}{j} = 0$, and in $\idxcolname{pos}{i}$ otherwise. The phase concludes with \cscheme{Dictionary} decoding of $\idxcolname{pos}{i}$ and $\colname{dictionary}_i$.

  An initial implicit column, $\colname{residual}_0$, is the identity column for $\zeroupto{n}$, i.e. initially, all elements are still not finalized. The disjoint subcolumn union $\Union_{i=1}^{k}\idxcolname{sc}{i}$ constitutes the decompressed column in default subcolumn representation.
\end{cschemes}

\begin{figure}[H]
  % -----------------------------------
  %  CascadedDictionaries}_k decoding
  % -----------------------------------
  \centering
  \begin{subfigure}[b]{0.25\linewidth}
    \centering
    \includegraphics[scale=0.6]{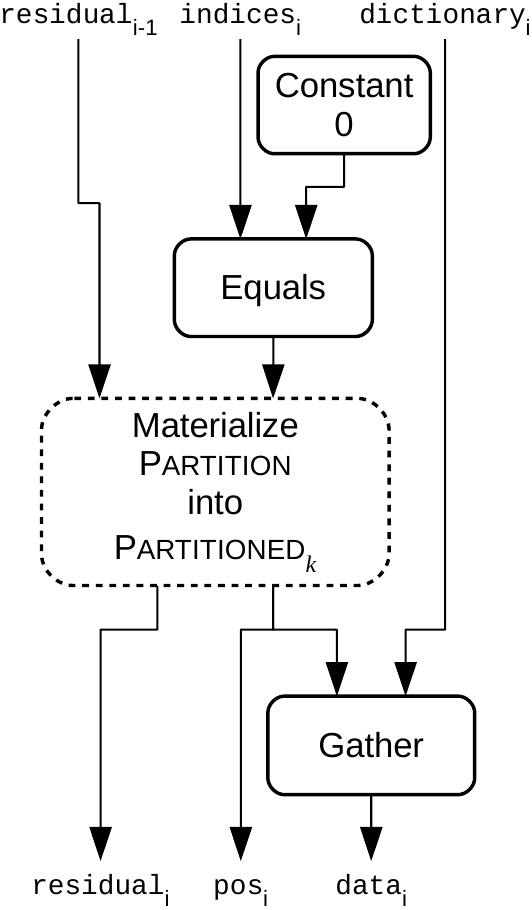}
    \subcaption{A single phase of the decoder (see \autoref{subsec:partitions} regarding partition materialization.) }
    \label{subfig:scheme:column:decoder:cascaded-dictionary:decoder-phase}
  \end{subfigure}
%  \hspace{0.05\linewidth}
  \begin{subfigure}[b]{0.7\linewidth}
    \centering
    \includegraphics[scale=0.6]{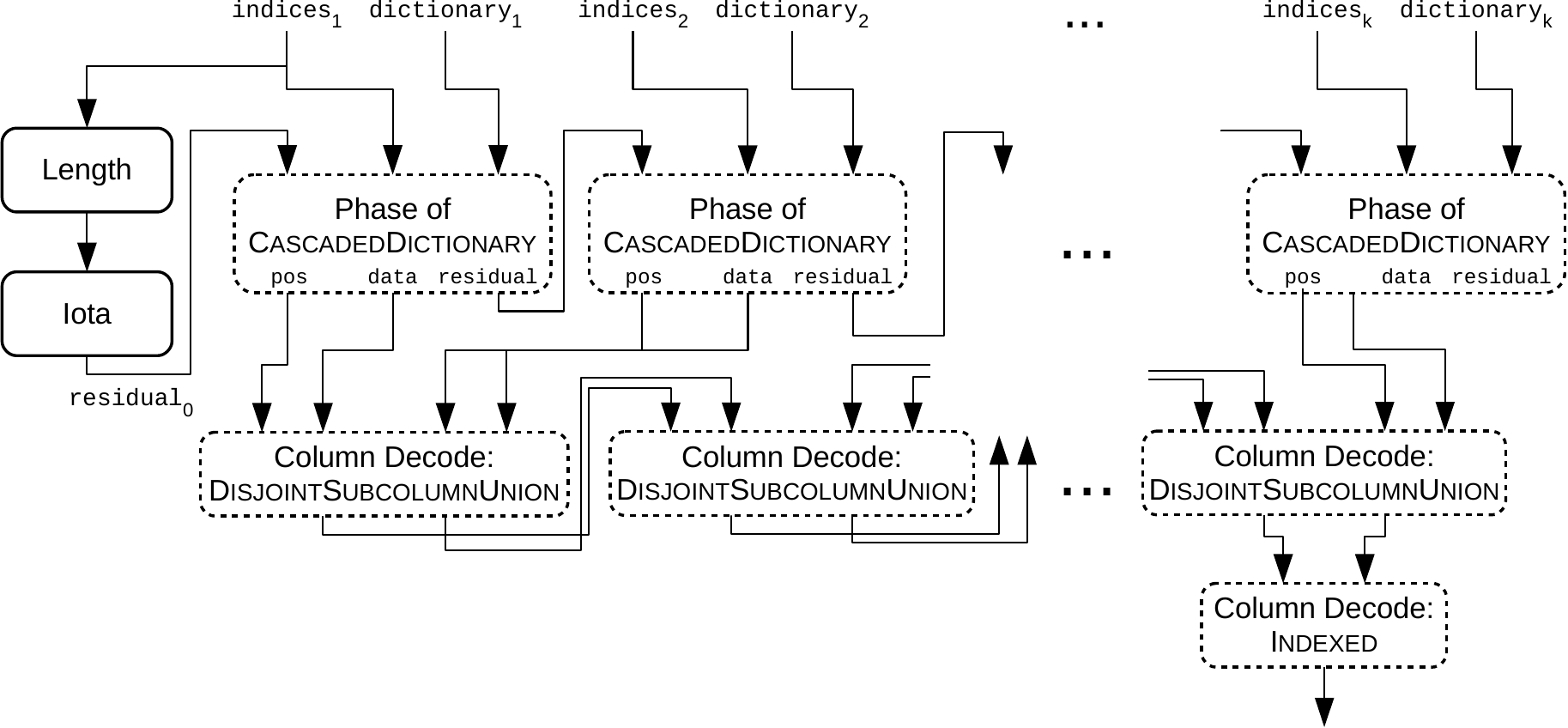}
    \subcaption{The full-blown ($k$-phase) decoder \\ \mbox{} \\ \mbox{}}
    \label{subfig:scheme:column:decoder:cascaded-dictionary:full-decoder}
  \end{subfigure}
  \caption{Decoder and subcircuit implementations for the \capucscheme{CascadedDictionaries} compression scheme}
%  \caption{A full $k$-phase decoder for the \capucscheme{CascadedDictionaries} compression scheme}
  \label{fig:scheme:column:decoder:cascaded-dictionary}
\end{figure}

We have not described an exact procedure for \emph{partitioning} the elements into $k$ subsets. Naturally, the earlier dictionaries involve the more frequent elements; but the number of dictionaries and their index sizes in bits must still be determined --- a process in which one may trade off a better compression ratio for fewer dictionaries and thus shallower dependencies of the decompressed column.

\begin{note}
The idea of partitioning columns by element frequency, to be able to use shorter-width dictionary indices for more frequent elements has already been adopted in lightweight compression in a DBMS context: IBM has used a ``frequency partition''-based compression scheme in its BlinkDB system\cite[\S III]{IBM2008}. The scheme is rather different from \cscheme{CascadedDictionaries}: BlinkDB is a row store, so the scheme there is applied to tuples; also, it compresses without preserving tuple order, so that the relative position of the various tuples is not encoded. With this said --- it is an interesting composite scheme also involving \cscheme{Delta}, which could be expressed in terms of the basic schemes and composition patterns presented in this chapter; and the discussion of partition formation \cite[\S 3.b]{IBM2008} is also of interest.
\end{note}

\subsection{Small-dictionary fitting}
\label{subsec:small-dictionary-fitting}

While the \ucscheme{CascadedDictionaries} scheme discussed in the previous section, is certainly of some utility, its definition involves something possibly even more useful: The single phase of its decompression, with a decoder as per \autoref{subfig:scheme:column:decoder:cascaded-dictionary:decoder-phase}. Instead of its repeated application on residual columns, let us apply it just once, and assume the decoded values for the residual are available explicitly:

\begin{cschemes}
 \csitem[SubcolumnDictionary]
  \begin{cslisting}
     $\colname{dictionary}$          & $\tau$            & up to $2^b$                                          & Dictionary entries \\
     $\colname{indices}$             & $b$-bit integers  & $n$                                                  & Indices into \colname{dictionary} or $0$ if uncovered \\
     $\colname{residual_data}$       & $\tau$            & $\setsize{\finverse{\colname{indices}}\paren{0}}$    & data for elements in $\finverse{\colname{indices}}\paren{0}$
  \end{cslisting}
  When decompressing,  $\colname{dictionary}$ and $\colname{indices}$ are treated just as in the \cscheme{Dictionary} scheme: $\colname{decompressed} \leftarrow \colname{dictionary} \circ \colname{indices}$ --- except for the elements of $\finverse{\colname{indices}}\paren{0}$; these have individual values provided, in order of their position, via the $\colname{residual_data}$ column.
\end{cschemes}
We can indeed formulate $\cscheme{CascadedDictionaries}_k$ as a repeated composition of \cscheme{SubcolumnDictionary} with itself, $k$ times --- with the appropriate subcolumns and bit widths $\vec{b}$), and dropping the empty final residual subcolumn.

\medskip

\cscheme{SubcolumnDictionary} alone will not yield a good compression ratio unless the dictionary covers almost all column elements. But even if it leaves out a sizable fraction, it is a prime candidate for composing with other encoding schemes, a procedure we refer to as \emph{small-dictionary fitting}. Such a separation of a column into an ``easier'' and ``harder'' part immediately reminds us of the patching procedure (described in \autoref{subsec:patching}); in some senses, they can be seen as each other's converse:

\begin{center}
%\begin{table}[H]
% \centering
 \resizebox{0.9\linewidth}{!}{%
 \setlength\extrarowheight{2pt}
% \label{tbl:comparison-of-patching-and-small-dict-fitting}
 \begin{tabu}{   >{\bfseries\raggedright}p{0.38\linewidth}   >{\raggedright}p{0.35\linewidth} >{\raggedright}p{0.35\linewidth}  }
 \toprule
 \rowfont{\bfseries}
                                                            & Patching                                      & Small-dictionary fitting \\
 \midrule
 Rationale for use                                          & Remove the infrequent, ``difficult'' elements & Remove the frequent elements \\
 Intended scope of application                              & Minimum necessary                             & Maximum possible \\
 Number of elements remaining to compress after application & The full column's length                      & Less than the full column length (hopefully much less) \\
 % Elements inconveient to compress reside in               & the smaller subcolumn                         & the smaller subcolumn \\
 Compression ratio of affected data                         & Pretty bad (typically worse than uncompressed) & Pretty good (typically better than rest of data) \\
 Overhead likely proportional to                            & Number of exceptional elements                & Total number of column elements \\
 Repeat application                                         & Useless                                       & Often relevant \\
 Scheme used for composition                                & \cscheme{SubcolumnOverlay}                    & \cscheme{SubcolumnDictionary} \\
 \bottomrule
\end{tabu}
}
\end{center}
Small-dictionary fitting can therefore take its place alongside the fundamental, or commonplace, scheme composition and transformation patterns in \autoref{sec:compression-scheme-composition-patterns}.

\section{Further compression of scheme-generated auxiliary columns}
\label{subsec:reducing-redundancy}

\subsection{Patch positions --- sparse index subsets}
\label{subsec:encoding-patch-positions-with-less-redundancy}

Patching, described in \autoref{subsec:patching} and used in several concrete schemes presented in this chapter, has a very steep space overhead per patched element: We need to store not only the full-size uncompressed element, but also an index into the uncompressed column. And the more effective the main compression scheme, and the longer the column --- the more ``painful'' this extra overhead of specifying the patch position becomes. As a consequence, naive patching --- a subcolumn of patched value in \cscheme{SparseIndexSet} representation --- is only beneficial with very low fractions of values to patch in a column.

Luckily, index subsets which aren't extremely sparse tend to be amenable to some compression, due to two characteristics: They can be arranged in monotone-increasing order; and the denser they are, the lower the arithmetic differences between subsequent indices are on the average, even well below the density at which one might consider switching to a dense representation.
\medskip

To simplify the exploitation of these two properties, we focus on \emph{long common prefixes} of indices' binary representation --- since if subsequent indices are close together, then on the average (though with slightly lower probability) they share a long common prefix. Indeed, we can define a family of representation schemes concretizing the prefix-agreement by grouping together the representation of identical prefixes:

\begin{operators}
  \opitem*[\planop{Carve}_{w,p}]
    Let $\kappa_{w, p}$ be the function carving $w$-bit values into their higher and lower $p$ and $(w-p)$ bits respectively --- that is, defined by $x \mapsto \paren{x \bmod{2^p}, \floor{x / 2^p} }$. $\uplanop{Carve}_{w,p}$ is an alias for $\uplanop{Elementwise}_f$ with $f = \kappa_{w,p}$.
\end{operators}

\begin{cschemes}
 \csitem*[\cscheme{CommonPrefix}_{w,p}]
  \begin{cslisting}
    $\colname{prefix}$        & unsigned integer, $p$ bits   &  $m'$  &  each to be prepended to multiple suffixes  \\
    $\colname{suffix_count}$  & unsigned integer, $w-p$ bits &  $m'$  &  Number of suffixes for each common prefix  \\
    $\colname{suffix}$        & unsigned integer, $w-p$ bits &  $m$   &  each preceded by the appropriate prefix    \\
  \end{cslisting}
  With this scheme, (encodings of) indices appear in order of index value. As consecutive indices tend to share their higher (more-significant) bits, we apply a $\uplanop{Carve}_{w,p}$ operator, carving each width-$w$ integer into its higher (or prefix) $p$ bits and its lower (or suffix) $w-p$ bits is performed. The column of $p$ high bits of patch positions exhibits (hopefully long) sequences of identical values, so we apply \cschemesh{RLE}{RunLengthEncoding} to it. The lower bits column remains untouched, completing the encoding. This procedure is clearly reversible, up to the sorting --- but sorting maintains the equivalence of the standard representations of index sets, so it unnecessary when decoding.
\end{cschemes}

\begin{note}The reader may notice the similarity between the $\cscheme{CommonPrefix}_{w,p}]$ scheme and \cscheme{FrameOfReference}. But the two schemes are fundamentally different in that the former only represents \emph{index sets}, or ordered column of integers --- not arbitrary columns. Another difference is the non-uniform number of represented indices with each prefix, while in \cscheme{FrameOfReference} that number is fixed.
\end{note}

An index subset representation in this family may take up as many as $\paren*{p + (w-p) + (w-p)} = 2w - p$ bits per element. This worst-case occurs when $m = m'$, i.e. when each element has a different prefix: In this case, each element requires $w$ bits for the actual carved-up element and $w-p$ bits for representing the length of the one-element subset. This fact seems to motivate a shorter prefix, to ensure more indices end up in the same set --- contrary to the basic motivation of sharing longer prefixes among many index set elements.

%On the other hand, as more of the elements share the same prefix, the burden of representing the \colname{prefix} and \colname{suffix_count} on each of these elements decreases. 
%As the size of these two values can never exceed $2^p \* paren{p + (w-p)} = 2^p w$
%Thus, for $2^p = \littleo{m}$ (both as functions of $n$), we have a size per element of $\paren{1+ \littleo{1}} \* (w-p)$ --- mostly just the suffix.

%The former, extra-sparse case motivates choosing a higher value of $p$, as for $p = 0$ we again reach the sparse representation; but the less-sparse case motivates a lower value of $p$. 
We can balance the contradicting motivations for a shorter and longer prefix by using multiple distinct (and disjoint) index sets: Each encoded in $\cscheme{CommonPrefix}$ (or naively), with the same value of $w$ but different values of $p$. Choose, with foresight, two $p$ values: $0$ and $w/2$. This choice results in the following compression scheme for index subsets:

\begin{cschemes}
 \csitem*[\cscheme{CommonUpperHalf}_w]
 Each element of the subset to be compressed is placed in one of two index subsets. The first of these is in the naive representation (and corresponds to no common prefix), while the second is represented using $\cscheme{CommonPrefix}_{w,w/2}$ (we assume $w$ is even for simplicity). The choice of placement for the different values depends on the density of the original index set's intersection with each consecutive $2^{w/2}$-aligned sequences of $2^{w/2}$ elements (i.e. the subsets of elements which all share the same $w-w/2$ prefix): If an index value is the single one within its $2^{w/2}$-aligned sequence, it is placed in the first index subset; and if there are multiple elements, they are all placed in the common-$w/2$-prefix index subset. The subsets are disjoint, so a decoder can simply concatenate the decoded standard representations of both subsets to obtain the original one.
\end{cschemes}
For each one of the $2^p = \sqrt{n}$ possible prefixes, we have either: No elements or one element in the first index subset, with width $w$; or at least two corresponding elements in the \colname{prefix} and \colname{suffix_count} columns of the second index subset, whose widths are $w - w/2 = w/2$ each, i.e. $w$ overall; or none of the above. Thus, a total size of $\sqrt{n}  \* w$ bits. This does not account for the elements in \colname{suffixes} over the first two elements for each suffix; their size is bounded by $\colsize{suffixes} = m \* w/2$. The overall size per element is therefore $m^{-1} \* \paren{m/2 + \sqrt{n}} \* w = \paren{\half + \sqrt{n}} \* w$. If $m = \bigOmega{\sqrt{n}}$, the size per element is $\paren{1+\littleo{1}} \* w/2$ --- asymptotically half of the naive representation.

\medskip

Returning finally to the context of patching, suppose a column of length $n$, has $\eps n$ of its values requiring patches. With $\cscheme{CommonUpperHalf}_{\ceil{\log{n}}}$, the amortized per-element overhead (in bits) of storing patch indices is $\eps \* \paren*{\half+\littleo{1}} \* \ceil{\log{n}}$ --- so that, asymptotically, the patches require about half as much space as their naive representation. To make this more concrete, suppose that that $\eps n > 10 \sqrt{n}$ (which, for $n > 10^6$, means $\eps > 1\%$); the index representation size per patch element is then under $0.6 \ceil{\log{n}}$, rather than the full $\ceil{\log{n}}$ for naively-represented indices.

\begin{note}
Common prefixes are a sub-optimal form of clustering the indices together; and so is the choice of only length-0 and lengh-$w/2$ prefixes in \cscheme{CommonUpperHalf}. One could, instead, undertake the computation of an actually optimal choice of cluster reference points, and assignment of indices to the clusters. This would be a non-trivial special case of the min-max facility location problem: Even if the optimal number of clusters $k_\text{opt}$ is known in advance, the best known upper bound on the time complexity seems to be $\bigO{k_\text{opt} \* n^2}$ \cite{WS2011} --- which is extremely prohibitive for long columns. If we also allow for the possibility of switching from a sparse to a dense regime locally in some clusters (as in Roaring bitmap compression \cite{CLKG2016}), the time complexity would increase even further.
\end{note}

\subsection{Element length specifiers}
\label{subsec:less-redundant-variable-element-length-encoding}

Using a variable-length representation of a column's elements often yields significant compression ratio improvement; but in the standard representation, and other compression schemes, this is offset by the significant overhead of storing elements' start positions. In the most useful cases of variable-width columns, the frequent elements only use a few bits --- in which case the starting positions takes up more space than the data itself. Thus, to effectively compress variable-width columns, they themselves must be compressed.

Fortunately, the element lengths, taken as a column, are often amenable at least to some of the simpler compression schemes, or a combination thereof: The range of extant element length is usually limited, and only as many bits as the logarithm of its size is required. Achieving this might requires using something like \cscheme{Dictionary} in the general case, but often that's not necessary, since the possible lengths are very dense in a certain range, thus \cscheme{FrameOfReference} (see \autoref{par:segmentized-frame-of-reference}) followed by \cscheme{DiscardHighBits} produce a near-optimal compression. The segmentization of the \cscheme{FrameOfReference} scheme may also come in useful, if the length distribution changes gradually.

%. That might also be useful if the local range of values (and lengths) changes gradually. Thus composing some of these schemes over the standard representation (defined in \autoref{subsec:variable-width-columns}) and dropping its redundant \colname{start_positions} column, is not half bad already, especially when nothing is known about the data and its distribution.

\subsubsection{Shaping element lengths by padding}

Beyond the straightforward approach above, there's a special feature of some variable-length representations which may be utilized to further improve the compression of a column of their length indicators:
\begin{itemize}
 \item For \emph{unsigned integers}, in the ubiquitous base-2 bitwise representation, a representation can always be padded with 0's as the top bits without changing the value.
 \item The case of \emph{signed integers} with two's-complement representation is similar, but using the sign bit for padding.
 \item \emph{Text strings} are usually represented in character set encoding schemes which allow for indicating the end of the string with a special value; this is commonly the ``null character'', or the numeric value of $0$. It can be repeated past the end of the string for padding.
\end{itemize}
For these kinds of data, the representation of an element can always be extended to a longer length without altering the value. We can thus trade off extra bytes in the element data column for a better fit to compressible patterns in the element lengths column.

This upwards-flexibility of element lengths is exploited in several well-known, non-columnar lightweight compression schemes: The 4-Gamma and 4-Wise NS schemes element widths for consecutive sequences of 4 elements (to the maximum of the actual widths); SIMD-128 and SIMD-FastPFOR go as far as setting a single length for sequences of 128 elements \cite{DHHL2017}. These all incur a significant penalty in compression ratio, however, when the compressed data is not locally uniform in width. Combining this with the fact that variable-length representation only makes sense to begin with when the element distribution is somewhat power-law-like, one is led conclude that the approach of these schemes is somewhat questionable in terms of achievable compression ratio. As an example, consider a column in the distribution of length exhibits a Geometric distribution with parameter $p$ (with minimum value $1$ occurring with probability $p$); and that we set the maximum length in bits every $k$ consecutive elements. Letting $W_\text{max}$ be the maximum width of some sequence $k$ elements and $\idxrng{W}{1}{k}$ be their individual widths, we have

\begin{align*}
\expect{W_i} &= \expect{\text{Geom}(p)} = \expect{\text{Geom}(p)} + 1 = 1/p \\
%Note: The \mathstrut is intended for vertical spacing between the overset text and the = sign
\expect{W_\text{max}}
 &= \sum_{j=0}^{\infty} j \* \prob{W_\text{max} = j}
 \overset{\substack{\text{Fubini's} \\ \text{theorem\mathstrut}}}{=} \sum_{j=1}^{\infty} \prob{W_\text{max} \geq j} \\
 &= \sum_{j=1}^{\infty} \paren*{1 - \prob{W_\text{max} < j} }
 \overset{\footnotesize\text{independence}}{=} \sum_{j=1}^{\infty} \paren*{1 - \prod_{i=1}^{k}{\prob{W_i < j}} } \\
 &=  \sum_{j=1}^{\infty} \paren*{1- \paren*{\prob{\text{Geom}(p) < j }}^k }
  =  \sum_{j=1}^{\infty} \paren*{1- \paren*{1 - (1-p)^{j-1}}^k } \\
 &=  \sum_{j=0}^{\infty} \paren*{1- \paren*{1 - (1-p)^j }^k }
\end{align*}
Taking $k = 4$ (as in 4-Gamma, 4-Wise NS etc.) and $p = (1 - \sqrt[8]{0.1}) \abouteq 1 - 0,749894 \abouteq 0.25$ (i.e. 90\% of the data takes up at most one byte), we would have an expected $W_\text{max}$ of $\sim 7.738$ and expected single-element length of about $3$. That would be worse than doubling the amount of space used for the actual compressed elements. This is rather dismal! ... of course if the element lengths are more likely to be identical to the predecessors or successors, such compression schemes can make sense.

Let us, therefore, describe a columnar compression scheme in the same vein as 4-Gamma, 4-Wise NS, SIMD-128 and SIMD-PFOR \cite{DHHL2017}:

\begin{cschemes}
 \csitem[PeriodicallyVariableWidth]
  \begin{cslisting}
     $\colname{period}$   & $\inttype$   & $1$            & The period of element length setting \\
     $\colname{widths}$   & $\inttype$   & $\ceil{\frac{\colname{length}}{\colname{period}}}$
                                                          & \makecell{The length (in units of $\tau$) of the representation of each\\
                                                                      of \colname{period} subsequent elements of the variable-width \\
                                                                      column }\\
     $\colname{length}$   & $\inttype$   & $1$            & \makecell{Number of elements in the decoded variable-width \\ column } \\
     $\colname{data}$     & $\tau$       & $\sum_i{\colval{widths}{i}}$
                                                          & \makecell{Concatenation of the sequences of $\tau$ values \\
                                                                      constituting the data of each of the elements} \\
  \end{cslisting}
  Each sequence of $\colname{period}$ consecutive elements $\itemrng{\colname{period}\*i}{\colname{period}\*(i+1)-1}$ of the encoded variable-width column have the same size: They each take up $\colval{widths}{i}$ type-$\tau$ elements in $\colname{data}$. The same is true for the last few elements (which may be less than $\colname{period}$ in number). \colname{total_length} indicates how many variable-size elements the decoded ``column'' actually has. %(Shorthand name: \IntroduceCompressionSchemeShorthand{PVL}{PeriodicallyVariableWidth}.)
\end{cschemes}
%A decompression circuit for \cscheme{PeriodicallyVariableWidth} would likely use
%(requiring an elementwise padding operator; call it \planop{PadToUniformLength})

\paragraph{Uniform element lengths via separate subcolumns} The performance pitfalls of the schemes mentioned above (including \cscheme{PeriodicallyVariableWidth}), along with their requirement of computing a prefix sum over element lengths, motivates the consideration of alternative approaches. One that immediately comes to mind when we recall \autoref{subsec:cascading-dictionaries} is the ``transposition'' from variable-length elements to multiple (sub)columns of uniform width. In our case, we can reduce the number necessary columns by padding elements of less-frequent widths --- while elements of frequent lengths should not require such padding. The overall number of columns should thus be limited. The down-side of such a scheme is the need to store the position of all these elements. This detriment may also be reduced by some creative use of padding: It is cheaper (possibly much cheaper) to pad an element by another bit (or basic column element type) than to relegate it to a subcolumn of patches.

\section{Multi-column compression}
\label{sec:multicolumn-compression}

This chapter has mostly been concerned with compression schemes for a single column (with the half-exception of variable-width columns). Using single-column schemes on a table yields one separate and independent compressed representation for each one of the columns; but this separation is often wasteful in terms of space, as some table columns are usually very well correlated. Also, a column store may have use for a copy of several columns in lexicographic order of the tuples they form (see \autoref{sec:choosing-between-schemes} below). It is thus worthwhile to consider intricating the compression schemes for different columns, to avoid some of the redundancy. Following is a scale of the intensity of such potential intrication:

\begin{enumerate}
 \item[0.] (\emph{Independent representations}: Compression and decompression of each column completely disregards other columns. This is the naive approach, directly following our discussion in this chapter thus far.)
 \item \emph{Decoder fusion}: When columns known to be related are decoded (decompressed), the disjoint union of their decoding circuits is considers as a decoding circuit for a column pair. In this circuit, one may opportunistically locate and remove identical, duplicate operators shared by the different decoders (or apply more involved optimizing transformations). Encoding may or may not be performed jointly also (``enconder fusion''), but that is of relatively minor concern.
 \item \emph{Derivative representation}: Some columns are compressed under the assumption that certain other related, uncompressed columns are available. At worst, this results in the independent compression scheme; hopefully, only little additional data is necessary over the pre-existing column. When decompressing, the base columns for the derivation, are decompressed first, realizing the assumption.
 \item \emph{Co-representation}: One chooses compression schemes for multi-column structures, e.g. pairs or tuples of several types --- from the outset. The columns are also decompressed all together.
\end{enumerate}
As one proceeds along this scale, compression ratio and speed for related columns \emph{together} drops, but for the individual columns it mostly rises.

%This scale constitutes yet another trade-off between, roughly, simplicity and speed of decompression o vis-a-vis smaller compressed size: When performing a columnar computation (e.g. executing an analytic database query) with independently-compressed columns, a column store will read and decompress exactly those columns necessary for the execution plan; and have the maximum flexibility in scheduling the work of reading, decoding and the rest of the execution, as would best fit that plan. With the other options --- additional reads are necessary, additional computation is necessary, and the data dependencies may also introduce constraints on scheduling and the distribution of work.

% On the other hand, the more the decompression of columns is allowed to depend on other columns and their decompression process, the better the potential of representing them with less specific information, and of avoiding repeated work during decompression.

\medskip

So as to limit the breadth of this monograph, we will not explore concrete multi-column schemes and different patterns of combining single-column schemes, except for a single simple case, immediately below. The reader is reminded that the general problem of optimally combining compression schemes for multiple columns is computationally intractable --- just like determining the optimal compression scheme for a single column: These are the problems of computing the plain, conditional and/or shared Kolmogorov complexity of columns as words over a finite alphabet. For an in-depth (abstract) treatment of these fundamental theoretical problems, see \cite{LV2008}.

\paragraph{Full derivation} Real-life data tables very often have columns which can be perfectly inferred from other columns --- even at the single tuple level. For example, a time-series table may have a \colname{date} column, alongside \colname{day_of_week}, or possibly \colname{year}, \colname{month} and \colname{month} as separate columns (e.g. in \cite{percona-blog-comparison}). %Another example would be a columnar representation of a dense matrix or a tensor: The axis coordinates are likely to each have a column of its own; and if the record index corresponds to a lexicographic ordering of the matrix or tensor cells, these columns' values can be
More generally, such columns tend to show up in databases not in the Third Normal Form (3NF, see \cite{Kent1983}).
%or with multiple candidate keys (as defined in \cite{Codd1972, LO1978}).
A column $\colname{c}$ is \emph{fully-derivable} from a family $B$ of columns if there is a columnar circuit which, for some assignment of columns from $B$ to its inputs, produces $\colname{c}$. If, the circuit is just a lifted $\planop{Elementwise}_f$ operator, the column is \emph{elementwise-derivable from $B$}.

The notion of full derivability can be abused if arbitrary operators are allowed in the circuit, as one can simply define an operator which outputs the desired column; but an appropriate constraint for excluding this case is difficult to formulate, seeing how column stores may generate operators dynamically. Elementwise derivability is a useful restriction of the general definition, both because the prevalence of such columns and because the elementwise-applied function is less amenable to such edge-case abuse. Typically, these functions correspond to multivariate arithmetic expressions with some deterministic mathematical functions available in SQL (such as \sqlcode{SIN()}, \sqlcode{SQRT()} and the like).

% ---------------------------------------
% ---------------------------------------
% ---------------------------------------
% ---------------------------------------
% ---------------------------------------

\chapter{Applying the model: Prospective notes}
\label{chap:applying-the-model-notes}
%\addcontentsline{toc}{chapter}{\nameref{chap:applying-the-model-notes}}

The columnar circuit model of computation we have presented in this work is intended for analysis of existing column stores systems --- but also as a foundation for developing new ones. This chapter focuses on what a column store might look like when designed with the model of computation in mind, and which aims to support and benefit from features more consistently and fully. \Autoref{chap:columnar-compression} explored one aspect of these prospects --- the use of compression --- in greater detail: This was a demonstration of how the model applies to a subject whose study so far has been somewhat fragmented and piecemeal; and which is possible to present in a a relatively self-contained manner. Other aspects of a column store system: Compilation, plan optimization, choices between representation schemes, mutability, execution flow and so on --- would require further, more concrete, work to be explored with confidence. Instead, this chapter will touch on these aspects more speculatively, and in brief;
%, focusing on the effect of intensive to adoption of the columnar circuit computation model.
these prospective notes may hopefully indicate paths towards actual implementation.

\section{Avoiding materializations: Operator fusion via JIT compilation}
\label{sec:jit}

A reader concerned with optimizing column store performance may become worried as they browse the previous chapters of this work: Ostensibly, this model suggests a column store materialize results after each and every operator in a circuit. Doing so is highly detrimental to query processing performance, and was a main motivation for the evolution of Vectorwise (initially ``X100'') out of MonetDB \cite{ZBNH2005}. To make matters worse, the columnar circuits tend to be composed of simpler operators than in existing column stores --- increasing the number of intermediate materializations even further.

This stands in contrast to the increasing prominence in recent years of DBMSes effecting compilition of query execution plans into native code : \cite{Neumann2011} and the implementation in HyperDB \cite{KN2011}; MS SQL Server's Hekaton engine \cite[\S 2.1.3]{Microsoft2013}, the MemSQL DBMS \cite[\S 1.1]{MemSQL2016} and others (see also the discussion in \cite{TER2018}). While these are not column stores, the motivation is similar: Avoiding a multitude of separate operators applied to ever element or tuple, with unnecessary data movement back and forth. A larger plan made of more operators seems to be the opposite of desirable.

Paradoxically, the copiousness intermediate materialization is intended to achieve the opposite of its immediate effect: The adherence to simple, uniform data structures and the simplicity of the operators maximize execution plans' amenability to optimizing transformations. Many of these let a column store avoid some computation altogether, or use alternative approaches to the computation of a subcircuit; but afer the system is done with these kinds of optimizations, numerous materializations still remain. It is important to have some final ``bake-in'' transformations of the resulting circuit: Inverting the decomposition impetus of our approach in previous chapters, these will take subcircuits made up of multiple simple operations and fuse them into single, complex, customized operators, which typically materializes only its final result.

We have, in fact, already described the formal machinery for effecting the above. In \autoref{sec:circuit-composition-and-transformation}, we formalized the transformation of circuits by replacing an induced subcircuit with another circuit having the same function. We noticed the special case of ``lifted operator'' single-node circuits, and  defined the fusion of a subcircuit into a lifted operator. There remains the question of where the implementations of these complex operators are to be found.

The simple, naive approach is for a large set of operators compiled into (optimized) machine-executable form to be determined \emph{statically}, apriori: The DBMS' executable files have their final implemented forms before any queries arrive; and complexities such as SQL UDFs are be handled by an interpreter (or an external compiler run \cite{c-cpp-udfs-in-monetdb}). Indeed, pre-compilation of operators is the approach column stores have taken, by and large, thus far: MonetDB, Vectorwise and C\=/Store have all of their machine code compiled apriori, with no additional (``Just-in-Time'') compilation for individual queries. %The author's personal impression of the source code of MonetDB and VectorWise is that doing so is somewhat cumbersome; and of course, the materialization

But while for some column stores apriori-compiled operators may be a baseline option --- with the columnar circuit model, and the compositional approach previous chapters have taken --- this approach is a non-sequitor: The variety of potentially fused subcircuits is enormous. Furthermore, transformation rules and even some representation schemes may be devised with \emph{reliance} on the elision of materializations. Thus it is absolutely necessary to generate operator implementations --- by JIT compilation of a higher or lower-leve language, or by using compilation results of different pieces of code.

%so it is not merely a matter of further optimization using JIT compilation, but effectively a necessesity.

%with JIT compilation as a potential feature to consider

%The author's personal impression of the source code of MonetDB and VectorWise is that doing so is somewhat cumbersome. But even had it been simpler --- for supporting subcircuit fusion in the columnar circuit model, and with
%, even under stricter constraints than those we've imposed.
%In defining our model, we have not even fixed a set of available operators --- a decision which would also determine which subcircuits can be fused. We have also not provided any description of whence such operators, or operator implementations, are to originate in a real-life system.
%(also refered to as \emph{``Just-In-Time''} compiling query)
%Instead,
%a column store needs to generate or instantiate operators \emph{dynamically} --- either by JIT compilation per se,
%compiling higher-level program code, or by fusing compiled or semi-compiled code from different sources. That is, to fuse subcircuits, a column store needs to perform of (sub)circuits into operators, or rather operator implementations.
%This capability is almost inherent to systems implemented over a JIT-employing virtual-machine (such as the Java VM \cite{PVC2001}); but is available, with some effort, in statically-compiled languages as well, using established compilation frameworks such as LLVM \cite{LA2004}.
\medskip

Let us consider a concrete example. A simple, archetypical scenario for compiling execution-plan-specific code is treated in \cite{Neumann2011} (in a non-columnar setting; and it is implemented by the DBMS, HyperDB \cite{KN2011}): A circuit includes two consecutive elementwise operators, $\planop{Elementwise}_f$ and  $\planop{Elementwise}_g$; and the column store intending to execute this circuit has compiled implementations for both individual operators, not of $\planop{Elementwise}_{\fcompose{f}{g}}$;
but it is able to fuse the machine code for both operators, with the result being an operator applying $\fcompose{f}{g}$ to the input. Or perhaps --- neither $\planop{Elementwise}_f$ nor $\planop{Elementwise}_g$  is available, but the code for $f$ and code for $g$ on individual elements is, and the column store can generate a parallelized ``for all elements'' operator from the code for a scalar. Banking on this ability, the column store may replace the subcircuit containing $\planop{Elementwise}_f$ and $\planop{Elementwise}_g$ with the lifted single operator $\planop{Elementwise}_{\fcompose{f}{g}}$, maintaining semantic equivalence and circuit output correctness.

There are, however, several important differences between compilation-based operator fusion in the columnar circuit model, and the JIT compilation in aHyPerDB-like system approach:

\begin{itemize}
 \item \textbf{Non-tree circuits}: Columnar circuits are not, in the general case, decomposable into compilable pipeline-segments in the sense of \cite[\S 3.1]{Neumann2011}; specifically, columnar circuit layouts are DAGs rather than trees, so that the same intermediate column(s) may be used multiple times for different purposes.
 \item \textbf{Fusion after parallelization}: In HyPerDB, operator fusion is a per-record level concept (as in the example of elementwise operators above; although operators like \planop{Gather} can also be thought of as being ``per-record''). With our model, operators can have complex parallelization in their implementation, while still being relevant for fusion. We must therefore allow for more complex memory access patterns in ``non-pipeline-breaking'' operators (again in the sense of \cite[\S 3.1]{Neumann2011}) --- as otherwise, many of the transformations presented in previous sections would preclude JITing: We would not be able to replace a non-pipeline-breaking operator with an implementation containing ``pipeline-temporary-breakers'' .
 \item Perhaps most importantly: In existing systems employing operator fusion (and in HyPerDB particularly), much of the computational work is not represented in execution plans: Application of statistical meta-data, combination of hot and cold structures into a single column, specifics of decompression, etc.; such work is therefore not amenable to fusion or JITing. Our model significantly increases the potential for fusion and JIT compilation of these kinds of work, especially as parts of overall query execution plans.
\end{itemize}

A related, but different, approach to operator fusion can be found in the more recent \cite{FBNMT2018}. The differences are fundamental enough to make a comparison of this approach to HyperDB-style JIT'ing exceed the scope of this monograph (albeit interesting to consider). But this approach is less general than HyPerDB's, which in turn is less general than fusion in our model.

\medskip

Finally, recall that both before and after subcircuit contraction, a column store has a columnar circuit to execute. While such circuits expose the potential for parallelism relatively well, the exploitation of this potential on the hardware side is ever more difficult as one considers operators of increasing complexity, on different processing devices, with multiple devices operating at once and cluster-level parallelism. The more sophisticed this exploitation becomes, the more challenging it is to support effective operator fusion. Unlike the trivial case of elementwise work, it may require adherence to certain programming conventions; and perhaps also instrumentation of the compilation chain to keep track of how pieces of code can fit together. Consider the following prototypical example, regarding the complication of even simple considerations: Suppose an operator $\uplanop{Op}$ has a GPU implementation involves the execution of two distinct kernels (which cannot be fused together), while a previous $\uplanop{Op}'$ in the execution plan, to be fused with $\uplanop{Op}$, is implemented using just one kernel. Should the fused operator employ three kernels, or two? Or maybe it's even just one, if the fusion obviates somehow the need for two kernels for $\uplanop{Op}$? And if it's two kernels, which of them will actually get the first operator's code fused? Such complications abound.

\section{The choice of compression schemes as a nexus of preprocessing effort}
\label{sec:choosing-between-schemes}

A column store with elaborate machinery for representation and compression of columnar data faces a challenge in choosing (or devising) compression schemes for its columns. Even for some single criterion for evaluating schemes, and even when capping the complexity of the scheme (e.g. in terms of compression and decompression circuit sizes and available operator set) --- the task is already rather daunting. This is both due to the size of the search space --- thousands upon thousands of potential schemes already with the above restrictions --- and the difficulty of the very choice between pairs of candidate schemes.

For example, consider an column \colname{c} of element type $\inttype$, which is close, but not identical, to an affine function, both in average distance and the maximum difference betwee elements --- i.e. both by the $L^1$ and by the $\infinitynorm$ metrics; and suppose we wish to optimize the compression ratio, disregarding all other scheme features.
%(i.e. $\mathop{\text{max}}_{i \in \fdomain{c}}{\abs*{\colval{c}{i} - a\*i + b}}$ for constants $a,b$.
Which compression scheme would yield a better ratio for column $\colname{c}$: \cscheme{Delta}, or a \cscheme{GeneratedPolynomial} with degree 1 (modeling $i \mapsto a\*i +b$) ,with elementwise offsets? If \colname{c} is perfectly affine, both schemes would have essentially the same compression ratio: An all-zero column of elementwise differences vs. an all-zero column of elementwise offsets; both could use an extra composition with the \cscheme{Constant} scheme, which would make the representation ``infinitely'' efficient. However, we \emph{do} have some differences from the affine function; which should we choose? The best ratio (ignoring the application of further compression schemes) would depend on the extremal offset values (for the noisy affine model scheme), and on the maximum second discrete derivative (for the \cscheme{Delta} scheme). And even realizing that these parameters are to be examined --- the dilemma is not nearly resolved, since composition of further compression must be considered. Specifically, if one examines the support sets for the first descrete derivative (i.e. the \colname{delta} column) and for the offsets in the \cscheme{GeneratedPolynomial} scheme, respectively --- one of them may be small enough to make an application of \cscheme{Dictionary} worthwhile; or the support set may mostly have weight on a small set of values, allowing for one of the approaches described in \autoref{sec:exploiting-non-uniform-frequencies} for exploiting the distribution skew? And so on.

This confounding state of affairs, already within limited search spaces for schemes, has led other researchers to approach the selection of a compression scheme as DBMSes typically approach the choice of a query plan: Rule-based pruning and strategic selection of composition steps for improvement of metrics or achievement of features; see \cite[\S 4.1]{FHL2010}.

% These dilemmas seem to justify the approach discussed briefly in \cite[\S 4.1]{FHL2010} for choosing a (parallelism-friendly) compression scheme, involving the computation of various properties of the column for estimating the efficacy of different compression schemes: sortedness (or Hamming-closeness to being sorted); support set size; distribution of values over the domain, and distributions of low-order derivatives; run lengths; and so on. Also, the benefits we have observed arising from composing schemes are often \emph{non-gradual}, i.e. taking a front of the compression circuit and considering the size of the intermediate results it produces, one often exceeds the size of the original data. Consequently, a combination of the following should be necessary for choosing an effective compression scheme:

\medskip

But Let us venture even farther, and lift the caps on scheme complexity and consideration criteria. In previous chapters of this work we've described quite a few (structural) features of a column store which, without the restriction, may induce additional dimensions of complexity to the the choice of compression schemes for the columns in a store:

\paragraph{Compressed-form execution} A key desirable aspect of compression schemes used in DBMSes is their being ``lightweight'' --- easy to decompress despite exhibiting attractive compression ratios. This has guided the entirely of our exploration of columnar compression schemes in \autoref{chap:columnar-compression} above. We have, however, hinted --- and made fully explicit in \autoref{sec:jit} above --- that we hope to avoid materializing intermediate results as much as possible; and decompressed columns are intermediate results of our columnar circuits when the inputs are compressed. In other words, we would rather \emph{avoid} actually decompressing our columns.

%to actually \emph{avoid} much (or most) of the work of materializing columns. When a column is already materialized, the fusion of subcircuits taking it as an input (see \autoref{sec:jit}) helps avoid further materializing intermediates of similar size; but for a compressed column, we would hope to \emph{entirely} avoid materialization of the decompressed form.

Several existing column stores have a partial capability to do this, via the ``push-down'' of operators into decompression routines --- so that what eventually gets materialized may be far smaller than the full decompressed column. Selection predicates (and, perhaps, aggregation) are ``pushed down'' in C\=/Store/Vertica \cite[\S 5.1]{AMF2006}, Vectorwise \cite[\S 2]{ZB2012}, MS SQL Server \cite{Microsoft2011}, and likely also other systems. This is usually (perhaps always) achieved via custom decompression routines supporting such push-down.

%; and a column store may also apply aggregations directly to its columns, rather than materializing the results of a selection. % but who does this? probably most of them...

The expressivity of the columnar circuit model allows column stores to surpass push-down optimizations. Since decompression is merely than another part of the larger circuit; and since the compression schemes devised in \autoref{chap:columnar-compression} are decomposable, and use simple, non-idiosyncratic operators as building blocks --- it is often be able to conceptually commute the decompression with the following (parts of) the execution: Simple transformations will get us to a point where we apply a circuit to the compressed form, whose output is the compressed form of applying a subsequent subcircuit in the original plan; and it is the result of this application that we then decompress.

% DIAGRAM here - push-down, simpler work on compressed form

A few examples: If a column is compressed in \cscheme{RunEncoding} (or \cschemesh{RLE}{RunLengthEncoding}, or \cschemesh{RPE}{RunPositionEncoding}), any elementwise operation can be applied to the \colname{run_values} column only, typically much shorter than the entire column; and a selection by predicate can be applied to the \colname{values}, \colname{positions} and \colname{lengths} columns, which, after having some corresponding elements removed, will now describe a subcolumn, compressed in \cscheme{RunEncoding}. Two columns compressed using \cscheme{Dictionary} can be \planop{Join}'ed by computing their individual histograms, \uplanop{Join}'ing their dictionaries (resulting in two columns of indices into the original dictionaries, a-la-MonetDB), and using the result to generate the appropriate subcolumn of the cartesian-product histogram. These three columns constitute the \planop{Join} result --- compressed in the \cscheme{Dictionary} scheme and with the original dictionary for each column.

\medskip
Unlike in existing systems, the examples above will not result in convoluted custom operators, which are nearly impossible to involve in further optimization. Instead, it will likely be possible to repeat the commutative process, pushing the decompression further and further into the circuit. Eventually, it may pass some vertex-cut in the circuit, beyond which a full materialization of the original column is no longer necessary. When this happens, the execution plan will have avoided decompression altogether.

Expecting to perform this kind of compressed-form execution, a column store considering compression schemes may assign less significance to maximal decompression bandwidth of a candidate scheme --- in favor of, say, better compression ratio.
% , or with decompression circuits which are more ``permeable'' to various operations
On the other hand, not every scheme is so ``permeable'' to execution order switching; so that the \emph{amenability to compressed-form execution} itself becomes a desirable metric in the choice among schemes.

\paragraph{Multi-column decoder fusion consideration in scheme choice} In \autoref{sec:multicolumn-compression} we listed several benefits of using joint compressed representations of correlated columns. Even if we refrain from considering them proper, the choice of individual column compression scheme may need to take inter-column correlation into consideration: The less the schemes chosen for two related columns have in common, the less likely the column store is to benefit from potential ``decoder fusion'', i.e. avoiding the repetition of similar or identical work when decoding both correlated columns for the same query. Also, given a choice of scheme, we may prefer sub-optimal settings of its parameters --- if these benefit decoder fusion with other columns. Examples could be a choice of segment length to fit the segmented representation of another column; choice of base functions and/or their coefficients for a \cscheme{NoisyGenerated} representation; or reuse of another column's dictionary despite it not being minimal for the reusing column.

\paragraph{Multiple (redundant) schemes for individual columns} When faced with several alternative compression schemes for a column, each with different relative advantages and none being overall-optimal --- why should a column store only choose one of them? True, holding redundant copies of a column goes against the basic rationale of compression --- reducing the amount of memory necessary for the column; but we have already noted this is just one of multiple qualities of a given scheme. Moreover --- while more memory would be needed for holding the column, the time required for reading the data from memory by a processor, or loading segments of it into smaller memory spaces, is \emph{not} increased by the existence of a redundant copy. This possibility expands the representation/compress scheme search space further, now with combinations of multiple schemes translating into different sets of parameters of scheme desirability.

Use of multiple schemes is even more relevant in a distributed setting, where each node only holds some of the columns (or some segments of some columns). As the number of nodes in such a system increases relative to the overall data size, columns will gradually become \emph{replicated} among the nodes --- improving both the availability of nodes with relevant information (needing less initial disk or network I/O), and the reliability / fault-tolerance of the system overall. When this is the case, it becomes even more attractive to choose different schemes for different copies of the column: Such \emph{asymmetric replication} allows the system to use different copies differently depending on the incoming queries and the availability of computational, memory and network bandwidth resources --- without paying the single-node premium of holding multiple copies of the same column: The risk is having the distribution of column replica schemes not match the distribution of ``desirable'' schemes w.r.t. the query workload.

Such a choice is described in \cite[\S 3]{RDS2002}, albeit with regards to a secondary disk array: Instead of employing RAID mirroring as protecting from hard disk faults, one stores different representation schemes for the same data (in the case of \cite{RDS2002} it was DSM vs NSM representation of tables; see the discussion in \cite[\S 2.2]{ABHIM2013}).

It should be noted that assymetric replication is potentially less robust as a fault tolerance measure: In the symmetric case, we simply need to get the same piece of data from a different replica; if replication is assymetric, however, restoring the part of a column representation stored at a certain node requires not just extra computational work, but possibly data from several distinct nodes.

\paragraph{Multi-column schemes: Derivation, co-encoding} As long as a column store has only one single representation per column, it is likely this will be a single-column, independent scheme; derivative or joint representation would be risky choices, considering the overhead of processing extra columns' data when they are not even used in a query. But once we have opened the door to holding multiple redundant representations, or if we're willing to adapt our schemes to the projected workload (see below), It becomes much more realistic for multi-column schemes to be used. This again expands the search space for compression schemes: In our description so far there has been a separate search spaces for each column (perhaps with some allowance for decoder fusion considerations); now it is a single search space, a choice of multiple schemes over all subsets of columns, with the constraint of their union covering every column.

% \paragraph{Write/Ingress-optimized schemes} Most of the representation schemes presented in this monograph do not agree completely, if at all, with established memory layout specifications and file formats for columnar or tabular data (such as CSV, Parquet, Orc, Arrow, MyISAM and so on). Most schemes depend on features of the data which may not be known in advance, before some pre-processing of the loaded data. While a column store is busy selectin
%
% Many schemes --- especially the  do not even of them lend themselves to being generated immediately while parsing secondary storage files in these formats.

\paragraph{Indices: Ordered multi-column representations} In general, table indices \cite[\S 11]{SKS2010} are data structures which make it easy to access table records according to a certain order of those records; and they are a key feature in non-columnar DBMSes, both for transactional and analytical queries.

In a column store, an index is essentially a copy of a sequence of columns, permuted into a lexicographic order of the induced records. But --- that definition is only true for the uncompressed representation of the index; it would be quite wasteful to actually store a full copy: First, the permuted copy of a column shared its support and frequency distribution with the original column --- which may well be enough to consider co-encoding them. Second, within every single column in the index, the lexicographic order induces long sorted stretches of elements, even if not all-column sortedness. Typically, this makes index columns compress very well with schemes such as \cschemesh{RLE}{RunLengthEncoding}/\cschemesh{RPE}{RunPositionEncoding}/\cscheme{RunEncoding}, or something more complex --- thus the overhead of keeping such sorted copies is limited. % And if the original column uses a dictionary, there is even more potential for co-encoding. Under some conditions it could be possible to even drop the \colname{values} column of this \cschemesh{RLE}{RunLengthEncoding}/\cschemesh{RPE}{RunPositionEncoding}/\cscheme{RunEncoding} representation altogether; the careful reader should be able to work out the details.
A specific (and useful) case of such index structures is the self-indexing of individual columns, i.e. keeping an extra, sorted copy of a column irrespective of any other column. Compression schemes used on a column's self-index provide information about a column which may be put to good use when planning queries, even if the sorted copy is not used instead of the column itself. When a self-index is compressed using \cscheme{RunEncoding} or \cschemesh{RLE}{RunLengthEncoding}, it provides us with a (sorted) representation of the column's support set. This fact was used for query plan optimization in \cite{AMPRR2016}.

Columnar DBMSes differ on the extent of the use of indices: MonetDB only offers support for self-indexing of a single column , and ignores other SQL statements involving table indexes \cite{MonetDBIndexDefinitions}). C\=/Store (and possibly, Vertica) takes the diametrically opposite approach: Its integration of indices is so deep, that in the papers introducing C\=/Store, the storage scheme is not even presented in terms of columns, but rather in terms of multiple (relational algebra) \emph{projections} of tables onto subsets of columns \cite[\S 3]{AMF2006}, each with some specific sort order. When processing a query, C\=/Store determines which are the relevant projections to decompress (often only a single one), according to the scan orders useful for the query's execution plan. These projections typically decompose into a ``key'' and a ``value'', in which case the ``key'' columns are a (multi-column) self-index, and the ``value'' column is indexd by the multi-column ones.
all but the last of the ``key'' columns in the projection typically exhibit long runs of identical values, due to the lexicographic sorting, and are thus efficiently compressed. However, C\=/Store also always keeps the compressed permutation of the records relative to some basic order, offseting the saving in space somewhat. Also, other than the indices, the columns are compressed independently, so that correlation among the ``non-key'' columns of the projection is under-exploited \cite{AMF2006}.

\begin{note}Some non-columnar DBMSes now utilize the column-store form of an index as their own index, to speed up analytic work, employing some upcasting and/or a \cscheme{Dictionary}-like compression scheme followed by an \ucscheme{RLE}-like scheme \cite{Microsoft2011}. The acceleration of non-columnar DBMSes by columnar indices is also explored in \cite{HRBLM2011,JSDKB2013} and in \cite[\S 4.8]{ABHIM2013}.
\end{note}

\paragraph{Workload dependence} Adptation of the physical layout of data to the workload a system expects is both common and useful in row-oriented DBMSes, e.g. in the form of materializing table views and choosing indices/sort orders to index by \cite[\S 24.1.6, \S 24.1.7]{SKS2010}. Existing column stores can generally benefit from auxiliary, workload-tuned indexes, and some, like C\=/Store/Vertica, also have ``column groups'' \cite{AMF2006} which can be seen as a limited materialized views, and can be chosen according to an expected workload. This can well be done with a larger space of compression and representation schemes of (permuted sets of)  columns: One alters the search or selection process to give more weight features which will benefit the processing of the expected queries with the available data (or expected future data), rather than trying to satisfty abstract general benefits.

A column store could also gradually adapt its choice of schemes to its workload, noticing when an alternative scheme it is now missing would better serve it, and switching at some threshold. This is easier if a column store holds multiple representations of columns, in which case it could ``evaluate'' schemes on actual data rather than only use analytic assumptions; and it does not have to discard all of its previously-held schemes, so that the threshold for adopting a scheme should not be as high. The practice of column cracking \cite{IKM2007} can be cast in terms of gradually refining a representation scheme of a column as multiple disjiont subcolumns (satisfying value range constraints), based on incoming range queries (and the prediction that future queries would be similar).

% Leaving this out:
%\paragraph{Intra-query-execution compression}

\section{Port digraph grammars for optimization}
\label{sec:graph-grammars}
%\addcontentsline{toc}{subsection}{\nameref{subsec:graph-grammars}}

A typical DBMS, columnar or otherwise, generates an initial execution plan from an SQL query, then proceeds to a distinct phase of plan optimization (in some systems these are two separated into two phases, optimization of the ``logical'' / relational algebra plan, and then the selection/optimization of a ``physical'' plan). \cite[\S 12.1, \S 13]{SKS2010}. These optimizations regard a set of operators which is fixed at compile-time, rather than variable and discovered at run-time; and so is the set of attributes, tags or meta-data which the system may compute.  Under these circumstances, optimizers (especially for ``physical plans'') tend to be implemented as general functions, implemented as part of the DBMS' codebase, in the same programming language as the rest of its code, and with wide or unrestricted access to DBMS internals.% --- rather than computationally-constrained rules.
For concrete examples in column stores, consider MonetDB's default optimizer pipeline \cite{monetdb-optimizer-pipelines}; or the compression-related optimizers of C\=/Store \cite[Fig. 3]{AMF2006}.

An alternative approach is exemplified by Spark SQL's Catalyst optimization engine \cite{Spark2015}. Spark SQL is not a column store, and its data model is quite dissimilar to the column-based model in this work; but the developers of Spark SQL's optimization engine (named Catalyst) prioritized for extensibility and external-developer involvement. They therefore designed the engine to expose the query execution plans to user-supplied optimization rules. Catalyst takes ``rules'' which interact with the plan in a rather domain-specific fashion \cite[\S 4]{Spark2015}; and the results of rule application is eventually JITed along with the Scala code in which the engine is written. However, these ``rules'' are still really full-blown Scala functions \cite{Spark2015, M2015} which the column store executes.

\medskip

In a column store following the columnar circuit execution model, execution plans are much more expressive, both in capabilities and in practical use: previous chapters (and sections in this chapter) demonstrate how various structural features of today's column store can ``implemeneted'' through columnar circuit execution plans. This implies more (or most) optimization work will involve straightward circuit transformations; it is partly for this reason that we lengthily and carefully exposited circuit transformations in \autoref{chap:columnar-computational-model}, even beyond what later chapters required. If we also allow for decorating vertices (operators), ports and edges with some property labels that optimization rules can match, the expressivity, and the coverage of optimization rules expands even further.

It is at this point that the need for utilizing a more general-purpose port-graph rewriting system, or (port-)graph \emph{grammar}, becomes apparent: Without apriori knowledge of the port-digraphs we need to work on, nor of the set of usable vertices (operators), nor of the set of usable transformation rules --- we require a mechanism for repeated rule selection and application, coupled with an abstract, general representation of what a transformation rule constitutes. With our model lending itself towards larger graphs and further decomposition than other column stores, match patterns will require some flexibility and may be larger sometimes than two or three vertices. The automation of rule consideration, prioritization, matching and application should prove to be quite challenging to implement efficiently (despite being computationally tractable). The choice of an appropriate formalism for is not immediately obvious, especially considering the variety of established models (see \cite[Vol. 1]{Rozenberg1999}) on one hand, and with our interest in rewriting rather than derivating from scratch or identifying a language. Going into any detail on this matter, however, would be well beyond the scope of this work.

\section{Mutable columns and column store write support}
\label{sec:mutable-columns}

The columnar circuit model regards columns as immutatble. Rather than columns changing, ``new'' columns are the result of an application of operators to existing ones, within a circuit. % When we consider circuits as being applied temporaly --- the application of a circuit actually creates new columns; and as the Turing machines corresponding to operators are executed, they mutate their state internally.
This restriction makes for a stark contrast with the extensive work on mutable, transactional DBMSes, both as practical implemented systems and as objects of theoretical study.

Could we complicate the model to also allow for column mutation? This is, perhaps, conceivable: Changes to input columns would propagate into the circuit (e.g. through an overlaying standard-representation subcolumn, see \autoref{subsec:subcolumns}) --- with operators having to either re-execute, or have a ``delta version'' execute on the original inputs and outputs, along with the changes. However, such a complication is probably too unwieldy to be worthwhile.

A more palatable alternative is for a column store utilizing the columnar circuit model to allow mutation, but have circuits isolated (in the ACID sense \cite{HR1983}) from mutations' effects. Such isolation would be achieved if representations of mutated columns each contained a complete representation of the original column, which can remain intact (and in-place in physical memory), even as the column mutates. A system applying this principle to each and every change a column undergoes must necessarily utilize \emph{versioned data structures} (a.k.a. \emph{persistent data structure} for its columns --- as defined in \cite{DSST1986}, and used extensively in some non-columnar DBMSes, such as LogicBlox \cite{ACGKOPVW2015}.

For the purpose of an analytics-focused column store, however, we can make do with much weaker ``isolative'' nature of a column's representation --- and we need look no further for it than the schemes already presented in previous chapters. Recalling the discussion in \autoref{subsec:combining-subcolumns}, we can adapt \cscheme{SubcolumnOverlay}, so that an overlaying subcolumn holds appended (or modified) elements; and deleted elements can be represented using an additional index subset representation (see \autoref{subsec:index-subsets}). Alternatively, the deletions could also be represented by a subcolumn --- to be ``subtracted'' from the \cscheme{SubcolumnOverlay}-encoded column; this approach would be similar to that of MonetDB \cite{Boncz2002} (see also the concise description in \cite[\S 2]{HZNSB2010}): Separate the main read-only data structure for a column (or a projection of columns) from an auxiliary but still columnar structure for insertions and deletions. Occasionally --- either every period of time or when changes exceed a certain size --- column data will be re-integrated by reconstructing a complete and up-to-date column with no mutations, overlaid by empty insertions and deletions subcolumns.

\medskip

Other existing column stores use non-columnar representations (\emph{delta-structures}) for column mutations. SAP HANA, for example, uses two-level structures situated in CPU cache: A row-major delta-table in L1 cache, and a column-major delta table in L2 cache; when the former fills up, it is merged into the latter, and when the latter fills up, it's merged into the main columns in main memory \cite{SFLCPB2012}. Another prominent delta-structure design are the B-tree-like positional delta trees (PDTs) used in Vectorwise \cite{HZNSB2010}, which take more after index structures in row-oriented DBMSes rather than actual tables. While these may be faster for transaction processing workloads, they would be more difficult (if not impossible) to make usable with columnar circuits for analytic queries. This difficult is also an explanation of C\=/Store's choice: Its columns are a Read Store, and a sepate Write-Store --- for inserts, updates and deletions --- is handled by a separqte non-columnar DBMS --- external to C\=/Store proper --- which uses B-trees under-the-hood \cite[\S 6]{SABCCFLLMOo2005}.

%(in C\=/Store, this is the ``merge-out'' process; see \cite[\S 7]{SABCCFLLMOo2005}).

%\erc{Consider looking for other options, e.g. in ``In-Memory Big Data Management and Processing: A Survey'', ``Enhancements to SQL Server Column Stores'', ``Fast Updates on Read-Optimized Databases Using Multi-Core CPUs''.}

% Talk about persistence here? : Write-supporting column stores must also provide persistence to non-volatile media (e.g. from DRAM to magnetic or flash disks)... etc. etc.

\section{Execution progression and residual plans}
\label{sec:representing-partial-execution-state}

The columnar circuit model abstracts away --- at least prima facie --- most aspects of execution on real machines. The following, at least, ones must somehow be accounted for when actually computing the output of a circuit (``executing'' the circuit):

\paragraph{Operator execution status} Columnar circuits do not define an order of execution for their constituent operators. Operator execution is not instantaneous, and multiple operators may execute concurrently. How should a column store, adhering to the columnar circuit model, keep track of which operators have concluded execution; which have not started execution; and which are currently executing?

\paragraph{Multiple finite memory spaces} Memory is a limited resource, and the faster it is, the less of it real-life machines have. Even if we choose to treat faster memory as cache, and consider secondary storage (e.g. magnetic or flash drives) to be ``large enough'' to hold all relevant data --- we would still not be able to fit the memory spaces on co-processors into a uniform, single-space view of memory. Also, performance considerations demand that column stores take great care to effectively utilize their small, faster memory. How would a column store reconcile this necessity with the hughe uniform columns in the circuits it is to execute/simulate?

\paragraph{Sub-operator-resolution scheduling} Simpler column store execution engines use the \emph{operator-at-a-time} model for plan execution \cite{MKB2009}: A plan operator is executed as a whole; there are no multiple recognized states before its execution completes. A prominent example is MonetDB\cite{monetdb}, which executes MAL \cite{MonetDBMAL} instructions in sequence (see \autoref{subfig:columnar-circuit:monetdb-equivalent}; but note that the multi-threaded capability makes MonetDB more of a per-thread operator-at-a-time system \cite[\S 3]{IKG2012}). This can indicated trivially by decorating the nodes of a columnar circuit.

Other column stores use finer-resolution scheduling: They operate on relatively small chunks of a column (called ``vectors'' in Vectorwise \cite{ZB2012}; and ``morsels'' in a further development of this notion in \cite{LBKN2014}). For multiple reasons, this choice of chunk-at-a-time execution is quite beneficial, performance-wise, at least on CPUs. However --- it is unreasoable to make column-chunks into ``first-class citizens'' of execution plans, as that would mean plan sizes would be huge --- linear in the size of the data. In Vectorwise, and likely in other chunk-at-a-time systems, chunks are not represented in the plan; instead, the column store's execution engine has the ``hard-wired'' capability of conceptually breaking columns up into chunks, keeping track of the different chunks, and recombining them after processing into a full column when necessary. % Chunks can also reflect the physical layout of the data, especially when it is compressed.: The boundaries of units of compression/representation align with boundaries of chunks for the purposes of scheduling

More generally, chunk-at-a-time execution seems to be paired with some idiosyncratic internal representation of query plan execution state. This implies that complex patterns of composition and de-composition of execution --- the very ones which are core to the columnar circuit model --- do not apply to this finer-resolution representation. The idiosyncracy also implies that very little optimization is possible on the currently-executing parts of the plan --- as the column store's optimizer mechanism is not aware (or not more than superficially aware) of the out-of-plan execution state information.

\subsection{Splitting operators and circuits}
\label{subsec:splits}

It so happens, that all three challenges described above can be addressed using a single ``silver bullet'': A well-developed capacity for \emph{splitting}. The possibility of splitting an operator and its inputs is not fundamental to the columnar circuit model, and we have barely used it so far in this work (see \autoref{subsec:segmentized-differences} for an exception); but no column store implementation can avoid it.

Splitting operator input columns is less trivial than it seems. First note that a split of a column can result in two or many (subcolumn) parts; the part sizes can be uniform or small ``chunk'' vs large ``rest''; and the resulting subcolumns can be contiguous or form a complex non-contiguous pattern. For a column unto itself, this is still perfectly trivial; and so is the case for splitting, say, an $\planop{Elementwise}_f$ operator. But as the operator's semantics are more complex, so does the splitting of its input become. A few examples:

\begin{itemize}
 \item $\planop{PrefixAggregate}_\oplus$ (see \autoref{subsec:differentiation}): A typical implementation of this operation --- and any reduction/fold-like operator --- is itself based on breaking up the column into smaller chunks: Compute a sum of the elements in the chunk; (recursively) compute a prefix sum of the chunk-sums, then, for each chunk add the sum of preceding chunks to the prefix sum within the single chunk. Partitioning the input column ``outside'' the operator merely adds another level of this procedure.
 \item \planOp{Sort} (see \autoref{subsec:column-support-set-compaction-1}): Similarly to $\uplanop{PrefixAggregate}_\oplus$, the computation on each part is a smaller \planop{Sort}; however, the combination of the parts is more involved --- a merging of sorted sequences. Alternatively, one could go through making the two parts into bitonic sequences, then finally merging them into a proper sorted sequence (see \cite{SHG2009} regarding the relevance of bitonic sorting in parallel settings).
 \item \planop{Gather} (see \autoref{subsec:column-support-set-compaction-1}): While this is not an elementwise operator, it \emph{is} elementwise operator with respect to its \colname{pos} input, but not with respect to the \colname{data} input. If we were to split up \colname{data} --- say, into the first and second halves --- we could not then simply employ two \planop{Gather}'s with the two halves ($\colname{data}_1$, $\colname{data}_2$), since some of the indices in $\colname{pos}$ may be out-of-range; and for the second half, all indices would need to be adjusted. One straightforward approach to handling these difficulties would be splitting \colname{pos} into two as well, using a filter computed by $\planop{Elementwise}_{\geq \colsize{data} / 2}$ (the split may not be into halves of course); we then have to subtract $\abs{\colname{data}_1}$, elementwise, from the elements of the second part of $\colname{pos}$. Finally, the two $\planop{Gather}$-resulting columns need to be re-integrated using the filter vector. (Note this approach is not the only one possible.)
\end{itemize}

How far do can these split recipes generalize? While they cannot exist for an arbitrary choice of input of any conceivable operator --- they do exist essentially for all inputs of the operators we have been using in this work (see the \hyperref[idx:operators]{Index of Operators}). Moreover, these splits are nothing other than columnar circuit transformation rule with certain specific features; and a simple, localized rule at that. Of course the question of \emph{when} these transformations are to be applied is a different matter: Some are relevant before execution has occured, but many become relevant \emph{during} execution, repeatedly. Also, the decision to apply one of these requires additional information besides the circut itself, e.g. information regarding actual column length, execution status and so on. More on this in \namecref{subsec:using-splits} below.

%to apply such a rule requires more than just the columnar circuit, leading us in the direction discussed in \autoref{sec:graph-grammars}.

\medskip

Splitting single operators also generalizes to subcircuits, i.e. instead of duplicating a single operator, the split inputs of a subcircuit are fed to two, potentially-but-not-necessarily-identical, circuits. This kind of subcircuit-split was used in the system described in \cite{AMPRRT2015} for schleduing parts of the data on different hardware devices and/or different threads. But perhaps a more prominent example is  MonetDB's mechanism for multi-threaded execution, named ``Mitosis'': Some (easy to split) columns are split up, one part per each working thread, while others are unotuched; and the MAL execution plan (see the example in \autoref{subfig:columnar-circuit:monetdb-equivalent}) has some of its instructions replicated for each of the  split column parts. This is performed as one of MonetDB's plan optimizers \cite{monetdb-optimizer-pipelines}, \cite[\S 3]{IKG2012}.

%
% Can these examples be generalized to any operator? From a complexity-theoretic perspective, the answer is clearly negative: The function computed by an operator for certain input lengths might not have any meaningful rule for derivation from related functions on shorter inputs. However, when this is the case, a column store cannot apply such an operator using chunk-at-a-time execution, or any other kind of decomposition --- whether it uses the columnar circuit model or not.

%We therefore note that operator splits are another ``structural'' feature or capability of a column store that is naturally expressible through the columnar-circuit execution plan itself; whenever splitting or taking chunks is possible for a certain input port of a certain operator, a staightforward transformation rule exists representing such a split: Instead of the single operator node with a large column feeding the relevant input port, a subcircuit beginning with a trivial split of the column (which doesn't require any actual computation on the entire column); all other operators using ``smaller'' inputs; and of course with the circuit outputs being the same as for the original operator (so that length may finally increase into an undesirable value). The notion of ``smaller'' inputs is rather vague, and would be made specific according to the expections for applying such a rule repeatedly.

\smallskip

%We find, therefore, that most of a column store's operators must be made available to it with associated input-splitting transformation rules; and that those rules themselves depend on other operators, so that a digraph (probably not even a DAG) of operator dependencies forms. With those being available for all operators in a circuit, it is often possible to combine them into transformation rules for splitting inputs of multi-operator circuits, together.

%\medskip

A precusor to the suggestions here is the effect by the author and others of splits and merges via execution plan transformations in \cite{AMPRRT2015,AMPRR2016}. It is also worth noting that MonetDB's multi-threading is heavily based on involves splitting columns, with the parts visible in the execution plan --- a procedure named ``mitosis'' and constituting an execution plan optimizer \cite[\S 3]{IKG2012}, \cite{monetdb-optimizer-pipelines}.

\subsection{Putting splits to use}
\label{subsec:using-splits}

With subcircuit splits achievable through a circuit transformation (and no ``unsplittable'' operators in use) --- any circuit is just a few transformations away from having an operator with small enough inputs to schedule a chunk for execution: At most one split transformation per input of the chosen operator. Similarly, the circuit is a few transformations away from having chunks of input columns which can be transmitted to a different memory space where another processor can access them and start working on a chosen operator. In both cases, the splitting can be expanded so that a larger subcircuit may execute using the split-off inputs; also in both cases, the actions above can be repeated several times to allow for concurrent execution on multiple chunks of the data, either using the same initial memory space or in different spaces. An example of both of the above would be concurrent execution of an operator on different chunks by different CPU threads and by a discrete GPU accessing chunks placed in its global memory. (A more simplistic form of such a mechanism was used in the AXE heterogeneous execution framework \cite{AMPRRT2015}).

In other words: Using splits, we reduce the challenge of executing circuits a-chunk-at-a-time to the more straightforward operator-at-a-time execution: Operators are only ever executed on input columns which are small enough not to merit further breakup, or tracking of the execution beyond the ``awaiting-/during-/after- execution'' trichotomy of operator-at-a-time column stores.

It should be stressed that the splitting described above, even when applied repeatedly, does not make the circuit balloon in size, to comprise the huge number of small column-chunks and small operators which end up being scheduled distinctly for execution: Only the single small part which is to be executed next is split-off; most of each column remains in a single large (sub)column; and the chunks already processed are also not kept separate --- they too are consolidated occasionally into larger partial-result columns.

%Also, the distinction between ``awaiting processing'', ``being processed'' and ``already processed'' parts of columns can be made more explicit with appropriate tagging of vertices and/or ports in the circuit (see \autoref{sec:graph-grammars}.
Finally, to address the remaining challenges listed at the beginning of this \namecref{sec:representing-partial-execution-state}, we can decorate the nodes and ports involved in splits, to indicate which column lies in which memory space, as well as the ``awaiting-/during-/after- execution'' status of each operator. Without the splits, each node and port would require complex information regarding what has happened to which part of it; the splits allow for the subsumption of execution engine idiosynctatic complexity by the complexity of of representation schemes for columns and subcolumns, which a column store must already contend with.

\subsection{Residual state and a streamlined execution flow}

A side-effect of represeting ongoing execution state within the execution plans (with the modification of allowing vertex and port tags) is that doing so is essentially maintaining an \emph{residual plan}: A plan whose execution, given the various columns currently available to a column store, produces the same results as the original query. The initial residual plan when execution commences is the execution plan originally drawn-up; while the final residual plan simply has the final results as an input with no further execution necessary.

Residual plans, being (decorated) columnar circuits, are subject to the same kind optimizing transformations as initial execution plans (and see also \autoref{sec:graph-grammars}). Now, this is also the case, essentially, for operator-at-a-time column stores like MonetDB; but with our model, plan optimizers or optimizing transformation grammars must already be tolerant of columns being being in a broken-up state, represented as multiple subcolumns with their own potentially different representation scheme, and requiring a circuit of computations just to re-materialize again. Thus there is more potential for optimization to be applicable during execution.

Considering our discussion above of the progression of execution using splits, and the potential for dynamic reisudal-plan optimization, we can conceive of the following rough outline for a column store's plan execution step:

\medskip

\noindent\begin{minipage}[c]{0.95 \textwidth}
\small
\setitemize{itemsep=1pt,topsep=3pt,parsep=0pt,partopsep=3pt}
\begin{itemize}
  \item \strong{Feeding the processing hardware:} If some processor (CPU, GPU, FPGA etc.) is running low on scheduled work,
  \begin{itemize}
    \item Find an unscheduled operator with small enough input \& output sizes, which are all accessible by said processor, and schedule it.
    \item If none was found, find such an operator but without the restriction on input \& output sizes, which could be split up and scheduled
    \item If one was found, apply a split transformation.
    \item If none was found, search for a set of columns to move into to a memory space directly accessible to the proessor, and which would be schedulable
    \item If no such set of columns was found, search for a set of columns which could be split up (or their generating operators split up) so that it would constitute a set of inputs for an operator schedulable on the processor
    %    consider operators with small-enough input located in other memory spaces, and determine whether ir is cheape
  \end{itemize}
  \item \strong{Feeding communication channels:} If some channel or bus is running low on scheduled transfers:
  \begin{itemize}
    \item Find a column which one can estimate would could be copied or moved directly over the channel into a more beneficial location.
    \item If one was found, schedule it for transfer
  \end{itemize}
  \item \strong{Acknowledging conclusion of computational work:} If some operator has completed its execution,
  \begin{itemize}
    \item Remove the operator its outputs become assignmed inputs of the circuit. Unused inputs of the deleted operator are marked unused (a later optimization may remove them).
    \item Create new circuit inputs and assign the operator's outputs to them.
  \end{itemize}
  \item \strong{Optimization:} Search for an optimizing transformation which could apply to the (residual) circuit. (The search could be longer and/or produce multiple optimizations; or may be timed and shorter, so as not to neglect minding the hardware, and continue the search at a later iteration.)
\end{itemize}
\end{minipage}

\medskip

This procedure is already attractively succinct, considering how it covers decompression, use of index structures, parallel processing on heterogenous hardware, dynamic reoptimization, some memory management (through the deletion of unused nodes) and so on. We can go even further, however, in streamlining the execution step: Recall that most of the conditions-and-actions above are nothing but applications of specific (sub)circuit transformation rules. They may be more ``structural'' than ``algorithmic''; there may be some importance to their relative order, or they may otherwise inter-relate --- but then, this can be true for other transformations. We only need to extend our node and port tag sets, to cover not only execution status but location: Which processor is executing an operator, in which memory space a column is located and where it is being trasmitted. With this modification, the execution step can be described much more generically:

%, it may be possible (and perhaps worthwhile) to streamline the execution step into the following:

%For example: when an operator is decorated with ``scheduled for execution'' and the identifier of a relevant processor, the column store must ensure it is indeed thus scheduled right after this marking takes place by a transformation.
%It may thus be worthwhile to streamline this execution loop by the more general abstraction, into the following:

\medskip

\noindent\begin{minipage}[c]{0.95 \textwidth}
\small
\setitemize{itemsep=1pt,topsep=3pt,parsep=0pt,partopsep=3pt}
\begin{enumerate}
 \item Determine out-of-circuit system changes not yet reflected in the plan (e.g. an operator having completed its execution or buffer transfer complete).
 \item Update the residual plan (= residual circuit) to reflect the changes.
% \item Schedule an operator execution or a column transfer if relevant.
 \item Update the (possibly-implicit) space of applicable circuit transformations, including transformations scheduling operator execution or transfers. The update includes the prior probability distributions of benefit for these transformations.
 \item Search, or continue a search, within the space of applicable circuit transformations (including unused input removal, execution or transfer scheduling, a split of an operator or a subcircuit, etc.), and possibly choose to apply one to the plan.
 \item Schedule executions or transfers according to the changes in the residual plan.
\end{enumerate}
\small
\end{minipage}

\medskip

In this formulation, multiple generic steps will be taken to cover the different actions in one single step of the longer, more specific one formulated above.

\clearpage
\bibliographystyle{alpha}
{\small
\bibliography{back_matter/hetero-computing,back_matter/dbmses,back_matter/compression,back_matter/tcs-general,back_matter/dataflow,back_matter/dbms-benchmarks,back_matter/programming-languages}
}

\clearpage
\listoffigures

\clearpage
\printindex[operators]
\noindent This is not a minimal set: Some of these operators can be implemented using other operators on the list, or are simply important special-cases of other operators.

%\indexprologue{\noindent Listed here are representation schemes for column, subcolumns and index subsets.}
\printindex[schemes]

\end{document}